# Renormalization group flows in Gauge-Gravity Duality

Arvind Murugan

A Dissertation
Presented to the Faculty
of Princeton University
in Candidacy for the Degree
of Doctor of Philosophy

Recommended for Acceptance
by the Department of
Physics
Adviser: Igor R. Klebanov

September 2009

// Abstract

We study aspects of Gauge-Gravity duality for theories with reduced supersymmetry and without conformal symmetry. We construct and study a few examples of renormalization group flow in gauge theories in $3+1$ and $2+1$ spacetime dimensions and the corresponding physics of the gravity dual. In the $3+1$ dimensional flow, we study how phenomena such as global symmetry breaking, large anomalous dimensions and moduli parameterized by baryonic operators, familiar in field theories, have corresponding elements in the gravity dual. In the $2+1$ dimensional theory, we find new phenomena due to a Chern-Simons term such as symmetry enhancement due to monopoles. We also use group theory to get surprisingly detailed information on anomalous dimensions of operators that helps check the gauge-gravity correspondence in this case.

Inspired by duality, we propose a new test of confinement in gauge theories based on entanglement entropy, analogous to the area law for the Wilson loop. When applied to the gravitational dual, the entanglement measure provides a particularly quick and yet delicate probe for the presence of a Hagedorn growth in the glueball spectrum of the gauge theory. Hence we propose the entanglement entropy transition as a simple test of confinement, similar in spirit to the thermal deconfinement phase transition.

Finally, we investigate a class of possible non-supersymmetric gauge-gravity duals for perturbative and non-perturbative instabilities. We find that while the former are not present in generic non-supersymmetric $AdS_4$ compactifications of 11 dim M-theory, the latter do destabilize most non-supersymmetric theories.

# Acknowledgements

I thank my advisor, Prof. Igor R. Klebanov for guiding me through these last five years. While I have learned much physics from him and from his papers before I even came to Princeton, I am most grateful for having had a patient mentor from whom I could learn how to conduct research by asking the right questions in incremental steps. Without his perceptive mix of latitude and guidance at different stages, I would have long tired of research or been lost in the wilderness.

I am also grateful to Prof. A. M. Polyakov for generously and patiently sharing his ideas with me. The questions he proposed for me to work on and the nature of the answers he sought were a constant reminder of the reasons I was first attracted to physics, before any formal education in it.

Much of my physics education in real terms came from the endless hours spent discussing the simple questions in physics that are often without answers, both at Princeton with L. Alday, M.Amarie, D.Baumann, A.Dahlen, T.Klose, M.Kulaxizi, D.Malyshev, A.Murugan, S.Pufu, M.Rechtsman, D.Rodriguez-Gomez and T.Tesilieanu and at Caltech with W.Cottrell, M.Solomon and S.Thomson. Along with M.Benna, M.Buican, S.Goyal and D.Hofman, I'd like to think our interactions helped maintain our collective sanity (to an extent) over the years.

I thank my collaborators D.Baumann, A.Dymarsky, L.McAllister, D.Kutasov, T.Klose, J.Maldacena, D. Rodriguez-Gomez and J. Ward on the research presented in this thesis. I have learned much physics from faculty members at Princeton and the IAS, Prof. Gubser, Prof. Herzog, Prof. Maldacena, Prof. Verlinde and others.

My years at Princeton, from the barbecue on the first day I arrived on campus, would have been very different without Ingrid. I thank my family for their support and am grateful to my parents and my grandfather for always encouraging me to pursue any interest (and subsequent obsession) of mine freely.

*Grown-ups like numbers. When you tell them about a new friend, they never ask questions about what really matters. They never ask: "What does his voice sound like?" "What games does he like best?" "Does he collect butterflies?" They ask:"How old is he?" "How many brothers does he have?" "How much does he weigh?" "How much money does his father make?" Only then do they think they know him.*

The Little Prince

Antoine De Saint-Exupery

For my parents.

# Contents

# List of Figures

# Chapter 1

# Introduction

Duality in physics refers to two seemingly different physical phenomena having descriptions that are mathematically equivalent to each other. The phenomena themselves can appear to have little to do with each other but there might be a correspondence that relates and identifies all elements of their descriptions to each other. We can identify the two phenomena as one in an abstract sense if we wish, with two differing descriptions.

There are several known examples of duality in various areas of physics, from statistical physics where the Kramers-Wannier duality relates high temperature properties of model magnets to the low temperature properties of such magnets to condensed matter and particle physics where certain two dimensional systems of particles with rather different properties and interactions have been shown to be mathematically equivalent.

The power of duality in physics emerges when behavior obscure in one description translates into transparent and tractable behavior in the other and vice-versa – i.e the two dual descriptions complement each other.

This thesis is based on a duality between a theory of quantum gravity (string theory) and a theory of particle physics (gauge theory) that has attracted much attention over the last dozen years under the name of gauge-gravity duality or more commonly, the Anti de-Sitter / Conformal Field Theory (or '$AdS/CFT$') duality. On the face of it, these two theories describe phenomena in spacetimes of *different dimension* and hence have little to do with each other. And yet their content is the same. Further, non-trivial quantum effects in one description are captured by the simple classical physics of the other and vice-versa, making this duality very powerful. Viewed in one direction, one can use elementary classical string theory to understand the strong interactions between particles such as quarks and gluons which have resisted many direct attempts at calculation. In the other direction, in the limit of weak particle interactions, quarks and gluons might teach us about strong quantum effects in gravity.

This thesis focuses on extensions and aspects of the gauge-gravity or the $AdS/CFT$ duality without conformality and with reduced or no supersymmetry. In this thesis, we construct and study a few examples of gauge-gravity duality with non-trivial renormalization flows, propose a general property of confining theories inspired by duality and finally investigate the possibility of completely non-supersymmetric duality.

Supersymmetry and conformality are simplifying features that allowed concrete basic examples of $AdS/CFT$ duality to be formulated. Supersymmetry is a simplifying property of certain theories which calls for a delicate arrangement among parameters such that many hard-to-compute physical effects present in conventional theories are absent. Conformality refers to the property of a system being self-similar at different scales, much like a fractal which looks the same when looked at at any magnification. Classical gauge theories with only massless matter are conformal and approximately conformal even after certain weak quantum effects are taken into account. Thus conformality and supersymmetry serve as bright lamp posts to perform calculations under. Such analytic calculations are essential for gaining physical insight, before tackling more complicated cases.

However, supersymmetry is not a known symmetry of any experimentally verified theory of particle physics and the physics of supersymmetry, while not simple, is considerably simpler than that of non-supersymmetric theories. Further, the world certainly looks very different at different length scales i.e it is not conformal. Much of the non-trivial physics of particles is related to the existence of special distance scales – for example, sub-atomic forces do not act over super-atomic distances and composite particles like protons and neutrons are of characteristic sizes and masses. In fact, one of the primary hopes of gauge-gravity duality is that it would shed light on quark confinement, the phenomenon by which quarks are never found separated from each other by more than sub-nuclear distances. Quark confinement is thus intimately related to the existence of special distance scale.

The goal of bringing the tools of gauge-gravity duality to bear on such theories with reduced supersymmetry and conformality is the underlying theme of this thesis. At the very least, one would like to at least be aware which of the lessons that one learns under the assumptions of supersymmetry and conformality are valid more generally and which lessons are artifacts of the simplification. Several gravity duals to gauge theories that exhibit various real-world non-conformal phenomena have been constructed and generalizing gauge-gravity duality to cases where the gauge theory and hence the gravitational background have lower supersymmetry has been an active area of research. (Attempts to describe non-supersymmetric theories have been fewer and less successful.) In this thesis, we test new ideas on such existing constructions while also advancing some simple new constructions in this spirit.

The plan of this thesis is as follows. We start in this chapter with a short introduction to gauge-gravity duality and its realization using D-branes of string theory, emphasizing themes and ideas most related to the work in this thesis.

In Chapter 2, we report on work on the resolved conifold, which provides an example of $AdS/CFT$ duality which is not conformal and has reduced supersymmetry. The resolved conifold is a 6 dimensional space on which one needs to place certain 4 dimensional objects, $D3$-branes, to obtain 10 dimensional spacetimes which can then be studied using gauge-gravity duality. We construct the first such non-singular 10 dim spacetimes using the resolved conifold, called the warped resolved conifold. Such a gravitational solution has a simple interpretation in terms of a renormalization group flow that starts from the Klebanov-Witten gauge theory at high energies and ending with the more (super)symmetric $\mathcal{N} = 4$ $SU(N)$ super Yang-Mills theory at low energies.

In Chapter 3, we propose a new theoretical test for confinement based on entanglement entropy. Confinement refers to very non-trivial property of the real universe in which quarks are always found tightly bound together and never individually like electrons or protons. It is a long-standing problem of particle physics to understand, through analytic methods, the emergence of such a property in a given theory of quarks and their interactions. Motivated by calculations using gauge-gravity duality, we propose an order parameter that can serve as an indicator of whether a given theory of quarks and gluons does lead to confinement.

In Chapter 4, we study a renormalization group flow of a $2+1$ dimensional Chern-Simmons theory. This theory arises from considering $M2$-branes of 11 dimensional M-theory. Such a theory in $2+1$ dimensions shows novel features compared to the usual gauge theories on $D3$-branes due to the Chern-Simmons term, non-dynamical gauge fields and the central role played by magnetic monopoles. In our study, we use group theory to resolve certain subtleties and hence properly identify the gravity dual to an RG flow of a gauge theory.

In Chapter 5, we investigate the possibility of going beyond reduced supersymmetry to having entirely non-supersymmetric examples of gauge-gravity duality. Non-supersymmetric spacetimes are plagued by instabilities and many have been discussed for duals to $3+1$ dimensional gauge theories. We consider duals to $2+1$ dimensional theories i.e non-supersymmetric vacua of 11 dimensional M-theory of the form $AdS_4 \times X^7$. We show that while the perturbative stability picture is better for $AdS_4$ vacua compared to $AdS_5$, non-perturbative instabilities are rather generic.

**Gauge-Gravity duality – origins**

Gauge theories are a class of theories of particle physics, first proposed in the 1950s based on purely mathematical motivation [1]. It was realized beginning in the the late 1960s and early 1970s, that different examples of these theories might successfully explain much of the particle physics that had been tested in particle accelerators. This included the physics of weak interactions which exist between many particles and the physics of strong interactions which exists between quarks due to particles called gluons. The gauge theory of strong interactions, named Quantum Chromodynamics (QCD), could make testable predictions of the forces between quarks when they were close together while the theory of weak interactions, named Electro-weak theory, successfully predicted and explained properties of new particles that were discovered in collider experiments.

The experimental success of Quantum Chromodynamics (QCD) and Electro-weak theory established these gauge theories firmly as the dominant paradigm, now known as the Standard Model of particle physics. However, while calculations in electro-weak theory were relatively straight-forward, deriving predictions from Quantum Chromodynamics (QCD) has proven a much harder mathematical task.

The confidence in QCD derives from the predictions that apply to quarks when they are close together. In this regime, the forces between them are weak (a fact known as asymptotic freedom) and theoretical predictions of QCD can be computed perturbatively and compared with experiment. On the other hand, many other experimentally verifiable aspects of quark interaction such as their confinement involve related to strong interactions between the quarks and gluons. Such effects are mathematically difficult to understand in QCD due to the absence of a perturbative scheme and the complexity of the equations. In later years, with the help of computer simu-lations, these mathematically difficult aspects could also be understood within QCD but an analytic mathematical understanding is still lacking.

The prototypical characteristic example of such mathematically intractable features is quark confinement. Quark confinement is a proposed feature of quark interaction that explains why quarks are never observed individually but always found as a bound state with a few other quarks. Such a confinement of quarks can result from a strong growing force that appears between quarks as one attempts to pull them apart. However, computational methods typically rely on having weak forces and computing quantities perturbatively in such weak interactions. Hence it is difficult to establish analytically that the theory of QCD indeed predicts a growing strong force between quarks.[1]

It was in this context that several researchers in the 1970s began looking for alternative descriptions of QCD that might shed light on the strongly interacting aspects. One idea, based on Wilson's linear potential between quarks, was that it might be reasonable to think of quarks interacting strongly as quarks with a string between them. The idea was appealing since such strings might naturally explain features such as quark confinement and also open up the possibility of using string theory to understand strong interactions. This proposal is called gauge-string (or gauge-gravity) duality since it proposed that strings could be an *effective* way – an alternative language to QCD – to describe the interaction between quarks. [2] This idea was given further impetus by 't Hooft [2] who proposed using Feynman diagrams of the form shown in Figure 1.1, quark interactions might indeed be modelled by string worldsheet and that in certain limits (where the number of colors $N \to \infty$), the string theory simplifies to that of non-interacting strings.

Using the theoretical developments in string theory of the early 1980s, Polyakov further studied the problem of modeling quark interactions by strings. He realized that since quarks move in 3+1 dimensional spacetime, any theory of strings connecting the quarks must be a non-critical string theory [3, 4]. Non-critical strings were known to generate an additional spacetime dimension that they could move in, due to a quantum anomaly that results in a degree of freedom not found classically. Hence Polyakov proposed that the strings describing quark interaction should really be in 4+1 dimensions, even though the quarks themselves move in 3+1 dimensional spacetime. Building on this in 1996-97, he described the 5 dimensional spacetime generated by the non-critical strings to be of the form [3, 4],

$$S \sim \int d^2\sigma \left[(\partial\phi)^2 + \alpha^2(\phi)(\partial X^\mu \partial X_\mu) + \text{vertex operators}\right] \tag{1.0.1}$$

where $\phi$ is the extra Liouville dimension. Spacetime $X_\mu$ is warped by a corresponding warp-factor $\alpha(\phi)$ which depends only on the Liouville coordinate. Viewing gauge-string duality in this setting as a case of open-closed string duality, Polyakov proposed that the gauge theory modes (or open string modes) live at a value of $\phi^*$ such that $\alpha^2(\phi^*) = \infty$. The warp-factor $\alpha(\phi)$ was interpreted as a scale-dependent string tension which encodes the non-trivial RG dynamics of the dual gauge theory. These ideas were made precise using a string prescription for the Wilson loop presented in [3, 4].

[1]It has been possible to investigate a discretized version of QCD through numerical simulations on computers and all indications are that quarks are indeed confined.

[2]This would complete a historical circle since string theory was originally (unsuccessfully) proposed in the 1960s precisely as a theory of quark interactions. When QCD proved to be the right theory of quarks, string theory was found to contain spin-2 particles and was hence reborn as a possible theory of quantum gravity.

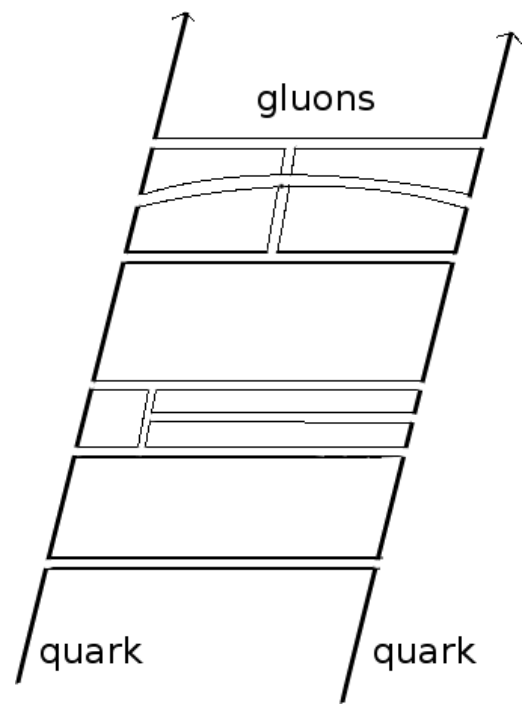

Figure 1.1: 't Hooft's proposal to view many-gluon exchange between quarks as being represented by a continuous string worldsheet.

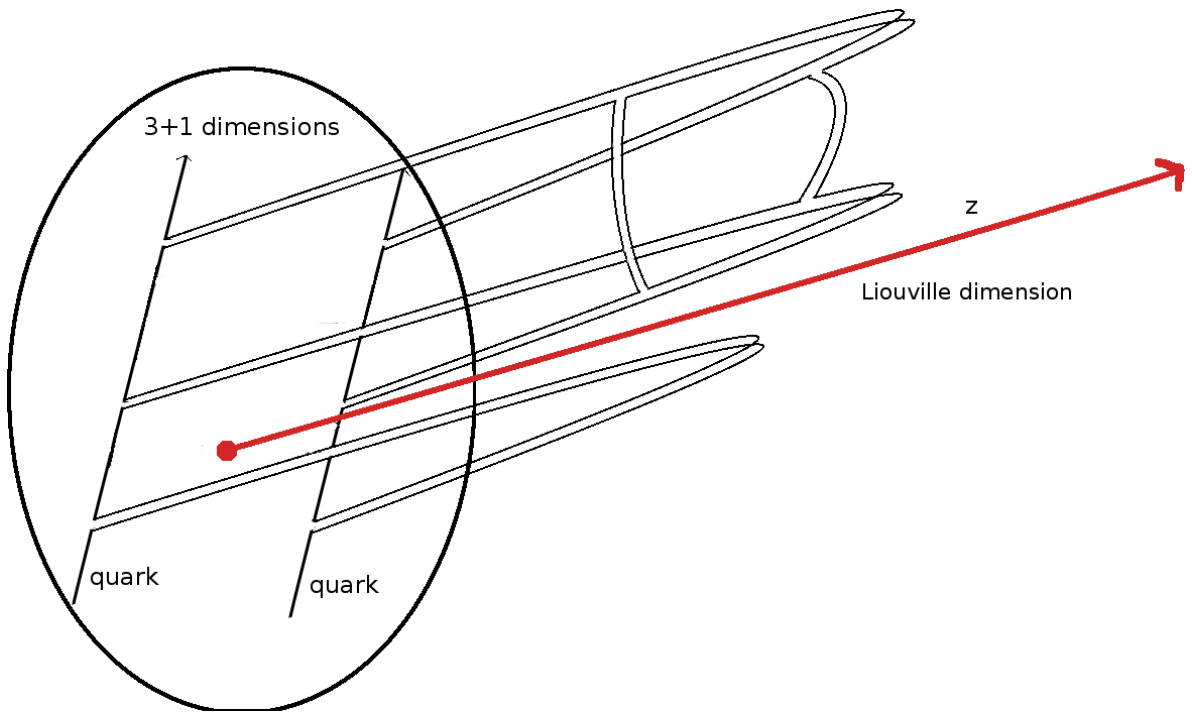

Figure 1.2: Polyakov's proposal of an extra Liouville (fifth) dimension generated by non-critical strings between 3+1 dimensional quarks

Despite these ideas, there was not much activity on gauge-gravity duality until the realization of such a duality using D-branes in 1997.

### Gauge-Gravity duality realization through D-branes

The general ideas and expectations of gauge-gravity duality over a couple of decades were realized in the first concrete example in 1997. This case, using D-branes to relate a certain special gauge theory to strings on Anti de-Sitter space has, along with some generalizations, come to be popularly known as the AdS/CFT correspondence.

D-branes are objects that naturally arise in the quantization of open strings in spacetime. Open strings, which satisfy free (or Neumann) boundary conditions in empty spacetime, can end on D-branes which effectively serve as Dirichlet boundary conditions for them. In the seminal work [5], Polchinski showed that such D-branes are charged massive dynamical objects of their own right in string theory.

Through such a connection, it was soon realized that a stack of a large number of D-branes in particular could be studied easily within the low-energy approximation to string theory, supergravity. Within supergravity, a stack of D-branes is a massive object that can be thought of as a charged black hole, generalized to have several internal dimensions called the world-volume. As charged massive objects in string theory, they interact with strings in a definite way. Open strings move with their ends constrained to D-brane worldvolume. This interaction with strings can be captured by an effective world-volume field theory for the dynamics of the D-branes. D-branes also interact with closed strings as they are massive objects that create a gravitational field. Much of the activity in the area of using D-branes to understand gauge-gravity duality arises from comparing their properties as seen by open strings and those seen by closed strings.

To understand the gravitational description of D-branes, we start with the low energy effective action for gravity obtained from string theory (in string frame) [6],

$$S = \frac{1}{(2\pi)^7 \alpha'^4} \int d^{10}x \sqrt{-g} \left[ e^{-2\phi} \left( R + 4(\nabla\phi)^2 \right) - \frac{2}{(8-p)!} F_{p+2}^2 \right] \tag{1.0.2}$$

where $\alpha'$, the string tension, sets the scale of the Newton constant and $F_{p+2} = dA_{p+1}$ is the field strength of a $p+1$ form. (For $p = 3$, the form is self-dual $F_5 = *F_5$ but this has to be imposed directly as an equation of motion.)

Stacks of D-branes of various dimensions can be identified with certain solutions of the action above. The number of such D-branes can be counted using their $p$-form charge by integrating the flux $F$ on the transverse sphere $S^{8-p}$,

$$\int_S^{8-p} *F_{p+2} \sim N \tag{1.0.3}$$

Focussing on D3-branes for concreteness, they correspond to solutions of the form (cf. [6, 7]),

$$\begin{aligned} ds^2 &= h^{-\frac{1}{2}} \left( -dt^2 + dx_1^2 + dx_2^2 + dx_3^2 \right) + h^{\frac{1}{2}} \left( dr^2 + r^2 d\Omega_5^2 \right), \\ h &= 1 + \frac{L^4}{r^4}, \end{aligned} \tag{1.0.4}$$

where $L$ is some length scale. This is a 10 dimensional metric where $t, x_1, x_2, x_3$ are the dimensions along the worldvolume of the D3-branes while the 6 dimensional space $R^6$ transverse to the D3-branes is split into the radial direction $r$ and a 5-sphere $\Omega_5$. D3-branes are special since this geometry has no regions of high curvature or any curvature singularities. For large $r \gg L$, the spacetime is flat 10 dimensional spacetime since $A \approx 1$ in this regime. For small $r \ll L$, we find $A \approx L^4/r^4$ and hence find a product of spacetimes with finite constant curvature,

$$ds^2 = \frac{r^2}{L^2}\left(-dt^2 + dx_1^2 + dx_2^2 + dx_3^2\right) + \frac{L^2}{r^2}dr^2 + L^2 d\Omega_5^2 \tag{1.0.5}$$

The last term is simply a 5-sphere of radius $L$. The first two terms combine to give a five dimensional space whose metric can be written in the following way by defining $z = L^2/r$,

$$ds^2_{AdS} = \frac{L^2}{z^2}\left(-dt^2 + dx_1^2 + dx_2^2 + dx_3^2 + dz^2\right) \tag{1.0.6}$$

which is known as Anti-de Sitter space (or Lobachevsky space if the space is made Euclidean $t \to i\tau$ ) and is the negative curvature analog of the round sphere – it is a space of constant negative curvature (of order $L^2$ in this case). Henceforth, we refer to it as AdS space (or $AdS_4$, $AdS_5$ etc to also specify the dimension). Note that the metric is of a form similar to Polyakov's warped form [4, 3] discussed in the last section with $z = Le^{-\phi/L}$ being the Liouville coordinate and $a^2 \sim e^{2\phi/L}$. This is no accident and it is the reason that Anti-de Sitter space is at the heart of the AdS/CFT correspondence.

The parameter $L$ above determines the ADM mass of the gravitational solution above. If this solution is to correspond to a stack of $N$ D3-branes placed at a point, we can equate the ADM mass to $N$ times the tension of a single brane which gives [7],

$$L^4 = 4\pi g_{st} N \alpha'^2 \tag{1.0.7}$$

where we have used the expression for the 10-d Newton constant $\kappa = 8\pi^{7/2} g_{st}\alpha'^2$.

This class of D-branes with $3+1$ dimensional worldvolumes were investigated in particular detail in the years preceeding 1997. In a typical study, the absorption cross-section of particles of various spin by D3-branes was computed by considering the D3-branes as gravitational objects in a 10 dimensional spacetime and by a corresponding calculation in the $3+1$ dimensional world-volume field theory.

One can compute the absorption of a scalar field $\phi(\omega r)$ incident on such objects – a scattering / absorption problem in the six transverse dimensions to the D3-brane (see Figure 1.3). For this, one needs to solve Laplace's equations on the transverse space for the scalar. In the simplest case of s-wave scattering, one finds for the absoprtion cross-section (per unit volume of the D3-brane) [8],

$$\sigma_{3-brane} = \frac{\pi^4}{8}\omega^3 L^8 \tag{1.0.8}$$

One can try to reproduce this result from the open-string properties of D3-branes. D3-branes have transverse and longtitudinal degrees of freedom. Using the coupling to open-strings, one can show that the low-energy dynamics of such D3-branes is captured by a gauge theory where the gauge fields capture the longtitudinal motions and matter fields capture the transverse movements. With the particular case of a stack of $N$ D3-branes placed at a point in flat space, one finds that the world-volume theory is $\mathcal{N}=4$ $SU(N)$ super Yang-Mills theory. In any case, the worldvolume theory of $D3$-branes must couple to background fields of the spacetimes the branes are placed in, in order to reflect the coupling of the branes themselves to such fields.

One can show that this includes couplings of the form [7],

$$S_{int} = \frac{\sqrt{\pi}}{\kappa}\int\left[\frac{1}{4}tr\left(e^{-\phi}F^2_{\mu\nu} - CF_{\mu\nu}\tilde{F}^{\mu\nu}\right) + h^{\mu\nu}T_{\mu\nu}\right] \tag{1.0.9}$$

where $\mu, \nu$ are coordinates along the brane, $\phi, C, h$ are the background dilaton, RR scalar and transverse graviton whose value at the position of the branes enters the action while $F_{\mu\nu}, T_{\mu\nu}$ are worldvolume fields.

Hence the scattering experiment where a scalar such as the dilaton is sent in can be understood in this open-string picture as a dilaton quantum being converted into two worldvolume photons according to the above action. By such a calculation, [8] showed that the absorption cross-section computed from the above couplings completely agrees with the closed string result,

$$\sigma_{3-brane} = \frac{\kappa^2 \omega^3 N^2}{32\pi} \tag{1.0.10}$$

using the relations $L^4 = 2\pi g_{st} N \alpha'^2, \kappa = 8\pi^{7/2} g_{st} \alpha'^2$ noted earlier. Such an exact agreement is surprising since the field theory analysis is inherently weakly coupled, valid only for small coupling $\lambda = g_{YM}^2 N$ while the SUGRA background is weakly curved and can be trusted only in the opposite limit $\lambda \gg 1$. It was later shown that the high level of supersymmetry leads to non-renormalization theorems for 2-point functions and hence for absorption cross-sections like that computed.

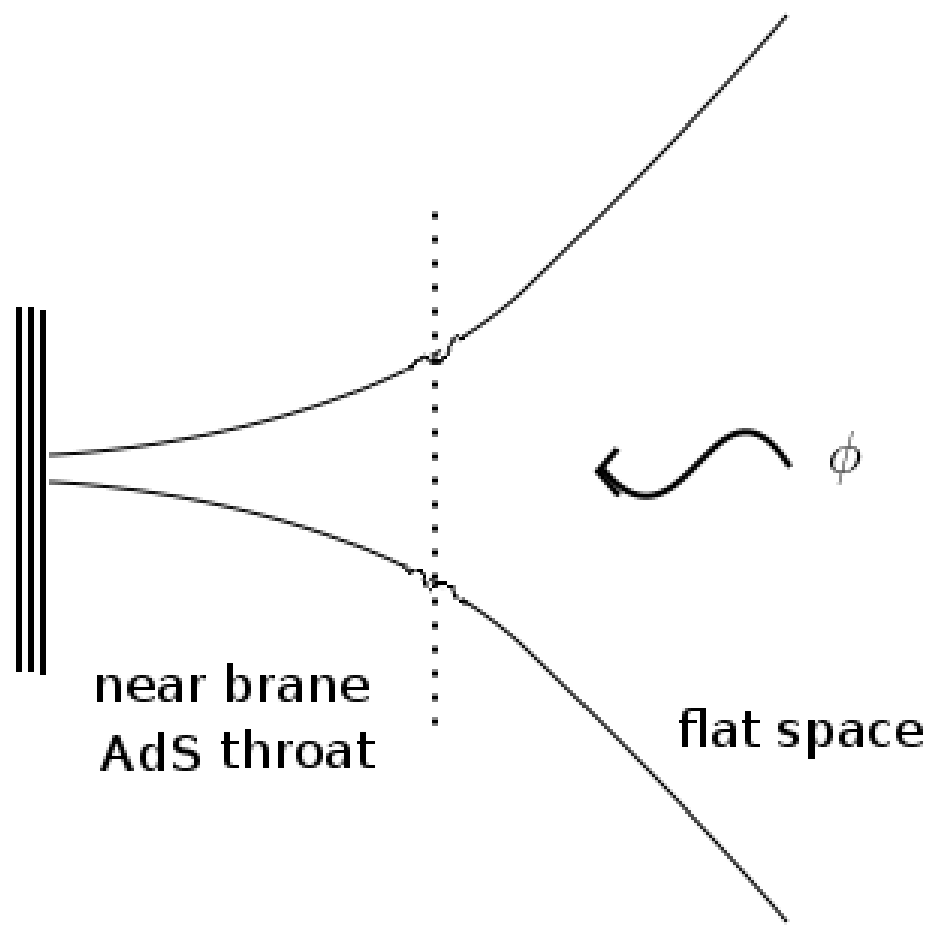

Figure 1.3: Absorption of an incident scalar by a stack of D-branes. The scalars are incident from flat space, onto the near-brane $AdS$ region and can hence be viewed as disturbances at the boundary of the AdS throat.

Beyond such scattering "experiments", computations of entropy of the low-energy theory at finite temperatures $T > 0$ were also done and matched up to numerical factors with a corresponding gravitational calculation of the entropy using D3-branes with a black hole horizon (which arises when the mass and charge are mismatched). Such calculations also gave similar striking results for the M2- and M5-branes of M-theory.

Thus there was accumulating evidence for a non-trivial connection between gravitational aspects of D-branes and the world-volume field theory living on them. For a precise identification, however, some limits were required where some sector of the gravitational dynamics could be exactly identified with the low-energy sector of the world-volume theory which is a gauge theory. This identification was made by Maldacena in 1997 in the paper that started a huge flurry of activity in an area that has come to be known as AdS/CFT duality.

The central insight leading to the AdS/CFT conjecture in [9] was that it is the near-brane (or near-horizon) region of spacetime, $AdS_5 \times S^5$, found in the limit $r \ll L$, which is relevant for the earlier gauge-gravity comparisons of absorption, entropy etc. One way to motivate this is by observing that the gauge theory in question is only the low-energy theory of the dynamics of the D3 branes due to its interactions with open strings in flat space, where there also exist closed strings due to gravity.

Maldacena argued that one could take such a low-energy limit directly in the supergravity solution which results in focussing on the $AdS_5 \times S^5$ near-horizon region. In the SUGRA brane solution Eq.1.0.4 which includes the $AdS$ region as well as the flat space region, any object at some finite radial position $r$ appears red-shifted to an observer at infinity by the warp factor, $h^{-1/4}$. Hence an energy packet of any energy, when at sufficiently small $r \ll L$, appears to be of low energy to an observer at infinity. These small $r$ low energy objects are in addition to waves of very large wavelength propagation through all of the warped spacetime. These latter bulk modes are the same as the closed strings that appear in the earlier description of D3-branes in flat space, interacting with open and closed strings. Hence, the excitations that are low energy by virtue of being at small $r \ll L$ must be identified with the gauge theory that describes the low-energy excitations of the worldvolume of D3-branes. Taking the limit $r \ll L$ precisely results in the $AdS_5 \times S^5$ part of the geometry as was seen after Eq.1.0.4.

This conjecture, while relying on calculations performed at large $\lambda = g_{YM}^2 N$ and large $N$, also goes beyond those limits. To have a gravity dual to a field theory with finite 't Hooft coupling $\lambda$ (with $N \to \infty$) would require us to correct supergravity with classical string scale corrections in powers of $\alpha'/L^2$. This amounts to studying the physics of classical strings moving on the AdS spacetime, as described by the 2 d sigma model. To also have finite $N$, one needs to take into account string loops and consider full interacting quantum string theory on the $AdS$ background (cf. [10]).

### The GKPW dictionary

Another way to understand the $AdS/CFT$ conjecture is by viewing the absorption calculations as computing the effect of perturbing the boundary of $AdS$ space. Scalars are incident on the branes from far away i.e from the flat space region of the geometry as shown in Figure 1.3. Such a scalar can be viewed as a disturbance at the boundary of the $AdS$ throat around $r \sim L$ which can then tunnel past the energy barrier into the region $r \sim (\omega L)L \ll L$.

Such a picture leads to a concrete proposal [11, 12] for computing correlation functions in the gauge theory by considering certain perturbations of the boundary of $AdS$ space. The proposal made was that the generating functional for the correlation functions of an operator in the gauge theory could be obtained from the classical string theory action on $AdS$ space with specific boundary conditions for the fields dual to the operators in question. See [11, 12, 6] for details.

An $AdS$ scalar field $\phi$ has two independent modes near the boundary $z = 0$ of $AdS$

space [13],

$$\phi(z,\vec{x}) \sim z^{\Delta}\left(A(\vec{x}) + O(z^2)\right) + z^{d-\Delta}\left(\phi_0(\vec{x}) + O(z^2)\right). \tag{1.0.11}$$

Here $\phi_0(\vec{x})$ is the source field or boundary value problem one is trying to solve – for example, to compute n-point functions, one would set $\phi_0$ to be a set of $\delta(\vec{x} - \vec{x_i})$ sources. $A$ is then determined from the classical equations for $\phi$ with such sources. It was argued in [11, 12] that the SUGRA action with such boundary conditions should be identified with the generating function of the gauge theory with sources $\phi_0$ for the operator dual to the field $\phi$. Namely, the action of the dual gauge theory is modified by the addition of the term,

$$\int d^4x \phi_0(\vec{x})\mathcal{O}(\vec{x}) \tag{1.0.12}$$

where $\mathcal{O}$ is the operator dual to the field $\phi$. Hence the $\phi_0(\vec{x})$ term in the asymptotic behavior of $\phi$ corresponds to modifying the lagrangian of the dual gauge theory by an operator. Generalizing this, it was argued in [13] that the right interpretation of the $A(\vec{x})$ term is a choice of *vacuum* in which the corresponding operator $\mathcal{O}$ has a vacuum expectation value (VEV) given by $A(\vec{x})$ (up to normalization). This relationship plays an important role in our construction of an RG flow in Chapter 2.

It was shown in [11,12] that $\Delta$ in the expansion of $\phi$ above is precisely the dimension of the dual operator $\mathcal{O}$. The relation between $\Delta$ and the mass of the field $\phi$ is among the most commonly used elements of the AdS/CFT dictionary,

$$\Delta_{\pm} = \frac{d}{2} \pm \sqrt{\frac{d^2}{4} + m^2L^2} \tag{1.0.13}$$

$\Delta_+$ is the usually the only meaningful dimension (i.e above the unitarity bound of $(d-2)/2$) in this formula but for the subtle case where both dimensions are valid,

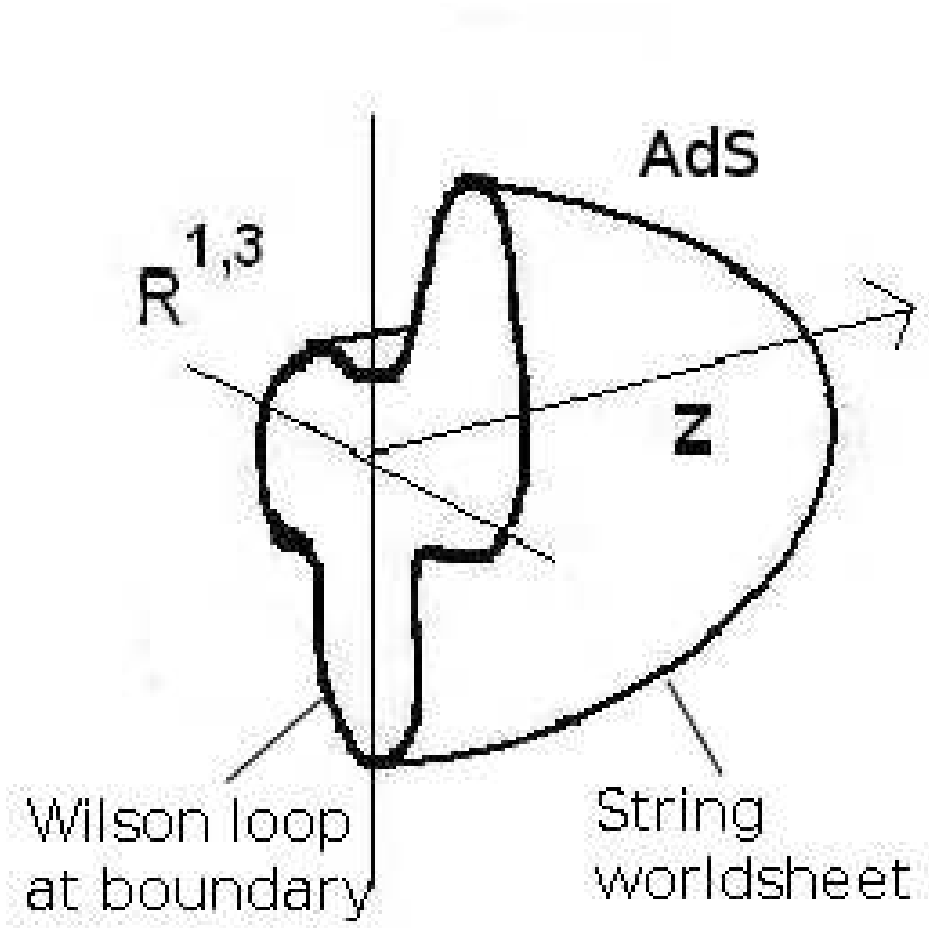

Figure 1.4: Holographic prescription for the Wilson loop. It is given by the action of the classical string worldsheet in $AdS$ space ending on the Wilson loop which is placed at the boundary.

see [7, 13].

A simple proposal for holographic computation of the non-local observables, Wilson loops, was made in [14, 15]. A Wilson loop can be thought of as the phase in the path-integral associated with the motion of massive non-dynamical external quarks (introduced as probes) along the path given by the Wilson loop. Since this can be interpreted as the energy of interaction, the Wilson loop provides a useful probe of the properties of the theory such as confinement. By separating a single $D3$-brane from a stack of $N$ $D3$-branes and viewing the massive stretched strings as the external quarks, [14, 15] proposed that the Wilson loop can be holographically computed by considering the minimal 2-d surface in $AdS$ that ends on the Wilson loop placed at the boundary of $AdS$ space, as illustrated in Fig 1.4.

### 1.0.1 Reducing SUSY

The original example of $AdS/CFT$ [9] duality related theories with a large amount of symmetry and supersymmetry. $\mathcal{N} = 4$ $SU(N)$ super Yang-Mills has the maximal

**Table 1**

Known Compactifications of Chiral N=2 d=10 Supergravity to d=5

| $M^5$ | Isometry | Supersymmetry $N$ = | $M^4$ | Symmetry of P.W. Solution |
|---|---|---|---|---|
| $S^5$ | SU(4) | 8 | $CP^2$ | SU(3) |
| $T^{11}$ | SU(2)×SU(2)×U(1) | 2 | $S^2 \times S^2$ | SU(2)×SU(2) |
| $S^3 \times S^3$ | [SU(2)]$^3$ | 0 | - | - |
| other $T^{pq}$ | SU(2)×SU(2)×U(1) | 0 | - | - |
| $\frac{SU(3)}{SO(3)}$ | SU(3) | 0 | - | - |
| K3×U(1) | U(1) | 4 | - | - |

Figure 1.5: Table compiled by Romans(1984) on compactifications with different amount of symmetry and supersymmetry.

supersymmetry a field theory in $3+1$ dimensions with spins $\leq 1$ can have. To generalize $AdS/CFT$ to describe gauge theories with less supersymmetry, one would need to consider a stack of D3-branes not in flat space but in a space with reduced symmetry. One might expect that the world-volume theory of such D3-branes would be less (super)symmetric than $\mathcal{N}=4$ $SU(N)$ super Yang-Mills.

The first simple generalizations of this form were obtained from placed branes at orbifold singularities [16, 17]. Later, Klebanov and Witten [18] constructed a gauge theory dual to the conifold. They considered branes placed at a conical singularity – a spacetime that is generally flat except at the tip of a cone where the curvature is singular due to a deficit angle. The motivation is that the immediate neighborhood (in the 6 transverse dimensions) of the stack of D3-branes placed at a singularity does not look like flat space and hence when one takes the near horizon limit described earlier, one finds a general spacetime of the form $AdS_5 \times X^5$ where $X^5$ is the base of the cone and is generally different from $S^5$.

Through this method, one can generalize the basic $AdS/CFT$ correspondence to a duality for string theory on spacetimes of the form $AdS_5 \times X^5$ where $X^5$ is a 5 dimensional manifold with less symmetry and supersymmetry than $S^5$. Several such spaces were already classified in the 1980s for completely unrelated purposes by Romans and others – a table from a paper of Romans [19] summarizing the symmetries and supersymmetries of these spaces is presented in Fig 1.5. The example worked out in Klebanov and Witten [18], $T^{11}$ can be seen on the second line.

The dual gauge theory has a gauge group and matter content determined by the details of $X^5$. For example, the global symmetry group of the gauge theory – which usually consists of the flavor symmetries and certain bosonic symmetries arising from the supersymmetry algebra – is exactly the global symmetry of $X^5$. The amount of supersymmetry the dual gauge theory has also corresponds exactly the supersymmetry preserved by the manifold $X^5$. For example, from the table of Romans in Fig 1.5, we see that $S^5$ has $SU(4)$ isometry and $\mathcal{N} = 8$ supersymmetry which is consistent with the $SU(4)$ R-charge symmetry and $\mathcal{N} = 4$ supersymmetry[3] of the dual super Yang-Mills theory. On the other hand, the conifold studied by Klebanov and Witten, $T^{11}$, has less symmetry $SU(2) \times SU(2) \times U(1)$ and only $\mathcal{N} = 2$ supersymmetry. This is consistent with the gauge theory proposed by Klebanov and Witten [18] which is reviewed in Chapter 2.

Thus we see a successful generalization of the $AdS/CFT$ duality to pairs of theories, each with fewer supersymmetries. Such theories share more non-trivial physics with conventional non-supersymmetric theories than $\mathcal{N} = 4$ super Yang-Mills. Looking at Fig 1.5, one might tempted to go further and use one of the non-supersymmetric spaces listed there such as $T^{pq}$ to construct duals to fully non-supersymmetric gauge theories. However, many difficulties associated with stability arise. While the real world at least appears to be both non-supersymmetric and stable, within the context of string theory, non-supersymmetric spacetimes appear to generically suffer various instabilities that need to be investigated on a case-by-case basis. There were many negative results derived over the years for non-supersymmetric $X^5$ [20–23]. In Chapter 5 of this thesis, we study non-supersymmetric $AdS_4 \times X^7$ spaces, showing that the situation is superficially better but has other non-perturbative problems.

[3]Note that the conventions for the amount of supersymmetry in supergravity is twice that of supersymmetric field theories. Hence $\mathcal{N} = 2$ supergravity has the same number of supercharges as a $\mathcal{N} = 1$ field theory.

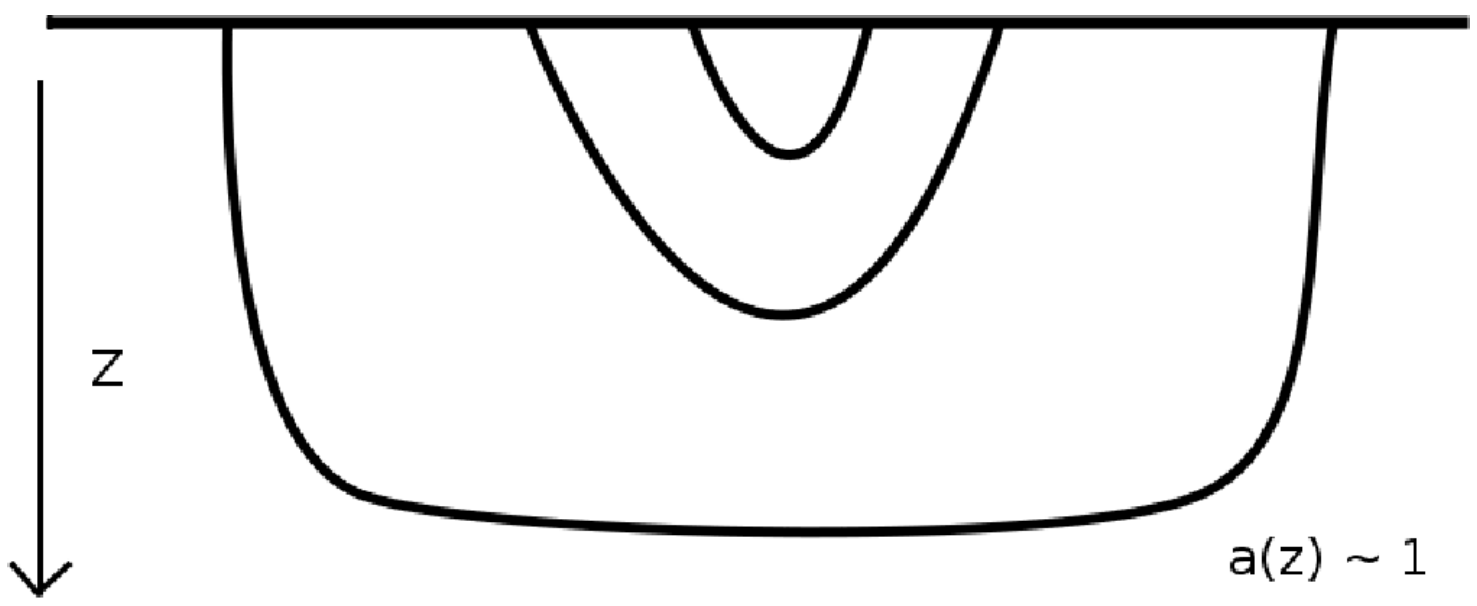

Figure 1.6: The Wilson loop begins to flatten out as the quarks get widely separated in a confining theory, leading to the area law.

### 1.0.2 Breaking conformality

The motivation to look for gauge-gravity duality from the 1970s was always to understand strongly coupled phenomena of QCD such as confinement and chiral symmetry breaking. These are by definition highly non-trivial phenomena that arise at particular energy scales. In this sense, the physics of conformal field theories, even if strongly coupled, is orthogonal to such phenomena.

The extension of $AdS/CFT$ to non-CFTs relies on deforming $AdS$ space. In Polyakov's picture of non-critical strings with a Liouville direction, the metric of a general string background with a gauge dual can be written in the form,

$$ds^2 = a^2(z) \left(-dt^2 + d\vec{x}^2 + dz^2\right) \tag{1.0.14}$$

where the Liouville direction $z$ has the interpretation of an energy scale and $a^2(z)$ is the related warp factor.

$AdS$ space alone has the warp factor $a^2(z) \sim \frac{1}{z^2}$ which makes the metric invariant under the scaling of the energy coordinate $z$ and spacetime,

$$z \to \lambda z, \quad \vec{x} \to \lambda \vec{x}, \quad t \to \lambda t. \tag{1.0.15}$$

As a result, the dual theory is at a conformal fixed point.

One can generalize the warp factor to describe an RG flow starting from a UV conformal fixed point and flowing to an IR conformal fixed point. Such a warp factor $a^2(z)$ would need to be in the conformal form $1/z^2$ in the UV and in the IR (though about a different point $1/|z-z_0|^2$). In between, the warp factor can fluctuate describing the non-trivial dynamics of an RG flow as shown in Figure 1.7. The size of the internal cycle, not shown in the figure, measures the central charge and breaks the apparent symmetry between the UV and the IR fixed points of the figure. Such RG flows have been constructed in the literature (cf. [24]) and Chapters 2 and 4 involve such constructions as well.

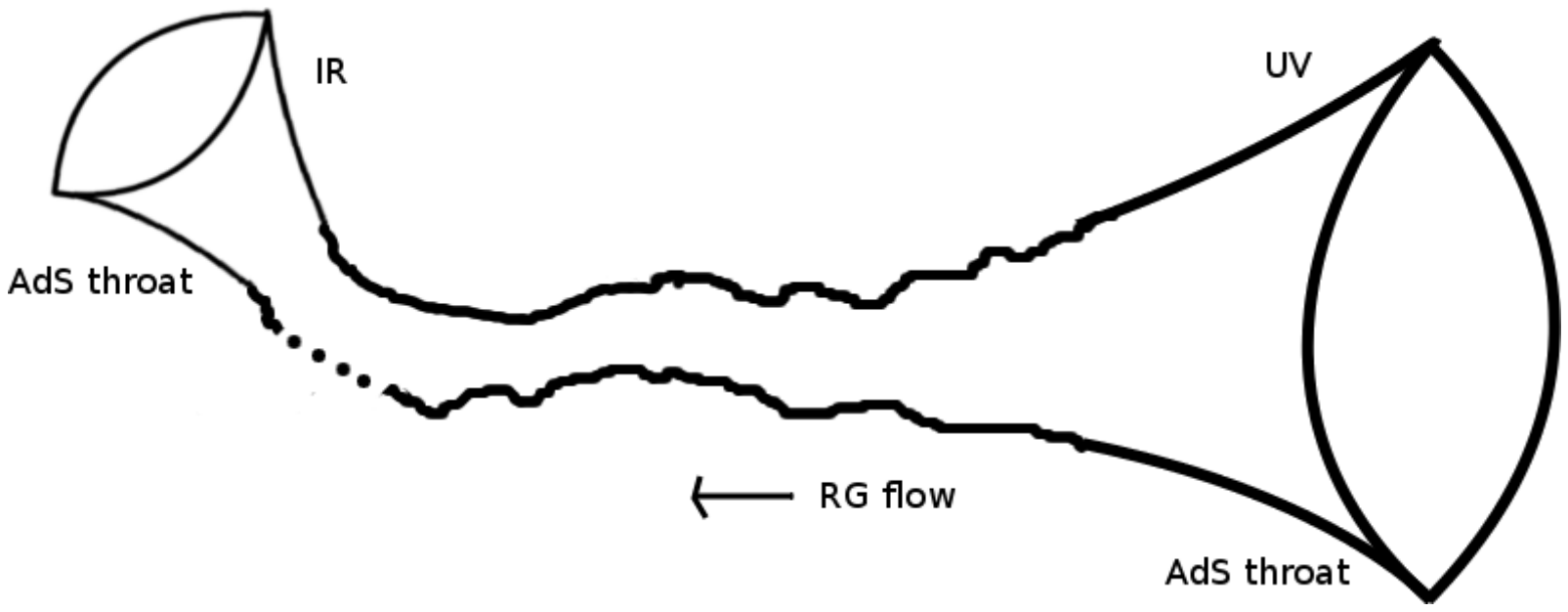

Figure 1.7: Schematic representation of the geometry dual to an RG flow between two fixed points. We have $AdS$ geometries in the UV and IR where the warp factor is $a^2(z) \sim \frac{1}{z^2}$ and deviations in between. The IR throat is show smaller since the internal space shrinks in accordance with the c-theorem for central charge.

If one wants to construct a string dual a theory like QCD, which is conformal at high energies and dynamically develops scales in the infrared, one would again need an $a^2(z)$ that tends to the conformal form $\frac{1}{z^2}$ near $z \sim 0$ (the UV) but deviates from that for large $z$ (the IR). Further, if one wants confinement as measured by an area law for the Wilson loop, one can show that we need $a^2(z) \to 1$ for large $z$ [25, 26]. This would lead to an area law for the Wilson loop in such a background when the quarks are well separated while it would produce the Coulomb force for small separations as shown in Fig. 1.6. This can be understood intuitively as minimal surface in a $z$-dependent gravitational field determined by $a^2(z)$. If $a^2(z) \to 0$ for large $z$ like in the case of $AdS$ space, the minimal surface drops further and further towards large $z$ and does not lead to an area law.

The Klebanov-Strassler solution [25], also known as the warped deformed conifold, was the first version of a complete string model of confinement with such a non-trivial warp factor and has turned out to be a cornerstone in studying confinement through gauge-string duality. There are several references and reviews which detail the physics of the conifold [27, 28, 25, 29]. Here we only outline the some of the properties most relevant to the discussion above. The Klebanov-Strassler (KS) solution is based on a deformation of the singular cone over $T^{11}$ with reduced supersymmetry discussed in the last section. By turning on fluxes $H_3, F_3$ and $F_5$, the KS solution is able to support a non-trivial warp factor $h_{KS}(z)$ that does go to a non-zero constant in the IR. The metric of the KS solution is of the form [25, 27, 28],

$$\begin{aligned} ds^2 &= h^{-1/2}(\tau)(-dt^2 + d\vec{x}^2) + h^{1/2}(\tau)ds_6^2 \qquad (1.0.16) \\ h &\sim \int_\tau^\infty dx \frac{x \coth x - 1}{\sinh^2 x}(\sinh(2x) - 2x)^{1/3} \to a_0 + O(\tau^2) \quad \text{for small } \tau \end{aligned}$$

where $h$ is the warp factor and $ds_6^2$ are internal coordinates of a modified $T^{1,1}$ along with the $\tau$ direction. See for eg. [27] for more details. Here $\tau$ plays the role of the Liouville/energy direction.

The KS warp factor $h(\tau)$ does go to a constant at the tip of the conifold, $\tau = 0$ (the IR). Hence the Wilson loop exhibits the area law. However $\sqrt{h}$ does not reduce to the conformal $\frac{1}{r^2}$ limit in the UV but rather to $\frac{\sqrt{\log r}}{r^2}$. This is a reflection of the fact that the KS solution does not have a fixed point in the UV but rather shows a very interesting cascade of ever-growing number of degrees of freedom. This has been interpreted as a sequence of Seiberg dualities, with the theory best described as an $SU(N) \times SU(N+M)$ for different $N$ at different scales. For more details, see [25].

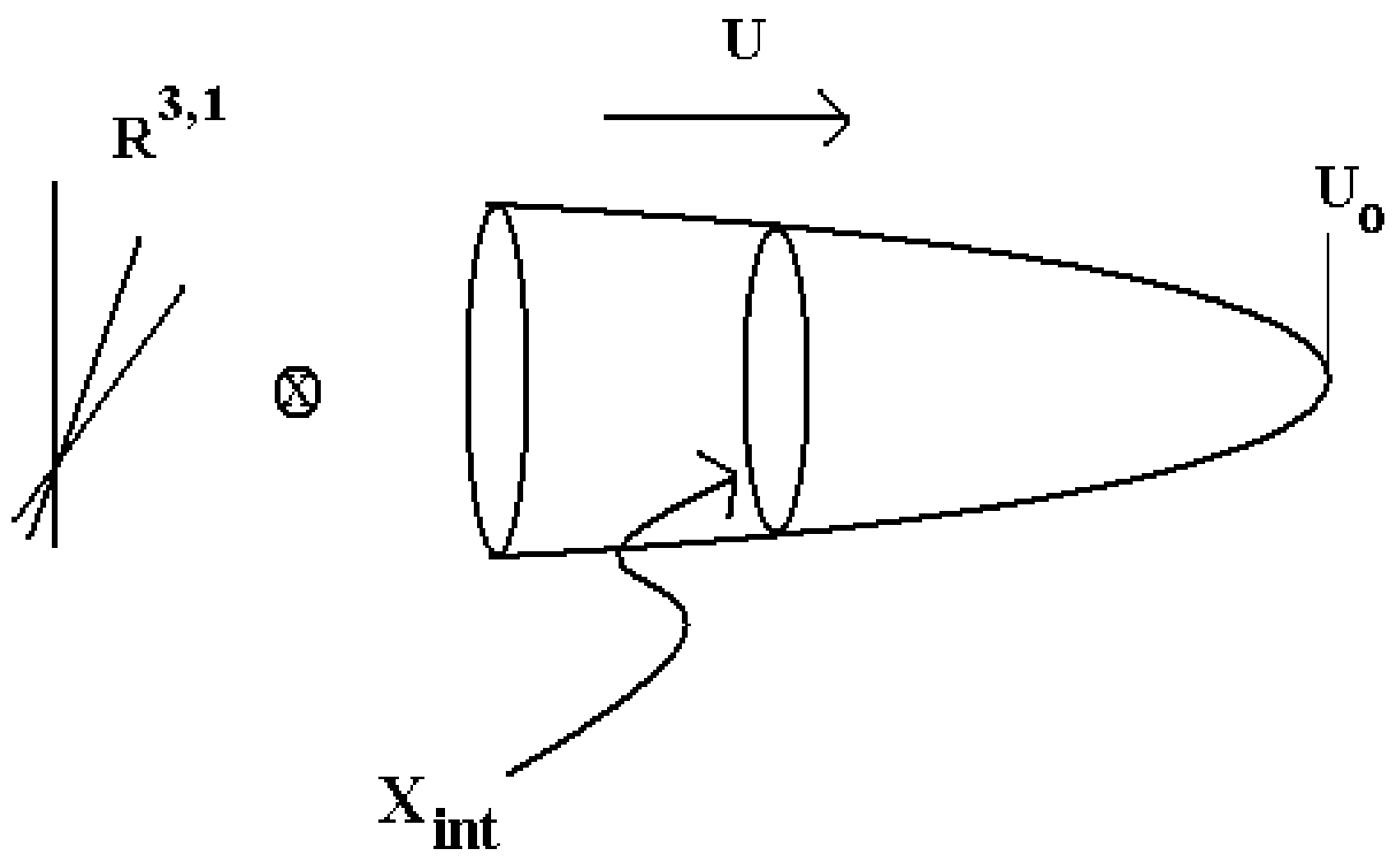

Figure 1.8: Schematic representation of the geometry dual to a confining gauge theory. The internal cycle $X_{int}$ shrinks along the Liouville/energy direction $U$ and vanishes at a point $U_0$. The transverse spacetime $R^{3,1}$ is also shown.

Note that the internal space $T^{1,1}$ shrinks to zero size at the tip where $h_{KS}(z) \sim 1$ and hence the geometry caps off. This phenomena where the internal cycle changes radius (unlike in the conformal case of $AdS_5 \times X^5$ where $X^5$ has fixed radius) along the Liouville energy direction and capps off the space by shrinking to zero size in the IR will be seen to be intimately related to confinement in Chapter 3.

Since the KS solution was found, several generalizations in related contexts have been made [30]. Brane constructions in other versions of string theory have also proven useful [31, 32]. Each model typically reproduces some non-trivial IR physics shared by QCD like chiral symmetry breaking but have their own peculiar behavior in the UV.

Other phenomenological models of confinement have also been cooked up, purely by manipulating the warp factor to get desirable properties without looking for solutions of string theory, such as the soft wall [33] and hard wall [26] models. Probe branes have been used to introduce quarks in the fundamental representation of the gauge groups in the dual theory [34]. Thus much progress has been made in modeling aspects of QCD-like physics within the classical SUGRA framework.

# Chapter 2

# Resolved conifold

This chapter is based on the paper 'Gauge/Gravity Duality and Warped Resolved Conifold' written in collaboration with I.R. Klebanov [35]. We study supergravity backgrounds encoded through the gauge/string correspondence by the $SU(N) \times SU(N)$ theory arising on $N$ D3-branes on the conifold. As discussed in hep-th/9905104, the dynamics of this theory describes warped versions of both the singular and the resolved conifolds through different (symmetry breaking) vacua. We construct these supergravity solutions explicitly and match them with the gauge theory with different sets of vacuum expectation values of the bi-fundamental fields $A_1, A_2, B_1, B_2$. For the resolved conifold, we find a non-singular $SU(2) \times U(1) \times U(1)$ symmetric warped solution produced by a stack of D3-branes localized at a point on the blown-up 2-sphere. It describes a smooth RG flow from $AdS_5 \times T^{1,1}$ in the UV to $AdS_5 \times S^5$ in the IR, produced by giving a VEV to just one field, e.g. $B_2$. The presence of a condensate of baryonic operator $\det B_2$ is confirmed using a Euclidean D3-brane wrapping a 4-cycle inside the resolved conifold. The Green's functions on the singular and resolved conifolds are central to our calculations and are discussed in some detail.

## 2.1 Introduction

In this chapter, we study elements of the AdS/CFT dictionary that relate a gauge theory and its gravitational dual when the former is not conformal. We induce a renormalization group flow in the gauge theory by giving an expectation value to certain operators and find that the gauge theory starts from a conformal UV fixed point and flows to another conformal IR fixed point. Hence the gravitational dual is of the form described in the Chapter 1, where there are two $AdS$ throats in the geometry (cf. Figure 1.7).

For the UV fixed point, we choose the simplest generalization of the AdS/CFT correspondence with lower amounts of supersymmetry. This generalization to $\mathcal{N} = 1$ superconformal symmetry was made in [18, 36] by considering a stack D3-branes, not in flat space, but placed at the tip of a 6d Calabi-Yau cone $X_6$. The near horizon limit in this case turns out to be $AdS_5 \times Y_5$ where $Y_5$ is the compact 5 dimensional base of $X_6$ and is a Sasaki-Einstein space.

Among the simplest of these examples is $Y_5 = T^{1,1}$, corresponding $X_6$ being the conifold. It was found that the low-energy gauge theory on the D3-branes at the tip of the conifold is a $\mathcal{N} = 1$ supersymmetric $SU(N) \times SU(N)$ gauge theory with bi-fundamental chiral superfields $A_i$, $B_j$ $(i, j = 1, 2)$ in $(N, \bar{N})$ and $(\bar{N}, N)$ representations of the gauge groups, respectively [18, 36]. The superpotential for this gauge theory is $W \sim \mathrm{Tr}\det A_i B_j = \mathrm{Tr}\ (A_1B_1A_2B_2 - A_1B_2A_2B_1)$. The continuous global symmetries of this theory are $SU(2) \times SU(2) \times U(1)_R \times U(1)_B$ where the $SU(2)$ factors act on $A_i$ and $B_j$ respectively, $U(1)_B$ is a baryonic symmetry, and $U(1)_R$ is the R-symmetry with $R_A = R_B = \frac{1}{2}$. This assignment ensures that $W$ is marginal, and one can also show that the gauge couplings do not run. Hence this theory is superconformal for all values of gauge couplings and superpotential coupling [18, 36].

When the above gauge theory is considered with no vacuum expectation values (VEV's) for any of the fields, we have a superconformal theory with the $AdS_5 \times T^{1,1}$ dual. In [13], more general vacua of this theory were studied. It was argued that moving the D3-branes off the tip of the singular conifold corresponds to a symmetry breaking in the gauge theory due to VEV's for the $A, B$ matter fields such that the VEV of operator

$$\mathcal{U} = \frac{1}{N}\mathrm{Tr}\left(|B_1|^2 + |B_2|^2 - |A_1|^2 - |A_2|^2\right) \tag{2.1.1}$$

vanishes. Further, more general vacua exist for this theory in which this operator acquires a non-zero VEV.[1] It was pointed out in [13] that these vacua cannot correspond to D3-branes on the singular conifold. Instead, such vacua with $\mathcal{U} \neq 0$ correspond to D3-branes on the resolved conifold. This "small resolution" is a motion along the Kähler moduli space where the singularity of the conifold is replaced by a finite $S^2$. Thus the $SU(N) \times SU(N)$ gauge theory was argued to incorporate in its different vacua both the singular and resolved conifolds. On the other hand, the deformation of the conifold, which is a motion along the complex structure moduli space, can be achieved through replacing the gauge theory by the cascading $SU(N) \times SU(N+M)$ gauge theory (see [25]).

One of the goals of this chapter is to construct the warped SUGRA solutions corresponding to the gauge theory vacua with $\mathcal{U} \neq 0$. Our work builds on the earlier resolved conifold solutions constructed by Pando Zayas and Tseytlin [37], where additional simplifying symmetries were sometimes imposed. Such solutions corresponding to D3-branes "smeared" over a region were found to be singular in the IR [37]. We will instead look for "localized" solutions corresponding to the whole D3-brane stack located at one point on the (resolved) conifold. This corresponds to giving VEV's to the fields $A_i, B_j$ which are proportional to $1_{N\times N}$. We construct the duals of these gauge theory vacua and find them to be completely non-singular. The solution acquires a particularly simple form when the stack is placed at the north pole of the blown up

[1] As was pointed out in [18], no D-term equation constrains this operator since the $U(1)$ gauge groups decouple in the infrared.

2-sphere at the bottom of the resolved conifold. It corresponds to the simplest way to have $\mathcal{U} \neq 0$ by setting $B_2 = u 1_{N\times N}$ while keeping $A_1 = A_2 = B_1 = 0$.

Following [38, 13], we also interpret our solutions as having an infinite series of VEV's for various operators in addition to $\mathcal{U}$. For this, we rely on the relation between normalizable SUGRA modes and gauge theory VEV's in the AdS/CFT dictionary. When a given asymptotically AdS solution has a (linearized) perturbation that falls off as $r^{-\Delta}$ at large $r$, it corresponds to assigning a VEV for a certain operator $\mathcal{O}$ of dimension $\Delta$ in the dual gauge theory [38,13]. The warp factor produced by a stack of D3-branes on the resolved conifold is related to the Green's function on the resolved conifold. This warp factor can be expanded in harmonics and corresponds to a series of normalizable fluctuations as above, and hence a series of operators in the gauge theory acquire VEV's.[2] For this purpose, we write the harmonics in a convenient set of variables $a_i, b_j$ that makes the link with gauge theory operators built from $A_i, B_j$ immediate. Due to these symmetry breaking VEV's, the gauge theory flows from the $SU(N)\times SU(N)$ $\mathcal{N}=1$ theory in the UV to the $SU(N)$ $\mathcal{N}=4$ theory in the IR, as one would expect when D3-branes are placed at a smooth point. The SUGRA solution is shown to have two asymptotic AdS regions – an $AdS_5 \times T^{1,1}$ region in the UV, and also an $AdS_5 \times S^5$ region produced in the IR by the localized stack of D3-branes. This can be considered an example of holographic RG flow. The Green's functions determined here might also have applications to models of D-brane inflation, and to computing 1-loop corrections to gauge couplings in gauge theories living on cycles in the geometry [42,43].

When the branes are placed on the blown up 2-sphere at the bottom of the resolved conifold, this corresponds to $A_1 = A_2 = 0$ in the gauge theory. Hence no chiral mesonic operators, such as $\mathrm{Tr} A_i B_j$, have VEV's, but baryonic operators, such as $\det B_2$, do acquire VEV's. Therefore, such solutions, parametrized by the size of the resolution and position of the stack on the 2-sphere, are dual to a "non-mesonic" (or "baryonic") branch of the $SU(N) \times SU(N)$ SCFT (see [44] for a related discussion). These solutions have a blown up $S^2$. On the other hand, the solutions dual to the baryonic branch of the cascading $SU(N) \times SU(N+M)$ gauge theory were constructed in [30, 29] (for an earlier linearized treatment, see [45]) and have a blown up $S^3$ supported by the 3-form flux.

[2] In the $\mathcal{N}=4$ SUSY example, the normalizations of the VEV's have been matched with the size of the SUGRA perturbations around $AdS_5 \times S^5$ (see [39–41]). In this chapter we limit ourselves to a more qualitative picture where the precise normalizations of the VEV's are not calculated.

The chapter is organized as follows. In Section 2.2, we review and establish notation for describing the conifold, its resolution, its symmetries and coordinates that make the symmetries manifest. We also review the metric of the resolved conifold and the singular smeared solution found in [37]. In Section 2.3, as a warm up, we study the simple example of moving a stack of D3-branes away from the tip of the singular conifold. We present the explicit supergravity solution for this configuration by determining the Green's function on the conifold. We interpret the operators that get VEV's and note that in general, chiral as well as non-chiral operators get VEV's. In Section 2.4, we determine the explicit SUGRA solution corresponding to a heavy stack of D3-branes at a point on the resolved conifold, again by finding the Green's function on the manifold. We find a non-singular solution with an $AdS_5 \times S^5$ region and interpret this construction in gauge theory. We consider a wrapped Euclidean D3-brane to confirm the presence of baryonic VEVs and reproduce the wavefunction of a charged particle in a monopole field from the DBI action as a check on our calculations. We make a brief note on turning on a fluxless NS-NS $B_2$ field on the warped resolved conifold in Section 2.5. In Appendix A we discuss the harmonics on $T^{1,1}$ in co-ordinates that make the symmetries manifest. We then classify operators in the gauge theory by symmetry in an analogous way to enable simple matching of operator VEV's and normalizable fluctuations.

## 2.2 The Conifold and its Resolution

The conifold is a singular non-compact Calabi-Yau three-fold [46]. Its importance arises from the fact that the generic singularity in a Calabi-Yau three-fold locally looks like the conifold. This is because it is given by the quadratic equation,

$$z_1^2 + z_2^2 + z_3^2 + z_4^2 = 0. \tag{2.2.1}$$

This homogeneous equation defines a real cone over a 5 dimensional manifold. For the cone to be Ricci-flat the 5d base must be an Einstein manifold ($R_{\mu\nu} = 4g_{\mu\nu}$). For the conifold [46], the topology of the base can be shown to be $S^2 \times S^3$ and it is called $T^{1,1}$ with the following Einstein metric,

$$\begin{aligned} d\Omega^2_{T^{1,1}} &= \frac{1}{9}\left(d\psi + \cos\theta_1 d\phi_1 + \cos\theta_2 d\phi_2\right)^2 \\ &\quad + \frac{1}{6}(d\theta_1^2 + \sin^2\theta_1 d\phi_1^2) + \frac{1}{6}(d\theta_2^2 + \sin^2\theta_2 d\phi_2^2). \end{aligned} \tag{2.2.2}$$

The metric on the cone is then $ds^2 = dr^2 + r^2 d\Omega^2_{T^{1,1}}$. As shown in [46] and earlier in [19], $T^{1,1}$ is a homogeneous space, being the coset $SU(2) \times SU(2)/U(1)$ and the above metric is the invariant metric on the coset space.

We may introduce two other types of complex coordinates on the conifold, $w_a$ and $a_i, b_j$, as follows,

$$\begin{aligned} Z = \begin{pmatrix} z^3 + iz^4 & z^1 - iz^2 \\ z^1 + iz^2 & -z^3 + iz^4 \end{pmatrix} = \begin{pmatrix} w_1 & w_3 \\ w_4 & w_2 \end{pmatrix} = \begin{pmatrix} a_1 b_1 & a_1 b_2 \\ a_2 b_1 & a_2 b_2 \end{pmatrix} \\ = r^{\frac{3}{2}} \begin{pmatrix} -c_1 s_2\, e^{\frac{i}{2}(\psi+\phi_1-\phi_2)} & c_1 c_2\, e^{\frac{i}{2}(\psi+\phi_1+\phi_2)} \\ -s_1 s_2\, e^{\frac{i}{2}(\psi-\phi_1-\phi_2)} & s_1 c_2\, e^{\frac{i}{2}(\psi-\phi_1+\phi_2)} \end{pmatrix} \end{aligned} \tag{2.2.3}$$

where $c_i = \cos\frac{\theta_i}{2}$, $s_i = \sin\frac{\theta_i}{2}$ (see [46] for other details on the $w, z$ and angular coordi-nates.) The equation defining the conifold is now $\det Z = 0$.

The $a, b$ coordinates above will be of particular interest in this chapter because the symmetries of the conifold are most apparent in this basis. The conifold equation has $SU(2) \times SU(2) \times U(1)$ symmetry since under these symmetry transformations,

$$\det LZR^T = \det e^{i\alpha} Z = 0. \tag{2.2.4}$$

This is also a symmetry of the metric presented above where each $SU(2)$ acts on $\theta_i, \phi_i, \psi$ (thought of as Euler angles on $S^3$) while the $U(1)$ acts by shifting $\psi$. This symmetry can be identified with $U(1)_R$, the R-symmetry of the dual gauge theory, in the conformal case. The action of the $SU(2) \times SU(2) \times U(1)_R$ symmetry on $a_i, b_j$ (defined in (2.2.3)):

$$\begin{aligned} SU(2) \times SU(2) \text{ symmetry} \quad &: \quad \begin{pmatrix} a_1 \\ a_2 \end{pmatrix} \to L \begin{pmatrix} a_1 \\ a_2 \end{pmatrix}, \qquad \begin{pmatrix} b_1 \\ b_2 \end{pmatrix} \to R \begin{pmatrix} b_1 \\ b_2 \end{pmatrix} \\ \text{R-symmetry} \quad &: \quad (a_i, b_j) \to e^{i\frac{\alpha}{2}} (a_i, b_j) \ , \end{aligned} \tag{2.2.5}$$

i.e. $a$ and $b$ transform as $(1/2, 0)$ and $(0, 1/2)$ under $SU(2) \times SU(2)$ with R-charge $1/2$ each. We can thus describe the singular conifold as points parametrized by $a, b$ but from (2.2.3), we see that there is some redundancy in the $a, b$ coordinates. Namely, the transformation

$$a_i \to \lambda\, a_i \quad , \quad b_j \to \frac{1}{\lambda}\, b_j \quad (\lambda \in \mathbf{C}) \tag{2.2.6}$$

give the same $z, w$ in (2.2.3). We impose the constraint $|a_1|^2 + |a_2|^2 - |b_1|^2 - |b_2|^2 = 0$ to fix the magnitude in the above transformation. To account for the remaining phase, we describe the singular conifold as the quotient of the $a, b$ space with the above constraint by the relation $a \sim e^{i\alpha} a, b \sim e^{-i\alpha} b$.

One simple way to describe the resolution is as the space obtained by modifying the above constraint to,

$$|b_1|^2 + |b_2|^2 - |a_1|^2 - |a_2|^2 = u^2 \tag{2.2.7}$$

and then taking the quotient, $a \sim e^{i\alpha}a, b \sim e^{-i\alpha}b$. Then $u$ is a measure of the resolution and it can be seen that this space is a smooth Calabi-Yau space where the singular point of the conifold is replaced by a finite $S^2$. The complex metric on this space is given in [46] while an explicit metric, first presented in [37], is:

$$\begin{aligned} ds_6^2 &= \kappa^{-1}(r)dr^2 + \frac{1}{9}\kappa(r)r^2\left(d\psi + \cos\theta_1 d\phi_1 + \cos\theta_2 d\phi_2\right)^2 \\ &\quad + \frac{1}{6}r^2(d\theta_1^2 + \sin^2\theta_1 d\phi_1^2) + \frac{1}{6}(r^2 + 6u^2)(d\theta_2^2 + \sin^2\theta_2 d\phi_2^2) \end{aligned} \tag{2.2.8}$$

where

$$\kappa(r) = \frac{r^2 + 9u^2}{r^2 + 6u^2} , \tag{2.2.9}$$

where $r$ ranges from 0 to $\infty$. Note that the above metric has a finite $S^2$ of radius $u$ at $r = 0$, parametrized by $\theta_2, \phi_2$. Topologically, the resolved conifold is an $\mathbf{R}^4$ bundle over $S^2$. The metric asymptotes to that of the singular conifold for large $r$.

Now we consider metrics produced by D3-branes on the conifold. As a warm-up to the case of the resolved conifold, we consider the example of placing a stack of D3-branes away from the apex of the singular conifold. As in [13], the corresponding supergravity solution is

$$ds^2 = \sqrt{H^{-1}(y)}\,\eta_{\mu\nu}dx^\mu dx^\nu + \sqrt{H(y)}\left(dr^2 + r^2 d\Omega_{T^{1,1}}^2\right) , \tag{2.2.10}$$

$$F_5 = (1 + *)dH^{-1} \wedge dx^0 \wedge dx^1 \wedge dx^2 \wedge dx^3, \qquad \Phi = \text{const} \tag{2.2.11}$$

where $\mu, \nu = 0, 1, 2, 3$ are the directions along the D3-branes. $H(y)$ is a solution of the

Green's equation on the conifold

$$\begin{aligned} \Delta H(r,Z;r_0,Z_0) = \frac{1}{\sqrt{g}}\partial_m(\sqrt{g}g^{mn}\partial_n H) &= -\mathcal{C}\frac{1}{\sqrt{g}}\delta(r-r_0)\delta^5(Z-Z_0)\ , \\ \mathcal{C} = 2\kappa_{10}^2 T_3 N &= (2\pi)^4 g_s N(\alpha')^2\ , \end{aligned} \tag{2.2.12}$$

where $(r_0, Z_0)$ is the location of the stack ($Z$ will represent coordinates on $T^{1,1}$) and $T_3 = \frac{1}{g_s(2\pi)^3(\alpha')^2}$ is the D3-brane tension.

When the stack of D3-branes is placed at $r_0 = 0$, the solution is $H = L^4/r^4$ where $L^4 = \frac{27\pi g_s N(\alpha')^2}{4}$. This reduces the metric to ($z = L^2/r$),

$$ds^2 = \frac{L^2}{z^2}\left(dz^2 + \eta_{\mu\nu}dx^\mu dx^\nu\right) + L^2 d\Omega^2_{T^{1,1}} \tag{2.2.13}$$

This is the $AdS_5 \times T^{1,1}$ background, which is dual to the superconformal $SU(N) \times SU(N)$ theory without any VEV's for the bifundamental superfields. More general locations of the stack, corresponding to giving VEV's that preserve the condition $\mathcal{U} = 0$, will be considered in section 4.

Now consider the case of resolved conifold. With D3-branes placed on this manifold, we get the warped 10-d metric,

$$ds_{10}^2 = \sqrt{H^{-1}(y)}dx^\mu dx_\mu + \sqrt{H(y)}ds_6^2 \tag{2.2.14}$$

where $ds_6^2$ is the resolved conifold metric (2.2.8) and $H(y)$ is the warp factor as a function of the transverse co-ordinates $y$, determined by the D3-brane positions. The dilaton is again constant, and $F_5 = (1+*)dH^{-1} \wedge dx^0 \wedge dx^1 \wedge dx^2 \wedge dx^3$.

In [37], the warped supergravity solution was worked out assuming a warp factor with only radial dependence (i.e no angular dependence on $\theta_2, \phi_2$):

$$H_{PT}(r) = \frac{2L^4}{9u^2r^2} - \frac{2L^4}{81u^4}\log\left(1 + \frac{9u^2}{r^2}\right)\ . \tag{2.2.15}$$

The small $r$ behavior of $H_{PT}$ is $\sim \frac{1}{r^2}$. This produces a metric singular at $r=0$ since the radius of $S^2(\theta_2, \phi_2)$ blows up and the Ricci tensor is singular. Imposing the symmetry that $H$ has only radial dependence corresponds not to having a stack of D3-branes at a point (which would necessarily break the $SU(2)$ symmetry in $\theta_2, \phi_2$) but rather having the branes smeared out uniformly on the entire two sphere at the origin. The origin of this singularity is precisely the smearing of the D3-brane charge. In Section 2.4, we confirm this by constructing the solution corresponding to localized branes and find that there is no singularity.

## 2.3 Flows on the Singular Conifold

Let us consider the case when the stack of D3-branes is moved away from the singular point of the conifold. Since the branes are at a smooth point on the conifold, we expect the near brane geometry to become $AdS_5 \times S^5$ and thus describe $\mathcal{N}=4$ $SU(N)$ SYM theory. The warp factor $H(r, Z)$ can be written as an expansion in harmonics on $T^{1,1}$ starting with the leading term $1/r^4$ followed by higher powers of $1/r$. Thus, the full solution still looks like $AdS_5 \times T^{1,1}$ at large $r$, but further terms in the expansion of the warp factor change the geometry near the branes to $AdS_5 \times S^5$. Such a SUGRA solution describes the RG flow from the $\mathcal{N}=1$ $SU(N) \times SU(N)$ theory in the UV to the $\mathcal{N}=4$ $SU(N)$ SYM in the IR. We will confirm this explicitly through the computation of the general Green's function on the conifold. We display the series of perturbations of the metric and interpret these normalizable solutions in terms of VEVs in the gauge theory for a series of operators using the setup of Appendix A. This was originally studied in [13] where a restricted class of chiral operators was considered.

Let us place the stack at a point $(r_0, Z_0)$ on the singular conifold. We rewrite

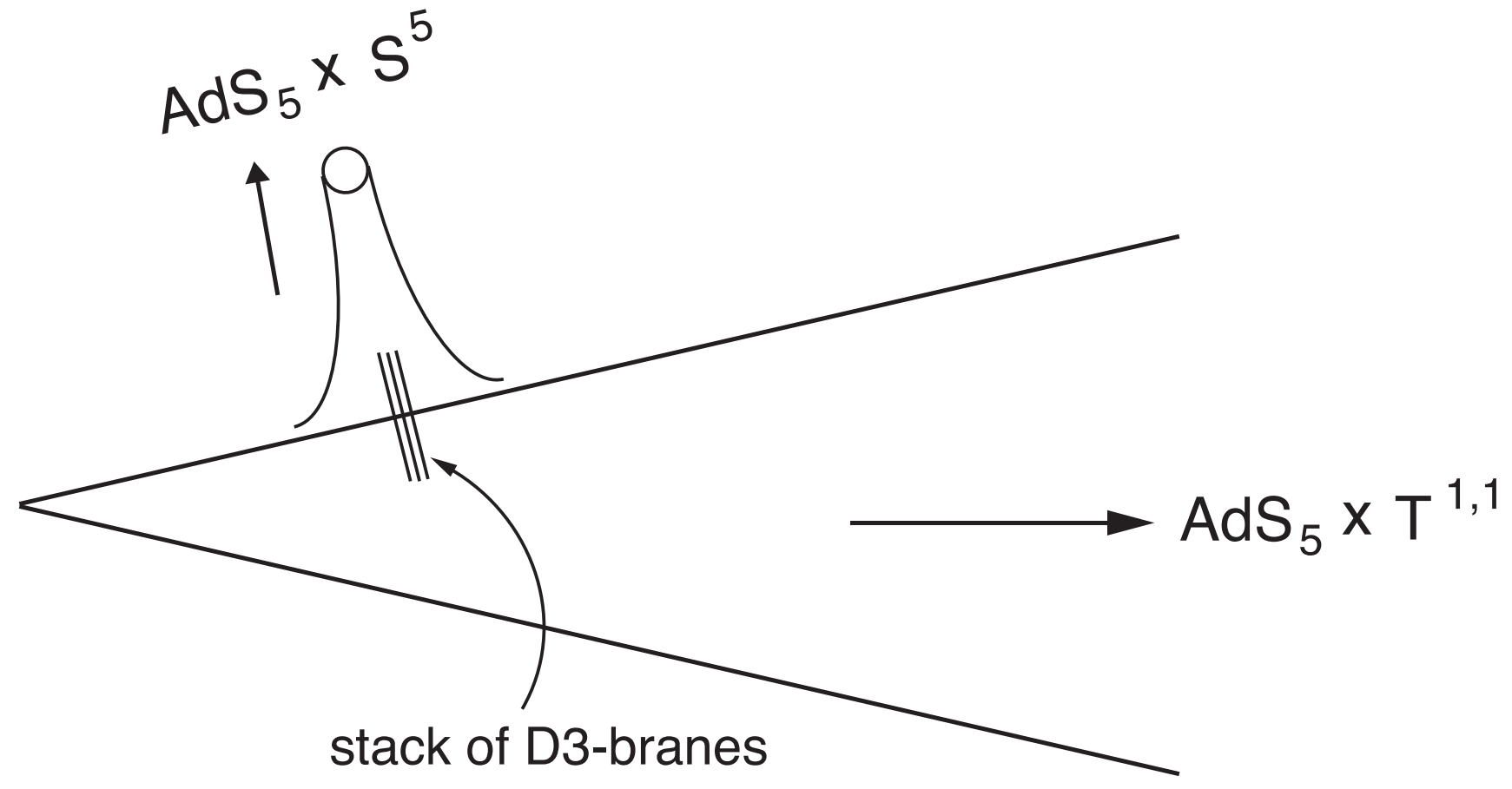

Figure 2.1: A stack of D3-branes warping the singular conifold

(2.2.12) as

$$\begin{aligned} \mathit{\Delta} H = \mathit{\Delta}_r H + \frac{\mathit{\Delta}_Z}{r^2} H &= -\frac{\mathcal{C}}{\sqrt{g}} \delta(r - r_0) \mathit{\Pi}_i \delta^5(Z_i - Z_{0i}) \\ &\equiv -\frac{\mathcal{C}}{\sqrt{g_r}} \delta(r - r_0) \delta_A(Z - Z_0) \end{aligned} \tag{2.3.1}$$

where $\mathit{\Delta}_r = \frac{1}{\sqrt{g}} \partial_r \left( \sqrt{g}\, \partial_r \right)$ is the radial Laplacian, $\mathit{\Delta}_Z$ the remaining angular laplacian. In the second line, $g_r$ is defined to have the radial dependence in $g$ and the angular delta function $\delta_A(Z - Z_0)$ is defined by absorbing the angular factor $\sqrt{g_5} = \sqrt{g/g_r}$. In this section, we have $\sqrt{g} = \frac{1}{108} r^5 \sin\theta_1 \sin\theta_2$ and we take $\sqrt{g_r} = r^5$.

The eigenfunctions $Y_I(Z)$ of the angular laplacian on $T^{1,1}$ can be classified by a set $I$ of symmetry charges since $T^{1,1}$ is a coset space [47, 48]. The eigenfunctions $Y_I$ are constructed explicitly in the appendix, including using the $a_i, b_j$ coordinates which will facilitate the comparison with the gauge theory below. If we normalize these angular eigenfunctions as,

$$\int Y^*_{I_0}(Z) Y_I(Z) \ \sqrt{g_5}\, d^5\varphi_i \ = \ \delta_{I_0, I} \tag{2.3.2}$$

we then have the complementary result,

$$\sum_I Y_I^*(Z_0)Y_I(Z) = \frac{1}{\sqrt{g_5}}\,\delta(\varphi_i - \varphi_{0i}) \equiv \delta_A(Z - Z_0). \tag{2.3.3}$$

We expand the $\delta_A(Z - Z_0)$ in (2.3.1) using (2.3.3) and see that the Green's function can be expanded as,

$$H \;=\; \sum_I \, H_I(r, r_0)\; Y_I(Z)\; Y_I^*(Z_0) \tag{2.3.4}$$

which reduces (2.3.1) to the radial equation,

$$\frac{1}{r^5}\frac{\partial}{\partial r}\left(r^5 \frac{\partial}{\partial r} H_I\right) - \frac{E_I}{r^2} H_I = -\frac{\mathcal{C}}{r^5}\delta(r - r_0) \tag{2.3.5}$$

where $\Delta_Z Y_I(Z) = -E_I Y_I(Z)$ (see appendix A for details of $E_I$.)

As is easily seen, the solutions to this equation away from $r = r_0$ are

$$H_I \;=\; A_\pm\, r^{c_\pm}, \text{ where } \;\; c_\pm = -2 \pm \sqrt{E_I + 4}.$$

The constants $A_\pm$ are uniquely determined integrating (2.3.5) past $r_0$. This determines $H_I$ and we put it all together to get the solution to (2.3.1), the Green's function on the singular conifold

$$H(r, Z; r_0, Z_0) = \sum_I \frac{\mathcal{C}}{2\sqrt{E_I + 4}}\; Y_I^*(Z_0) Y_I(Z) \;\times \begin{cases} \dfrac{1}{r_0^4}\left(\dfrac{r}{r_0}\right)^{c_I} & r \le r_0 \\[2ex] \dfrac{1}{r^4}\left(\dfrac{r_0}{r}\right)^{c_I} & r \ge r_0 \;, \end{cases} \tag{2.3.6}$$

where $c_I = c_+$. The term with $E_I = 0$ gives $L^4/r^4$ where

$$L^4 = \frac{\mathcal{C}}{4\mathbf{Vol}(T^{1,1})} = \frac{27\pi g_s N (\alpha')^2}{4}\ . \tag{2.3.7}$$

Since $E_I = 6(l_1(l_1+1) + l_2(l_2+1) - R^2/8)$, there are $(2l_1+1) \times (2l_2+1)$ terms with the same $E_I$ and hence powers of $r$ and factors. Also note that when $l_1 = l_2 = \pm\frac{R}{2}$, $c_I$ is a rational number and these are related to (anti) chiral superfields in the gauge theory.

We can argue that the geometry near the stack (at $r_0, Z_0$) is actually a long $AdS_5 \times S^5$ throat. We observe that $H$ must behave as $L^4/y^4$ near the stack (where $y$ is the distance between $(r, Z)$ and $(r_0, Z_0)$) since it is the solution of the Green's function and locally, the manifold looks flat and is 6 dimensional. This leads to the usual $AdS_5 \times S^5$ throat. We show this explicitly in Appendix B. The complete metric thus describes holographic RG flow from $AdS_5 \times T^{1,1}$ geometry in the UV to $AdS_5 \times S^5$ in the IR. Note, however that this background has a conifold singularity at $r = 0$.

## Gauge theory operators

Let the stack of branes be placed at a point $a_i, b_j$ on the conifold. Then consider assigning the VEVS, $A_i = a_i^* 1_{N\times N}, B_j = b_j^* 1_{N\times N}$, i.e the prescription

$$Z_0 = \begin{pmatrix} a_1 b_1 & a_1 b_2 \\ a_2 b_1 & a_2 b_2 \end{pmatrix} \quad \Longleftrightarrow \quad \begin{array}{c} A_1 = a_1^* 1_{N\times N}, \quad A_2 = a_2^* 1_{N\times N}, \\ B_1 = b_1^* 1_{N\times N}, \quad B_2 = b_2^* 1_{N\times N}. \end{array} \tag{2.3.8}$$

In the appendix, we construct operators $\mathcal{O}_I$ transforming with the symmetry charges $I$. From the similar construction of the operator $\mathcal{O}_I$ and $Y_I(Z)$ (compare (2.A.9) and (2.A.11)), this automatically leads to a VEV proportional to $Y_I^*(Z_0)$ for the operator $\mathcal{O}_I$.

Meanwhile, the linearized perturbations of the metric are determined by binomially expanding $\sqrt{H}$ in (2.2.10) and considering terms linear in $Y_I(Z)$. These are easily seen to be of the form $Y_I^*(Z_0) Y_I(Z) \left(\frac{r_0}{r}\right)^{c_I}$. From its form and symmetry properties, we conclude that it is the dual to the above VEV,

$$Y_I^*(Z_0)Y_I(Z) \times \left(\frac{r_0}{r}\right)^{c_I} \quad \Longleftrightarrow \quad \langle \mathcal{O}_I \rangle \propto Y_I^*(Z_0)\, r_0^{c_I}. \tag{2.3.9}$$

This is the sought relation between normalizable perturbations and operator VEV's. For a general position of the stack $(r_0, Z_0)$, all $Y_I^*(Z_0)$ are non-vanishing. Being a coset space, we can use the symmetry of $T^{1,1}$, to set the D3-branes to lie at any specific point without loss of generality. For example, consider the choice

$$Z_0 = \begin{pmatrix} a_1b_1 & a_1b_2 \\ a_2b_1 & a_2b_2 \end{pmatrix} = \begin{pmatrix} 1 & 0 \\ 0 & 0 \end{pmatrix} \qquad \Rightarrow \qquad a_1 = b_1 = 1,\ a_2 = b_2 = 0. \tag{2.3.10}$$

Using (2.A.9) and (2.A.8) for $Y_I$, we find that $Y_I(Z_0) = 0$ unless $m_1 = m_2 = R/2$ and for these non-vanishing $Y_I$ we get,

$$Y_I(Z_0) \sim a_1^{l_1+\frac{R}{2}} \bar{a}_1^{l_1-\frac{R}{2}} b_1^{l_2+\frac{R}{2}} \bar{b}_1^{l_2-\frac{R}{2}} \tag{2.3.11}$$

If we give the VEVs $A_1 = B_1 = 1_{N\times N}, A_2 = B_2 = 0$, we get $\langle TrA_1B_1\rangle \neq 0$ and all other $\langle TrA_iB_j\rangle = 0$. In fact, by this assignment, the only gauge invariant operators with non-zero vevs are the $\mathcal{O}_I$ with $m_1 = m_2 = R/2$. These are precisely the operators dual to fluctuations $Y_I(Z)$ that have non-zero coefficient $Y_I^*(Z_0)$ as was seen in (2.3.11).

The physical dimension of this operator (at the UV fixed point) is read off as $c_I$ from the metric fluctuation - a supergravity prediction for strongly coupled gauge theory. (Above, $r_0$ serves as a scale for dimensional consistency.) In [13], the (anti) chiral operators were discussed $(l_1 = l_2 = \pm\frac{R}{2})$ . These have rational dimensions but as we see here, for any position of the stack of D3-branes, other operators (with generically irrational dimensions) also get vevs. For example, the dimension of the simplest non-chiral operator $(I \equiv l_1 = 1, l_2 = 0, R = 0)$ is 2 but when $I \equiv l_1 = 2, l_2 = 0, R = 0,$ $\mathcal{O}_I$ has dimension $2(\sqrt{10}-1)$. This interesting observation about highly non-trivial scaling dimensions in strongly coupled gauge theory was first made in [47].

When operators $A_i, B_j$ get vevs as in (2.3.8), the $SU(N) \times SU(N)$ gauge group is broken down to the diagonal $SU(N)$. The bifundamental fields $A, B$ now become adjoint fields. With one linear combination of fields having a VEV, we can expand the superpotential $W \sim \mathrm{Tr}\det A_i B_j = \mathrm{Tr}(A_1 B_1 A_2 B_2 - A_1 B_2 A_2 B_1)$ of the $SU(N) \times SU(N)$ theory to find that it is of the form $\mathrm{Tr}(X[Y, Z])$ in the remaining adjoint fields [18]. This is exactly $\mathcal{N} = 4$ $SU(N)$ super Yang-Mills, now obtained through symmetry breaking in the conifold theory. This corresponds to the $AdS_5 \times S^5$ throat we found on the gravity side near the source at $r_0, Z_0$.

Thus we have established a gauge theory RG flow from $\mathcal{N} = 1$ $SU(N) \times SU(N)$ theory in the UV to $\mathcal{N} = 4$ $SU(N)$ theory in the IR. The corresponding gravity dual was constructed and found to be asymptotically $AdS_5 \times T^{1,1}$ (the UV fixed point) but developing a $AdS_5 \times S^5$ throat at the other end of the geometry (the IR fixed point). The simple example is generalized to the resolved conifold in the next section.

## 2.4 Flows on the Resolved Conifold

In this section we use similar methods to construct the Green's function on the resolved conifold and corresponding warped solutions due to a localized stack of D3-branes. We will work out explicitly the $SU(2) \times U(1) \times U(1)$ symmetric RG flow corresponding to a stack of D3-branes localized on the finite $S^2$ at $r = 0$. Such a solution is dual to giving a VEV to just one bi-fundamental field, e.g. $B_2$, which Higgses the $\mathcal{N} = 1$ $SU(N) \times SU(N)$ gauge theory theory to the $\mathcal{N} = 4$ $SU(N)$ SYM. We also show how the naked singularity found in [37] is removed through the localization of the D3-branes.

The supergravity metric is of the form (2.2.14). The stack could be placed at non-zero $r$ but in this case, the symmetry breaking pattern is similar in character to the singular case discussed above. The essence of what is new to the resolved conifold is best captured with the stack placed at a point on the blown up $S^2$ at $r = 0$; this breaks the $SU(2)$ symmetry rotating $(\theta_2, \phi_2)$ down to a $U(1)$. The branes also preserve the $SU(2)$ symmetry rotating $(\theta_1, \phi_1)$ as well as the $U(1)$ symmetry corresponding to the shift of $\psi$. On the other hand, the $U(1)_B$ symmetry is broken because the resolved conifold has no non-trivial three-cycles [13]. Thus the warped resolved conifold background has unbroken $SU(2) \times U(1) \times U(1)$ symmetry.

To match this with the gauge theory, we first recall that in the absence of VEV's we have $SU(2) \times SU(2) \times U(1)_R \times U(1)_B$ where the $SU(2)$'s act on $A_i, B_j$ respectively, the $U(1)_R$ is the R-charge ($R_A = R_B = 1/2$) and $U(1)_B$ is the baryonic symmetry, $A \to e^{i\theta} A, B \to e^{-i\theta} B$. As noted above, the VEV $B_2 = u 1_{N\times N}, B_1 = A_i = 0$ corresponds to placing the branes at a point on the blown-up 2-sphere. This clearly leaves one of the $SU(2)$ factors unbroken. While $U(1)_R$ and $U(1)_B$ are both broken by the baryonic operator $\det B_2$, their certain $U(1)$ linear combination remains unbroken. Similarly, a combination of $U(1)_B$ and the $U(1)$ subgroup of the other $SU(2)$, that rotates the $B_i$ by phases, remains unbroken. Thus we again have $SU(2) \times U(1) \times U(1)$ as the unbroken symmetry, consistent with the warped resolved conifold solution. Since the baryon operator $\det B_2$ acquires a VEV while no chiral mesonic operators do (because $A_1 = A_2 = 0$), the solutions found in this section are dual to a "baryonic branch" of the CFT (see [44] for a discussion of such branches).

## Solving for the warp factor

Since the resolution of the conifold preserves the $SU(2)_L \times SU(2)_R \times U(1)_\psi$ symmetry (where $U(1)_\psi$ shifts $\psi$), the equation for Green's function $H$ looks analogous to (2.3.1) for the resolved conifold,

$$\frac{1}{r^3(r^2+6u^2)}\frac{\partial}{\partial r}\left(r^3(r^2+6u^2)\kappa(r)\frac{\partial}{\partial r}H\right) \quad + \quad \mathbf{A}H = \\ -\frac{\mathcal{C}}{r^3(r^2+6u^2)} \quad \times \quad \delta(r-r_0)\;\delta_{T^{1,1}}(Z-Z_0) \quad (2.4.1)$$

where

$$\mathbf{A}H = 6\frac{\mathit{\Delta}_1}{r^2}H + 6\frac{\mathit{\Delta}_2}{r^2+6u^2}H + 9\frac{\mathit{\Delta}_R}{\kappa(r)r^2}H \quad (2.4.2)$$

and $\mathit{\Delta}_i$ , $\mathit{\Delta}_R$ are defined in the appendix. ($\mathit{\Delta}_i$ are $S^3$ laplacians and $\mathit{\Delta}_R = \partial_\psi^2$. Note that $6\mathit{\Delta}_1 + 6\mathit{\Delta}_2 + 9\mathit{\Delta}_R = \mathit{\Delta}_{T^{1,1}}$).

This form of the $\mathbf{A}$ is fortuitous and allows us to use the $Y_I$ from the singular conifold, since $Y_I$ are eigenfunctions of each of the three pieces of $\mathbf{A}$ above. We could solve it for general $r_0$, but $r_0 = 0$ is a particularly simple case that is of primary interest in this chapter.

Since (2.4.1) involves the same $\delta_{T^{1,1}}(Z-Z_0)$ as the singular case, we can expand $H$ again in terms of the angular and radial functions as $H = \sum_I H_I(r,r_0)Y_I(Z)Y_I^*(Z_0)$ to find the radial equation,

$$-\frac{1}{r^3(r^2+6u^2)} \quad \frac{\partial}{\partial r}\left(r^3(r^2+6u^2)\kappa(r)\frac{\partial}{\partial r}H_I\right) \\ + \quad \left(\frac{6(l_1(l_1+1)-R^2/4)}{r^2} + \frac{6(l_2(l_2+1)-R^2/4)}{r^2+6u^2} + \frac{9R^2/4}{\kappa(r)r^2}\right)H_I \\ = \frac{\mathcal{C}}{r^3(r^2+6u^2)}\delta(r-r_0). \quad (2.4.3)$$

This equation can be solved for $H_I(r)$ exactly in terms of some special functions. If we place the stack at $r_0 = 0$, i.e at location $(\theta_0, \phi_0)$ on the blown up $S^2$, then an additional simplification occurs. The warp factor $H$ must be a singlet under the $SU(2) \times U(1)_\psi$ that rotates $(\theta_1, \phi_1)$ and $\psi$ since these have shrunk at the point where the branes are placed. Hence we only need to solve this equation for $l_1 = R = 0$, $l_2 = l$.

The two independent solutions (with convenient normalization) to the homogeneous equation in this case, in terms of the hypergeometric function ${}_2F_1$, are

$$\begin{aligned} H_l^A(r) &= \frac{2}{9u^2}\frac{C_\beta}{r^{2+2\beta}}\ {}_2F_1\left(\beta, 1+\beta; 1+2\beta; -\frac{9u^2}{r^2}\right) \\ H_l^B(r) &\sim {}_2F_1\left(1-\beta, 1+\beta; 2; -\frac{r^2}{9u^2}\right) \end{aligned} \tag{2.4.4}$$

where

$$C_\beta = \frac{(3u)^{2\beta}\Gamma(1+\beta)^2}{\Gamma(1+2\beta)}\ , \qquad \beta = \sqrt{1+(3/2)l(l+1)}\ . \tag{2.4.5}$$

These two solutions have the following asymptotic behaviors,

$$\frac{2}{9u^2r^2} + \frac{4\beta^2}{81u^4}\ln r + \mathcal{O}(1) \quad \overset{0\leftarrow r}{\longleftarrow} \quad H_l^A(r) \quad \overset{r\to\infty}{\longrightarrow} \quad \frac{2C_\beta}{9u^2r^{2+2\beta}} \tag{2.4.6}$$

$$\mathcal{O}(1) \quad \overset{0\leftarrow r}{\longleftarrow} \quad H_l^B(r) \quad \overset{r\to\infty}{\longrightarrow} \quad \mathcal{O}\left(r^{-2+2\beta}\right) \tag{2.4.7}$$

To find the solution to (2.4.3) with the $\delta(r-r_0)$ on the RHS, we need to match the two solutions at $r=r_0$ as well as satisfy the condition on derivatives obtained by integrating past $r_0$. Since we are interested in normalizable modes, we use $H_l^A(r)$ for $r>r_0$ and $H_l^B(r)$ for $r<r_0$. Finally, we take $r_0=\epsilon$ and take the limit $\epsilon\to 0$ (since the stack of branes is on the finite $S^2$). We find simply that $H_l(r) = \mathcal{C}H_l^A(r)$ due to the normalization chosen earlier in (2.4.4). Putting it all together, we find,

$$H(r, Z; r_0=0, Z_0) = \mathcal{C}\sum_I Y_I^*(Z_0)H_I^A(r)Y_I(Z) \tag{2.4.8}$$

where only the $l_1=0, R=0$ harmonics contribute since the stack leaves $SU(2)\times U(1)\times U(1)$ symmetry unbroken. In this situation, the $Y_I$ wavefunctions simplify to the usual $S^2$ spherical harmonics $\sqrt{\frac{4\pi}{\mathbf{Vol}(T^{1,1})}}\,Y_{l,m}$.

Let us take $b_i$ to describe the finite $S^2(\theta_2, \phi_2)$ while $a_j$ are associated with the $S^2$ that shrinks to a point. As reviewed in Section 2.2, the resolved conifold can be described with $a, b$ variables governed by the constraint (2.2.7), where $u$ is the measure of resolution, the radius of the finite $S^2$. The position of the branes on the finite sphere can be parametrized as $b_1 = u \sin\frac{\theta_0}{2} e^{-i\phi_0/2}, b_2 = u\cos\frac{\theta_0}{2} e^{i\phi_0/2}$ and $a_1 = a_2 = 0$ (since the branes do not break the $SU(2)$ symmetry rotating the $a$'s). Then,

$$H(r, Z; r_0 = 0, Z_0 = (\theta_0, \phi_0)) = 4\pi L^4 \sum_{l,m} H_l^A(r)\, Y^*_{l,m}(\theta_0, \phi_0) Y_{l,m}(\theta_2, \phi_2). \qquad (2.4.9)$$

Without a loss of generality, we can place the stack of D3-branes at the north pole ($\theta_0 = 0$) of the 2-sphere. Then (2.4.9) simplifies further: only $m = 0$ harmonics contribute and we get the explicit expression for the warp factor which is one of our main results,

$$H(r, \theta_2) = L^4 \sum_{l=0}^{\infty} (2l+1) H_l^A(r) P_l(\cos\theta_2). \qquad (2.4.10)$$

Now the two unbroken $U(1)$ symmetries are manifest as shifts of $\phi_2$ and $\psi$.

The 'smeared' singular solution found in [37] corresponds to retaining only the $l = 0$ term in this sum. Indeed, we find that

$$H_0^A(r) = \frac{2C_1}{9u^2 r^4}\ {}_2F_1\left(1, 2; 3; -\frac{9u^2}{r^2}\right) = \frac{2}{9u^2r^2} - \frac{2}{81u^4}\log\left(1 + \frac{9u^2}{r^2}\right) \qquad (2.4.11)$$

in agreement with [37]. Fortunately, if we consider the full sum over modes appearing in (2.4.12), the geometry is no longer singular. The leading term in the warp factor (2.4.10) at small $r$ is

$$\frac{2L^4}{9u^2r^2} \sum_{l=0}^{\infty} (2l+1) P_l(\cos\theta_2) = \frac{4L^4}{9u^2r^2}\delta(1 - \cos\theta_2) \qquad (2.4.12)$$

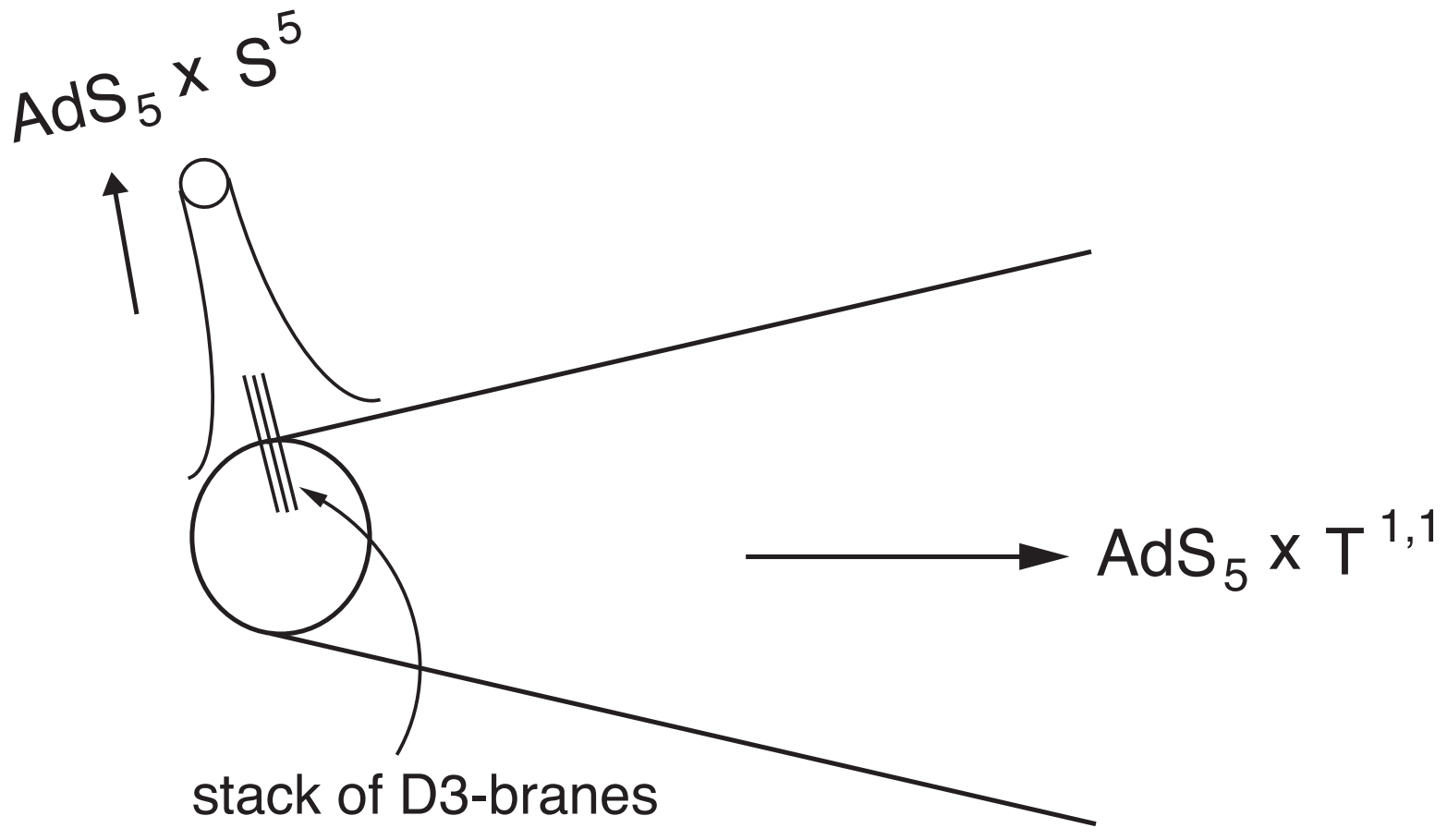

Figure 2.2: A stack of D3-branes warping the resolved conifold

This shows that away from the north pole the $1/r^2$ divergence of the warp factor cancels. Similarly, after summing over $l$ the term $\sim \ln r$ cancels away from the north pole. This implies that the warp factor is finite at $r = 0$ away from the north pole. However, at the north pole it diverges as expected. Indeed, since the branes are now localized at a smooth point on the 6-manifold (all points on the resolved conifold are smooth), very near the source $H$ must again be of the form $L^4/y^4$ where $y$ is the distance from the source. This is shown explicitly in Appendix B. Writing the local metric in the form $dy^2 + y^2 d\Omega^2_{S^5}$ near the source, we get the $AdS_5 \times S^5$ throat, avoiding the singularity found in [37].

## Gauge theory operators

With the branes placed at the point $b_1 = u \sin\frac{\theta_0}{2} e^{-i\frac{\phi_0}{2}}, b_2 = u \cos\frac{\theta_0}{2} e^{i\frac{\phi_0}{2}}$ , $a_1 = a_2 = 0$ on the finite $S^2$, consider the assignment of VEVs, $B_1 = u \sin\frac{\theta_0}{2} e^{i\frac{\phi_0}{2}} 1_{N\times N}, B_2 = u \cos\frac{\theta_0}{2} e^{-i\frac{\phi_0}{2}} 1_{N\times N}, A_1 = A_2 = 0$.

The linearized fluctuations compared to the leading term $1/r^4$ ($l_2 = 0$) are of the form,

$$Y_I^*(Z_0)Y_I(Z)r^4H_I(r) \to Y_I^*(Z_0)Y_I(Z)(\frac{1}{r})^{c_I} \quad (r \gg a) \tag{2.4.13}$$

where as earlier, $c_I = 2\sqrt{1+(3/2)l_2(l_2+1)} - 2$.

With this assignment of VEVs above, operators $\mathcal{O}_I$ with $l_1 = R = 0$ acquire VEVs. In the notation $|l_2, m_2; R>$ of Appendix A,

$$< \mathcal{O}_I > \; = \; < Tr|l_2, m_2; 0\rangle_B > \neq 0. \tag{2.4.14}$$

For example, when $l_2 = 1, m_2 = 0$ above, $\langle \mathcal{O}_I \rangle = \langle Tr B_1\bar{B}_1 - B_2\bar{B}_2 \rangle = u^2(\sin^2\frac{\theta_0}{2} - \cos^2\frac{\theta_0}{2})$. By construction of $\mathcal{O}_I$ and $Y_I$, it is clear that $< \mathcal{O}_I > \sim Y_I^*(Z_0)$ and is dual to the metric fluctuation above. We can read off the dimensions of these operators as $c_I$ from the large $r$ behavior of the fluctuation (2.4.13). For the $l_2 = 1$ operator as above, the exact dimension is 2 (the classical value) because the operator is a superpartner of a conserved current [49]. Similarly, the dimension 2 of $\mathcal{U}$ is protected against quantum corrections because of its relation to a conserved baryonic current [13]. When one expands $H_I(r)$ at large $r$, one finds sub-leading terms in addition to $1/r^{c_I}$ shown above. These terms, which do not appear for the singular conifold, increase in powers of $1/r^2$ and hence describe a series of operators with the same symmetry $I$ but dimension increasing in steps of 2 from $c_I$. These modes appear to correspond to VEV's for the operators $\mathrm{Tr}\mathcal{O}_I\mathcal{U}^n$. It would be interesting to investigate such operators and their dimensions further.

Hence we have an infinite series of operators that get VEV's in the gauge theory dual to the warped resolved conifold. These are in addition to the basic operator $\mathcal{U}$ which gets a VEV due to the asymptotics of the unwarped resolved conifold metric itself [13]. The operator $\mathcal{U}$ would get the same VEV of $u^2$ for any position of the brane on the $S^2$ while the VEV's for the infinite series of operators $\mathcal{O}_I$ depend on the position. We also note that $\mathcal{U} = u^2$ is the gauge dual of the constraint $|b_1|^2 + |b_2|^2 - |a_1|^2 - |a_2|^2 = u^2$ defining the resolved conifold in Section 2.2.

Lastly, we verify that the gauge theory does flow in the infrared to $\mathcal{N} = 4$ $SU(N)$ SYM. Without loss of generality, we can take the stack of branes to lie on the north pole of the finite sphere ($B_2 = u1_{N\times N}, B_1 = 0$). As in the singular case, $B_2 = u1_{N\times N}$ breaks the $SU(N) \times SU(N)$ gauge group down to $SU(N)$, all the chiral fields now transforming in the adjoint of this diagonal group. Consider the $\mathcal{N} = 1$ superpotential $W \sim \mathrm{Tr}\det A_i B_j = Tr(A_1B_1A_2B_2 - A_1B_2A_2B_1)$. When $B_2 \propto 1_{N\times N}$, the superpotential reduces to the $\mathcal{N} = 4$ form,

$$W = \lambda \mathrm{Tr}(A_1B_1A_2 - A_1A_2B_1) = \lambda \mathrm{Tr}(A_1[B_1, A_2]). \tag{2.4.15}$$

This confirms that the gauge theory flows to the $\mathcal{N} = 4$ $SU(N)$ SYM theory in the infrared.

## Baryonic Condensates and Euclidean D3-branes

Here we present a calculation of the baryonic VEV using the dual string theory on the warped resolved conifold background.[3] A similar question was addressed for the cascading theories on the baryonic branch where the baryonic condensates are related to the action of a Euclidean D5-brane wrapping the deformed conifold [50, 51]. In this section we present an analogous construction for the warped resolved conifolds, which are asymptotic to $AdS_5 \times T^{1,1}$.

The objects in $AdS_5 \times T^{1,1}$ that are dual to baryonic operators are D3-branes wrapping 3-cycles in $T^{1,1}$ [52]. Classically, the 3-cycles dual to the baryons made out of the $B$'s are located at fixed $\theta_2$ and $\phi_2$ (quantum mechanically, one has to carry out collective coordinate quantization and finds wave functions of spin $N/2$ on the 2-

[3] We are indebted to E. Witten for his suggestion that led to the calculation presented in this section.

sphere). To calculate VEV's of such baryonic operators, we need to consider Euclidean D3-branes which at large $r$ wrap a 3-cycle at fixed $\theta_2$ and $\phi_2$. In fact, the symmetries of the calculation suggest that the smooth 4-cycle wrapped by the Euclidean D3-brane is located at fixed $\theta_2$ and $\phi_2$, and spans the $r$, $\theta_1$, $\phi_1$ and $\psi$ directions. In other words, the Euclidean D3-brane wraps the $\mathbf{R}^4$ fiber of the $\mathbf{R}^4$ bundle over $S^2$ (recall that the resolved conifold is such a bundle).

The action of the D3-brane will be integrated up to a radial cut-off $r_c$, and we identify $e^{-S(r_c)}$ with the classical field dual to the baryonic operator. The Born-Infeld action is

$$S_{BI} = T_3 \int d^4\xi \sqrt{g} \ , \tag{2.4.16}$$

where $g_{\mu\nu}$ is the metric induced on the D3 world volume. We find

$$S_{BI} = \frac{3N}{4L^4} \int_0^{r_c} dr r^3 H(r,\theta_2) = \frac{3N}{4} \int_0^{r_c} dr r^3 \sum_{l=0}^{\infty} (2l+1) H_l^A(r) P_l(\cos\theta_2) \ . \tag{2.4.17}$$

The $l = 0$ term (2.4.11) needs to be evaluated separately since it contains a logarithmic divergence:[4]

$$\int_0^{r_c} dr r^3 H_0^A(r) = \frac{1}{4} + \frac{1}{2} \ln\left(1 + \frac{r_c^2}{9u^2}\right) \ . \tag{2.4.18}$$

For the $l > 0$ terms the cut-off may be removed and we find a nice cancellation involving the normalization (2.4.5):

$$\int_0^{\infty} dr r^3 H_l^A(r) = \frac{2}{3l(l+1)} \ . \tag{2.4.19}$$

[4]A careful holographic renormalization of divergences for D-brane actions was considered in [53]. We leave a similar construction in the present situation for future work.

Therefore,

$$\int_0^\infty drr^3 \sum_{l=1}^{\infty}(2l+1)H_l^A(r)P_l(\cos\theta_2) = \frac{2}{3}\sum_{l=1}^{\infty}\frac{2l+1}{l(l+1)}P_l(\cos\theta_2)$$
$$= \frac{2}{3}(-1-2\ln[\sin(\theta_2/2)]) \ . \quad (2.4.20)$$

This expression is recognized as the Green's function on a sphere. Combining the results, and taking $r_c \gg u$, we find

$$e^{-S_{BI}} = \left(\frac{3e^{5/12}u}{r_c}\right)^{3N/4} \sin^N(\theta_2/2) \ . \quad (2.4.21)$$

In [52] it was argued that the wave functions of $\theta_2, \phi_2$, which arise though the collective coordinate quantization of the D3-branes wrapped over the 3-cycle $(\psi, \theta_1, \phi_1)$, correspond to eigenstates of a charged particle on $S^2$ in the presence of a charge $N$ magnetic monopole. Taking the gauge potential $A_\phi = N(1+\cos\theta)/2,\ A_\theta = 0$ we find that the ground state wave function $\sim \sin^N(\theta_2/2)$ carries the $J = N/2, m = -N/2$ quantum numbers.[5] These are the $SU(2)$ quantum numbers of $\det B_2$. Therefore, the angular dependence of $e^{-S}$ identifies $\det B_2$ as the only operator that acquires a VEV, in agreement with the gauge theory.

The power of $r_c$ indicates that the operator dimension is $\Delta = 3N/4$, which again corresponds to the baryonic operators. The VEV depends on the parameter $u$ as $\sim u^{3N/4}$. This is not the same as the classical scaling that would give $\det B_2 = u^N$. The classical scaling is not obeyed because this is an interacting theory where the baryonic operator acquires an anomalous dimension.

The string theoretic arguments presented in this section provide nice consistency checks on the picture developed in this chapter, and also confirm that the Eucldean 3-brane can be used to calculate the baryonic condensate.

[5] In a different gauge this wave function would acquire a phase. In the string calculation it comes from the purely imaginary Chern-Simons term in the Euclidean D3-brane action.

## 2.5 $B$-field on the Resolved Conifold

Our warped resolved conifold solution written with no NS-NS $B$ field corresponds to a special isolated point in the space of gauge coupling constants. From [28, 27], the relation between coupling constants and the SUGRA background is known to be,

$$\frac{4\pi^2}{g_1^2} + \frac{4\pi^2}{g_2^2} = \frac{\pi}{g_s e^{\Phi}} \tag{2.5.1}$$

$$\frac{4\pi^2}{g_1^2} - \frac{4\pi^2}{g_2^2} = \frac{1}{g_s e^{\Phi}} \left( \frac{1}{2\pi\alpha'} \int_{S^2} B_2 - \pi \right) \tag{2.5.2}$$

where $\Phi$ is the dilaton. Hence when $B = 0$, $g_1$ is infinite.

Since the resolved conifold has a topologically non-trivial two cycle and we could turn on a $B$-field proportional to the volume of this cycle [18]:

$$B_2 \sim \sin\theta_2 d\theta_2 \wedge d\phi_2. \tag{2.5.3}$$

Such a $B$-field would have no flux, $H = dB_2 = 0$, while still being non-trivial ($\int_{S^2} B_2 \neq 0$). Since there is no flux, the rest of the SUGRA solution remains untouched and we have a description of the gauge theory at generic coupling.

When the resolved conifold is warped by a stack of branes as we have in this chapter, the argument of [13] continues to hold. A new $AdS_5 \times S^5$ throat branches out at the point where the stack is placed. This modifies the topology by introducing a new non-trivial 5-cycle. However, the earlier two-cycle is untouched and does not become topologically trivial. One way to see this is to note that the new 5-cycle was the trivial cycle that could shrink to a point at the place where the stack is placed. But the finite two cycle of the resolution is topologically distinct from the cycles that shrink here and hence it obviously survives the creation of a new 5-cycle. Hence the fluxless NS-NS $B_2$ field above that naturally exists on such a space can be used to describe the gauge theory at generic coupling.

Had we considered a stack of D3-branes on the deformed conifold, the situation would have been quite different, as emphasized in [13]. In that case, a fluxless $B_2$ field cannot be turned on; therefore, there is no simple $SU(N) \times SU(N)$ gauge theory interpretation for backgrounds of the form (2.2.14) with $ds_6^2$ being the deformed conifold metric, and $H$ the Green's function of the scalar Laplacian on it. Of course, the deformed conifold with a different warp factor created by self-dual 3-form fluxes corresponds to the cascading $SU(kM) \times SU(k(M+1))$ gauge theory [25, 54].

## 2.6 Conclusions

We have constructed the SUGRA duals of the $SU(N) \times SU(N)$ conifold gauge theory with certain VEV's for the bi-fundamental fields. As discussed in [13], the different vacua of the theory correspond to D3-branes localized on the singular as well as resolved conifold. Vacua with $\mathcal{U} = 0$ describe the singular conifold with a localized stack of D3-branes; vacua with $\mathcal{U} \neq 0$ instead describe D3-branes localized on the conifold resolved through blowing up of a 2-sphere. We constructed explicit SUGRA solutions corresponding to these vacua. In particular, the solution corresponding to giving a VEV to only one of the fields in the gauge theory, $B_2 = u 1_{N\times N}$, while keeping $A_i = B_1 = 0$, corresponds to a certain warped resolved conifold. In this case the warp factor is given by the Green's function with a source at a point on the blown-up 2-sphere at $r = 0$. The baryonic operator $\det B_2$ gets a VEV while no chiral mesonic operator does. This background is thus dual to a non-mesonic, or baryonic, branch of the CFT. To confirm this, we used the action of a Euclidean D3-brane wrapping a 4-cycle in the resolved conifold, to calculate the VEV of the baryonic operator.

The explicit SUGRA solution was determined and found to asymptote to $AdS_5 \times T^{1,1}$ in the large $r$ region. When one approaches the blown-up 2-sphere, the warp factor causes an $AdS_5 \times S^5$ throat to branch off at a point on the 2-sphere. Our calculation makes use of the explicit metric on the resolved conifold found in [37]. Our warped solution, with a localized stack of D3-branes, is completely non-singular in contrast to the smeared-brane solution obtained in [37].

The Green's functions on the singular and resolved conifolds were determined in detail for the purpose of constructing the SUGRA solutions. These Green's functions are also useful in brane models of inflation where they play a role in computing the one-loop corrections to non-perturbative superpotentials (see [42, 43] for such an application). The Green's functions were written using harmonics on $T^{1,1}$ in the $a, b$ variables on the conifold (instead of the usual angular variables or the $z, w$ co-ordinates). This facilitated the comparison with the explicit gauge theory operators that acquire VEVs.

We see a number of possible extensions of our work. One of them deals with the AdS/CFT dualities based on the Sasaki-Einstein spaces $Y^{p,q}$ [55–57]. Calculations similar to ours can be performed for the resolved cones over $Y^{p,q}$ manifolds (for recent work, see [58,59]). Harmonics in convenient co-ordinates similar to the ones constructed here could perhaps be constructed using the bifundamental fields of these quiver gauge theories. Again, the basic non-singular solutions will correspond to a stack of branes at a point, and it would be interesting to solve for the corresponding warp factors. One could also study the resolved cone versions of the solutions found in [60], which correspond to cascading gauge theories. It is also possible to consider Calabi-Yau cones with blown-up 4-cycles [61–63, 57, 64]. In [44], the gauge theory operator whose VEV corresponds to blown-up 4-cycles of certain cones was identified. Perhaps the Green's function could be determined for a stack of branes on such 4-cycles, giving the non-singular SUGRA dual of corresponding non-mesonic branches in the gauge theory.

## 2.A Eigenfunctions of the Scalar Laplacian on $T^{1,1}$

The main emphasis of this Appendix is on writing the harmonics on $T^{1,1}$ in a way that makes the connection with the dual gauge theory operators most transparent. The eigenfunctions of the scalar Laplacian on $T^{1,1}$ have been worked out in [47, 48]. We first review this calculation and present the harmonics in angular variables on $T^{1,1}$. This form of the harmonics is useful for some purposes, such as in [43] where it was used to find the potential generated for a D3-brane moving on the conifold due to a wrapped D7. We then write the harmonics using the complex $a_i, b_j$ coordinates, generalizing the $z_i$ construction of [13], that makes the connection with the gauge theory manifest. We also construct the operators using $A_i, B_j$ with given symmetry charges, related to the harmonics through the AdS/CFT correspondance.

Since $T^{1,1}$ is a product of two 3−spheres divided by a $U(1)$, the eigenfunctions are simply products of harmonics on two 3−spheres, restricted by the fact that the two spheres share an angle $\psi$. The laplacian (defined by $\mathit{\Delta}_Z H = \frac{1}{\sqrt{g}}\partial_m(g^{mn}\sqrt{g}\partial_n H)$) on $T^{1,1}$ can be written in the following form,

$$\mathit{\Delta}_Z = 6\mathit{\Delta}_1 + 6\mathit{\Delta}_2 + 9\mathit{\Delta}_R \tag{2.A.1}$$

where

$$\mathit{\Delta}_i = \frac{1}{\sin\theta_i}\partial_{\theta_i}\left(\sin\theta_i\,\partial_{\theta_i}\right) + \left(\frac{1}{\sin\theta_i}\partial_{\phi_i} - \cot\theta_i\partial_\psi\right)^2 \tag{2.A.2}$$

$$\mathit{\Delta}_R = \partial_\psi^2 \tag{2.A.3}$$

We can solve for the eigenfunctions through separation of variables,

$$Y_I(Z) \sim J_{l_1,m_1,R}(\theta_1)\; J_{l_2,m_2,R}(\theta_2)\; e^{im_1\phi_1 + im_2\phi_2}\; e^{\frac{iR\psi}{2}}$$

This leads to

$$\frac{1}{\sin\theta}\partial_\theta\left(\sin\theta\,\partial_\theta\, J_{lmR}(\theta)\right)-\left(\frac{1}{\sin\theta}m-\cot\theta\frac{R}{2}\right)^2 J_{lmR}(\theta)=-EJ_{lmR}(\theta) \qquad (2.A.4)$$

for both sets of angles. When $R=0$, this reduces to the equation for harmonics on $S^2$. For general integer $R$, this is closely related to the harmonic equation on $S^3$ in Euler angles $(\theta,\phi,\psi)$. The eigenvalues $E$ are $l(l+1)-\frac{R^2}{4}$ as can be seen by comparing with Laplace's equation on $S^3$.

The solutions for $J_{lmR}$ are,

$$J^A_{lmR}(\theta)=\sin^m\theta\;\;\cot^{\frac{R}{2}}\frac{\theta}{2}\;\;{}_2F_1\left(-l+m,1+l+m;1+m-\frac{R}{2};\sin^2\frac{\theta}{2}\right) \qquad (2.A.5)$$

$$J^B_{lmR}(\theta)=\sin^{\frac{R}{2}}\theta\;\;\cot^m\frac{\theta}{2}\;\;{}_2F_1\left(-l+\frac{R}{2},1+l+\frac{R}{2};1-m+\frac{R}{2};\sin^2\frac{\theta}{2}\right) \qquad (2.A.6)$$

Here ${}_2F_1$ is the hypergeometric function. If $m\le R/2$, solution B is non-singular. If $m\ge R/2$, solution A is non-singular. (The solutions coincide when $m=R/2$).

Putting together these solutions, the spectrum is of the form

$$E_I=6\left(l_1(l_1+1)+l_2(l_2+1)-\frac{R^2}{8}\right)$$

with eigenfunctions that transform under $SU(2)_A\times SU(2)_B$ as the spin $(l_1,l_2)$ representation and under the shift of $\psi/2$ (which is $U(1)_R$ in the UV) with charge $R$. Here $I$ is a multi-index with the data:

$$I\;\equiv\;(l_1,m_1),\;(l_2,m_2),\;R$$

with the following restrictions coming from existence of single valued regular solutions:

- $l_1$ and $l_2$ both integers or both half-integers
- $R \in \mathbf{Z}$ with $\frac{R}{2} \in \{-l_1, \cdots, l_1\}$ and $\frac{R}{2} \in \{-l_2, \cdots, l_2\}$
- $m_1 \in \{-l_1, \cdots, l_1\}$ and $m_2 \in \{-l_2, \cdots, l_2\}$

As above $(l_1, l_2), R$ are the $SU(2) \times SU(2)$ spins and R-charge and $(m_1, m_2)$, the $J_z$ values under the two $SU(2)$s.

## Harmonics in the $a, b$ basis

In [13], the 'chiral' harmonics were constructed using the complex $z_i$ coordinates. We generalize this to construct harmonics by using the $a_i, b_j$ coordinates which facilitates the comparison with the gauge . We form the eigenfunction $Y_I$ in the $a, b$ basis by tensoring representations. As we wish to construct harmonics on the base $T^{1,1}$, we fix the radius $r$ of the conetheory by setting $|a_1|^2 + |a_2|^2 = |b_1|^2 + |b_2|^2 = 1$. Since we are dealing with commuting functions (or symmetric tensors), only the highest total spin survives the tensor product. First we introduce the products,

$$\begin{aligned} \sqrt{\frac{n!}{(2m)!(n-2m)!}} a_1^{\frac{n}{2}+m} a_2^{\frac{n}{2}-m} &\equiv |\frac{n}{2}, \; m\rangle \qquad \left(n \in \mathbf{Z}, m \in \mathbf{Z} - \frac{n}{2}\right) \\ \sqrt{\frac{n!}{(2m)!(n-2m)!}} \bar{a}_2^{\frac{n}{2}+m} \bar{a}_1^{\frac{n}{2}-m} &\equiv \overline{|\frac{n}{2}, \; m\rangle} \qquad \left(n \in \mathbf{Z}, m \in \mathbf{Z} - \frac{n}{2}\right) \qquad (2.A.7) \end{aligned}$$

which are states of definite $SU(2)$ spin $n$ and $R$ charge $\pm n/2$, since the product of $n$ commuting $a$'s and $\bar{a}$'s automatically has only spin $n/2$ states. We combine these to form a state of arbitrary $SU(2)$ spin and $R$ charge using Clebsch-Gordon coefficients,

[6] by,

$$
\begin{aligned}
|l_1, m_1; R/2\rangle_a &= \sum_{\substack{k,\tilde{k} \\ k+\tilde{k}=m_1}} {}_R C^{l_1,m_1}_{k\,;\,\tilde{k}} |\frac{l_1}{2}+\frac{R}{4}, k\rangle \overline{|\frac{l_1}{2}-\frac{R}{4}, \tilde{k}\rangle} \\
&= (a_1 a_2)^{\frac{l_1}{2}+\frac{R}{4}} (\bar{a}_1 \bar{a}_2)^{\frac{l_1}{2}-\frac{R}{4}} \sum_{k+\tilde{k}=m_1} {}_R C^{l_1,m_1}_{k\,;\,\tilde{k}} \; a_1^k a_2^{-k} \bar{a}_2^{\tilde{k}} \bar{a}_1^{-\tilde{k}} \quad (2.A.8)
\end{aligned}
$$

where we have introduced $|l_1, m_1; R/2\rangle_a$ to denote the wavefunctions with $SU(2)$ spin $(l_1, m_1)$ and $U(1)_R$ charge $R/2$ constructed from $a_i$ variables.

Using the same notation for $b_i$, $|l_2, m_2; R/2\rangle_b$ is the state with the required symmetry charges. To construct an eigenfunction $Y_I$ on $T^{1,1}$, we must have equal $R$ charge for the $a$ and $b$ states above in order to have invariance under the transformation $a \to e^{i\alpha} a, b \to e^{-i\alpha} b$ explained earlier (see (2.2.6)). Hence, $Y_I$ is simply a product of the $a$ and $b$ states constructed above,

$$
Y_I \sim |l_1, m_1; R/2\rangle_a |l_2, m_2; R/2\rangle_b \,. \quad (2.A.9)
$$

For example, some of the wavefunctions for $l_1 = l_2 = 1, R = 0$ are :

$$
\begin{aligned}
&a_1 \bar{a}_2 \, b_1 \bar{b}_2 && (m_1, m_2) = (1, 1) \\
&(a_1 \bar{a}_1 - a_2 \bar{a}_2) \, b_1 \bar{b}_2 && (m_1, m_2) = (0, 1) \\
&a_2 \bar{a}_1 \, b_2 \bar{b}_1 && (m_1, m_2) = (-1, -1)
\end{aligned}
$$

While the harmonics (2.A.9) are obviously relevant to the singular conifold, it was also shown in Section 2.4 that the Laplacian on the resolved conifold (see (2.4.2))

[6] We are only using the 'top-spin' Clebsch Gordon coefficients. The notation here is:

$$
{}_R C^{l_1,m_1}_{k;\tilde{k}} = \langle l_1, m_1 | \frac{l_1}{2}+\frac{R}{4}, k \,;\, \frac{l_1}{2}-\frac{R}{4}, \tilde{k}\rangle \times (-1)^{\frac{l_1}{2}-\frac{R}{4}-\tilde{k}}
$$

We need this extra $-1$ factor because we tensoring conjugate representations of $SU(2)$ : $J_- a_1 \sim a_2$ but $J_- \bar{a}_2 \sim -\bar{a}_1$ factors in a form that allows one to use the same angular functions. This is because the resolution of the conifold preserves the $SU(2)_L \times SU(2)_R \times U(1)_R$ symmetry.

### Construction of the dual operators

The above construction of eigenfunctions is useful primarily because of their one-to-one correspondence with (single trace) operators in the guage theory. Our stragey is to replace $a_i, b_j$ in the eigenfunctions by the chiral superfields $A_i, B_j$. However, since $A_i, B_j$ are non-commuting operators in the gauge theory, we need to modify the procedure of the previous section to obtain an operator $\mathcal{O}_I$ of a given symmetry.

We may start with (2.A.7), and symmetrize the product of $A_1, A_2$'s (and $\bar{A}_1, \bar{A}_2$'s) by hand (the gauge index structure seems ill defined but this will be fixed when the total operator is put together.) So we could now write instead of (2.A.7) (with a different normalization factor),

$$\frac{1}{\sqrt{\frac{n!}{(2m)!(n-2m)!}}} \sum_{\substack{\frac{n}{2}+m=\sum i \\ \frac{n}{2}-m=\sum j}} A_1^{i_1} A_2^{j_1} A_1^{i_2} \cdots A_2^{j_k} \equiv |\frac{n}{2}\,,\ m\rangle \qquad \Big(n \in \mathbf{Z}, m \in \mathbf{Z} - \frac{n}{2}\Big)$$

The same symmetrization applies to $\bar{A}$'s as well. With this modified definition of $|\frac{n}{2}, m\rangle$, we can write down the equation analogous to (2.A.8) with no change in the form,

$$|l_1, m_1; R/2\rangle_A = \sum_{\substack{k,\tilde{k} \\ k+\tilde{k}=m_1}} {}_R C^{l_1,m_1}_{k\,;\,\tilde{k}} |\frac{l_1}{2} + \frac{R}{4}, k\rangle \overline{|\frac{l_1}{2} - \frac{R}{4}, \tilde{k}\rangle} \tag{2.A.10}$$

We make the analogous definitions for $B$. Finally, we can write down dual operator $\mathcal{O}_I$ as,

$$\mathcal{O}_I = \mathrm{Tr}\left(|l_1, m_1; R/2\rangle_A |l_2, m_2; R/2\rangle_B\right) \tag{2.A.11}$$

The product of the operators $|l_1, m_1; R/2\rangle_A$ and $|l_2, m_2; R/2\rangle_B$ is taken in the following way. All the terms are multiplied out and in each term, one is free to move operators in the $(N, \bar{N})$ rep of the gauge group (i.e $A, \bar{B}$) past $(\bar{N}, N)$ (i.e $B, \bar{A}$) but no rearrangement among themselves is allowed. We shuffle them past each other until they alternate and so we can contract gauge indices properly and take the trace. It is easy to verify that the numbers of fields of each type are equal and so there is always one essentially unique way of doing this. By construction, this operator has the specified symmetry $I$ under the global symmetry group.

## 2.B $AdS_5 \times S^5$ Throats in the IR

Here we show explicitly that the Green's function on the resolved conifold reduces to the form $\frac{1}{y^4}$ near the source as it must ($y$ here is the physical distance from the source on the transverse space). This leads to the usual near-horizon limit when the branes are at a smooth point and hence an $AdS_5 \times S^5$ throat. This is of course to be expected since close to the source, we can find coordinates in which the space looks flat at leading order and hence the Green's function must behave as $\frac{1}{y^4}$. But it is instructive to see how the series does add up to such a divergence while each individual term has a different kind of divergence that gives a singular geometry in the case of the resolved conifold.

We focus on the resolved conifold and consider $\theta_0 = 0 = \phi_0$, i.e set the stack on the 'north pole' of the finite $S^2$. Also, we approach the singularity by first setting $\theta_2 = 0$ and taking the $r \to 0$ limit. Now $r$ is the physical distance and from (2.4.12), we would like to show that,

$$\sum_l (2l+1) H_I^A(r) \sim \frac{1}{r^4} \quad \text{while} \quad H_I^A(r) \sim \frac{1}{r^2} \quad \text{as } r \to 0 \tag{2.B.1}$$

Consider the regulation of the sum of squares of integers,

$$\sum_{n=0}^{\infty} n^2 \to \sum_{n=0}^{\infty} n^2 R(n\epsilon) \tag{2.B.2}$$

where $R(x)$ is a regulator such as $R(x) = e^{-x}$ with the property $R(x) \to 0$ (fast enough in a sense to be seen below) as $x \to \infty$. As $\epsilon \to 0$, the sum diverges and this allows one to approximate the sum by an integral in this limit. Further, only $0 \leq n \leq 1/\epsilon$ will contribute. Hence we find,

$$\int_0^{\frac{1}{\epsilon}} n^2 R(n\epsilon) dn \quad \sim \quad \frac{1}{\epsilon^3} \int_0^1 y^2 R(y) dy \quad (\epsilon \to 0) \tag{2.B.3}$$

Note that the above argument just amounts to dimensional analysis. To cast the given expression (2.B.1) in the above form with $H_I^A(r)$ playing the role of a regulator, we note that $H_I^A(r)$ can be approximated for $r \ll a$ by $(\sqrt{l(l+1)}/r)K_1(\sqrt{l(l+1)}r)$. [7] Hence we have for $r \ll a$,

$$\begin{aligned} \sum_l (2l+1) H_I^A(r) \quad &\sim \quad \sum_l (2l+1) \frac{\sqrt{l(l+1)}}{r} K_1(\sqrt{l(l+1)}r) \sim \frac{1}{r} \int_n (2n+1) n K_1(nr) \\ &\sim \quad \frac{1}{r} \int_0^{1/r} n^2 K_1(nr) dn \sim \frac{1}{r} \times \frac{1}{r^3} \quad \int_0^1 y^2 K_1(y) dy \end{aligned}$$

where we have kept track of only the leading order singularity. We have identified $R(y) = K_1(y)$ despite the fact $K_1(y) \sim 1/y$ for small y. This is allowed here because $\int_0^1 dy y^2 K_1(y)$ converges.

Hence we see that indeed, $H(r, \theta_2 = 0) \sim \frac{1}{r^4}$ near $r = 0$ and hence the geometry is non-singular (though each term in the expansion of $H$ behaves as $\frac{1}{r^2}$ giving a singular geometry by itself). The result essentially follows from dimensional analysis in (2.B.3).

[7] We mean this in the sense that $K_1(y)$ is the solution to the differential equation obtained by applying $r \ll a$ to (2.4.3) whose exact solution was obtained as $H_I^A(r)$. We are interested in how $r$ scales with $l$ to keep $H_I^A(r)$ constant for very small $r$, since this determines the leading order singularity through essentially dimensional analysis in (2.B.3). Approximating $H_I^A$ by $K_1$ is valid in this sense.

# Chapter 3

# Entanglement entropy

This chapter is based on the paper 'Entanglement as a Probe of Confinement' written in collaboration with I.R. Klebanov and D. Kutasov [65]. We investigate the entanglement entropy in gravity duals of confining large $N_c$ gauge theories using the proposal of [66, 67]. Dividing one of the directions of space into a line segment of length $l$ and its complement, the entanglement entropy between the two subspaces is given by the classical action of the minimal bulk hypersurface which approaches the endpoints of the line segment at the boundary. We find that in confining backgrounds there are generally two such surfaces. One consists of two disconnected components localized at the endpoints of the line segment. The other contains a tube connecting the two components. The disconnected surface dominates the entropy for $l$ above a certain critical value $l_{\rm crit}$ while the connected one dominates below that value. The change of behavior at $l = l_{\rm crit}$ is reminiscent of the finite temperature deconfinement transition: for $l < l_{\rm crit}$ the entropy scales as $N_c^2$, while for $l > l_{\rm crit}$ as $N_c^0$. We argue that a similar transition should occur in any field theory with a Hagedorn spectrum of non-interacting bound states. The requirement that the entanglement entropy has a phase transition may be useful in constraining gravity duals of confining theories.

## 3.1 Introduction

Consider a $d+1$ dimensional quantum field theory (QFT) on $\mathbb{R}^{d+1}$ in its vacuum state $|0\rangle$. Divide the $d$ dimensional space into two complementary regions,

$$\begin{aligned} A &= \mathbb{R}^{d-1} \times I_l \ , \\ B &= \mathbb{R}^{d-1} \times (\mathbb{R} - I_l) \ , \end{aligned} \tag{3.1.1}$$

where $I_l$ is a line segment of length $l$. The entanglement entropy between the regions $A$ and $B$ is defined as the entropy seen by an observer in $A$ who does not have access to the degrees of freedom in $B$, or vice versa (see e.g. [68] for a recent review and references to earlier work). It can be calculated by tracing the density matrix of the vacuum, $\rho_0 = |0\rangle\langle 0|$, over the degrees of freedom in $B$ and forming the reduced density matrix

$$\rho_A = \mathrm{Tr}_B \rho_0 \ . \tag{3.1.2}$$

The quantum entanglement entropy $S_A$ is then given by

$$S_A = -\mathrm{Tr}_A \rho_A \ln \rho_A \ . \tag{3.1.3}$$

The above construction can be generalized in a number of ways. In particular, one can replace the vacuum state $|0\rangle$ by any other pure or mixed state, and choose the submanifold of $\mathbb{R}^d$, $A$, to be different than (3.1.1). In this chapter we will restrict to the choices above, which are sufficient for our purposes.

The entanglement entropy (3.1.3) is in general UV divergent. To leading order in the UV cutoff $a$ it scales like [69, 70]

$$S_A \simeq \frac{V_{d-1}}{a^{d-1}} \tag{3.1.4}$$

where $V_{d-1}$ is the volume of $\mathbb{R}^{d-1}$ in (3.1.1). Note that (3.1.4) is independent of $l$. This turns out to be a general feature – the entropy is defined up to an $l$ independent (infinite) additive constant. In particular, $\partial_l S_A$ and differences of entropies at different values of $l$ approach a finite limit as $a \to 0$. In $d+1$ dimensional CFT with $d > 1$, the finite $l$-dependent part of the entropy is negative and proportional to $V_{d-1}/l^{d-1}$, while for $d = 1$ it goes like $\ln l$.

If the QFT in question has a gravity dual [9, 11, 12], it is natural to ask whether the entanglement entropy can be calculated using the bulk description. This problem was addressed in [66, 67]. For the case of $d+1$ dimensional large $N_c$ conformal field theories with $AdS_{d+2}$ gravity duals, the authors of [66, 67] proposed a simple geometric method for computing the entanglement entropy and subjected it to various tests. This method is to find the minimal area $d$-dimensional surface $\gamma$ in $AdS_{d+1}$ such that the boundary of $\gamma$ coincides with the boundary of $A$, which in the case (3.1.1) consists of two copies of $\mathbb{R}^{d-1}$ a distance $l$ apart. The quantum entanglement entropy between the regions $A$ and $B$ is proportional to the classical area of this surface,

$$S_A = \frac{1}{4G_N^{(d+2)}} \int_\gamma d^d\sigma \sqrt{G_{\text{ind}}^{(d)}} \, , \tag{3.1.5}$$

where $G_N^{(d+2)}$ is the $d+2$ dimensional Newton constant and $G_{\text{ind}}^{(d)}$ is the induced string frame metric on $\gamma$. Note that the surface $\gamma$ is defined at a fixed time and (3.1.5) gives the entanglement entropy at that time. For static states, such as the vacuum, the resulting entropy is time independent.[1] Also, since $\gamma$ is extended in the transverse $\mathbb{R}^{d-1}$ (3.1.1), the entropy (3.1.5) is proportional to its volume $V_{d-1}$. Thus, in this case it is better to consider the entropy per unit transverse volume.

In non-conformal theories, the volume of the $8-d$ compact dimensions and the dilaton are in general not constant. A natural generalization of (3.1.5) to the corresponding

[1] Generalizations of the proposal of [66, 67] to time dependent states were discussed in [71].

ten dimensional geometries is [66, 67, 72]

$$S_A = \frac{1}{4G_N^{(10)}} \int d^8\sigma e^{-2\phi} \sqrt{G_{\rm ind}^{(8)}} \ . \tag{3.1.6}$$

The entropy is obtained by minimizing the action (3.1.6) over all surfaces that approach the boundary of $A$ (3.1.1) at the boundary of the bulk manifold, and are extended in the remaining spatial directions. Since $G_N^{(10)} = 8\pi^6\alpha'^4 g_s^2$, this gives an answer of order $N_c^2$ in the 't Hooft limit $N_c \to \infty$ with $g_s N_c$ held fixed.

It was shown in [66,67] that for $AdS_3$ the prescription (3.1.5) successfully reproduces the known form of the entanglement entropy in two-dimensional conformal field theory, and that it gives sensible results when applied to some higher dimensional vacua, such as $AdS_5 \times S^5$. Nevertheless, some aspects of the proposal are not well understood. In particular, it is not clear how to extend it beyond leading order in $1/N_c$.

In this chapter we apply the proposal of [66, 67] to some confining backgrounds [31, 25]. One of the motivations for this investigation is to subject the proposal (3.1.6) to further tests. Another is to study the $l$ dependence of the entanglement entropy, which is in general difficult to determine in strongly coupled field theories.

Gravitational backgrounds dual to confining gauge theories typically have the following structure. As one moves in the radial direction away from the boundary, an internal cycle smoothly contracts and approaches zero size at the infrared (IR) end of space. The radial direction together with the shrinking cycle make a type of cigar geometry, with the IR end of space corresponding to the tip of the cigar.

We will see that in such geometries there are in general multiple local minima of the action (3.1.6) for given $l$. One of those is a disconnected surface, which consists of two cigars extended in $\mathbb{R}^{d-1}$ and separated in the remaining direction in $\mathbb{R}^d$ by the distance $l$. A second one is a connected surface, in which the two cigars are connected by a tube whose width depends on $l$. Since the two geometries are related by a continuous deformation, there is a third extremum of the action between them, which is a saddle point of (3.1.6). A natural generalization of the proposal of [66, 67] to this case is to identify the entanglement entropy with the absolute minimum of the action. We will see that this leads to a phase transition in the behavior of the entanglement entropy as a function of $l$.

For the disconnected solution, $S_A$ (3.1.6) does not depend on $l$. As mentioned above, the actual value of $S_A$ depends on the UV cutoff, but if we are only interested in differences of entropies, or the derivative of the entropy with respect to $l$, we can set it to zero. For the connected solution, $S_A$ depends non-trivially on $l$. For small $l$, it is smaller than that of the disconnected one. Thus, it dominates the entropy (3.1.6). For $l > l_{\text{crit}}$ the action of the connected solution becomes larger than that of the disconnected one, and it is the latter that governs the entropy. Thus, in going from $l < l_{\text{crit}}$ to $l > l_{\text{crit}}$, $\partial_l S_A$ goes from being of order $N_c^2$ to being of order $N_c^0$. One can think of this change of behavior as a phase transition which, as we show, is typical in large $N_c$ confining theories.

Similar transitions between connected and disconnected D-brane configurations play a role in other contexts. In [73] an analogous transition is responsible for screening of magnetic charges in confining gravitational backgrounds; in [74] it governs the pattern of metastable supersymmetry breaking vacua in a brane construction of supersymmetric QCD. An important difference is that in all these cases the transitions involve the rearrangement of real branes, whereas the hypersurface whose area is being minimized here does not seem to have such an interpretation.

The plan of the rest of the chapter is as follows. In section 2 we present a general analysis of a class of gravitational backgrounds that arises in the construction of holographic duals of confining gauge theories. We show that in this class there are multiple local minima of the action (3.1.6), as discussed above. With some mild assumptions, we also show that for small $l$ the global minimum of the action corresponds to a con-nected solution, while for large $l$ it corresponds to a disconnected one. We also show that the connected solution does not exist for sufficiently large $l$.

In sections $3-5$ we illustrate the discussion of section 2 with a few examples. Section 3 contains an analysis of the geometry of $N_c$ $D4$-branes wrapped around a circle with twisted boundary conditions for the fermions. For $g_sN_c \ll 1$ this system reduces at low energies to pure Yang-Mills (YM) theory, while for $g_sN_c \gg 1$ it can be analyzed using the near-horizon geometry of the $D4$-branes [31]. In section 4 we describe the analogous $D3$-brane system, which for $g_sN_c \ll 1$ gives rise to YM in $2+1$ dimensions. Section 5 contains an analysis of the warped deformed conifold (KS) background [25], which corresponds to a cascading, confining $SU(M(k+1)) \times SU(Mk)$ supersymmetric gauge theory. This theory approaches pure $SU(M)$ SYM theory in the limit $g_sM \ll 1$, while the dual supergravity description is reliable in the opposite limit, $g_sM \gg 1$.

In section 6 we connect the results of sections $2-5$ to large $N_c$ confining field theories such as pure YM. To leading order in $1/N_c$, such theories are expected to reduce to free field theories of the gauge singlet bound states. The latter are expected to have a Hagedorn density of states at high mass, $\rho(m) \sim m^\alpha \exp(\beta_H m)$. The entanglement entropy in such theories can be calculated by summing the contributions of the individual states. We show that this sum over states has a very similar character to the finite temperature partition sum, with $l$ playing the role of the inverse temperature $\beta$. It converges for sufficiently large $l$ and diverges below a critical value of $l$, since the large entropy overwhelms the exponential suppression of the contribution of a given state of large mass. In the thermodynamic case, this phenomenon is related to the appearance of a deconfinement transition. By analogy, it is natural to expect that here it signifies a transition between an entropy that goes like $N_c^0$ at large $l$, and one that goes like $N_c^2$ below a critical value. Since the gravitational analysis reproduces this feature of the dynamics, we conclude that the system with $g_sN_c \gg 1$ is in the same universality class as the one with $g_s N_c \ll 1$.

In section 7 we comment on our results and discuss other systems which one can analyze using similar techniques. We also point out some general issues related to the proposal of [66, 67].

## 3.2 Holographic computation of entropy

The gravitational backgrounds we will consider have the string frame metric

$$ds^2 = \alpha(U)\left[\beta(U)dU^2 + dx^\mu dx_\mu\right] + g_{ij}dy^i dy^j \tag{3.2.1}$$

where $x^\mu$ ($\mu = 0, 1, \ldots, d$) parametrize $\mathrm{I\!R}^{d+1}$, $U$ is the holographic radial coordinate, and $y^i$ ($i = d+2, \cdots, 9$) are the $8-d$ internal directions. The volume of the internal manifold,

$$V_{\text{int}} = \int \prod_{i=1}^{8-d} dy^i \sqrt{\det g}\ , \tag{3.2.2}$$

and the dilaton, $\phi$, are taken to depend only on $U$.

The radial coordinate $U$ ranges from a minimal value, $U_0$, to infinity. As $U \to U_0$, a $p$-cycle in the internal $(8-d)$-dimensional space shrinks to zero size, so $V_{\text{int}}(U_0) = 0$. The vicinity of $U = U_0$ looks locally like the origin of spherical coordinates in $\mathrm{I\!R}^{p+1}$ (times a compact space), and we assume that all the supergravity fields are regular there. In particular, $\alpha(U)$ and $\phi(U)$ approach fixed finite values as $U \to U_0$. The fact that $\alpha(U_0) > 0$ implies that the string tension is non-vanishing. This is the gravitational manifestation of the fact that the dual gauge theory is confining.

Examples of backgrounds in this class that will be discussed below are the geometries of coincident $D3$ and $D4$-branes on a circle with twisted boundary conditions [31], in which the shrinking cycle is a circle ($p = 1$), and the KS geometry [25] in which it is a two-sphere ($p = 2$). In the $D3$-brane and KS cases, the dilaton is independent of $U$.

We would like to use the proposal (3.1.6) to calculate the entanglement entropy of $A$ and $B$ (3.1.1) in the geometry (3.2.1). Denoting the direction along which the line segment $I_l$ in (3.1.1) is oriented by $x$, the entropy per unit volume in the transverse $\mathbb{R}^{d-1}$ is given by

$$\frac{S_A}{V_{d-1}} = \frac{1}{4G_N^{(10)}} \int_{-\frac{l}{2}}^{\frac{l}{2}} dx \sqrt{H(U)} \sqrt{1 + \beta(U)(\partial_x U)^2} \tag{3.2.3}$$

where we introduced the notation

$$H(U) = e^{-4\phi} V_{\text{int}}^2 \alpha^d \ . \tag{3.2.4}$$

Due to the shrinking of the $p$-cycle, we have $H(U_0) = 0$. Thus, as $U$ varies between $U_0$ and $\infty$, $H(U)$ varies between 0 and $\infty$. It provides a natural parametrization of the radial direction of the space (3.2.1). Near $U_0$, one has $H \sim r^{2p}$, where $r \in [0, \infty)$ is a natural radial coordinate, $dr = \sqrt{\beta(U)} dU$.

The quantity (3.2.4) is simply related to the warp factor we get upon dimensionally reducing on the $(8 - d)$-dimensional compact manifold. The resulting $(d + 2)$-dimensional Einstein frame metric may be written as

$$ds_{d+2}^2 = \kappa(U) \left[ \beta(U) dU^2 + dx^\mu dx_\mu \right] \ . \tag{3.2.5}$$

A standard calculation shows that $\kappa(U)^d = H(U)$. It is a common assumption that the warp factor $\kappa(U)$ is a monotonic function of the holographic radial coordinate. In particular, finiteness of the holographic central charge [75, 24],

$$c \sim \beta^{\frac{d}{2}} \kappa^{\frac{3d}{2}} \left( \kappa' \right)^{-d} \ , \tag{3.2.6}$$

requires $\kappa$ to be monotonic. Since it goes to zero as $U \to U_0$ and to infinity as $U \to \infty$, it must be that $\kappa' > 0$ for all $U$. This implies $H'(U) > 0$, a fact that will be useful below.

We need to find the shape $U(x)$ that minimizes the action (3.2.3) subject to the constraint $U(x \to \pm\frac{l}{2}) \to \infty$. Denoting by $U^*$ the minimal value of $U$ along this curve,[2] and using the fact that the action does not depend directly on $x$, its equation of motion can be integrated once and written in the form

$$\partial_x U = \pm \frac{1}{\sqrt{\beta}} \sqrt{\frac{H(U)}{H(U^*)} - 1} \ . \tag{3.2.7}$$

Integrating once more we find

$$l(U^*) = 2\sqrt{H(U^*)} \int_{U^*}^{\infty} \frac{dU \sqrt{\beta(U)}}{\sqrt{H(U) - H(U^*)}} \ . \tag{3.2.8}$$

Plugging (3.2.7) into (3.2.3) we find

$$\frac{S_A}{V_{d-1}} = \frac{1}{2G_N^{(10)}} \int_{U^*}^{U_\infty} \frac{dU \sqrt{\beta(U)} H(U)}{\sqrt{H(U) - H(U^*)}} \ . \tag{3.2.9}$$

In the examples we study below, and probably much more generally, the integral in (3.2.8) turns out to be convergent, while that in (3.2.9) is not. This is the reason for the appearance of the UV cutoff $U_\infty$ in the latter and its absence in the former.

As mentioned earlier, the entropy $S_A$ depends on the cutoff only via an $l$ independent constant, which cancels in differences of entropies. This can be seen from (3.2.9) as follows. Denoting by $U_1^*$ and $U_2^*$ the solutions of (3.2.8) for $l = l_1$ and $l = l_2$, respectively, we have

$$S_A(l_1) - S_A(l_2) \sim \int^{\infty} dU \sqrt{\beta(U) H(U)} \left[ \left(1 - \frac{H(U_1^*)}{H(U)}\right)^{-\frac{1}{2}} - \left(1 - \frac{H(U_2^*)}{H(U)}\right)^{-\frac{1}{2}} \right]$$

[2]If the curve is smooth, this value is attained at $x = 0$, where $\partial_x U = 0$.

where we omitted an overall multiplicative constant and focused on the behavior of the integral in the UV region $U \to \infty$. In that region $H(U) \to \infty$, and we can approximate the integrand in (3.2.10) by

$$S_A(l_1) - S_A(l_2) \sim (H(U_1^*) - H(U_2^*)) \int^{\infty} dU \sqrt{\frac{\beta(U)}{H(U)}} \ . \tag{3.2.10}$$

The integrand in (3.2.10) behaves as $U \to \infty$ in the same way as that in (3.2.8). Thus, if the latter is finite and does not require introduction of a UV cutoff, the same is true for the former.

To find the dependence of the entropy on $l$ we need to determine $U^*(l)$ by solving (3.2.8), and then use it in (3.2.9). In the next sections we will study specific examples of this procedure; here we would like to make some general comments on it.

Consider first the limit $U^* \to \infty$. Physically, one would expect $l(U^*)$ to go to zero in this limit since as $l \to 0$ the minimal action surface should be located at larger and larger $U$. In terms of (3.2.8) this means that although the prefactor $\sqrt{H(U^*)}$ goes to infinity, the integral goes to zero faster, such that the product of the two goes to zero as well. We will see below that this is indeed what happens in all the examples we will consider.

It turns out that $l$ (3.2.8) also goes to zero in the opposite limit $U^* \to U_0$. The prefactor $\sqrt{H(U^*)}$ goes to zero in this limit, and as long as the integral does not diverge rapidly enough to overwhelm it, $l \to 0$. Since any divergence of the integral as $U^* \to U_0$ must come from the region $U \simeq U^* \simeq U_0$, it is enough to estimate the contribution to it from this region. In terms of the coordinate $r$ defined above, one has

$$l(r_*) \sim r_*^p \int_{r_*} \frac{dr}{\sqrt{r^{2p} - r_*^{2p}}} \ . \tag{3.2.11}$$

For $p > 1$ one finds that for small $r_*$, $l(r_*) \sim r_*$; for $p = 1$, $l(r_*) \sim r_* \ln r_*$. In both cases, $l \to 0$ in the limit $r_* \to 0$ (or, equivalently, $U^* \to U_0$).

We see that for small $l$ the equation of motion (3.2.7) has two independent solutions, one with large $U^*$ and the other with $U^* \simeq U_0$. The former is a local minimum of the action (3.2.9) while the latter is a saddle point. We can interpolate between them with a sequence of curves which differ in the minimal value of $U$, such that the solution with large $U^*$ is a local minimum along this sequence, while the one with $U^* \simeq U_0$ is a local maximum.

This implies that there must be another local minimum of the effective action, with $U^*$ smaller than that of the saddle point. This solution cannot correspond to a smooth $U(x)$, since then it would be captured by the above analysis. Therefore, it must correspond to a disconnected solution, which formally has $U^* = U_0$, but is better described as two disconnected surfaces that are extended in all spatial directions except for $x$, and are located at $x = \pm\frac{l}{2}$.

The entropy corresponding to this solution is given by (see (3.2.9))

$$\frac{S_A}{V_{d-1}} = \frac{1}{2G_N^{(10)}} \int_{U_0}^{U_\infty} dU \sqrt{\beta(U)H(U)}\ . \qquad (3.2.12)$$

By the above analysis it must be smaller than that of the connected solution with $U^* \simeq U_0$, but may be larger or smaller than that of the connected local minimum with large $U^*$.

We saw before that $l(U^*)$, (3.2.8), goes to zero both at large $U^*$ and as $U^* \to U_0$. If the supergravity background is regular, one can show that between these two extremes $l$ is a smooth function of $U^*$, that remains finite everywhere. The simplest behavior it can have is to increase up to some point where $\partial l/\partial U^* = 0$, and then decline back to zero as $U^* \to \infty$. We will see that this is indeed what happens in all the examples we study below.

Denoting the value of $l(U^*)$ at the maximum by $l_{\max}$, this behavior implies that smooth solutions to the equation of motion (3.2.7) only exist for $l \leq l_{\max}$. As $l \to l_{\max}$ from below, the local minimum and saddle point discussed above approach each other, merge and annihilate for $l > l_{\max}$.

At first sight, the fact that there are no solutions to (3.2.7) for $l > l_{\max}$ may seem puzzling, but it is important to remember that this analysis only captures smooth connected solutions. As discussed above, for all $l$ we have in addition a disconnected solution for which $U'(x)$ is infinite. For $l > l_{\max}$ the entanglement entropy $S_A$ is governed by this solution and is given by (3.2.12). For $l < l_{\max}$ one needs to compare the entropies of the connected and disconnected solutions and find the smaller one. This difference can be written as

$$\frac{2G_N^{(10)}}{V_{d-1}}\left(S_A^{(\mathrm{conn})} - S_A^{(\mathrm{disconn})}\right) = \int_{U^*}^{\infty} dU\sqrt{\beta H}\left(\frac{1}{\sqrt{1-\frac{H(U^*)}{H(U)}}}-1\right) - \int_{U_0}^{U^*} dU\sqrt{\beta H}\ .$$

It is physically clear and easy to see from (3.2.13) that for small $l$ the connected solution with large $U^*$ has the lower entropy. As $l$ increases, there are in general two possibilities. The connected solution can remain the lower action one until $l = l_{\max}$, or there could be a critical value $l_{\mathrm{crit}} < l_{\max}$ above which the right hand side of (3.2.13) is positive, so that the disconnected solution becomes the dominant one. In the first case there would be a phase transition at $l = l_{\max}$; in the second, the transition would occur at $l_{\mathrm{crit}}$, and in the range $l_{\mathrm{crit}} < l < l_{\max}$, the connected solution would be a metastable local minimum. In all the examples we study below it is the second possibility, $l_{\mathrm{crit}} < l_{\max}$, that is realized: as we increase $l$, the transition occurs before the connected solution ceases to exist. This is similar to the first-order finite temperature deconfinement transitions found in gravitational duals of confining gauge theories [31, 76–79].

## 3.3 D4-branes on a circle

The low energy dynamics of $N_c$ $D4$-branes in type IIA string theory is governed by 4+1 dimensional supersymmetric Yang-Mills theory with gauge group $U(N_c)$ and ‘t Hooft coupling $\lambda = g_s N_c l_s$. In order to reduce to $3+1$ dimensions and break supersymmetry, we compactify one of the directions along the branes, $x^4$, on a circle of radius $R_4$, $x^4 \sim x^4 + 2\pi R_4$, with twisted boundary conditions for the fermions.

The low energy dynamics of this system, which was studied in [31], depends on the dimensionless parameter $\lambda_4 = \lambda/R_4$, and can be investigated using different tools in different regions of parameter space. For $\lambda_4 \ll 1$, it corresponds to pure Yang-Mills theory with gauge group $U(N_c)$ and ’t Hooft coupling $\lambda_4$ (at the scale $R_4$). In the opposite limit, $\lambda_4 \gg 1$, one can use a gravitational description in terms of the near-horizon geometry of the branes[3]

$$\begin{aligned} ds^2 &= \left(\frac{U}{R}\right)^{3/2}\left[\left(\frac{R}{U}\right)^3 \frac{dU^2}{f(U)} + dx^\mu dx_\mu\right] + R^{3/2}U^{1/2}d\Omega_4^2 + \left(\frac{U}{R}\right)^{3/2} f(U)(dx^4)^2 , \\ e^{-2\phi} &= \left(\frac{R}{U}\right)^{3/2} , \end{aligned} \tag{3.3.1}$$

where $R$ is related to the five dimensional ’t Hooft coupling via the relation $R^3 = \pi\lambda$, and

$$f(U) = 1 - \left(\frac{U_0}{U}\right)^3 , \qquad U_0 = \frac{4\pi}{9}\frac{\lambda}{R_4^2} . \tag{3.3.2}$$

As $U \to U_0$, the radius of the $x^4$-circle goes to zero; $(U, x^4)$ form together a cigar geometry of the type described in the previous sections.

[3]Here and below we set $\alpha' = 1$.

Comparing (3.3.1) to (3.2.1) we identify $\alpha, \beta, V_{\text{int}}$ as,

$$\begin{aligned}
\alpha &= \left(\frac{U}{R}\right)^{3/2} , \quad \beta = \left(\frac{R}{U}\right)^3 \frac{1}{f(U)} , \qquad (3.3.3) \\
V_{\text{int}} &= \frac{8\pi^2}{3}(R^3U) \times 2\pi R_4 \left(\frac{U}{R}\right)^{3/4} \sqrt{f(U)} = \frac{32\pi^3 R^{\frac{15}{4}}}{9U_0^{\frac{1}{2}}} U^{\frac{7}{4}} \sqrt{f(U)} . \qquad (3.3.4)
\end{aligned}$$

The combination (3.2.4) is given in this case by

$$H(U) = R^6 \left(\frac{32\pi^3}{9}\right)^2 \frac{U^2(U^3 - U_0^3)}{U_0} . \qquad (3.3.5)$$

Note that $H'(U) > 0$ for all $U \geq U_0$, as mentioned in the previous section.

The explicit form of the background can be used to verify the assertions of section 2 about the behavior of $l(U^*)$. In particular, it is easy to check that the integral (3.2.8) converges. For $U^* \gg U_0$ it is given by

$$l(U^*) = 2R^{3/2} \times 2\sqrt{\pi} \frac{\Gamma\left(\frac{3}{5}\right)}{\Gamma\left(\frac{1}{10}\right)} \frac{1}{\sqrt{U^*}} . \qquad (3.3.6)$$

We see that $l$ indeed goes to zero in the limit $U^* \to \infty$, as expected. Similarly, one can check that it goes to zero in the opposite limit $U^* \to U_0$. The full curve $l(U^*)$ can be computed numerically and is plotted in figure 1. It has the qualitative structure anticipated in section 2. The maximum of the curve occurs at $U^* \simeq 1.2U_0$, with

$$l_{\max} \simeq 1.418 R_4 . \qquad (3.3.7)$$

At larger values of $l$, there is no smooth solution to the equations of motion (3.2.7).

Turning to the entanglement entropy $S_A$, following the discussion of section 2 we need to calculate the entropies of the connected solution (3.2.9) and the disconnected one (3.2.12), and compare them. The calculations of the individual entropies must be done with the UV cutoff $U_\infty$ in place, but the difference of entropies is insensitive to

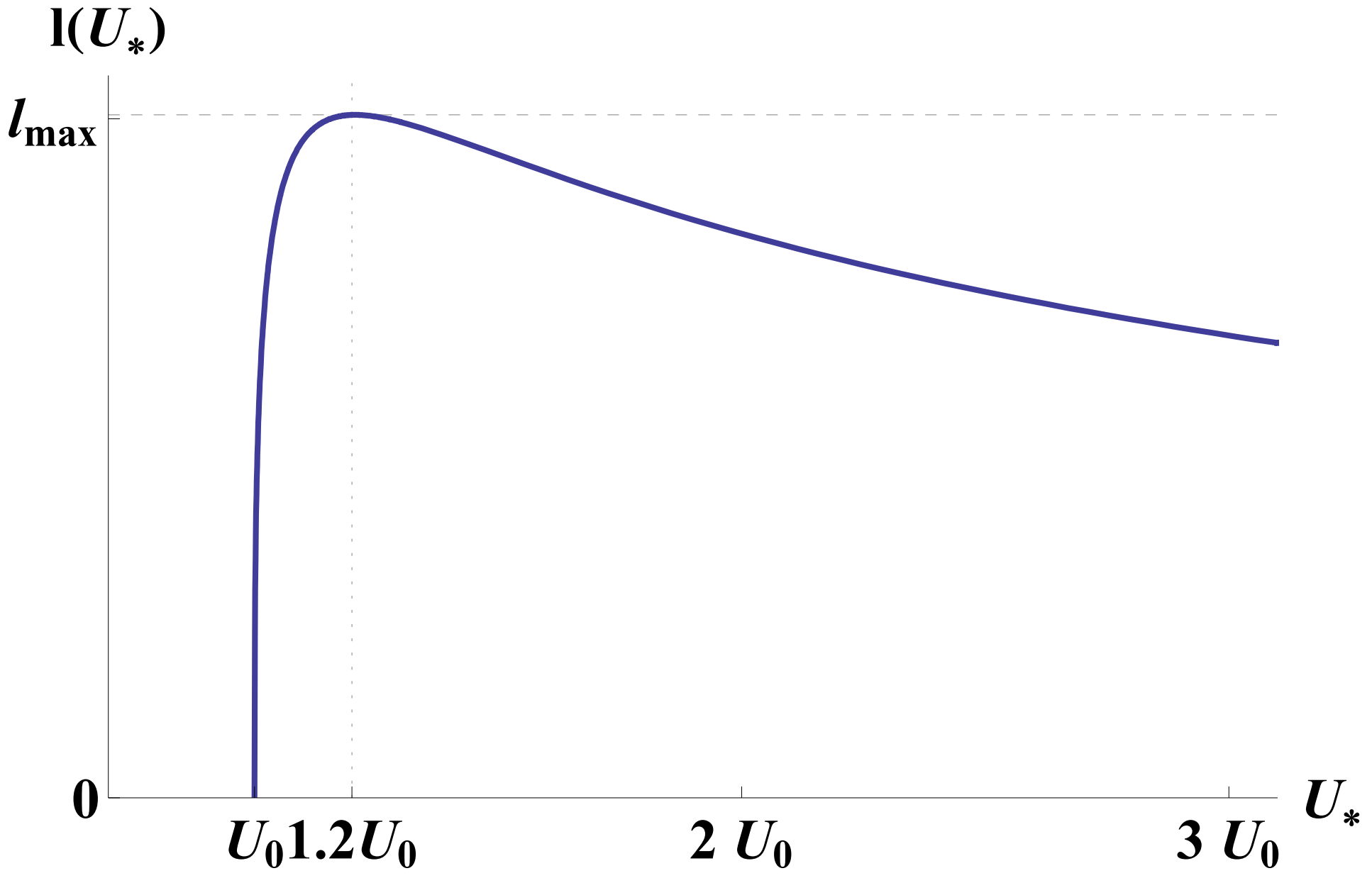

Figure 3.1: $l(U^*)$ for $D4$-branes on a circle.

it (see (3.2.10), (3.2.13)).

For the disconnected solution, the entropy can be calculated in closed form:

$$S_A^{(\text{disconn})} = \frac{8\pi^3}{9} \frac{V_2 R^{9/2}}{U_0^{1/2} G_N^{(10)}} \left( U_\infty^2 - U_0^2 \right) . \tag{3.3.8}$$

For the connected one it is given by (3.2.9), which in general has to be computed numerically. For small $l$ one can again perform the integral using the fact that in this case $U^* \gg U_0$. One finds

$$S_A^{(\text{conn})}(l) = \frac{8\pi^3}{9} \frac{V_2 R^{9/2}}{U_0^{1/2} G_N^{(10)}} \left( U_\infty^2 - 256 \left[ \frac{\sqrt{\pi}\Gamma\left(\frac{3}{5}\right)}{\Gamma\left(\frac{1}{10}\right)} \right]^5 \frac{R^6}{l^4} \right) . \tag{3.3.9}$$

Comparing to (3.3.8) we see that for small $l$ the connected solution has lower entropy, in agreement with the general discussion of section 2. The fact that the entropy (3.3.9) scales like $1/l^4$ at small $l$ is indicative of 5+1 dimensional scale invariant dynamics. This is what one expects, since at short distances the dynamics on the wrapped $D4$-branes is described by the $(2,0)$ superconformal field theory in $5+1$ dimensions. Indeed, we find

$$S_A^{(\mathrm{conn})}(l) - S_A^{(\mathrm{disconn})} = -V_2(2\pi R_4)(2\pi R_{10})\frac{32\sqrt{\pi}}{3}\left[\frac{\Gamma(\frac{3}{5})}{\Gamma(\frac{1}{10})}\right]^5 \frac{N_c^3}{l^4} + \dots \qquad (3.3.10)$$

which is precisely the entanglement entropy of $N_c$ coincident $M5$-branes compactified on a circle of radius $R_4$ and the M-theory circle of radius $R_{10} = g_s$ found in [66, 67].

A naive use of (3.3.8), (3.3.9) suggests that the disconnected solution becomes the lower entropy one at $l \sim R^{\frac{3}{2}}/U_0 \sim R_4$, not far from $l_{\max}$ (3.3.7). Of course, the small $l$ approximation leading to (3.3.9) is not valid there, and in order to determine the precise position of the transition we need to evaluate (3.2.9). The result of that evaluation is shown in figure 2, where we also exhibit the entropy of the disconnected solution and, for completeness, that of the saddle point discussed in section 2 as well.

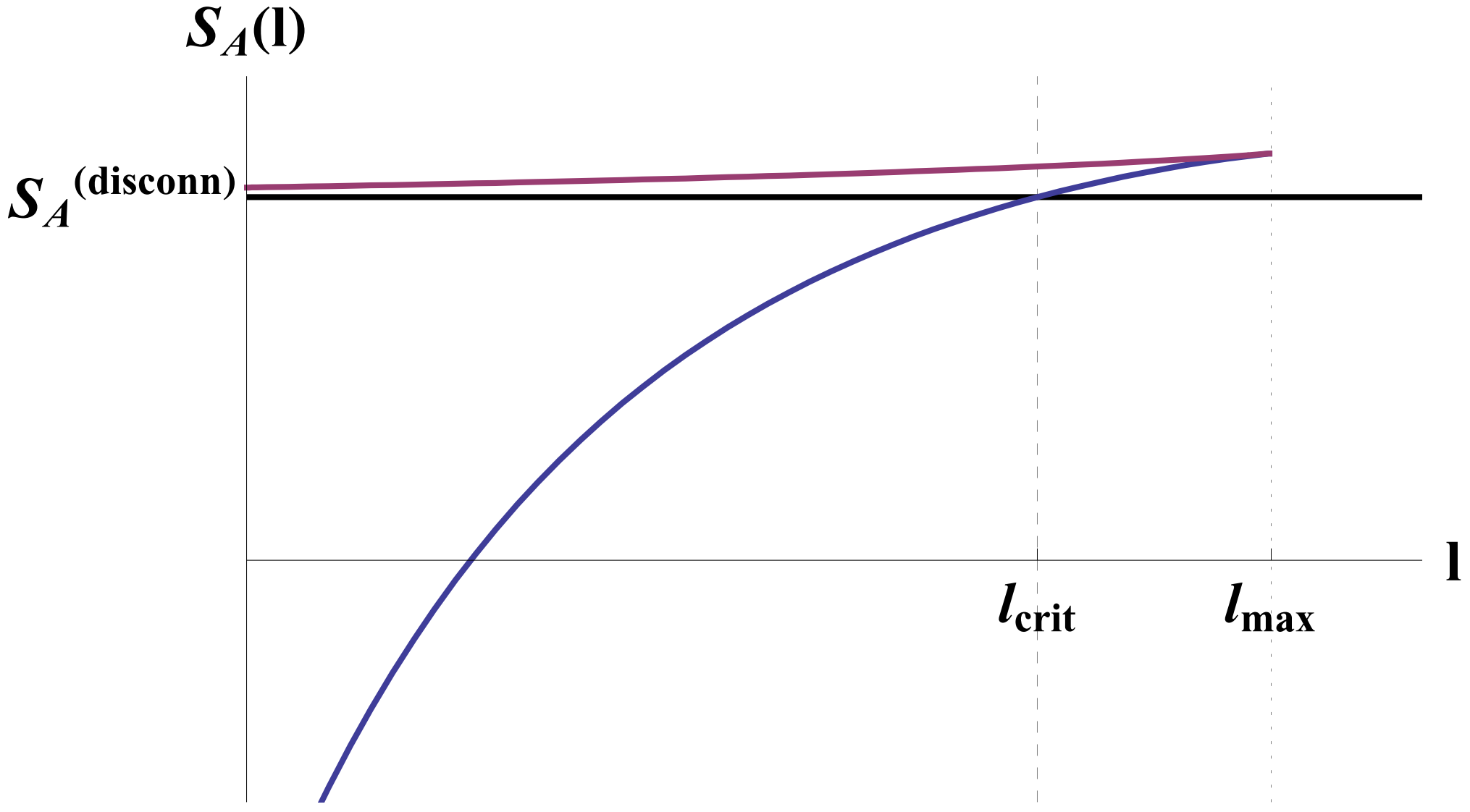

Figure 3.2: Entropies of the connected (blue and red) and disconnected (black) solutions for the wrapped $D4$-brane geometry.

We see that, as expected, the saddle point entropy is larger than that of the connected and disconnected local minima for all $l$. It approaches that of the connected one as $l \to l_{\max}$, and the disconnected one as $l \to 0$. The entropies of the connected and disconnected solutions cross at $l = l_{\rm crit} < l_{\rm max}$ given by

$$l_{\rm crit} \simeq 1.288 R_4 \ . \tag{3.3.11}$$

As explained in section 2, the entropy is governed by the connected solution and exhibits non-trivial dependence on $l$ for $l < l_{\rm crit}$, while for $l > l_{\rm crit}$ it is governed by the disconnected one and is $l$ independent (to leading order in $1/N_c$).

## 3.4 D3-branes on a circle

In this section we study the system of $N_c$ $D3$-branes wrapped around a circle of radius $R_3$ with twisted boundary conditions for the fermions. The discussion is largely parallel to that of the previous section, so we will be brief.

Before the compactification, the low energy theory on the $D3$-branes is $N = 4$ SYM with 't Hooft coupling $\lambda = g_s N_c$. For finite $R_3$ one finds at long distances a $2+1$ dimensional confining theory. For $\lambda \ll 1$ that theory is $2+1$ dimensional YM with 't Hooft coupling $\lambda_3 = \lambda/R_3$ [31]. For $\lambda \gg 1$ one can instead use a gravitational description in terms of the near-horizon geometry of the $N_c$ D3-branes,

$$ds_{10}^2 = \left(\frac{U}{L}\right)^2 \left[\left(\frac{L}{U}\right)^4 \frac{dU^2}{h(U)} + dx^\mu dx_\mu\right] + L^2 d\Omega_5^2 + \left(\frac{U}{L}\right)^2 h(U)(dx^3)^2 \ , \tag{3.4.1}$$

$$h(U) = 1 - \left(\frac{U_0}{U}\right)^4 \ , \tag{3.4.2}$$

where

$$L^4 = 4\pi\lambda \ , \qquad U_0^2 = \frac{\pi\lambda}{R_3^2} \ , \tag{3.4.3}$$

and the dilaton is constant, $\phi(U) = 0$. Comparing (3.4.1) to (3.2.1) we find

$$\alpha = \left(\frac{U}{L}\right)^2 , \quad \beta = \left(\frac{L}{U}\right)^4 \frac{1}{h(U)} , \quad V_{\rm int} = 2\pi^4 R_3 L^4 U \sqrt{h(U)} . \tag{3.4.4}$$

The combination (3.2.4) is given by

$$H(U) = (2\pi^4 R_3)^2 L^4 U^6 h(U) . \tag{3.4.5}$$

It is again monotonically increasing with $U$, as expected.

All the calculations of the previous section can be done in this case as well. The integral (3.2.8) is again convergent. For small $l$ (and large $U^*$) one finds

$$l(U^*) = 2\sqrt{\pi}\frac{\Gamma(\frac{2}{3})}{\Gamma(\frac{1}{6})}\frac{L^2}{U^*} . \tag{3.4.6}$$

The extension to all $U^*$ is plotted in figure 3. The qualitative shape of $l(U^*)$ is similar to the $D4$-brane case shown in figure 1. The maximum occurs at $U^* \simeq 1.113 U_0$, and

$$l_{\rm max} \simeq 1.383 R_3 . \tag{3.4.7}$$

The entropy of the disconnected solution is given by

$$S_A^{(\rm disconn)} = \frac{\pi^4 R_3 L^4 V_1}{2G_N^{(10)}} \left(U_\infty^2 - U_0^2\right) , \tag{3.4.8}$$

where $V_1$ is the length of the strip. The entropy of the connected solution is exhibited in figure 4. For small $l$ one has

$$S_A^{(\rm conn)}(l) = \frac{\pi^4 R_3 L^4 V_1}{2G_N^{(10)}} \left(U_\infty^2 - 4\left[\frac{\sqrt{\pi}\Gamma(\frac{2}{3})}{\Gamma(\frac{1}{6})}\right]^3 \frac{L^4}{l^2}\right) . \tag{3.4.9}$$

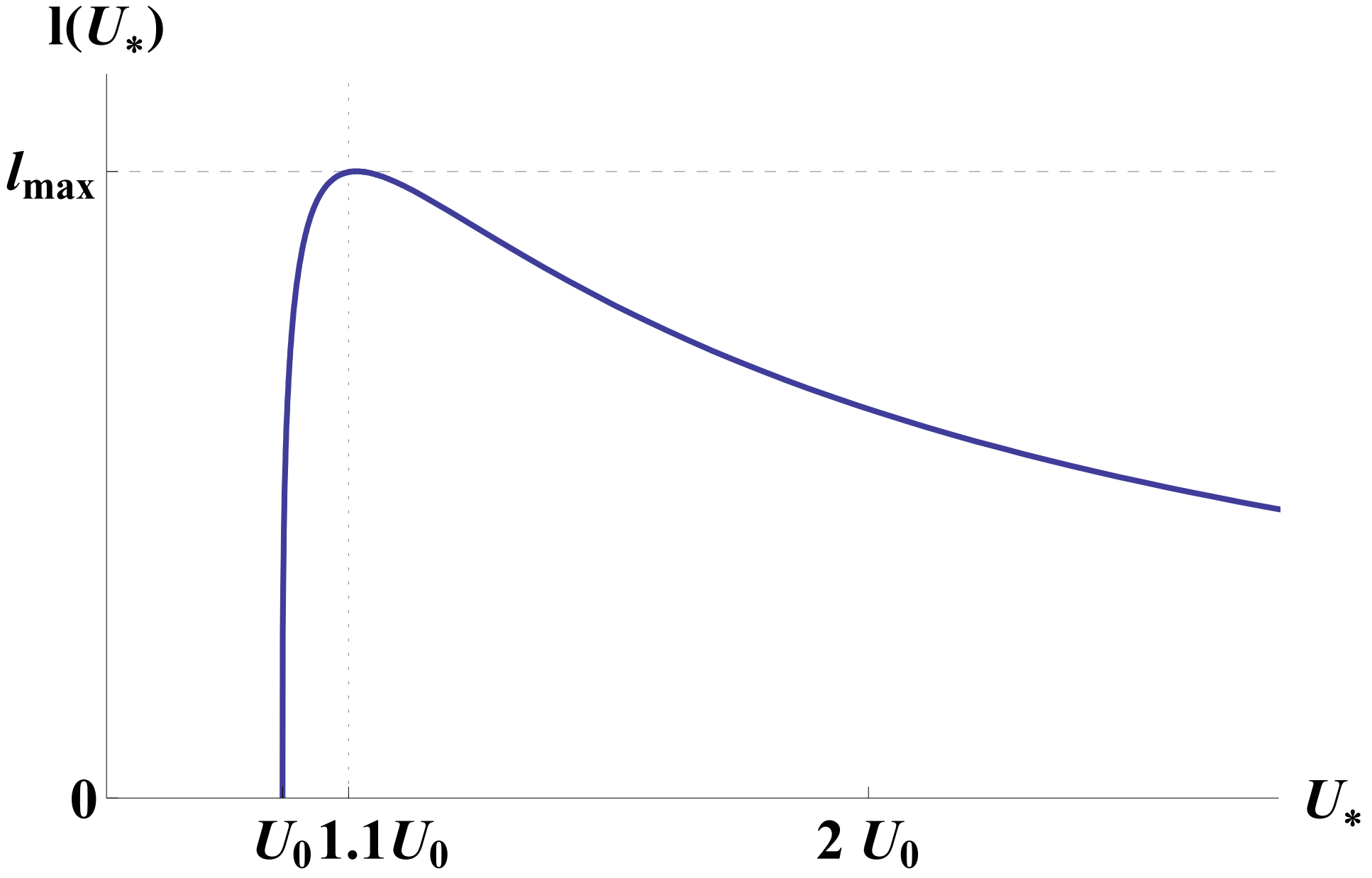

Figure 3.3: $l(U_*)$ for $D3$-branes on a circle.

Therefore, for small $l$ we find

$$S_A^{(\text{conn})}(l) - S_A^{(\text{disconn})} = -2\sqrt{\pi}\left[\frac{\Gamma(\frac{2}{3})}{\Gamma(\frac{1}{6})}\right]^3 V_1(2\pi R_3)\frac{N_c^2}{l^2} + \ldots \tag{3.4.10}$$

which is the entanglement entropy of the $3+1$ dimensional $\mathcal{N}=4$ SYM theory compactified on a circle of radius $R_3$ [66, 67].

As is clear from figure 4, the transition between the connected and disconnected solutions happens again at a value of $l$ smaller than $l_{\max}$. The numerical evaluation gives

$$l_{\text{crit}} \simeq 1.2376 R_3\ . \tag{3.4.11}$$

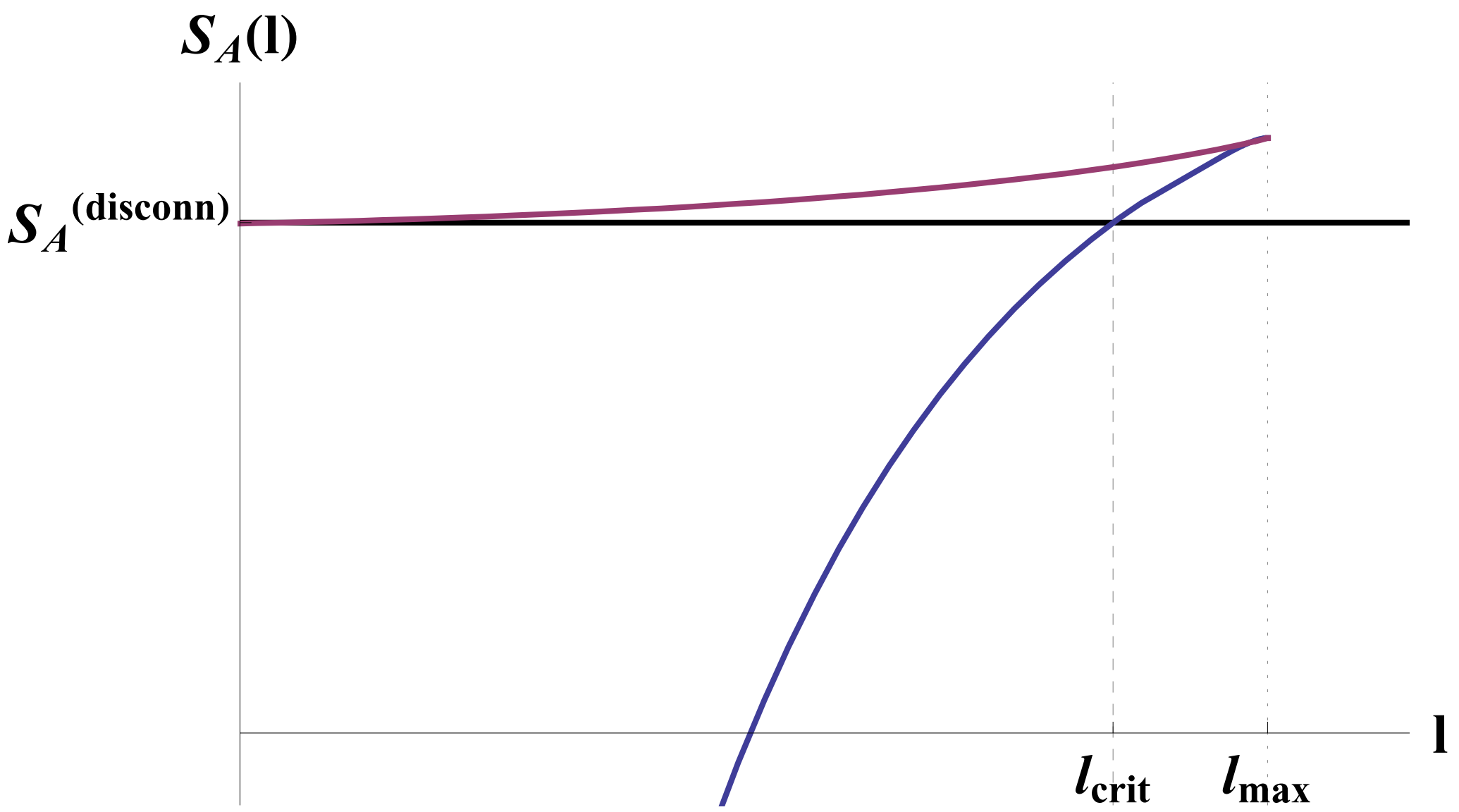

Figure 3.4: Entropies of the connected (blue and red) and disconnected (black) solutions for the wrapped $D3$-brane geometry.

## 3.5 Cascading Confining Gauge Theory

The background dual to the cascading $SU(M(k+1))\times SU(Mk)$ supersymmetric gauge theory is the deformed conifold $\sum_{i=1}^{4} z_i^2 = \epsilon^2$ warped by $M$ units of RR 3-form flux. The relevant metric is [25],

$$ds_{10}^2 = h^{-1/2}(\tau)dx^\mu dx_\mu + h^{1/2}(\tau)ds_6^2 \ , \tag{3.5.1}$$

where $ds_6^2$ is the metric of the deformed conifold

$$ds_6^2 = \frac{1}{2}\epsilon^{4/3}K(\tau)\Bigg[\frac{1}{3K^3(\tau)}(d\tau^2 + (g^5)^2) + \cosh^2\left(\frac{\tau}{2}\right)[(g^3)^2 + (g^4)^2] + \sinh^2\left(\frac{\tau}{2}\right)[(g^1)^2 + (g^2)^2]\Bigg] \ . \tag{3.5.2}$$

Here

$$K(\tau) = \frac{(\sinh(2\tau) - 2\tau)^{1/3}}{2^{1/3}\sinh\tau} \ , \tag{3.5.3}$$

and the warp factor is given by

$$h(\tau) = (g_s M\alpha')^2 2^{2/3}\epsilon^{-8/3}\int_\tau^\infty dx\frac{x\coth x - 1}{\sinh^2 x}(\sinh(2x) - 2x)^{1/3}\ . \tag{3.5.4}$$

The dilaton is constant and we set it to zero. For the details of the angular forms $g_i$, see [25, 27].

The cascading gauge theory has a continuous parameter, $g_s M$. The theory approaches the pure $SU(M)$ SYM theory in the limit $g_s M \to 0$, while the dual supergravity description is reliable in the opposite limit, $g_s M \to \infty$. In this limit the geometry describes a gauge theory with two widely separated scales: the scale of glueball masses,

$$m_{\rm glueball} = \frac{\epsilon^{2/3}}{g_s M\alpha'}\ , \tag{3.5.5}$$

and the scale of the string tension at the IR end of space (the tip of the cigar), $\sqrt{T_s} \sim \sqrt{g_s M} m_{\rm glueball}$.

The metric (3.5.1), (3.5.2) is of the form (3.2.1) with

$$\alpha \equiv h^{-1/2}, \quad \beta \equiv \frac{h(\tau)\epsilon^{4/3}}{6K^2(\tau)}\ . \tag{3.5.6}$$

Using $\int g_1 \wedge g_2 \wedge g_3 \wedge g_4 \wedge g_5 = 64\pi^3$, we get

$$V_{\rm int} = \frac{4\pi^3}{\sqrt{6}}\ h^{5/4}\epsilon^{10/3} K \sinh^2(\tau)\ . \tag{3.5.7}$$

Thus, all the general formulae of section 2 apply, with $U$ replaced by the standard deformed conifold radial variable $\tau$.

We find

$$H(\tau) = e^{-4\phi} V_{\rm int}^2 \alpha^3 = \frac{8\pi^6}{3}\epsilon^{20/3} h(\tau) K^2(\tau)\sinh^4(\tau)\ . \tag{3.5.8}$$

$H$ can be seen to be monotonically increasing with $\tau$ as noted in section 3.2 from general considerations. The general equation (3.2.8) with these identifications gives $l(\tau_*)$ for the KS background. As in the previous sections, the integral is convergent. For large $\tau^*$, we can approximate $l(\tau)$ using the asymptotic forms valid at large $\tau$,

$$\begin{aligned} h(\tau) &\rightarrow 2^{1/3}3\,(g_s M\alpha')^2\epsilon^{-8/3}\left(\tau-\frac{1}{4}\right)e^{-4\tau/3}\,, \qquad K\rightarrow 2^{1/3}e^{-\tau/3}\,, \\ H(\tau) &\rightarrow \pi^6\epsilon^4(g_s M\alpha')^2\left(\tau-\frac{1}{4}\right)e^{2\tau}\,, \quad \sqrt{\beta}\rightarrow 2^{-2/3}\epsilon^{-2/3}(g_s M\alpha')\sqrt{\tau-\frac{1}{4}}e^{-\tau/3}\,. \end{aligned} \tag{3.5.9}$$

This leads to the simplified expression,

$$l(\tau^*) = 2^{1/3}\epsilon^{-2/3}g_s M\alpha'\int_{\tau^*}^{\infty}\frac{\sqrt{\tau}e^{-\tau/3}d\tau}{\sqrt{\frac{\tau e^{2\tau}}{\tau^* e^{2\tau^*}}-1}}\,. \tag{3.5.10}$$

The main contribution is from the region $\tau\sim\tau^*$; shifting $\tau\rightarrow\tau_*+y$ and keeping the lowest order term in $y$ we conclude that for large $\tau^*$,

$$l(\tau^*) = \frac{2^{1/3}3\sqrt{\pi}\Gamma(2/3)}{\Gamma(1/6)}\epsilon^{-2/3}g_s M\alpha'\sqrt{\tau^*}e^{-\tau^*/3} \tag{3.5.11}$$

As earlier, $l$ goes to zero as $\tau^*\rightarrow\infty$. One can also verify, as outlined in section 3.2, that as $\tau^*\rightarrow 0$, $l$ goes to zero again. The full curve, computed numerically, is presented in figure 3.5. We see that it shows the same qualitative behavior as the other cases (figures 1,3). The maximum occurs at $\tau^*\approx 2.1$ with

$$l_{\max}\approx 1.00\ \ m_{\text{glueball}}^{-1}\,. \tag{3.5.12}$$

We now turn to the entanglement entropy $S_A$. As earlier, we have to calculate and compare the entropies of the connected (3.2.9) and disconnected (3.2.12) surfaces. As discussed in section 2, each of these entropies must be computed with a UV cut-off in place, but the difference of the entropies is UV finite. The entropy of the disconnected solution is found to be

$$S_A^{(\mathrm{disconn})} = V_2 \frac{M^2\epsilon^{4/3}}{2^{2/3}16\pi^3\alpha'^2} \left(\frac{3}{2}\tau_\infty e^{2\tau_\infty/3} - \frac{21}{8}e^{2\tau_\infty/3} + 2.194\right) \quad (3.5.13)$$

where the finite additive constant was computed numerically. For the connected solution, we first consider an analytic approximation valid for small $l$:

$$S_A(\tau^*) = V_2 \frac{M^2\epsilon^{4/3}}{2^{2/3}16\pi^3\alpha'^2} \int_{\tau^*}^{\tau_\infty} \frac{(\tau - 1/4)^{3/2} e^{5\tau/3} d\tau}{\sqrt{(\tau - 1/4)e^{2\tau} - (\tau^* - 1/4)e^{2\tau^*}}} . \quad (3.5.14)$$

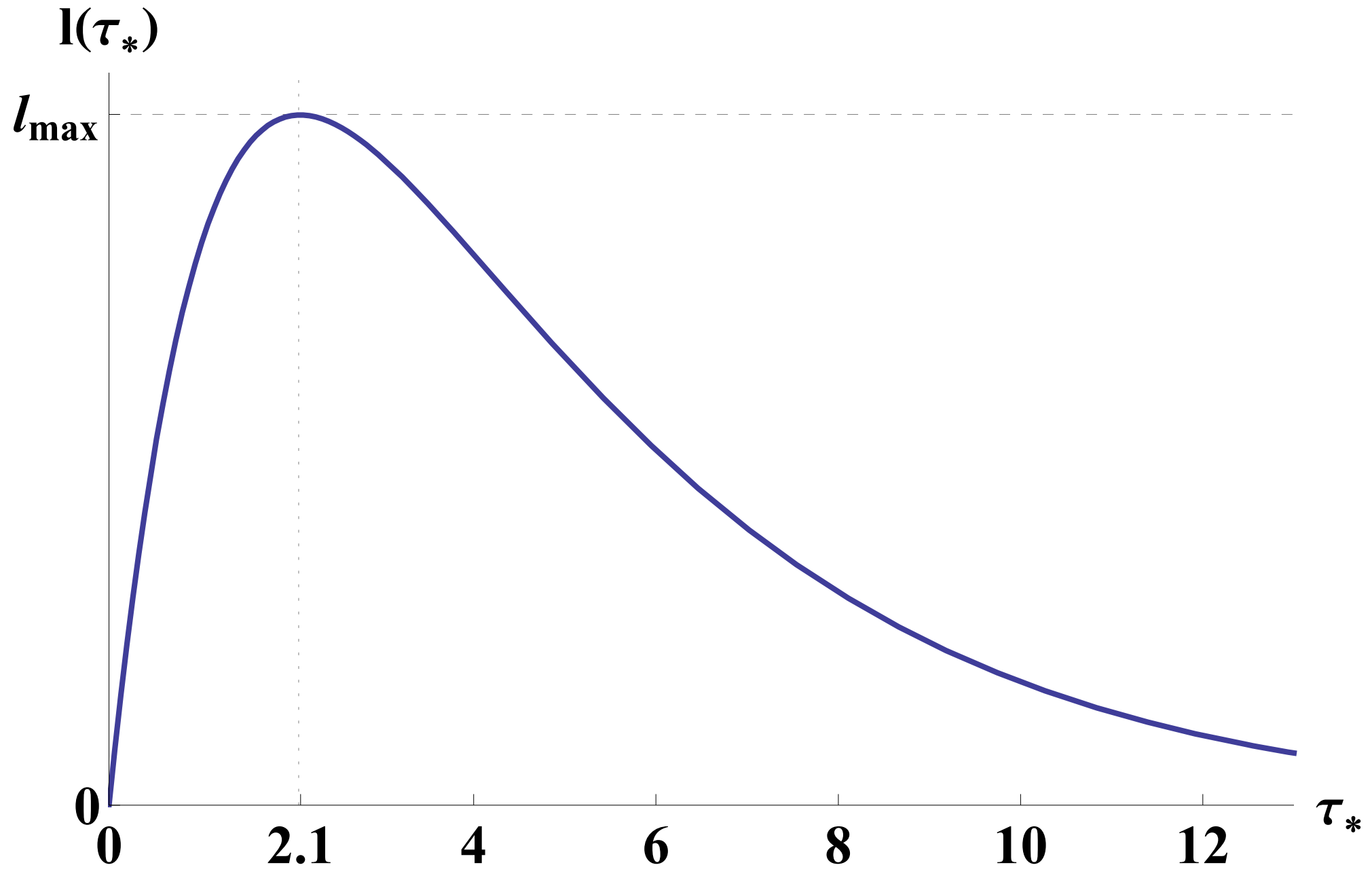

Figure 3.5: $l(\tau_*)$ for the KS geometry.

Approximating this integral as we did for $l(\tau^*)$, we find

$$S_A(\tau^*) = V_2 \frac{M^2\epsilon^{4/3}}{2^{2/3}16\pi^3\alpha'^2} M^2\epsilon^{4/3} \left(\frac{3}{2}\tau_\infty e^{2\tau_\infty/3} - \frac{21}{8}e^{2\tau_\infty/3} - \frac{3\sqrt{\pi}\Gamma(2/3)}{2\Gamma(1/6)}\tau^* e^{2\tau^*/3}\right).$$

Thus, for $l \ll 1/m_{\text{glueball}}$,

$$S_A^{(\text{conn})} - S_A^{(\text{disconn})} = -V_2 \frac{243\Gamma\left(\frac{2}{3}\right)^3}{32\pi^{3/2}\Gamma\left(\frac{1}{6}\right)^3} \frac{g_s^2 M^4}{l^2} \log^2(m_{\text{glueball}} l) + \dots \ . \tag{3.5.15}$$

In the cascading theory the effective number of colors is a logarithmic function of the distance scale [54, 25, 27]:

$$N_{\text{eff}}(l) = \frac{3}{2\pi} g_s M^2 \log(m_{\text{glueball}} l) + \dots \ . \tag{3.5.16}$$

We see that the finite piece of the entropy is

$$-V_2 \frac{27\sqrt{\pi}\Gamma\left(\frac{2}{3}\right)^3 N_{\text{eff}}^2(l)}{8\Gamma\left(\frac{1}{6}\right)^3 l^2} + \dots \ . \tag{3.5.17}$$

For a $3+1$ dimensional conformal gauge theory, the finite piece of the entanglement entropy is indeed of the form $N_c^2(V_2/l^2)$. Following [66, 67], we may use a minimal surface in $AdS_5 \times T^{11}$ to find the entanglement entropy in the dual $SU(N) \times SU(N)$ SCFT [18]:[4]

$$-V_2 \frac{27\sqrt{\pi}\Gamma\left(\frac{2}{3}\right)^3 N^2}{8\Gamma\left(\frac{1}{6}\right)^3 l^2} \ . \tag{3.5.18}$$

Hence, the result (3.5.17) we find for the cascading gauge theory is a reasonably modified form of the conformal behavior. The same distance-dependent effective number of colors was found in evaluation of correlation functions in the cascading theory [80, 81].

Going beyond the small $l$ limit, we present the result of the numerical evaluation of $S_A$ in figure 3.5, which compares the connected, disconnected and saddle point entropies. As expected, the saddle point entropy is always the largest and approaches the disconnected solution for small $l$ and the connected solution as $l \to l_{\max}$. The connected solution has the lowest entropy for small $l$ and is the dominant contribution in this regime. The point at which the connected and disconnected solutions cross is $l_{\rm crit} < l_{\rm max}$, which is found to be

$$l_{\rm crit} \approx 0.95 \ \ m_{\rm glueball}^{-1} \ . \tag{3.5.19}$$

[4]The extra factor of 27/16 compared to the result (3.4.10) for $AdS_5 \times S^5$ comes from the fact that $\text{vol}(T^{11}) = \frac{16}{27}\text{vol}(S^5)$.

For $l > l_{\rm crit}$, the $O(N_c^2)$ entropy is $l$-independent as explained in section 3.2.

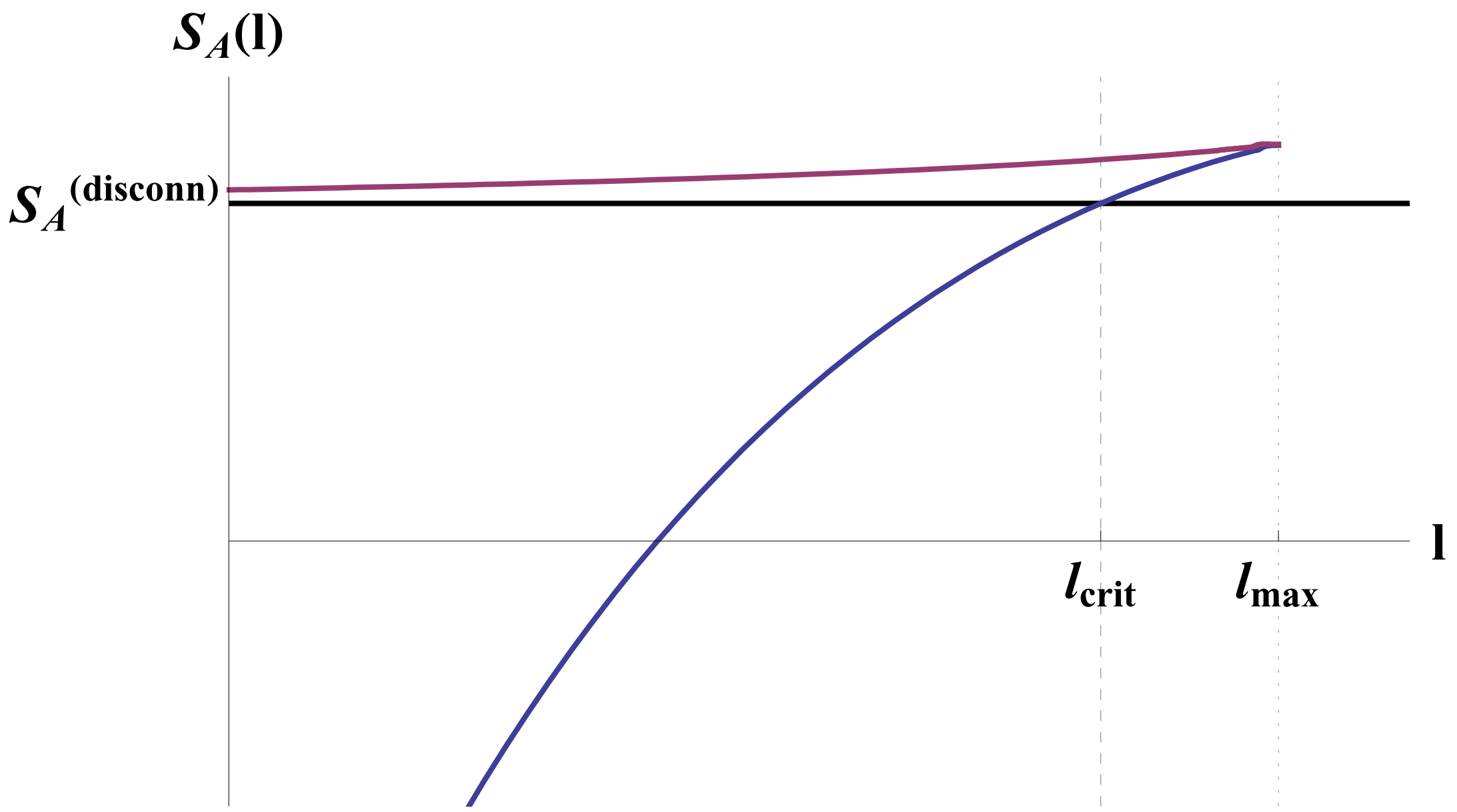

Figure 3.6: Entropies of the connected (blue and red) and disconnected (black) solutions for the KS geometry.

## 3.6 Comparison to field theory

It is natural to ask whether the transition at finite $l$ that we found in confining gravitational backgrounds also occurs in large $N_c$ asymptotically free gauge theories, such as pure YM or $\mathcal{N} = 1$ SYM with gauge group $SU(N_c)$. The location of such a transition would have to be around the QCD scale, $l_{\rm crit}\Lambda_{QCD} \sim 1$. At such scales the theory is strongly coupled and it is difficult to evaluate the entanglement entropy $S_A$ directly.

To proceed one can use the fact that, at large $N_c$, confining gauge theories are expected to reduce to free field theories of glueballs, whose density of states grows like

$$\rho(m) \simeq m^{\alpha} e^{\beta_H m} \tag{3.6.1}$$

at large mass $m$. The inverse Hagedorn temperature $\beta_H$ is of order $1/\Lambda_{QCD}$, and $\alpha$ is a constant. Both are difficult to calculate from first principles. Most of the states that contribute to (3.6.1) are unstable resonances whose width goes to zero as $N_c \to \infty$. More generally, all interactions between the glueballs go to zero in this limit. At finite $N_c$ the spectrum (3.6.1) is effectively cut off at some large mass scale.

We can use the above picture to calculate the entanglement entropy at large $N_c$, by summing the contributions of the glueballs. To avoid UV divergences, we will consider the quantity

$$C = l \frac{dS_A(l)}{dl} \tag{3.6.2}$$

which, as mentioned in the previous sections, does not depend on the UV cutoff. Consider, for example, a free scalar field of mass $m$. It is clear that the non-trivial dependence of (3.6.2) on $l$ is via the combination $ml$. We will be interested in the region $ml \gg 1$, where $C(ml)$ can be calculated as follows. In $1+1$ dimensions, the large $l$ form of $C(ml)$ has been obtained in [82]; it is given by

$$C_1(ml) = \frac{ml}{4} K_1(2ml) \simeq \frac{\sqrt{\pi m l}}{8} e^{-2ml} \tag{3.6.3}$$

A four-dimensional free scalar field can be thought of as an infinite collection of two-dimensional ones, labeled by the momentum in the transverse $\mathbb{R}^2$, $\vec{k}$, with mass $m(\vec{k}) = \sqrt{m^2 + \vec{k}^2}$. Summing over these momentum modes we find the 3+1 dimensional version

of (3.6.3),

$$C_3(ml) = \frac{V_2}{(2\pi)^2} \int d^2\vec{k} C_1(m(\vec{k})l) \simeq \frac{V_2}{32\pi} \frac{\sqrt{\pi} m^2}{\sqrt{ml}} e^{-2ml} \ . \qquad (3.6.4)$$

We see that the contribution of a single scalar field to the entanglement entropy is exponentially suppressed at large mass.[5] This is similar to the exponential suppression of its contribution to the canonical partition sum at finite temperature, with the role of the inverse temperature $\beta$ played here by $2l$.

For a theory with a Hagedorn spectrum (3.6.1) of bound states, the total entropy is obtained by summing over all states,

$$C_{\text{total}} = \int dm \rho(m) C_3(m) \sim \int^{\infty} dm m^{\beta} e^{(\beta_H - 2l)m} \qquad (3.6.5)$$

The integral converges for $l > \beta_H/2$ and diverges otherwise. This is the analog of the usual Hagedorn divergence of the canonical partition sum at the Hagedorn temperature. There, the physical picture is that for temperatures below some critical temperature, that is believed to be somewhat below the Hagedorn one [83, 84], the system is in the confining phase and the thermal free energy scales like $N_c^0$. Above that temperature, the system is in a deconfined phase and the free energy scales like $N_c^2$.

Similarly, for the entanglement entropy in gauge theory we expect that for $l$ above some $l_{\text{crit}}$ that is somewhat larger than $\beta_H/2$ the entanglement entropy is of order $N_c^0$ and is given by the convergent integral (3.6.5), while for $l < l_{\text{crit}}$ the entropy is of order $N_c^2$, in agreemeent with the divergence of (3.6.5).

The resulting picture is qualitatively similar to what we got from the gravity analysis in sections 2 – 5. Of course, as usual, the details are expected to differ because in the gravity regime the theory contains two widely separated scales. One is the scale of the lightest glueball masses, which goes like $1/R_4$ in the $D4$-brane analysis of section 3, like $1/R_3$ in that of section 4, and like $m_{\rm glueball}$ (3.5.5) in the KS geometry. The other is the scale of massive string excitations living near the tip of the cigar, $\sqrt{T_s}$, which is parametrically higher than the glueball scale. Since the exponential density of states comes from these string modes, we expect $\beta_H$ to be of order $T_s^{-1/2}$.

[5]The same is true for fermions and other higher spin fields.

The transition point $l_{\rm crit}$ in the gravity regime is instead determined by the inverse of the lightest glueball mass, and is parametrically larger than the Hagedorn scale $T_s^{-1/2}$. Thus, as we decrease $l$, the transition at $l = l_{\rm crit}$ to entangelement entropy of order $N_c^2$ happens long before $\beta_H$, $l_{\rm crit} \gg \beta_H$. For example, in the cascading theory $l_{\rm crit}/\beta_H \sim \sqrt{g_s M}$.

In the asymptotically free field theory regime, there is a single scale $\Lambda_{QCD}$ and everything happens around it. One can interpolate between the two regimes by tuning the 't Hooft coupling (e.g. making $g_s M$ small in the KS example). Our results suggest that no phase transition is encountered along such an interpolation – the two regimes are in the same universality class.

The arguments presented above apply directly to large $N_c$ theories. It would be interesting to investigate whether the phase transition we found continues to exist at finite $N_c$, and to characterize its order. Studying the entanglement entropy in pure glue $SU(N_c)$ lattice gauge theory would therefore be very interesting.

## 3.7 Discussion

In this chapter we applied the holographic method for calculating the entanglement entropy, introduced in [66, 67], to confining theories with gravity duals. In the simple case of entanglement between a strip of width $l$ and its complement, we found an interesting phase transition as a function of $l$: for $l < l_{\rm crit}$ the entropy is dominated by the action of a connected surface, while for $l > l_{\rm crit}$ by that of a disconnected one. After a subtraction of an $l$-independent UV divergent contribution, we conclude that the entropy is $O(N_c^2)$ for $l < l_{\rm crit}$ and $O(1)$ for $l > l_{\rm crit}$. This transition is qualitatively similar to the confinement/deconfinement transition at finite temperature.

Studying the thermal phase transition in confining gravitational backgrounds requires finding a SUGRA solution with an event horizon, and comparing its action with that of another solution which is horizon-free but has the Euclidean time periodically identified [31]. In general, these calculations are complicated and require a considerable amount of numerical work (see, for example, [76–79]). Studying the qualitatively similar transition for the entanglement entropy is much simpler; instead of finding new SUGRA solutions, one needs to find locally stable surfaces in previously known backgrounds.

We also argued that a transition similar to the one we observed using the methods of [66,67] should occur in any confining large $N_c$ gauge theory. This reasoning, and the several examples we have presented, make it plausible that any consistent gravity dual of a confining theory has to exhibit this phase transition. This is a useful prediction for any confining gauge/gravity dual pairs that remain to be discovered.

The existence of the transition in the cases we have discussed is linked to $p$-cycles of the internal geometry that shrink in the IR. One could ask if this is the most general situation that results in the phase transition. As we showed, the monotonic function $H(U) = e^{-4\phi} V_{\rm int}^2 \alpha^d$ has to vanish at the IR "end of space," $U = U_0$. On the other hand, $\alpha(U_0)$ should be non-vanishing for the string to retain its tension in the IR. This seems to restrict us to the vanishing of $e^{-2\phi} V_{\rm int}$. Thus, we should consider models where there are shrinking cycles and/or $\phi$ diverges in the IR.

Curiously, one of the most widely used gravitational models of confinement [26], $AdS_5$ with a hard IR wall at $U = U_0$, exhibits neither of these phenomena because both $\phi$ and $V_{\rm int}$ are assumed to be constant. Therefore, for such a model the transition of the entanglement entropy does not seem to occur. This is not surprising, since the notion of the disconnected solution wrapping the entire geometry is not *a priori* well-defined in this case. A related problem is that the equations of motion are not satisfied at $U = U_0$, hence the boundary conditions are ambiguous there.

There may exist a definition of the boundary conditions that allows the disconnected solution and produces a phase transition of the entanglement entropy (an encouraging sign is that the thermal deconfining transition does take place in the hard-wall model [85]). Indeed, when the hard wall model was considered in [66, 67] the contribution from the part of the minimal surface lying along the cutoff horizon was not included in the calculation; hence, it was treated as a disconnected surface. Justifying such a prescription may be a good problem for the future.

Another popular phenomenological model is the "soft wall" model where space-time has the geometry of $AdS_5$, while $\phi(U)$ blows up in the IR [33]:

$$ds_5^2 = U^2 \left( U^{-4} dU^2 + dx^\mu dx_\mu \right) , \qquad \phi(U) = U^{-2} . \tag{3.7.1}$$

Here, there is no shrinking internal cycle but the blow-up of the dilaton causes $H(U)$ to rapidly approach zero at $U = 0$.[6] In general, if $H(U) \sim U^p e^{-k/U^q}$ as $U \to 0$, one finds a finite $l_{\max}$ (above which the connected solution does not exist) provided $\beta(U)$ has a pole of order $2q + 2$ or less at $U = 0$. One can show this by similar means to those employed in section 3.2 where only shrinking cycles were considered. In all the models considered in this chapter so far, $q = 0$ and $\beta$ had a pole of order less than 2. On the other hand, the soft wall model corresponds to $q = 2$ while $\beta(U) = 1/U^4$ and hence still satisfies the criterion for the existence of a finite $l_{\max}$. For the soft-wall model one finds that there is indeed a transition between the disconnected solution stretching from $U = 0$ to $U = \infty$ and the connected one that becomes unstable for $l > l_{\text{crit}}$ and stops existing at $l_{\max}$.

We see that the entanglement entropy may be useful as a simple test of holographic models of confinement. More amibitiously, it would be nice to show that, if the confining background satisfies the supergravity equations of motion (neither the hard-wall nor the soft-wall do), then there is a phase transition of the entanglement entropy.

[6]For the soft-wall model $\alpha(U) = U^2$, hence the string loses its tension at $U = 0$. However, the model is typically treated as a five dimensional field theory, so it is not clear if the string tension requirement needs to be imposed.

Finally, it is important to understand the underlying reasons for the success of the geometric method of [66,67]. This prescription is designed to capture only the leading, $O(N_c^2)$, term in the entanglement entropy. While it has a superficial similarity to probe brane calculations, it does not seem to be consistent to think of the bulk surface that appears in the construction as a brane. Indeed a brane with the worldvolume action (3.1.6) would have tension proportional to $1/g_s^2$, and would back-react on the geometry at leading order in $g_s$. In any case, branes with the right properties do not seem to exist (see e.g. [86, 87]). We need to formulate the problem in semiclassical gravity whose solution to leading order in $G_N^{(10)}$ is the minimization problem proposed in [66, 67]. Hopefully, this can pave the way to finding the $O(N_c^0)$ corrections to the entanglement entropy and comparing them with field theory.

# Chapter 4

# M2 branes

This chapter is based on the paper '$AdS_4/CFT_3$ – Squashed, Stretched and Warped' written in collaboration with I.R. Klebanov and T. Klose [88]. We use group theoretic methods to calculate the spectrum of short multiplets around the extremum of $\mathcal{N} = 8$ gauged supergravity potential which possesses $\mathcal{N} = 2$ supersymmetry and SU(3) global symmetry. Upon uplifting to M-theory, it describes a warped product of $AdS_4$ and a certain squashed and stretched 7-sphere. We find quantum numbers in agreement with those of the gauge invariant operators in the $\mathcal{N} = 2$ superconformal Chern-Simons theory recently proposed to be the dual of this M-theory background. This theory is obtained from the U($N$) × U($N$) theory through deforming the superpotential by a term quadratic in one of the superfields. To construct this model explicitly, one needs to employ monopole operators whose complete understanding is still lacking. However, for the U(2) × U(2) gauge theory we make a proposal for the form of the monopole operators which has a number of desired properties. In particular, this proposal implies enhanced symmetry of the U(2)×U(2) ABJM theory for $k = 1, 2$; it makes its similarity to and subtle difference from the BLG theory quite explicit.

## 4.1 Introduction and Summary

The AdS/CFT correspondence has been a very active field of research since the original papers [9, 11, 12] appeared in late 1997/ early 1998. However most work has focussed on the version that involves D3-branes and relates a 3+1 dimensional CFT to Type II B string theory on an $AdS_5$ compactification from 10 dimensions.

Much less progress was made on investigating an alternative case of the conjecture, relating $2+1$ dimensional CFTs to M-theory compactifications on $AdS_4$. Such constructions parallel the more common version involving stacks of D3-branes in 10 d spacetime by replacing them with M2-branes in 11 d spacetime. While the gravitational background described by a heavy stack of such M2-branes was clear, the dual worldvolume CFT was not understood until the work of Bagger and Lambert [89–91], and by Gustavsson [92]. They conjectured a certain $2+1$ dimensional superconformal Chern-Simons theory with the maximal $\mathcal{N}=8$ supersymmetry and manifest SO(8) R-symmetry as the world-volume theory of a few M2-branes (these papers were inspired in part by the ideas of [93,94]). The Bagger-Lambert-Gustavsson (BLG) 3-algebra construction was, under the assumption of manifest unitarity, limited to the gauge group SO(4). This BLG theory is conveniently reformulated as an $\mathrm{SU}(2)\times\mathrm{SU}(2)$ gauge theory with conventional Chern-Simons terms having opposite levels $k$ and $-k$ [95, 96]. While the extension to more general gauge groups at first appeared to be difficult, major progress was eventually achieved by Aharony, Bergman, Jafferis and Maldacena (ABJM) [97] who proposed a $\mathrm{U}(N)\times\mathrm{U}(N)$ Chern-Simons gauge theory with levels $k$ and $-k$ as a dual description of $N$ M2-branes placed at the singularity of $\mathbb{R}^8/\mathbb{Z}_k$. The $\mathbb{Z}_k$ acts by simultaneous rotation in the four planes; for $k>2$ this orbifold preserves only $\mathcal{N}=6$ supersymmetry. ABJM gave strong evidence that their Chern-Simons gauge theory indeed possesses this amount of supersymmetry, and further work in [98,99] provided confirmation of this claim. Furthermore, for $k=1,2$ the supersymmetry of the orbifold, and therefore of the gauge theory, is expected to be enhanced to $\mathcal{N} = 8$. This is not manifest in the ABJM theory. Generally, inclusion of monopole operators is expected to play a crucial role both in the enhancement of the supersymmetry and in describing the full spectrum of gauge invariant operators. Explicit construction of these monopoles in ABJM theory was initiated in [100] but not all properties required here have been established. We will make some comments on these monopole operators, although our explicit calculations will mostly refer to the $\mathrm{U}(2) \times \mathrm{U}(2)$ case. Without the use of monopole operators one can make at most $\mathcal{N} = 6$ supersymmetry manifest in theories with higher rank gauge groups. These theories were classified in [101].

The explicit formulation of highly supersymmetric theories on M2-branes raises hope that one can also formulate $\mathrm{AdS}_4/\mathrm{CFT}_3$ dualities with reduced supersymmetry. To this end one may consider orbifolds or orientifolds of the BLG and ABJM theories [102,103,98,104–107]. But it is also interesting to look for gauge theories that are dual to backgrounds that do not locally look like $AdS_4 \times S^7$. Recent steps in this direction were made in [108] where a dual to the $\mathcal{N} = 1$ supersymmetric squashing of the $S^7$ was proposed, and in [109–115] where $S^7$ was replaced by manifolds preserving $\mathcal{N} = 2$ or $\mathcal{N} = 3$ supersymmetry. In the present paper we continue the program begun in [98] (see also [116]) where an $\mathcal{N} = 2$ superpotential deformation of $k = 1, 2$ ABJM theory by a term quadratic in one of the bi-fundamental superfields was shown to create an RG flow leading to a new Chern-Simons CFT with $\mathcal{N} = 2$ supersymmetry and SU(3) global symmetry. This CFT was conjectured to be dual to Warner's $\mathrm{SU}(3) \times \mathrm{U}(1)_R$ invariant extremum [117] of the potential in the gauged $\mathcal{N} = 8$ supergravity [118]. This extremum was uplifted to an 11-dimensional warped $AdS_4$ background containing a 'squashed and stretched' 7-sphere [119] (this terminology suggested the title of our paper). This background is of the Englert type in that there is a 4-form field strength turned on in the 7-sphere directions [120]. As a result, it breaks parity (reflection of one world volume direction accompanied by $C_{IJK} \to -C_{IJK}$) and we will show that the parity is also broken in the gauge theory. The $\mathcal{N} = 2$ superconformal symmetry of this background facilitates the comparison, via the AdS/CFT map [9, 11, 12], of the $SU(3) \times U(1)_R$ quantum numbers and energies of supergravity fluctuations with those of the gauge invariant operators in the Chern-Simons CFT. One interesting feature of the gauge theory is that far in the IR the effective superpotential is sextic in the bi-fundamental chiral superfields. The marginality then requires that their $U(1)_R$ charges equal 1/3.

On the supergravity side, the analysis of the $SU(3) \times U(1)_R$ quantum numbers was initiated in [121], where some low-lying supermultiplets were constructed. It was noted that $\mathcal{N} = 2$ supersymmetry allowed for two alternative ways of assigning $SU(3) \times U(1)_R$ quantum numbers; however, the two U(1) embeddings were found to be essentially equivalent at the lowest level [121]. Indeed, in App. 4.B we will show that there is no difference between the two choices in the values of $m^2$ in $AdS_4$ corresponding to the lowest hypermultiplet studied in [121]. The only difference concerns the choice of branches in the square root formula entering the operator dimensions. However, working only at the level of superconformal symmetry alone and not doing an explicit KK reduction, we show that these two choices of assigning $SU(3) \times U(1)_R$ quantum numbers lead to completely different spectra at higher levels. It should be stressed that even though such a group theoretical method does not necessarily lead to a unique answer, it is a rather efficient tool to gain insights into the spectrum. The first assignment of charges, which we will call Scenario I, produces agreement with the proposed gauge theory. The second one, Scenario II, which for the lowest hypermultiplet was spelled out in [121], turns out not to agree with our gauge theory proposal. In general, the mass spectra resulting from the two scenarios are distinct and hence an explicit KK reduction could agree with only one of them. We show that when considering higher massive multiplets, Scenario II does not appear to give a spectrum characteristic of

KK reduction.[1]

In Sec. 4.2 we review the gauged supergravity analysis of multiplets from [121], extend this work to higher levels and scrutinize the differences between Scenarios I and II for grouping fluctuations into supermultiplets. In Sec. 4.3 we review the ABJM theory and its relevant deformation, emphasizing the important role of monopole operators. We show how the expected symmetries of the theory emerge for $N = 2$ for a certain form of these operators. In Sec. 4.4 we analyze the short multiplets of chiral operators, demonstrating agreement with the gauged supergravity. The general structure of $\mathcal{N} = 2$ supermultiplets, and their specific examples occurring in this theory, as well as some comments on the monopole operators are left for the Appendices.

## 4.2 Supergravity side

The supergravity background proposed in [98] as a dual to the mass deformed ABJM gauge theory (a related yet somewhat different proposal independently appeared in [116]) was first found by Warner [117] as one of several non-trivial extrema of the gauged $\mathcal{N} = 8$ SUGRA potential [118]. The vacuum of interest preserves $\mathcal{N} = 2$ SUSY and the global symmetry SU(3) $\times$ U(1) (broken down from SO(8)) and corresponds to a scalar and a pseudo-scalar of $\mathcal{N} = 8$ gauge supergravity acquiring VEVs. As a consequence, this background does not preserve parity.

The 11d uplift of this $AdS_4$ vacuum was found more recently, in [119], and studied further in [123]. The solution is not a simple Freund-Rubin direct product $AdS_4 \times X_7$ but instead the metric of $AdS_4$ is warped by a function $f(y)$ of the coordinates $y$ on the internal manifold $X_7$. $X_7$ itself is a ‘squashed and stretched’ $S^7$ [119]. As noted earlier, this background has an Englert type flux in the $S^7$ directions [120], which is another way of seeing the breaking of parity.

[1] In fact, some evidence for the correctness of scenario I has recently been obtained also from solving the minimally coupled scalar equation in the background under consideration [122].

To determine the SUGRA spectrum, one could in principle perform the $11 \to 4$ dimensional KK reduction on this warped, squashed and stretched space. By performing the KK reduction for modes of various $AdS_4$ spin, one can group the resulting particles into $\mathcal{N} = 2$ supermultiplets of definite energy. Such an analysis was performed for example in [124, 125] for Freund-Rubin vacua with $X_7 = M^{111}, Q^{111}$. We can avoid such an involved calculation for this warped spacetime since it is obtained at the end of a SUSY preserving RG flow from the $\mathcal{N} = 8$ theory. A similar analysis has been performed earlier in [121] for the $\mathrm{SU}(3) \times \mathrm{U}(1)_R$ case at hand and in [24] for the analogous case in $\mathrm{AdS}_5/\mathrm{CFT}_4$ for a gauge theory with $\mathrm{SU}(2) \times \mathrm{U}(1)$ symmetry. However, here we go beyond gauged supergravity and study the rearrangement of the massive KK modes of the $\mathcal{N} = 8$ theory into $\mathcal{N} = 2$ supermultiplets. In doing so, we find that of the two alternative charge assignments, only one (referred to as "Scenario I" below) leads to agreement with the proposed gauge dual, while the other ("Scenario II") does not appear to be characteristic of a KK reduction. Hence while both assignments are consistent at the level of symmetry, only the former is likely to be reproduced through an explicit KK reduction from $11 \to 4$ dimensions.

| Spin | Field | SO(8) irrep | SO(8) Dynkin labels |
|---|---|---|---|
| 2 | ${e_\mu}^a$ | $\mathbf{1}$ | $[0,0,0,0]$ |
| $\frac{3}{2}$ | ${\psi_\mu}^I$ | $\mathbf{8}_s$ | $[0,0,0,1]$ |
| 1 | ${A_\mu}^{IJ}$ | $\mathbf{28}$ | $[0,1,0,0]$ |
| $\frac{1}{2}$ | $\chi^{IJK}$ | $\mathbf{56}_s$ | $[1,0,1,0]$ |
| $0^+$ | $S^{[IJKL]_+}$ | $\mathbf{35}_v$ | $[2,0,0,0]$ |
| $0^-$ | $P^{[IJKL]_-}$ | $\mathbf{35}_c$ | $[0,0,2,0]$ |

Table 4.1: **The massless $\boldsymbol{\mathcal{N}} = 8$ supermultiplet.** All degrees of freedom of 11d supergravity form one supermultiplet. When compactified on a round seven-sphere this supermultiplet splits into a series of Osp(8|4) supermultiplets. This table lists the components of the lowest supermultiplet in this series.

| Spin | Field | SO(8) Dynkin labels |
|---|---|---|
| 2 | ${e_\mu}^a$ | $[n,0,0,0]$ |
| $\frac{3}{2}$ | ${\psi_\mu}^I$ | $[n,0,0,1]+[n-1,0,1,0]$ |
| 1 | ${A_\mu}^{IJ}$ | $[n,1,0,0]+[n-1,0,1,1]+[n-2,1,0,0]$ |
| $\frac{1}{2}$ | $\chi^{IJK}$ | $[n+1,0,1,0]+[n-1,1,1,0]+[n-2,1,0,1]+[n-2,0,0,1]$ |
| $0^+$ | $S^{[IJKL]_+}$ | $[n+2,0,0,0]+[n-2,2,0,0]+[n-2,0,0,0]$ |
| $0^-$ | $P^{[IJKL]_-}$ | $[n,0,2,0]+[n-2,0,0,2]$ |

Table 4.2: **The massive $\boldsymbol{\mathcal{N}} = 8$ supermultiplet at level $\boldsymbol{n}$.** Representations with negative labels are absent. For $n = 0$ the massless $\mathcal{N} = 8$ supermultiplet from Tab. 4.1 is recovered.

### 4.2.1 Spectrum on the stretched and squashed seven-sphere

The spectrum of $\mathcal{N} = 8$ supermultiplets obtained by KK reduction on the round $S^7$ is well-known [126, 127]. All multiplets are shortened and have maximum spin 2. The massless multiplet is shown in Tab. 4.1 while the $SO(8)_R$ representations that higher massive multiplets transform in is presented in Tab. 4.2.

Now we would like to find the spectrum on the deformed $S^7$. We do this by exploiting the restrictions on the spectrum due to the symmetries of the background. The strategy for this derivation is summarized in the following diagram:

$$\begin{array}{ccc} \mathrm{Osp}(8|4) & \xrightarrow[\text{squashing of } S^7]{\text{stretching and}} & \mathrm{SU}(3)\times\mathrm{Osp}(2|4) \\ \begin{smallmatrix}\text{decompose } \mathcal{N}=8\\ \text{supermultiplets}\end{smallmatrix}\Big\downarrow & & \Big\uparrow\begin{smallmatrix}\text{assemble } \mathcal{N}=2\\ \text{supermultiplets}\end{smallmatrix} \\ \mathrm{SO}(8)_R\times\mathrm{SO}(3,2) & \xrightarrow{\text{RG flow}} & \mathrm{SU}(3)\times\mathrm{U}(1)_R\times\mathrm{SO}(3,2) \end{array} \tag{4.2.1}$$

The $\mathrm{Osp}(8|4)$ supermultiplets are decomposed into irreducible representations of the bosonic subgroup $\mathrm{SO}(8)_R \times \mathrm{SO}(3,2)$ as already given in Tab. 4.1 and 4.2. This set of representations is then further decomposed into irreducible representations of the bosonic symmetry group $\mathrm{SU}(3) \times \mathrm{U}(1)_R \times \mathrm{SO}(3,2)$ of the IR theory. Finally, we re-assemble these bosonic multiplets into supermultiplets of $\mathrm{Osp}(2|4)$ with definite $\mathrm{SU}(3)$ representations. This procedure is carried out for every level $n$ separately.

The described method is applicable because the RG flow preserves the $\mathrm{Osp}(2|4) \subset \mathrm{Osp}(8|4)$ supersymmetry. The only thing we do not know is how $\mathrm{Osp}(2|4)$ is embedded into $\mathrm{Osp}(8|4)$, or how $\mathrm{SU}(3) \times \mathrm{U}(1)_R \times \mathrm{SO}(3,2)$ is embedded into $\mathrm{SO}(8)_R \times \mathrm{SO}(3,2)$. Therefore we will make a general ansatz for the latter embedding:

$$[a,b,c,d] \to [f,g]_h \ , \tag{4.2.2}$$

where $f$, $g$, $h$ are linear functions of the $\mathrm{SO}(8)_R$ Dynkin labels $a,b,c,d$. The functions $f$ and $g$ represent the $\mathrm{SU}(3)$ Dynkin labels, and the function $h$ is the $\mathrm{U}(1)_R$ charge. The $\mathrm{SO}(3,2)$ labels are given by the spin $s$ and the energy $E$. While the spin is unaltered during the flow, the energy can in general not be determined by group theoretical arguments alone. We can only find the energy for short multiplets where it is fixed by the values of the other labels.

| Spin | SO(8) | | SU(3) |
|---|---|---|---|
| 2 | $\mathbf{1}$ | $\to$ | $\mathbf{1}$ |
| $\frac{3}{2}$ | $\mathbf{8}_s$ | $\to$ | $\mathbf{3}+\bar{\mathbf{3}}+2\cdot\mathbf{1}$ |
| 1 | $\mathbf{28}$ | $\to$ | $\mathbf{8}+3\cdot\mathbf{3}+3\cdot\bar{\mathbf{3}}+2\cdot\mathbf{1}$ |
| $\frac{1}{2}$ | $\mathbf{56}_s$ | $\to$ | $2\cdot\mathbf{8}+\mathbf{6}+\bar{\mathbf{6}}+4\cdot\mathbf{3}+4\cdot\bar{\mathbf{3}}+4\cdot\mathbf{1}$ |
| $0^+$ | $\mathbf{35}_v$ | $\to$ | $\mathbf{8}+\mathbf{6}+\bar{\mathbf{6}}+2\cdot\mathbf{3}+2\cdot\bar{\mathbf{3}}+3\cdot\mathbf{1}$ |
| $0^-$ | $\mathbf{35}_c$ | $\to$ | $\mathbf{8}+\mathbf{6}+\bar{\mathbf{6}}+2\cdot\mathbf{3}+2\cdot\bar{\mathbf{3}}+3\cdot\mathbf{1}$ |

Table 4.3: **Decomposition of the massless $\mathcal{N}=8$ supermultiplet under SU(3).**

The functions in the ansatz (4.2.2) are restricted in the following way. First of all there are only three choices of canonical embeddings of SU(3) into SO(8) which are given by $[f,g]=[a,b]$ or $[b,c]$ or $[b,d]$. All three choices lead to the same decomposition if the R-charge is ignored; the result for the massless level is printed in Tab. 4.3. We can now fix the $U(1)_R$ charges as follows. Fields of different spin but same SU(3) representation in the decomposition of the $\mathcal{N}=8$ supermultiplet must all recombine into various $\mathcal{N}=2$ supermultiplets which we list in Tab. 4.8 to 4.16 in App. 4.A. This is only possible when the R-charges of the states that go into one $\mathcal{N}=2$ supermultiplet are correlated in the way given in the tables.

For example, there are only three fields in Tab. 4.3 in the sextet **6** of SU(3) – a spin 1/2 field and two scalars. (Recall that the deformation and the IR background are not parity invariant and hence the UV parity assignments should be ignored.) The only supermultiplet they can form is a hypermultiplet described by Tab. 4.16. This requires an R-charge assignment of the form[2] $y_0 \mp 1$ for the spin 1/2 field when the scalars are assigned $y_0, y_0 \mp 2$. When we repeat this multiplet-forming exercise for the other fields, this further constrains the embedding of $U(1)_R$ into $SO(8)_R$ until we are left with exactly two possibilities consistent with supersymmetry.

Doing this in a systematic way, we find that the two choices can be described in terms of the Dynkin labels of SO(8) as,

$$[a,b,c,d] \to \begin{cases} [a,b]_{\left(\frac{a}{3}+\frac{2b}{3}+d\right)\varepsilon} & \text{Scenario I ,} \\ [a,b]_{-\left(\frac{2a}{3}+\frac{4b}{3}+c+d\right)\varepsilon} & \text{Scenario II ,} \end{cases} \tag{4.2.3}$$

where $\varepsilon = \pm 1$, the two integers $[a,b]$ give the SU(3) Dynkin labels and the subscript is the $\mathrm{U}(1)_R$ charge. The choice of $\varepsilon = \pm 1$ is simply a flip of the $\mathrm{U}(1)_R$ definition. We note that the SU(3) embedding $[b,c]$ and $[b,d]$ lead to no consistent regrouping into $\mathcal{N}=2$ supermultiplets.

[2]We adopt the usual notation where the upper sign applies if $y_0 > 0$ and the lower one if $y_0 < 0$.

With the $\mathrm{SU}(3) \times \mathrm{U}(1)_R$ charges of Scenario I above, we proceed to group the fields into $\mathcal{N}=2$ supermultiplets. The result for fields from the $\mathcal{N}=8$ massless sector is in Tab. 4.4. We find from the table that the massless $\mathcal{N}=8$ multiplet yields some familiar massless $\mathcal{N}=2$ multiplets such as the massless graviton multiplet in a singlet under SU(3) and a massless vector multiplet in the adjoint of SU(3). The former is expected in any theory of SUGRA while the latter contains the massless bosons gauging the SU(3) symmetry in the bulk. We also find several other massive multiplets that acquired mass in the breaking $\mathrm{SO}(8) \to \mathrm{SU}(3) \times \mathrm{U}(1)_R$. When massless particles of spin 1 or greater acquire a mass, they need to 'eat' spin 1/2 and spin 0 particles to furnish the extra polarizations. Hence when we find massive gravitinos in a **3** of SU(3) for example, we need to set aside some spin 1/2 triplets from Tab. 4.3 to be eaten and not group them into other multiplets. These are listed in the last column of Tab. 4.4.

Scenario II produces a different set of $\mathrm{U}(1)_R$ charges. The grouping of massless $\mathcal{N}=8$ fields into $\mathcal{N}=2$ multiplets is detailed in Tab. 4.5. The crucial differences between Tab. 4.5 and Tab. 4.4 are in the hyper- and long vector multiplets where a reassignment of R-charges leads to differing physical dimensions. For the hypermultiplet, Scenario II (Tab. 4.5) assigns the ground state a R-charge of $y_0 = -\frac{4}{3}$ and hence by Tab. 4.16, a dimension of $E_0 = |y_0| = \frac{4}{3}$. On the other hand, Scenario I (Tab. 4.4) results in the assignment $y_0 = \frac{2}{3}$ and $E_0 = |y_0| = \frac{2}{3}$. In App. 4.B we further observe that the energies of the hypermultiplet in the two scenarios can be related to the same mass spectrum when different dressings are used for the two different scenarios. However, this relationship holds only at level $n = 0$.

The massive multiplets of $\mathcal{N} = 8$, listed in Tab. 4.2 for $n = 1, 2, 3, \ldots$, are decomposed in a similar way for each of the two scenarios. We have delegated the details to the appendices in Tab. 4.17 through 4.23. We find several series of $\mathcal{N} = 2$ multiplets as we increase $n$ with different R-charges in the two different scenarios. To compare with the gauge theory, the short multiplets are the most interesting since their energy can be determined entirely from their R-charge. We collect the four distinct series of short multiplets that emerge from decomposing massive $\mathcal{N} = 8$ multiplets in Tab. 4.6 for the two scenarios. The SU(3) representations are given in terms of Dynkin labels, i.e. $[a, b]$ is the symmetric product of $a$ $\mathbf{3}$'s and $b$ $\bar{\mathbf{3}}$'s. The subscript again gives the $\mathrm{U}(1)_R$ charge. When $n = 0$, these multiplets are also found in Tables 4.5,4.4 discussed earlier.

We stress that the two scenarios arise as logical possibilities when one only works at the level of the symmetry breaking $\mathrm{SO}(8) \to \mathrm{SU}(3) \times \mathrm{U}(1)_R$ and one does not perform an explicit KK reduction to find the mass spectrum. The two scenarios correspond to two different embeddings of $\mathrm{U}(1)_R$ in SO(8). From Tab. 4.6, one can easily verify that set of masses resulting from the two scenarios are distinct (though this is not true when $n = 0$ as discussed in App. 4.B). For example, we note the unusual feature that in Scenario II, the short gravitons have the $n$-independent charge $[0, 0]_0$. This leads to $n$-independent mass of $m^2 = 0$ for the graviton. It would seem very unlikely that such an infinite sequence of zero masses can be obtained from a KK reduction. In contrast, Scenario I has masses that increase with $n$ for all the short series. Thus Scenario II is unlikely to be obtained from an explicit KK reduction and we conjecture that it is Scenario I that will agree with such a direct computation. Hence we will primarily

| Spin | SO(8) | $\mathrm{SU}(3)_{\mathrm{U}(1)}$ | | | | | | | |
|---|---|---|---|---|---|---|---|---|---|
| 2 | $\mathbf{1}$ | $\mathbf{1}_0$ | | | | | | | |
| $\frac{3}{2}$ | $\mathbf{8}_s$ | $\mathbf{1}_{+1}$<br>$\mathbf{1}_{-1}$ | | $\mathbf{3}_{\frac{\varepsilon}{3}}$ | $\bar{\mathbf{3}}_{-\frac{\varepsilon}{3}}$ | | | | |
| 1 | $\mathbf{28}$ | $\mathbf{1}_0$ | $\mathbf{8}_0$ | $\mathbf{3}_{\frac{4\varepsilon}{3}}$<br>$\mathbf{3}_{-\frac{2\varepsilon}{3}}$<br>$\mathbf{3}_{-\frac{2\varepsilon}{3}}$ | $\bar{\mathbf{3}}_{-\frac{4\varepsilon}{3}}$<br>$\bar{\mathbf{3}}_{\frac{2\varepsilon}{3}}$<br>$\bar{\mathbf{3}}_{\frac{2\varepsilon}{3}}$ | | | $\mathbf{1}_0$ | |
| $\frac{1}{2}$ | $\mathbf{56}_s$ | | $\mathbf{8}_{+1}$<br>$\mathbf{8}_{-1}$ | $\mathbf{3}_{\frac{\varepsilon}{3}}$<br>$\mathbf{3}_{\frac{\varepsilon}{3}}$<br>$\mathbf{3}_{-\frac{5\varepsilon}{3}}$ | $\bar{\mathbf{3}}_{-\frac{\varepsilon}{3}}$<br>$\bar{\mathbf{3}}_{-\frac{\varepsilon}{3}}$<br>$\bar{\mathbf{3}}_{\frac{5\varepsilon}{3}}$ | $\mathbf{6}_{-\frac{\varepsilon}{3}}$ | $\bar{\mathbf{6}}_{\frac{\varepsilon}{3}}$ | $\mathbf{1}_{-1}$<br>$\mathbf{1}_{+1}$<br>$\mathbf{1}_{-1}$<br>$\mathbf{1}_{+1}$ | $\mathbf{3}_{\frac{\varepsilon}{3}}$<br>$\bar{\mathbf{3}}_{-\frac{\varepsilon}{3}}$ |
| $0^+$ | $\mathbf{35}_v$ | | $\mathbf{8}_0$ | $\mathbf{3}_{-\frac{2\varepsilon}{3}}$ | $\bar{\mathbf{3}}_{\frac{2\varepsilon}{3}}$ | $\mathbf{6}_{\frac{2\varepsilon}{3}}$ | $\bar{\mathbf{6}}_{-\frac{2\varepsilon}{3}}$ | $\mathbf{1}_2$<br>$\mathbf{1}_0$<br>$\mathbf{1}_{-2}$ | $\mathbf{3}_{\frac{4\varepsilon}{3}}$<br>$\bar{\mathbf{3}}_{-\frac{4\varepsilon}{3}}$ |
| $0^-$ | $\mathbf{35}_c$ | | $\mathbf{8}_0$ | | | $\mathbf{6}_{-\frac{4\varepsilon}{3}}$ | $\bar{\mathbf{6}}_{\frac{4\varepsilon}{3}}$ | $\mathbf{1}_0$<br>$\mathbf{1}_0$ | $\mathbf{3}_{-\frac{2\varepsilon}{3}}$<br>$\mathbf{3}_{-\frac{2\varepsilon}{3}}$<br>$\bar{\mathbf{3}}_{\frac{2\varepsilon}{3}}$<br>$\bar{\mathbf{3}}_{\frac{2\varepsilon}{3}}$<br>$\mathbf{1}_0$ |
| | | Massless graviton | Massless vector | Massive short gravitino | Massive short gravitino | Massive hyper | Massive hyper | Massive vector | eaten |

Table 4.4: **Decomposition of massless $\mathcal{N} = 8$ supermultiplet: Scenario I.** $\varepsilon$ can be set to $\pm 1$.

| Spin | SO(8) | $\mathrm{SU}(3)_{\mathrm{U}(1)}$ | | | | | | | |
|---|---|---|---|---|---|---|---|---|---|
| 2 | $\mathbf{1}$ | $\mathbf{1}_0$ | | | | | | | |
| $\frac{3}{2}$ | $\mathbf{8}_s$ | $\mathbf{1}_{+1}$<br>$\mathbf{1}_{-1}$ | | $\mathbf{3}_{\frac{\varepsilon}{3}}$ | $\bar{\mathbf{3}}_{-\frac{\varepsilon}{3}}$ | | | | |
| 1 | $\mathbf{28}$ | $\mathbf{1}_0$ | $\mathbf{8}_0$ | $\mathbf{3}_{\frac{4\varepsilon}{3}}$<br>$\mathbf{3}_{-\frac{2\varepsilon}{3}}$<br>$\mathbf{3}_{-\frac{2\varepsilon}{3}}$ | $\bar{\mathbf{3}}_{-\frac{4\varepsilon}{3}}$<br>$\bar{\mathbf{3}}_{\frac{2\varepsilon}{3}}$<br>$\bar{\mathbf{3}}_{\frac{2\varepsilon}{3}}$ | | | $\mathbf{1}_0$ | |
| $\frac{1}{2}$ | $\mathbf{56}_s$ | | $\mathbf{8}_{+1}$<br>$\mathbf{8}_{-1}$ | $\mathbf{3}_{\frac{\varepsilon}{3}}$<br>$\mathbf{3}_{\frac{\varepsilon}{3}}$<br>$\mathbf{3}_{-\frac{5\varepsilon}{3}}$ | $\bar{\mathbf{3}}_{-\frac{\varepsilon}{3}}$<br>$\bar{\mathbf{3}}_{-\frac{\varepsilon}{3}}$<br>$\bar{\mathbf{3}}_{\frac{5\varepsilon}{3}}$ | $\mathbf{6}_{-\frac{\varepsilon}{3}}$ | $\bar{\mathbf{6}}_{\frac{\varepsilon}{3}}$ | $\mathbf{1}_{-1}$<br>$\mathbf{1}_{+1}$<br>$\mathbf{1}_{-1}$<br>$\mathbf{1}_{+1}$ | $\mathbf{3}_{\frac{\varepsilon}{3}}$<br>$\bar{\mathbf{3}}_{-\frac{\varepsilon}{3}}$ |
| $0^+$ | $\mathbf{35}_v$ | | $\mathbf{8}_0$ | $\mathbf{3}_{-\frac{2\varepsilon}{3}}$ | $\bar{\mathbf{3}}_{\frac{2\varepsilon}{3}}$ | $\mathbf{6}_{-\frac{4\varepsilon}{3}}$ | $\bar{\mathbf{6}}_{\frac{4\varepsilon}{3}}$ | $\mathbf{1}_0$<br>$\mathbf{1}_0$ | $\mathbf{3}_{-\frac{2\varepsilon}{3}}$<br>$\bar{\mathbf{3}}_{\frac{2\varepsilon}{3}}$<br>$\mathbf{1}_0$ |
| $0^-$ | $\mathbf{35}_c$ | | $\mathbf{8}_0$ | | | $\mathbf{6}_{\frac{2\varepsilon}{3}}$ | $\bar{\mathbf{6}}_{-\frac{2\varepsilon}{3}}$ | $\mathbf{1}_{+2}$<br>$\mathbf{1}_0$<br>$\mathbf{1}_{-2}$ | $\mathbf{3}_{\frac{4\varepsilon}{3}}$<br>$\mathbf{3}_{-\frac{2\varepsilon}{3}}$<br>$\bar{\mathbf{3}}_{-\frac{4\varepsilon}{3}}$<br>$\bar{\mathbf{3}}_{\frac{2\varepsilon}{3}}$ |
| | | Massless graviton | Massless vector | Massive short gravitino | Massive short gravitino | Massive hyper | Massive hyper | Massive vector | eaten |

Table 4.5: **Decomposition of massless $\mathcal{N} = 8$ supermultiplet : Scenario II.** $\varepsilon$ can be set to $\pm 1$.

work with Scenario I in this paper and compare it with proposed dual gauge theory.

| | Scenario I | Scenario II |
|---|---|---|
| Hyper | $[n+2,0]_{\frac{n+2}{3}}$, $[0,n+2]_{-\frac{n+2}{3}}$ | $[n+2,0]_{-\frac{2n+4}{3}}$, $[0,n+2]_{\frac{2n+4}{3}}$ |
| Vector | $[n+1,1]_{\frac{n}{3}}$, $[1,n+1]_{-\frac{n}{3}}$ | $[n+1,1]_{-\frac{2n}{3}}$, $[1,n+1]_{\frac{2n}{3}}$ |
| Gravitino | $[n+1,0]_{\frac{n+1}{3}}$, $[0,n+1]_{-\frac{n+1}{3}}$ | $[n+1,0]_{-\frac{2n-1}{3}}$, $[0,n+1]_{\frac{2n-1}{3}}$ |
| Graviton | $[0,0]_n$, $[0,0]_{-n}$ | $[0,0]_0$, $[0,0]_0$ |

Table 4.6: **Series of short multiplets in the two scenarios.** These are four series of short multiplets labeled by $n = 0, 1, 2, \ldots$. When $n = 0$, there is only one $[1,1]_0$ vector and one $[0,0]_0$ graviton, both of which are massless.

## 4.3 Gauge theory side

In this section we discuss the conjectured gauge theory dual to the supergravity background described above, i.e. the warped product of $\mathrm{AdS}_4$ and a squashed and stretched $S^7$. We provide evidence that this gauge theory is the IR limit of the ABJM theory [97] with a superpotential mass term for one of the superfields, as conjectured in [98] (see also [116]).

### 4.3.1 Review of ABJM theory

We begin with a brief recap of ABJM theory [97] following the notation in [98]. The $\mathrm{U}(N) \times \mathrm{U}(N)$ gauge superfields are $\mathcal{V}^a{}_b$ and $\hat{\mathcal{V}}^{\hat{a}}{}_{\hat{b}}$. The matter superfields $(\mathcal{Z}^A)^a{}_{\hat{a}}$ and $(\mathcal{W}_A)^{\hat{a}}{}_a$ transform under gauge transformations in the representation $(\mathbf{N}, \bar{\mathbf{N}})$ and $(\bar{\mathbf{N}}, \mathbf{N})$, respectively. They also transform under two different global SU(2)'s in the $\mathbf{2}$ and $\bar{\mathbf{2}}$ indicated by the indices $A = 1, 2$. The action is given by standard Chern-Simons terms with level $k$ for $\mathcal{V}$ and level $-k$ for $\hat{\mathcal{V}}$. The matter action is given by the standard kinetic terms for $\mathcal{Z}$ and $\mathcal{W}$ minimally coupled to the gauge fields. Finally, the theory includes the SU(2) × SU(2) invariant superpotential [18]

$$\mathrm{W} = \frac{1}{4}\epsilon_{AC}\epsilon^{BD}\,\mathrm{tr}\,\mathcal{Z}^A\mathcal{W}_B\mathcal{Z}^C\mathcal{W}_D \ . \tag{4.3.1}$$

This gauge theory was conjectured to be the CFT dual of M-theory on $\mathrm{AdS}_4 \times (S^7/\mathbb{Z}_k)$ supported by $N$ units of 4-form flux [97].

Special attention needs to be paid to the U(1) × U(1) part of the gauge group. All matter fields are neutral under one linear combination of the U(1)'s, $c_\mu$, which therefore corresponds to the center of mass degree of freedom of the stack of M2-branes. The flux for this non-interacting U(1) is quantized, and it may be dualized into a periodic scalar. As a result the other linear combination, the 'baryonic' U(1) gauge field $b_\nu$, which enters the Chern-Simons action as

$$\frac{k}{2\pi}\epsilon^{\mu\nu\lambda}b_\mu\partial_\nu c_\lambda \ , \tag{4.3.2}$$

gets broken to $\mathbb{Z}_k$ [97] through a mechanism demonstrated in [128,129]. The generator of this group acts on the superfields as

$$\mathcal{Z}^A \to e^{2\pi i/k}\mathcal{Z}^A \ , \qquad \mathcal{W}_B \to e^{-2\pi i/k}\mathcal{W}_B \ . \tag{4.3.3}$$

This argument loosely suggests that for $k = 1$ the $\mathrm{U}(N) \times \mathrm{U}(N)$ gauge theory is simply equivalent to the $\mathrm{SU}(N) \times \mathrm{SU}(N)$. However, this is not quite correct since the moduli spaces of the two theories are different [128, 129, 97].[3] The ABJM theory with $k > 2$ has been demonstrated to possess the $\mathcal{N} = 6$ superconformal invariance, in agreement with that of its proposed M-theory dual. For $k = 1, 2$ the superconformal symmetry of the ABJM theory is expected to enhance to $\mathcal{N} = 8$ but this is yet to be demonstrated explicitly. An important manifestation of this enhancement is that, when

[3]For $k = 2$ and $N = 2$ the moduli spaces of the two gauge theories do coincide [128,129], so in this case they may be equivalent.

$\mathcal{N} = 2$ superspace is used, then the theory possesses $\mathrm{U}(1)_R \times \mathrm{SU}(4)$ global symmetry. Although the non-R SU(4) flavor symmetry is difficult to establish in general, in the next section we discuss how it may appear using some specific examples.

### 4.3.2 Towards establishing the $\mathbf{U(1)_R \times SU(4)}$ invariance

In [98] the BLG theory was reformulated using $\mathcal{N} = 2$ superspace where its has manifest $\mathrm{U}(1)_R \times \mathrm{SU}(4)$ global symmetry. This theory is exactly equivalent to the $\mathrm{SU}(2) \times \mathrm{SU}(2)$ version of the ABJM theory [97]. The four complex bi-fundamental superfields $\mathcal{Z}^A$ of the BLG theory, which transform in the fundamental of the global SU(4), are related to the fields entering (4.3.1) through

$$\mathcal{Z}^3 = \mathcal{W}_1^{\ddagger}, \qquad \mathcal{Z}^4 = \mathcal{W}_2^{\ddagger} \ . \tag{4.3.4}$$

This uses an operation special to the $\mathrm{SU}(2) \times \mathrm{SU}(2)$ gauge theory

$$\mathcal{W}^{\ddagger} := -\epsilon \mathcal{W}^T \epsilon \qquad \text{with } \epsilon = \begin{pmatrix} 0 & 1 \\ -1 & 0 \end{pmatrix} \tag{4.3.5}$$

because it relies on the invariant tensor $\epsilon_{ab}$. After this transformation the superpotential (4.3.1), which is manifestly only $\mathrm{SU}(2)^2$ invariant, acquires the SU(4) invariant form [98]

$$\mathrm{W} = \frac{1}{4!}\epsilon_{ABCD}\mathrm{tr}\mathcal{Z}^A \mathcal{Z}^{\ddagger B} \mathcal{Z}^C \mathcal{Z}^{\ddagger D} = -\frac{1}{8 \cdot 4!}\epsilon_{ABCD}\epsilon^{abcd}\mathcal{Z}_c^A \mathcal{Z}_b^B \mathcal{Z}_c^C \mathcal{Z}_d^D \tag{4.3.6}$$

where the relation to SO(4) notation $\mathcal{Z}_a^A$ is explained in [98].

Below we will suggest how the operation (4.3.4) may be generalized to $\mathrm{U}(N) \times \mathrm{U}(N)$ gauge theories. To accomplish this one likely has to invoke monopole operators, often called 't Hooft operators because of his pioneering work [130]. Such operators naturally carry the magnetic charge determined by the flux they insert at a point. In a Chern-Simons theory, they also carry an electric charge (or gauge representation) proportional to the Chern-Simons level $k$. We assemble some useful facts about these operators in App. 4.D. For a recent explicit study of monopole operators in the ABJM theory, see [100].

When $k = 1$, the simplest monopole operators are $(e^{\tau})^{a}_{\hat{a}}$, which transforms in the representation $(\mathbf{N}, \bar{\mathbf{N}})$, and its conjugate $(e^{-\tau})^{\hat{a}}_{a}$. They are obtained for the choices of flux described in App. 4.D. We can also construct the "double" monopole operators, $(e^{2\tau})^{ab}_{\hat{a}\hat{b}}$ and $(e^{-2\tau})^{\hat{a}\hat{b}}_{ab}$. They can be either symmetric or anti-symmetric under separate interchanges of upper or lower indices, but both choices have the symmetry under the interchange of both:

$$(e^{2\tau})^{ab}_{\hat{a}\hat{b}} = (e^{2\tau})^{ba}_{\hat{b}\hat{a}} \quad , \quad (e^{-2\tau})^{\hat{a}\hat{b}}_{ab} = (e^{-2\tau})^{\hat{b}\hat{a}}_{ba} \, . \tag{4.3.7}$$

These operators transform under $\mathrm{U}(N) \times \mathrm{U}(N)$ as indicated by their indices. In particular, they are charged under the baryonic U(1) gauge group, which is the interacting part of the $\mathrm{U}(1) \times \mathrm{U}(1)$. In our notation, $e^{n\tau}$ has charge $n$ under this baryonic U(1).

When $k = 2$, no choice of flux can give an operator of the form $(e^{\tau})^{a}_{\hat{a}}$ and the smallest operators one can form are $(e^{2\tau})^{ab}_{\hat{a}\hat{b}}$ and its conjugate, as discussed in the appendix.

Let us use the monopole operators to establish $\mathrm{U}(1)_R \times \mathrm{SU}(4)$ symmetry of the $\mathrm{U}(2) \times \mathrm{U}(2)$ ABJM theory, which has some subtle differences from the BLG theory. Inspired by (4.3.5), we propose to use the monopole operators in the ABJM gauge theory that are *anti-symmetric* in each set of indices,

$$(e^{2\tau})^{ab}_{\hat{a}\hat{b}} = -(e^{2\tau})^{ba}_{\hat{a}\hat{b}} \quad , \quad (e^{2\tau})^{ab}_{\hat{a}\hat{b}} = -(e^{2\tau})^{ab}_{\hat{b}\hat{a}} \, . \tag{4.3.8}$$

Thinking of the $N = 2$ ABJM gauge theory as $\mathrm{SU}(2) \times \mathrm{SU}(2) \times \mathrm{U}(1) \times \mathrm{U}(1)$, we can use the SU(2) invariant tensors to write this as,

$$(e^{2\tau})^{ab}_{\hat{a}\hat{b}} = T^2 \epsilon^{ab}\epsilon_{\hat{a}\hat{b}} \ , \quad (e^{-2\tau})^{\hat{a}\hat{b}}_{ab} = T^{-2}\epsilon_{ab}\epsilon^{\hat{a}\hat{b}} \ , \tag{4.3.9}$$

where $T^2$ is a monopole operator that creates two (one) units of magnetic flux when $k = 1$ ($k = 2$) for the decoupled U(1) field $c_\mu$ in the $U(1) \times U(1)$ Chern-Simons gauge theory (4.3.2) coupled to the charged matter. Due to the coupling (4.3.2), $T^2$ is doubly charged under the baryonic U(1) in both cases ($k = 1, 2$).

Using the expressions (4.3.9) valid when $N = 2$, the following invertibility identity can be verified,

$$(e^{2\tau})^{ab}_{\hat{a}\hat{b}}(e^{-2\tau})^{\hat{b}\hat{c}}_{bc} = \delta^a_c \delta^{\hat{c}}_{\hat{a}} \ , \tag{4.3.10}$$

where we have assumed that these monopole operators do not contribute to the scaling dimensions of gauge invariant operators. This is a non-trivial assumption, since in some theories where the monopole operators were constructed explicitly, their scaling dimensions are non-vanishing [131, 132]. The assumption that their scaling dimensions vanish in the ABJM theory was central in forming operators with the right dimension and R-charge for AdS/CFT duality [97], and that will be the case here as well. However a definitive proof of this has been lacking.

To search for a global symmetry enhancement in the superpotential (4.3.1), let us introduce a multiplet of superfields in the fundamental of SU(4)

$$\mathcal{Z}^A = (\mathcal{Z}^1, \mathcal{Z}^2, \mathcal{W}_1 e^{2\tau}, \mathcal{W}_2 e^{2\tau}) \ , \qquad A = 1, 2, 3, 4, \tag{4.3.11}$$

where the explicit index structure is

$$(\mathcal{Z}^3)^a{}_{\hat{a}} = (\mathcal{W}_1)^{\hat{b}}{}_b (e^{2\tau})^{ab}_{\hat{a}\hat{b}} \quad , \quad (\mathcal{Z}^4)^a{}_{\hat{a}} = (\mathcal{W}_2)^{\hat{b}}{}_b (e^{2\tau})^{ab}_{\hat{a}\hat{b}}. \tag{4.3.12}$$

We note that the fields $\mathcal{Z}^A, A = 1, \dots 4$ have the same baryonic charge, even though $\mathcal{W}_{1,2}$ have the opposite charge. With this definition the superpotential can be written as

$$\mathrm{W} = \frac{1}{2}(\mathcal{Z}^1)^a{}_{\hat{a}}(\mathcal{Z}^2)^b{}_{\hat{b}}(\mathcal{Z}^3)^c{}_{\hat{c}}(\mathcal{Z}^4)^d{}_{\hat{d}}\Big[(e^{-2\tau})^{\hat{a}\hat{c}}_{bc}(e^{-2\tau})^{\hat{b}\hat{d}}_{ad} - (e^{-2\tau})^{\hat{a}\hat{d}}_{bd}(e^{-2\tau})^{\hat{b}\hat{c}}_{ac}\Big] \ . \quad (4.3.13)$$

In the $\mathrm{U}(2) \times \mathrm{U}(2)$ ABJM theory, using the expressions (4.3.9), we find that the superpotential (4.3.13) has a close relation to that of the BLG theory, but also contains the abelian monopole operators needed for its $\mathrm{U}(1)_b$ gauge invariance:

$$\mathrm{W} = \frac{1}{4!}T^{-4}\epsilon_{ABCD}\mathrm{tr}\mathcal{Z}^A\mathcal{Z}^{\ddagger B}\mathcal{Z}^C\mathcal{Z}^{\ddagger D} \ . \quad (4.3.14)$$

It would be very interesting to extend the validity of the above arguments and expressions to $N > 2$, and to establish the SU(4) invariance of the superpotential in the $\mathrm{U}(N) \times \mathrm{U}(N)$ ABJM theory. This would provide a clear argument in favor of its $\mathcal{N} = 8$ supersymmetry.

### 4.3.3 Quadratic Deformations of the Superpotential

While for any $k$ we can add quadratic operators of the form $\mathrm{tr}\, \mathcal{Z}^A\mathcal{W}_B$, for $k = 1$ and $k = 2$ we can also deform the ABJM theory by a relevant operator which is quadratic in *just one* of the chiral superfields. To write these operators explicitly we need the monopole operators:

$$\Delta\mathrm{W} = m(\mathcal{Z}^4)^a{}_{\hat{a}}(\mathcal{Z}^4)^b{}_{\hat{b}}(e^{-2\tau})^{\hat{a}\hat{b}}_{ab} \quad (4.3.15)$$

This relevant operator creates RG flow. To find the effective superpotential of the infrared theory, we integrate out the massive field $\mathcal{Z}^4$ in the IR, leaving a sextic potential for the remaining fields. It is natural to conjecture [98] that this IR fixed point is dual to the warped $AdS_4$ background of M-theory containing a $\mathrm{U}(1)_R \times \mathrm{SU}(3)$ symmetric 'squashed and stretched' 7-sphere, whose original gauged supergravity formulation was found in [117]. In order to achieve the $\mathrm{U}(1)_R$ symmetry, the total R-charge of the superpotential should equal 2. In [97] it was assumed that all the necessary monopole operators have vanishing R-charge and dimension. We will assume the same here without a more detailed study involving matter fields to justify this. Then we can assign the following dimensions and R-charges:

$$\varDelta(\mathcal{Z}^A) = R(\mathcal{Z}^A) = \frac{1}{3} \qquad \text{for } A = 1,2,3 \quad , \quad \varDelta(\mathcal{Z}^4) = R(\mathcal{Z}^4) = 1 \; . \tag{4.3.16}$$

It is interesting that the U(1) symmetry with these charges holds not just in the IR, but along the entire RG flow. The M-theory dual of this RG flow was found in [117,123]. Remarkably, it possesses [133] a U(1) symmetry with the same charges as in the field theory.[4] This can be demonstrated by identifying the U(1) symmetry of the 3-form potential (see eq. (121), (122) of [133]) and showing that three of the complex coordinates of the 7-sphere transform with charge 1/3, and the fourth one with charge 1. This provides an immediate check of the gauge/gravity duality along the entire RG flow.

For general $N$ explicit demonstration of the SU(3) global symmetry of the superpotential remains a challenge, just like the SU(4) global symmetry of the ABJM superpotential (4.3.13). Fortunately, this symmetry is explicit for $\mathrm{U}(2) \times \mathrm{U}(2)$ ABJM theory, if we use our assumption (4.3.9) about the monopole operators. Then the quadratic superpotential deformation assumes the form

$$\varDelta \mathrm{W} = m T^{-2} \, \mathrm{tr} \, \mathcal{Z}^4 \mathcal{Z}^{4\ddagger} \; . \tag{4.3.17}$$

which is closely related to the deformation of the BLG theory proposed in [98]. Adding

[4] We thank Juan Maldacena for an enlightening discussion on this issue.

such a mass term and integrating out $\mathcal{Z}^4$, we find

$$\mathcal{Z}^4 = -\frac{T^{-2}}{12m}\,\epsilon_{ABC}\,\mathcal{Z}^A\mathcal{Z}^{\ddagger B}\mathcal{Z}^C \tag{4.3.18}$$

and hence the new superpotential,

$$\mathrm{W}_{\text{eff}} = \frac{T^{-6}}{144m}\,\epsilon_{ABC}\epsilon_{DEF}\ \mathrm{tr}\,\mathcal{Z}^A\mathcal{Z}^{\ddagger B}\mathcal{Z}^C\mathcal{Z}^{\ddagger D}\mathcal{Z}^E\mathcal{Z}^{\ddagger F}\ . \tag{4.3.19}$$

We conclude this section by making the breaking of parity invariance due to the deformation (4.3.17) more apparent[5]. The parity operation in the gauge theory sends $(x^0, x^1, x^2) \to (x^0, -x^1, x^2)$ [96]. The fermionic coordinates transform as $\theta_\alpha = -\gamma^1_{\alpha\beta}\theta^\beta$. These maps are accompanied by a transformation of the fields. In the $N = 2$ theory the superfield transforms as $\mathcal{Z}^A \to \mathcal{Z}^{\ddagger A}$ $(A = 1, \dots, 4)$, and the component fields as $Z^A \to Z^{\ddagger A}$, $\zeta^A_\alpha \to \gamma^1_{\alpha\beta}\zeta^{\ddagger A\beta}$, and $F^A \to -F^{\ddagger A}$ $(A = 1, \dots, 4)$. Now, consider the deformation (4.3.17) integrated over superspace

$$\begin{aligned}\mathit{\Delta}\mathcal{L}_{\text{pot}} &= mT^{-2}\int d^2\theta\ \mathrm{tr}\,\mathcal{Z}^4\mathcal{Z}^{\ddagger 4}\\ &= -mT^{-2}\,\mathrm{tr}\,\zeta^4\zeta^{\ddagger 4} + 2mT^{-2}\,\mathrm{tr}\,F^4 Z^{\ddagger 4}\\ &= -mT^{-2}\,\mathrm{tr}\,\zeta^4\zeta^{\ddagger 4} + T^{-4}\frac{mL}{3}\epsilon^{ABC}\,\mathrm{tr}\,\bar{Z}_A\bar{Z}^{\ddagger}_B\bar{Z}_C Z^{\ddagger 4}\ ,\end{aligned} \tag{4.3.20}$$

where in the last line we replaced the auxiliary field $F$ using its equation of motion. Any of these expressions makes it explicit that $\mathit{\Delta}\mathcal{L}_{\text{pot}}$ is parity odd and hence breaks the parity invariance of the original theory.

[5]We use the notation and conventions of [98].

## 4.4 Matching of short multiplets

Having described the field content of the IR fixed point, we can proceed to match gauge theory operators with the gravity multiplets found earlier. For every supermultiplet there is a superfield of the gauge theory. Long supermultiplets correspond to unconstrained superfields, short supermultiplets to constrained ones. We will focus on the four series of short multiplets (cf. Tab. 4.6) and show that with our assignment there are four corresponding series of gauge theory operators. For the duality to hold it is essential to assign the charges of the IR gravity states according to Scenario I.

To facilitate the comparison of the components of the gravity supermultiplets and the components of the gauge theory superfields, we summarize the charges of the component fields in Tab. 4.7.

| | $Z^A$ | $\zeta^A$ | $Z^\dagger_A$ | $\zeta^\dagger_A$ | $Z^4$ | $\zeta^4$ | $Z^\dagger_4$ | $\zeta^\dagger_4$ | $x$ | $\theta$ | $\bar\theta$ |
|---|---|---|---|---|---|---|---|---|---|---|---|
| SU(3) | **3** | **3** | $\bar{\mathbf{3}}$ | $\bar{\mathbf{3}}$ | **1** | **1** | **1** | **1** | **1** | **1** | **1** |
| Dimension | $\frac{1}{3}$ | $\frac{5}{6}$ | $\frac{1}{3}$ | $\frac{5}{6}$ | 1 | $\frac{3}{2}$ | 1 | $\frac{3}{2}$ | $-1$ | $-\frac{1}{2}$ | $-\frac{1}{2}$ |
| R-charge | $+\frac{1}{3}$ | $-\frac{2}{3}$ | $-\frac{1}{3}$ | $+\frac{2}{3}$ | $+1$ | 0 | $-1$ | 0 | 0 | $+1$ | $-1$ |

Table 4.7: **Dimensions and R-charges of building blocks.** The components of the superfields are $\mathcal{Z} = Z + \sqrt{2}\theta^\alpha\zeta_\alpha + \text{aux.}$ and $\bar{\mathcal{Z}} = Z^\dagger - \sqrt{2}\bar\theta^\alpha\zeta^\dagger_\alpha + \text{aux.}$

### Hypermultiplets

In Sec. 4.2 we found that, in Scenario I, the hypermultiplets come in the SU(3) representations $[n+2,0]$ where $n = 0, 1, 2, \dots$ (see left column of Tab. 4.6). They have R-charge $y_0 = \frac{n+2}{3}$ and dimension[6] $\Delta_0 = |y_0| = \frac{n+2}{3}$, both of which suggestively increase in steps of 1/3, the R-charge and dimension of the superfields $\mathcal{Z}^A, A = 1, 2, 3$.

[6] $y_0$ and $\Delta_0$ in this section must be compared to $y_0$ and $E_0$ in Tab. 4.8 to 4.16. Note that $\Delta_0 = E_0$ refers to the dimension of the ground state in a multiplet and the dimensions of the other components are related as shown in those tables.

Hence we write down a series of corresponding operators,

$$H^{(n)A_1\ldots A_{n+2}} \sim \mathcal{Z}^{(A_1}\mathcal{Z}^{A_2}\cdots\mathcal{Z}^{A_{n+2})}\ , \tag{4.4.1}$$

ignoring their gauge indices for the moment. We have symmetrized the SU(3) indices $A_i$ to obtain the $[n+2,0]$ representation. These operators are chiral, $\bar{D}_\alpha H^{(n)} = 0$, which implies that they have the structure of the $\mathcal{N} = 2$ hypermultiplet as given in Tab. 4.16. To see this explicitly in this simple example, we write out the components of this superfield:

$$\begin{aligned} H^{(n)} \sim\ & Z^{(A_1}\cdots Z^{A_{n+2})} \\ & + n\sqrt{2}\,\theta^\alpha\,\zeta_\alpha^{(A_1}Z^{A_2}\cdots Z^{A_{n+2})} \\ & - \tfrac{1}{2}n(n-1)\,\theta^2\,\zeta^{\alpha(A_1}\zeta_\alpha^{A_2}Z^{A_3}\cdots Z^{A_{n+2})}\ . \end{aligned} \tag{4.4.2}$$

Using the charges from Tab. 4.7, it is simple to verify that the dimensions and R-charges, as well as the spins, of the components match.

To render the schematic operator expression (4.4.1) gauge invariant, we need to make use of monopole operators. For even $n$, the natural expression is

$$H^{(n)A_1\ldots A_{n+2}} = \mathrm{tr}\,\mathcal{Z}^{(A_1}\mathcal{Z}^{A_2}e^{-2\tau}\,\mathcal{Z}^{A_3}\mathcal{Z}^{A_4}e^{-2\tau}\cdots\mathcal{Z}^{A_{n+1}}\mathcal{Z}^{A_{n+2})}e^{-2\tau}\ , \tag{4.4.3}$$

where the operator $e^{-2\tau}$ is contracted with the preceding field as

$$(\mathcal{Z}e^{-2\tau})^{\hat{a}}{}_a = \mathcal{Z}^b{}_{\hat{b}}(e^{-2\tau})^{\hat{a}\hat{b}}_{ab}.$$

For $N = 2$, where the form of the monopole operators simplifies, these operators

become

$$H^{(n)A_1\ldots A_{n+2}} = T^{-n-2}\,\mathrm{tr}\,\mathcal{Z}^{(A_1}\mathcal{Z}^{\ddagger A_2}\,\mathcal{Z}^{A_3}\mathcal{Z}^{\ddagger A_4}\cdots\mathcal{Z}^{A_{n+1}}\mathcal{Z}^{\ddagger A_{n+2})}\ . \tag{4.4.4}$$

They are generalizations of the $n = 0$ quadratic operator studied in [98]. In order to write down the operators for odd $n$, present for $k = 1$, we need to insert one monopole operator $(e^{-\tau})_a^{\hat{a}}$:

$$H^{(n)A_1\ldots A_{n+2}} = \mathrm{tr}\,\mathcal{Z}^{(A_1}\mathcal{Z}^{A_2}e^{-2\tau}\cdots\mathcal{Z}^{A_n}\mathcal{Z}^{A_{n+1}}e^{-2\tau}\mathcal{Z}^{A_{n+2})}e^{-\tau}\ . \tag{4.4.5}$$

For $k = 2$ the operator $e^{-\tau}$ is not available, and we can construct only the even operators. This is consistent with the supergravity side: when $n$ is odd and $k = 2$, the $Z_2$ orbifold action projects out the corresponding SUGRA mode.

**Short graviton multiplets**

From Tab. 4.6, we see that the short graviton multiplets are always SU(3) singlets. In Scenario I they possess R-charges $y_0 = n$ and dimensions $\Delta_0 = |y_0| + 2 = n + 2$ for $n = 0, 1, 2, \ldots$. When $n = 0$, this is actually the familiar massless graviton in $AdS$ and hence corresponds to the energy momentum tensor in the CFT. The other two massless components in this supermultiplet are the gravitino which is the SUSY generator and a massless vector boson which corresponds to the $\mathrm{U}(1)_R$ symmetry of the dual CFT.

The gauge theory operator dual to the massless graviton multiplet is given by the stress-energy superfield

$$\mathcal{T}^{(0)}_{\alpha\beta} = \mathrm{tr}\,\bar{D}_{(\alpha}\bar{\mathcal{Z}}_A D_{\beta)}\mathcal{Z}^A + i\,\mathrm{tr}\,\bar{\mathcal{Z}}_A\overleftrightarrow{\partial}_{\alpha\beta}\mathcal{Z}^A\ , \tag{4.4.6}$$

which satisfies the corresponding constraint $D^\alpha\mathcal{T}^{(0)}_{\alpha\beta} = \bar{D}^\alpha\mathcal{T}^{(0)}_{\alpha\beta} = 0$ and has protected classical dimension. For example the spin-two component has exact dimension 3 and the ground state component has dimension $\Delta_0 = 2$. For higher $n$ we expect the series to continue schematically as

$$\mathcal{T}^{(n)}_{\alpha\beta} \sim \mathcal{T}^{(0)}_{\alpha\beta}(\epsilon_{ABC}\mathcal{Z}^A\mathcal{Z}^B\mathcal{Z}^C)^n \qquad \text{for } n = 1, 2, 3, \dots , \tag{4.4.7}$$

where we again understand none of the gauge indices to be contracted yet. The anti-symmetric combination of three $\mathcal{Z}$s may be thought of as the field $\mathcal{Z}^4$ which was integrated out. For $n \geq 1$ these superfields satisfy only $\bar{D}^\alpha \mathcal{T}^{(n)}_{\alpha\beta} = 0$. Such a series has R-charge and dimension increasing in steps of 1 and in complete agreement with Scenario I in Tab. 4.6.

The fields (4.4.7) are again made gauge invariant by means of appropriate monopole operators. For even $n$ we insert a total of $3\frac{n}{2}$ monopole operators with two units of flux, $e^{-2\tau}$, and contract them with every other field as we described for the hypermultiplet. To find the superfield corresponding to the short graviton multiplet, one also needs to sum over all permutations of the fields. A typical term in such a sum is

$$\operatorname{tr} \mathcal{T}^{(0)}_{\alpha\beta} \left[ \left(\epsilon_{ABC}\mathcal{Z}^A e^{-2\tau}\mathcal{Z}^B\mathcal{Z}^C e^{-2\tau}\right)\left(\epsilon_{DEF}\mathcal{Z}^D\mathcal{Z}^E e^{-2\tau}\mathcal{Z}^F\right)\cdots\right] . \tag{4.4.8}$$

For odd $n$ we need to insert another monopole operator with one unit of flux, $e^{-\tau}$. If $k = 2$ we do not have such a monopole at our disposal and hence there are no gauge theory operators for odd $n$. This mirrors the fact that such modes are projected out by the orbifolding action on the gravity side, just as we saw for the hypermultiplets.

The dimensions and R-charge in Scenario II appear difficult to interpret in a CFT. The corresponding short graviton series has a fixed R-charge of 0 and dimension of 2 for all $n$. As remarked earlier, this does not seem characteristic of a KK reduction.

**Short gravitino multiplets**

The short gravitino multiplets come in the SU(3) representations $[n+1, 0]$ with R-charges $y_0 = \frac{n+1}{3}$ and dimensions $\Delta_0 = |y_0| + \frac{3}{2} = \frac{2n+11}{6}$ for $n = 0, 1, 2, \ldots$. Note that this is a massive multiplet even for $n = 0$. The existence of a massless gravitino multiplet would indicate enhancement of SUSY beyond $\mathcal{N} = 2$. Based on this data, we can write down the following candidate superfield,

$$\Lambda_\alpha^{(n)A_1 \ldots A_{n+1}} \sim \epsilon_{ABC} \mathcal{Z}^A \mathcal{Z}^B \mathcal{Z}^C D_\alpha \mathcal{Z}^{(A_1} \mathcal{Z}^{A_2} \cdots \mathcal{Z}^{A_{n+1})} , \tag{4.4.9}$$

where the derivative acts only onto the $\mathcal{Z}$ next to it. These fields are a fermionic superfields and satisfy $\bar{D}^\alpha \Lambda_\alpha = 0$. We can verify that (4.4.9) is the correct dual operator by checking the explicit components of this superfield against the known SUGRA multiplet. We show this for $n = 0$. Let us restrict ourselves to $N = 2$ where we can use the SO(4) notation $\mathcal{Z}_a^A$ that enables us to write the operator in the following gauge invariant way

$$\Lambda_\alpha^{(0)A_1} \sim \epsilon_{ABC} \epsilon^{abcd} \mathcal{Z}_a^A \mathcal{Z}_b^B \mathcal{Z}_c^C D_\alpha \mathcal{Z}_d^{A_1} \quad . \tag{4.4.10}$$

The component expansion of this superfield is (up to total derivatives)

$$\begin{aligned}\Lambda_\alpha^{(0)A_1} \sim \epsilon_{ABC}\epsilon^{abcd}\Big[ & -\sqrt{2}i\,(\theta\gamma^\mu\bar{\theta})\,\big(ZZZ\partial_\mu\zeta_\alpha + \epsilon_{\mu\nu\rho}ZZZ(\gamma^\nu\partial^\rho\zeta)_\alpha \\ & \qquad + 3\,\zeta_\alpha ZZ\partial_\mu Z - 3\,\epsilon_{\mu\nu\rho}(\gamma^\nu\zeta)_\alpha ZZ\partial^\rho Z\big) \\ & + 2i\,(\gamma^\mu\bar{\theta})_\alpha\, ZZZ\partial_\mu Z \\ & - 6i\,\theta^2(\gamma^\mu\bar{\theta})_\alpha\,\big(\zeta\zeta Z\partial_\mu Z + ZZ\zeta\partial_\mu\zeta\big) \\ & - 3\,(\gamma^\mu\theta)_\alpha\, ZZ\zeta\gamma_\mu\zeta \\ & - \sqrt{2}i\,\theta\bar{\theta}\,\big(ZZZ(\partial\!\!\!/\zeta)_\alpha + 3\,(\gamma^\mu\zeta)_\alpha ZZ\partial_\mu Z\big) \\ & + \sqrt{2}\,ZZZ\zeta_\alpha \\ & - 3\sqrt{2}\theta^2\, Z\zeta\zeta\zeta_\alpha \\ & - 3\,\theta_\alpha\, ZZ\zeta\zeta\Big]\,. \end{aligned} \tag{4.4.11}$$

To simplify the notation, we have omitted the SU(3) indices $ABCA_1$ and the SO(4) gauge indices $abcd$ from the fields on the right hand side. The dimensions, R-charge and spin of each component presented on distinct lines above match up with the components of the supermultiplet in Tab. 4.11.

The monopole operators required to make these operators gauge invariant for general $n$ are similar to those used for the hypermultiplets with $e^{-2\tau}$ inserted on every other $\mathcal{Z}$ and summing over all permutations. A typical term in such a sum (when $n$ is even) is,

$$\left(\epsilon_{ABC}\mathcal{Z}^A e^{-2\tau}\mathcal{Z}^B\mathcal{Z}^C e^{-2\tau}\right) D_\alpha \mathcal{Z}^{(A_1}\mathcal{Z}^{A_2}e^{-2\tau}\cdots\mathcal{Z}^{A_{n-1}}\mathcal{Z}^{A_n}e^{-2\tau}\mathcal{Z}^{A_{n+1})} \tag{4.4.12}$$

If $n$ is odd, we need an extra $e^{-\tau}$ monopole operator which is allowed only when $k=1$. This agrees with the fact that the corresponding SUGRA modes are projected out by the $k=2$ orbifold.

**Short vector multiplets**

The short vector multiplets come in the SU(3) representations $[n+1,1]$ with R-charges $y_0 = \frac{n}{3}$ and dimensions $\Delta_0 = |y_0| + 1 = \frac{n+3}{3}$ for $n = 0, 1, 2, \ldots$. When $n = 0$, this is in fact the conserved current multiplet $\mathcal{J}_A^{(0)B}$ corresponding to the SU(3) global symmetry of the CFT. This superfield satisfies the constraint $D^2 \mathcal{J}_A^{(0)B} = \bar{D}^2 \mathcal{J}_A^{(0)B} = 0$. Its highest spin component is the bosonic current

$$J_{\mu A}^{(0)B} = \bar{Z}_A \overleftrightarrow{\partial}_\mu Z^B - \frac{1}{3} \delta_A^B \bar{Z}_C \overleftrightarrow{\partial}_\mu Z^C \ . \tag{4.4.13}$$

It has the protected classical dimension of 2. For higher $n$ we expect the series to continue as

$$\mathcal{J}_{A_0}^{(n)A_1 \ldots A_{n+1}} \sim \mathcal{J}_{A_0}^{(0)(A_1} \mathcal{Z}^{A_2} \cdots \mathcal{Z}^{A_{n+1})} \qquad \text{for } n = 1, 2, 3, \ldots \ , \tag{4.4.14}$$

where we still have to deal with the gauge indices. For $n \geq 1$ these operators satisfy only the constraint $\bar{D}^2 \mathcal{J}^{(n)} = 0$.

To make these operators gauge invariant, we need $\lfloor \frac{n}{2} \rfloor$ monopole operators with two units of flux, $e^{-2\tau}$, and in case $n =$ odd another one with one unit of flux, $e^{-\tau}$. Since the latter ones do not exist for $k = 2$, there are no operators for odd $n$, just as the corresponding SUGRA mode is projected out by the $k = 2$ orbifold. The $e^{-2\tau}$ operators are inserted on every other $\mathcal{Z}$ just as for the hypermultiplet and summed over all possible permutations. One typical permutation is for example,

$$\operatorname{tr} \mathcal{J}_{A_0}^{(0)(A_1} \mathcal{Z}^{A_2} e^{-2\tau} \mathcal{Z}^{A_3} \cdots \mathcal{Z}^{A_n} e^{-2\tau} \mathcal{Z}^{A_{n+1})} \ . \tag{4.4.15}$$

## 4.A $\mathcal{N} = 2$ supermultiplets

In the main text we have used the knowledge of the structure of Osp(2|4) supermultiplets to constrain the spectrum of gravity states on the 'stretched and squashed' seven sphere. These supermultiplets have been worked out in the context of general $\mathcal{N} = 2$ compactifications in [134] (see also [124]). The short multiplets and their gauge theory interpretation in a general $AdS_4/CFT_3$ context were discussed in [135]. For the convenience of the reader we list the multiplets relevant to our discussion in this appendix.

The bosonic subgroup of Osp(2|4) is $SO(3,2) \times SO(2)$. The $SO(3,2)$ part is the conformal group in 2+1 dimensions or, equivalently, the isometry group of $AdS_4$. Unitary, positive energy representations of $SO(3,2)$ are labeled by spin $s$ and energy $E$ [136]. The SO(2) part is the R-symmetry and the representation label is the hypercharge $y$. An $\mathcal{N} = 2$ supermultiplet is a set of $SO(3,2) \times SO(2)$ representations which is obtained by acting with the fermionic raising operators of Osp(2|4) onto a chosen $SO(3,2) \times SO(2)$ with labels $(s_0, E_0, y_0)$, the so-called lowest bosonic submultiplet.

The total number of bosonic submultiplets within one Osp(2|4) representation depends on the relationships between the labels $(s_0, E_0, y_0)$:

- Long multiplets for $E_0 > |y_0| + s_0 + 1$:
  long graviton ($s_0 = 1$), long gravitino ($s_0 = \frac{1}{2}$), long vector ($s_0 = 0$),

- Short multiplets 'I' for $E_0 = |y_0| + s_0 + 1$:
  short graviton ($s_0 = 1$), short gravitino ($s_0 = \frac{1}{2}$), short vector ($s_0 = 0$),

- Short multiplets 'II' for $E_0 = |y_0| \geq \frac{1}{2}$:
  hypermultiplet ($s_0 = 0$),

- Ultrashort multiplets for $E_0 = s_0 + 1$, $y_0 = 0$:
  massless graviton ($s_0 = 1$), massless vector ($s_0 = 0$).

Note that there is no massless gravitino as its presence would enhance the supersymmetry to $\mathcal{N} > 2$.

| Spin | 2 | $\frac{3}{2}$ | $\frac{3}{2}$ | 1 |
|---|---|---|---|---|
| Energy | 3 | $\frac{5}{2}$ | $\frac{5}{2}$ | 2 |
| R-charge | 0 | $+1$ | $-1$ | 0 |

Table 4.8: **$\mathcal{N} = 2$ massless graviton multiplet (MGRAV).**

| Spin | 2 | $\frac{3}{2}$ | $\frac{3}{2}$ | $\frac{3}{2}$ | 1 | 1 | 1 | $\frac{1}{2}$ |
|---|---|---|---|---|---|---|---|---|
| Energy | $E_0+1$ | $E_0+\frac{3}{2}$ | $E_0+\frac{1}{2}$ | $E_0+\frac{1}{2}$ | $E_0+1$ | $E_0+1$ | $E_0$ | $E_0+\frac{1}{2}$ |
| R-charge | $y_0$ | $y_0\mp 1$ | $y_0+1$ | $y_0-1$ | $y_0\mp 2$ | $y_0$ | $y_0$ | $y_0\mp 1$ |

Table 4.9: **$\mathcal{N} = 2$ short graviton multiplet (SGRAV).** $E_0 = |y_0| + 2$

| Spin | 2 | $\frac{3}{2}$ | $\frac{3}{2}$ | $\frac{3}{2}$ | $\frac{3}{2}$ | 1 | 1 | 1 |
|---|---|---|---|---|---|---|---|---|
| Energy | $E_0+1$ | $E_0+\frac{3}{2}$ | $E_0+\frac{3}{2}$ | $E_0+\frac{1}{2}$ | $E_0+\frac{1}{2}$ | $E_0+2$ | $E_0+1$ | $E_0+1$ |
| R-charge | $y_0$ | $y_0-1$ | $y_0+1$ | $y_0-1$ | $y_0+1$ | $y_0$ | $y_0-2$ | $y_0+2$ |
| Spin | 1 | 1 | 1 | $\frac{1}{2}$ | $\frac{1}{2}$ | $\frac{1}{2}$ | $\frac{1}{2}$ | 0 |
| Energy | $E_0+1$ | $E_0+1$ | $E_0$ | $E_0+\frac{3}{2}$ | $E_0+\frac{3}{2}$ | $E_0+\frac{1}{2}$ | $E_0+\frac{1}{2}$ | $E_0+1$ |
| R-charge | $y_0$ | $y_0$ | $y_0$ | $y_0-1$ | $y_0+1$ | $y_0-1$ | $y_0+1$ | $y_0$ |

Table 4.10: **$\mathcal{N} = 2$ long graviton multiplet (LGRAV).**

| Spin | $\frac{3}{2}$ | 1 | 1 | 1 | $\frac{1}{2}$ | $\frac{1}{2}$ | $\frac{1}{2}$ | 0 |
|---|---|---|---|---|---|---|---|---|
| Energy | $E_0+1$ | $E_0+\frac{1}{2}$ | $E_0+\frac{1}{2}$ | $E_0+\frac{3}{2}$ | $E_0+1$ | $E_0+1$ | $E_0$ | $E_0+\frac{1}{2}$ |
| R-charge | $y_0$ | $y_0-1$ | $y_0+1$ | $y_0\mp 1$ | $y_0\mp 2$ | $y_0$ | $y_0$ | $y_0\mp 1$ |

Table 4.11: **$\mathcal{N} = 2$ short gravitino multiplet (SGINO).** $E_0 = |y_0| + \frac{3}{2}$

| | | | | | | | | |
|---|---|---|---|---|---|---|---|---|
| Spin | $\frac{3}{2}$ | 1 | 1 | 1 | 1 | $\frac{1}{2}$ | $\frac{1}{2}$ | $\frac{1}{2}$ |
| Energy | $E_0+1$ | $E_0+\frac{3}{2}$ | $E_0+\frac{3}{2}$ | $E_0+\frac{1}{2}$ | $E_0+\frac{1}{2}$ | $E_0+2$ | $E_0+1$ | $E_0+1$ |
| R-charge | $y_0$ | $y_0-1$ | $y_0+1$ | $y_0-1$ | $y_0+1$ | $y_0$ | $y_0-2$ | $y_0$ |
| Spin | $\frac{1}{2}$ | $\frac{1}{2}$ | $\frac{1}{2}$ | 0 | 0 | 0 | 0 | |
| Energy | $E_0+1$ | $E_0+1$ | $E_0$ | $E_0+\frac{3}{2}$ | $E_0+\frac{3}{2}$ | $E_0+\frac{1}{2}$ | $E_0+\frac{1}{2}$ | |
| R-charge | $y_0+2$ | $y_0$ | $y_0$ | $y_0-1$ | $y_0+1$ | $y_0-1$ | $y_0+1$ | |

Table 4.12: **$\mathcal{N}=2$ long gravitino multiplet (LGINO).**

| | | | | | |
|---|---|---|---|---|---|
| Spin | 1 | $\frac{1}{2}$ | $\frac{1}{2}$ | 0 | 0 |
| Energy | 2 | $\frac{3}{2}$ | $\frac{3}{2}$ | 2 | 1 |
| R-charge | 0 | $+1$ | $-1$ | 0 | 0 |

Table 4.13: **$\mathcal{N}=2$ massless vector multiplet (MVEC).**

| | | | | | | | |
|---|---|---|---|---|---|---|---|
| Spin | 1 | $\frac{1}{2}$ | $\frac{1}{2}$ | $\frac{1}{2}$ | 0 | 0 | 0 |
| Energy | $E_0+1$ | $E_0+\frac{3}{2}$ | $E_0+\frac{1}{2}$ | $E_0+\frac{1}{2}$ | $E_0+1$ | $E_0+1$ | $E_0$ |
| R-charge | $y_0$ | $y_0\mp 1$ | $y_0-1$ | $y_0+1$ | $y_0\mp 2$ | $y_0$ | $y_0$ |

Table 4.14: **$\mathcal{N}=2$ short vector multiplet (SVEC).** $E_0=|y_0|+1$

| | | | | | |
|---|---|---|---|---|---|
| Spin | 1 | $\frac{1}{2}$ | $\frac{1}{2}$ | $\frac{1}{2}$ | $\frac{1}{2}$ |
| Energy | $E_0+1$ | $E_0+\frac{3}{2}$ | $E_0+\frac{3}{2}$ | $E_0+\frac{1}{2}$ | $E_0+\frac{1}{2}$ |
| R-charge | $y_0$ | $y_0-1$ | $y_0+1$ | $y_0-1$ | $y_0+1$ |
| Spin | 0 | 0 | 0 | 0 | 0 |
| Energy | $E_0+2$ | $E_0+1$ | $E_0+1$ | $E_0+1$ | $E_0$ |
| R-charge | $y_0$ | $y_0-2$ | $y_0$ | $y_0+2$ | $y_0$ |

Table 4.15: **$\mathcal{N}=2$ long vector multiplet (LVEC).**

| Spin | $\frac{1}{2}$ | 0 | 0 |
|---|---|---|---|
| Energy | $E_0 + \frac{1}{2}$ | $E_0$ | $E_0 + 1$ |
| R-charge | $y_0 \mp 1$ | $y_0$ | $y_0 \mp 2$ |

Table 4.16: **$\mathcal{N} = 2$ hyper multiplet (HYP).** $E_0 = |y_0|$

## 4.B Choices of dressing for the lowest hypermultiplet

In this appendix we make a curious observation which relates the operator dimensions of the fields in the hypermultiplet in Scenario I to the ones in Scenario II at the massless level originally studied in [121]. Recall that in Scenario I the hypermultiplet contains scalar operators of dimension $\frac{2}{3}$ and $\frac{5}{3}$, and a fermionic operator of dimension $\frac{7}{6}$; in Scenario II it contains scalar operators of dimension $\frac{4}{3}$ and $\frac{7}{3}$, and a fermionic operator of dimension $\frac{11}{6}$. We show that the three mass-squared values of the fields comprising these hypermultiplets are the same for the two scenarios, but they differ only in the choice of the branches in the formulae for the dimension. For scalars in $AdS_4$ the corresponding operators have dimensions

$$\Delta_{\pm} = \frac{3}{2} \pm \sqrt{\frac{9}{4} + m^2} \qquad (4.B.1)$$

and both choices are allowed [13] for $-\frac{9}{4} < m^2 < -\frac{5}{4}$. For a scalar of $m^2 = -\frac{14}{9}$, we find that $\Delta_- = \frac{2}{3}$ giving the ground state of the Scenario I multiplet, while $\Delta_+ = \frac{7}{3}$ corresponding to the second scalar in the Scenario II multiplet. Similarly, for $m^2 = -\frac{20}{9}$, $\Delta_- = \frac{4}{3}$ giving the ground state scalar in Scenario II, while $\Delta_+ = \frac{5}{3}$ corresponding to the second scalar in Scenario I. For the fermionic operators the correct formula is [137]

$$\Delta_f = 1 + |m + \tfrac{1}{2}| \; . \tag{4.B.2}$$

We find that with $m^2 = \frac{1}{9}$ the two choices of sign, $m = \pm\frac{1}{3}$, reproduce dimensions $\frac{7}{6}$ and $\frac{11}{6}$. Thus, for this part of the spectrum the distinction between the two scenarios does not concern the $m^2$ spectrum in $AdS_4$ but only the boundary conditions. However, we note that this relationship does not persist to higher levels where completely different values of $m^2$ occur in the two scenarios.

## 4.C Supermultiplets at higher levels

In this appendix, we list the $\mathcal{N} = 2$ supermultiplets of gravity states at the first few Kaluza-Klein levels $n$. We group them according to the SU(3) representations $[a, b]$ under which they transform. One observes that at level $n$ exactly those SU(3) representations occur which satisfy $a + b \leq n + 2$. Furthermore, the supermultiplets with representation $[b, a]$ are conjugate to the ones in the representation $[a, b]$ in the sense that their R-charge is negated.

In the first subsection of this appendix we present the spectrum following from the embedding of SU(3) $\times$ U(1)$_R$ into SO(8) which yields agreement with the gauge theory (Scenario I). For comparison we also exhibit the first few levels of the spectrum resulting form Scenario II in the second subsection. The acronyms as MGRAV, SGINO, etc. refer to the $\mathcal{N} = 2$ supermultiplets defined in the tables 4.8 to 4.16 in App. 4.A. The numbers following the acronyms specify the R-charges of the supermultiplets of this kind.

Since parity is broken, there are some ambiguities for grouping the states into supermultiplets. For certain ranges of R-charges one finds SVEC$_y$ $\cup$ HYP$_{y+2}$ = LVEC$_y$ and SGRAV$_y$ $\cup$ SGINO$_{y+1}$ = LGRAV$_y$. In these cases we have noted the long multi-plets in the tables below. These ambiguities can only be resolved by an explicit KK reduction, but in any case they do not affect the four series of short operators which we are mainly interested in.

### 4.C.1 Scenario I

| | | |
|---|---|---|
| $[0,0]$<br>MGRAV 0<br>LVEC 0 | $[0,1]$<br>SGINO $-\frac{1}{3}$ | $[0,2]$<br>HYP $-\frac{2}{3}$ |
| $[1,0]$<br>SGINO $+\frac{1}{3}$ | $[1,1]$<br>MVEC 0 | |
| $[2,0]$<br>HYP $+\frac{2}{3}$ | | |

Table 4.17: **Multiplets of IR theory at level $n = 0$.**

| | | | |
|---|---|---|---|
| $[0,0]$<br>SGRAV $+1, -1$<br>LVEC $+1, -1$ | $[0,1]$<br>LGRAV $-\frac{1}{3}$<br>LGINO $+\frac{2}{3}$<br>LVEC $-\frac{1}{3}$ | $[0,2]$<br>SGINO $-\frac{2}{3}$<br>LVEC $+\frac{1}{3}$ | $[0,3]$<br>HYP $-1$ |
| $[1,0]$<br>LGRAV $+\frac{1}{3}$<br>LGINO $-\frac{2}{3}$<br>LVEC $+\frac{1}{3}$ | $[1,1]$<br>LGINO 0 | $[1,2]$<br>SVEC $-\frac{1}{3}$ | |
| $[2,0]$<br>SGINO $+\frac{2}{3}$<br>LVEC $-\frac{1}{3}$ | $[2,1]$<br>SVEC $+\frac{1}{3}$ | | |
| $[3,0]$<br>HYP $+1$ | | | |

Table 4.18: **Multiplets of IR theory at level $n = 1$.**

| | | | | |
|---|---|---|---|---|
| $[0,0]$<br>LGRAV 0<br>SGRAV $-2$, $+2$<br>LVEC $-2$, 0, $+2$ | $[0,1]$<br>LGRAV $-\frac{4}{3}$, $+\frac{2}{3}$<br>LGINO $-\frac{1}{3}$, $-\frac{1}{3}$, $+\frac{5}{3}$<br>LVEC $-\frac{4}{3}$, $+\frac{2}{3}$ | $[0,2]$<br>LGRAV $-\frac{2}{3}$<br>LGINO $+\frac{1}{3}$<br>LVEC $-\frac{2}{3}$, $-\frac{2}{3}$, $+\frac{4}{3}$ | $[0,3]$<br>SGINO $-1$<br>LVEC 0 | $[0,4]$<br>HYP $-\frac{4}{3}$ |
| $[1,0]$<br>LGRAV $-\frac{2}{3}$, $+\frac{4}{3}$<br>LGINO $-\frac{5}{3}$, $+\frac{1}{3}$, $+\frac{1}{3}$<br>LVEC $-\frac{2}{3}$, $+\frac{4}{3}$ | $[1,1]$<br>LGRAV 0<br>LGINO $-1$, $-1$, $+1$, $+1$<br>LVEC 0, 0 | $[1,2]$<br>LGINO $-\frac{1}{3}$, $-\frac{1}{3}$<br>LVEC $+\frac{2}{3}$ | $[1,3]$<br>SVEC $-\frac{2}{3}$ | |
| $[2,0]$<br>LGRAV $+\frac{2}{3}$<br>LGINO $-\frac{1}{3}$<br>LVEC $-\frac{4}{3}$, $+\frac{2}{3}$, $+\frac{2}{3}$ | $[2,1]$<br>LGINO $+\frac{1}{3}$, $+\frac{1}{3}$<br>LVEC $-\frac{2}{3}$ | $[2,2]$<br>LVEC 0 | | |
| $[3,0]$<br>SGINO $+1$<br>LVEC 0 | $[3,1]$<br>SVEC $\frac{2}{3}$ | | | |
| $[4,0]$<br>HYP $+\frac{4}{3}$ | | | | |

Table 4.19: **Multiplets of IR theory at level $n = 2$.**

| | | | | | |
|---|---|---|---|---|---|
| $[0,0]$<br>LGRAV $-1$, $+1$<br>SGRAV $-3$, $+3$<br>LVEC $-3$, $-1$, $-1$, $+1$, $+1$, $+3$ | $[0,1]$<br>conj. to $[1,0]$ | $[0,2]$<br>conj. to $[2,0]$ | $[0,3]$<br>conj. to $[3,0]$ | $[0,4]$<br>conj. to $[4,0]$ | $[0,5]$<br>conj. to $[5,0]$ |
| $[1,0]$<br>LGRAV $-\frac{5}{3}$, $+\frac{1}{3}$, $+\frac{7}{3}$<br>LGINO $-\frac{8}{3}$, $-\frac{2}{3}$, $-\frac{2}{3}$, $+\frac{4}{3}$, $+\frac{4}{3}$<br>LVEC $-\frac{5}{3}$, $-\frac{1}{3}$, $-\frac{1}{3}$, $+\frac{7}{3}$ | $[1,1]$<br>LGRAV $-1$, $+1$<br>LGINO $-2$, $-2$, $0$, $0$, $0$, $0$, $+2$, $+2$<br>LVEC $-1$, $-1$, $+1$, $+1$ | $[1,2]$<br>conj. to $[2,1]$ | $[1,3]$<br>conj. to $[3,1]$ | $[1,4]$<br>conj. to $[4,1]$ | |
| $[2,0]$<br>LGRAV $-\frac{1}{3}$, $+\frac{5}{3}$<br>LGINO $-\frac{4}{3}$, $+\frac{2}{3}$, $+\frac{2}{3}$<br>LVEC $-\frac{7}{3}$, $-\frac{1}{3}$, $-\frac{1}{3}$, $+\frac{5}{3}$, $+\frac{5}{3}$ | $[2,1]$<br>LGRAV $+\frac{1}{3}$<br>LGINO $-\frac{2}{3}$, $-\frac{2}{3}$, $+\frac{4}{3}$, $+\frac{4}{3}$<br>LVEC $-\frac{5}{3}$, $+\frac{1}{3}$, $+\frac{1}{3}$, $+\frac{1}{3}$ | $[2,2]$<br>LGINO $0$, $0$<br>LVEC $-1$, $+1$ | $[2,3]$<br>conj. to $[3,2]$ | | |
| $[3,0]$<br>LGRAV $+1$<br>LGINO $0$<br>LVEC $-1$, $+1$, $+1$ | $[3,1]$<br>LGINO $+\frac{2}{3}$, $+\frac{2}{3}$<br>LVEC $-\frac{1}{3}$ | $[3,2]$<br>LVEC $+\frac{1}{3}$ | | | |
| $[4,0]$<br>SGINO $+\frac{4}{3}$<br>LVEC $+\frac{1}{3}$ | $[4,1]$<br>SVEC $+1$ | | | | |
| $[5,0]$<br>HYP $+\frac{5}{3}$ | | | | | |

Table 4.20: **Multiplets of IR theory at level $n = 3$.**

### 4.C.2 Scenario II

| | | |
|---|---|---|
| $[0,0]$<br>MGRAV 0<br>LVEC 0 | $[0,1]$<br>SGINO $-\frac{1}{3}$ | $[0,2]$<br>HYP $\frac{4}{3}$ |
| $[1,0]$<br>SGINO $+\frac{1}{3}$ | $[1,1]$<br>MVEC 0 | |
| $[2,0]$<br>HYP $-\frac{4}{3}$ | | |

Table 4.21: **Multiplets of Scenario II IR theory at level $n = 0$.**

| | | | |
|---|---|---|---|
| $[0,0]$<br>SGRAV $+0$, $-0$<br>LVEC 0, 0 | $[0,1]$<br>conj. to $[1,0]$ | $[0,2]$<br>conj. to $[2,0]$ | $[0,3]$<br>conj. to $[3,0]$ |
| $[1,0]$<br>SGRAV $-\frac{2}{3}$<br>LGINO $+\frac{1}{3}$<br>SGINO $+\frac{1}{3}$<br>LVEC $-\frac{2}{3}$ | $[1,1]$<br>LGINO 0 | $[1,2]$<br>conj. to $[2,1]$ | |
| $[2,0]$<br>SGINO $-\frac{1}{3}$<br>SVEC $-\frac{4}{3}$<br>HYP $-\frac{4}{3}$ | $[2,1]$<br>SVEC $-\frac{2}{3}$ | | |
| $[3,0]$<br>HYP $-2$ | | | |

Table 4.22: **Multiplets of Scenario II IR theory at level $n = 1$.**

| | | | | |
|---|---|---|---|---|
| [0, 0]<br>LGRAV 0<br>SGRAV $-0$, $+0$<br>LVEC 0, 0, 0, 0 | [0, 1]<br>conj. to [1, 0] | [0, 2]<br>conj. to [2, 0] | [0, 3]<br>conj. to [3, 0] | [0, 4]<br>conj. to [4, 0] |
| [1, 0]<br>LGRAV $-\frac{2}{3}$<br>SGRAV $-\frac{2}{3}$<br>LGINO $+\frac{1}{3}$, $+\frac{1}{3}$, $+\frac{1}{3}$<br>SGINO $+\frac{1}{3}$<br>LVEC $-\frac{2}{3}$, $-\frac{2}{3}$ | [1, 1]<br>LGRAV 0<br>LGINO $-1$, $+1$<br>SGINO $-1$, $+1$<br>SVEC $-0$, $-0$, $+0$, $+0$<br>HYP $-2$, $+2$ | [1, 2]<br>conj. to [2, 1] | [1, 3]<br>conj. to [3, 1] | |
| [2, 0]<br>SGRAV $-\frac{4}{3}$<br>LGINO $-\frac{1}{3}$, $-\frac{1}{3}$<br>LVEC $-\frac{4}{3}$, $+\frac{2}{3}$<br>HYP $-\frac{4}{3}$ | [2, 1]<br>LGINO $+\frac{1}{3}$<br>SGINO $-\frac{5}{3}$<br>LVEC $-\frac{2}{3}$<br>SVEC $-\frac{2}{3}$ | [2, 2]<br>LVEC 0 | | |
| [3, 0]<br>SGINO $-1$<br>SVEC $-2$<br>HYP $-2$ | [3, 1]<br>SVEC $-\frac{4}{3}$ | | | |
| [4, 0]<br>HYP $-\frac{8}{3}$ | | | | |

Table 4.23: **Multiplets of Scenario II IR theory at level $n = 2$.**

# 4.D Monopole Operators

The monopole (or 't Hooft) operators in $2+1$ dimensions can be viewed as changing the boundary conditions for fields in the path integral in a way that produces some specified magnetic flux through an $S^2$ around some point $x$. Hence these can also be called monopole creation operators [138] and are local.

We can classify the flux of magnetic monopoles in a 3d gauge theory using the scheme in [139][7]. We take the singularity to be of the form,

$$F \sim *d\left(\frac{1}{|x|}\right)M \tag{4.D.1}$$

where $M$ is some generator of the gauge group $G$. The generalized Dirac quantization condition is,

$$e^{2\pi i M} = 1 \tag{4.D.2}$$

By conjugation, $M$ can be brought to the form $\beta^a G_a$ where $G_a$ are the Cartan generators of $G$.

When there is a Chern-Simons term with level $k$, such monopoles transform in a representation of $G$. For example, consider an abelian theory on $S^2 \times R$ (i.e in the radial quantization picture) with the Chern-Simons term $k \int A \wedge dA$. With $n$ units of flux through the $S^2$, we can integrate the Chern-Simons term over $S^2$ to obtain $kn \int A_0 dt$ which is a coupling to a particle of charge $kn$.

In general, a monopole with flux $\beta^a$ transforms in the representation of $G$ with highest weight state given by $k\beta^a$. Let us illustrate this in the case of $\mathrm{U}(N)$. The quantization condition is solved (up to conjugation) by $M$ in the form of a diagonal matrix $\mathrm{diag}(m_1, m_2, \ldots, m_N)$ with $m_1 \geq m_2 \geq \ldots \geq m_N$ all being integers (cf. [140]). Such a monopole would transform in a representation of $\mathrm{U}(N)$ with the highest weight state given by $(km_1, km_2, \ldots, km_N)$. In the notation of [140], this corresponds to a Young tableaux with rows of length $km_1, km_2, \ldots, km_N$. We note that since we are interested in representations of $\mathrm{U}(N)$ and not $\mathrm{SU}(N)$, we must keep track of columns of length $N$ since they give the charge under the central $\mathrm{U}(1)$ subgroup of $\mathrm{U}(N)$.

[7] The theories of interest in [139] were 4d gauge theories but the monopoles were time-independent objects identical to what we wish to insert in our 3d gauge theory.

Turning our attention to the U($N$) × U($N$) gauge theory of interest, we will be interested in monopole operators of the form $(e^{n\tau})^{a_1\ldots a_n}_{\hat{a}_1\ldots\hat{a}_n}$ which transform in conjugate representations of the two gauge groups. Hence we give the choice of flux $M$ in the first group alone. The conjugate representation is understood to be chosen in the other U($N$).

**$k = 1$**

The basic monopole operator for $k = 1$ transforms in the bi-fundamental representation with the simplest choice of flux,

$$M = \mathrm{diag}(1, 0, 0, \ldots, 0) \quad (e^{\tau})^{a}_{\hat{a}} \tag{4.D.3}$$

It can be used to render operators with odd powers of $\mathcal{Z}$ gauge-invariant [97]. For operators with two indices in each group, we have the following choices for the flux giving symmetric and anti-symmetric operators,

$$M = \mathrm{diag}(2, 0, 0, \ldots, 0) \qquad (e^{2\tau})^{ab}_{\hat{a}\hat{b}} = (e^{2\tau})^{ba}_{\hat{a}\hat{b}} = (e^{2\tau})^{ba}_{\hat{b}\hat{a}} \,, \tag{4.D.4}$$

$$M = \mathrm{diag}(1, 1, 0, \ldots, 0) \qquad (e^{2\tau})^{ab}_{\hat{a}\hat{b}} = -(e^{2\tau})^{ba}_{\hat{a}\hat{b}} = (e^{2\tau})^{ba}_{\hat{b}\hat{a}} \,. \tag{4.D.5}$$

The symmetric operators were used in [97] while the anti-symmetric operators are important in writing down the mass deformation discussed in this paper. Note that both choices are symmetric under the simultaneous interchange of both sets of indices.

When $N = 2$, the anti-symmetric operator can also be viewed as an abelian monopole operator creating flux for $\mathrm{U}(1)_{\mathrm{diag}}$ of U(2) × U(2) which hence carries $\mathrm{U}(1)_b$ charge due to the Chern-Simons term of ABJM theory as explained in [97, 128, 129]. Hence it was denoted $(e^{2\tau})^{ab}_{\hat{a}\hat{b}} = T^2\epsilon^{ab}\epsilon_{\hat{a}\hat{b}}$ in this paper where $T^2$ is the abelian operator with two units of $\mathrm{U}(1)_b$ charge and creates two units of flux for $\mathrm{U}(1)_{\mathrm{diag}}$.

### $\boldsymbol{k = 2}$

When $k = 2$, one cannot construct a monopole operator with the indices $(e^{\tau})^{a}_{\hat{a}}$. The smallest choice of flux $M = \text{diag}(1, 0, \ldots, 0)$, after multiplying by $k = 2$, already corresponds to an operator with two pairs of indices, $(e^{2\tau})^{ab}_{\hat{a}\hat{b}}$ symmetric in upper and lower indices separately,

$$kM = \text{diag}(2, 0, 0, \ldots, 0) \qquad\qquad (e^{2\tau})^{ab}_{\hat{a}\hat{b}} = (e^{2\tau})^{ba}_{\hat{a}\hat{b}} = (e^{2\tau})^{ba}_{\hat{b}\hat{a}} \,. \tag{4.D.6}$$

Trying to form an anti-symmetric operator with 2 indices fails since we would need $kM = \text{diag}(1, 1, 0, \ldots, 0)$ but such a $M$ would not obey the Dirac quantization above for general $N$. However, when $N = 2$, we can effectively create an anti-symmetric operator by using an abelian monopole operator charged under $\text{U}(1)_b$ as in the $k = 1$ case. Such an operator can again be written as

$$(e^{2\tau})^{ab}_{\hat{a}\hat{b}} = T^2 \epsilon^{ab} \epsilon_{\hat{a}\hat{b}} \,. \tag{4.D.7}$$

$T^2$ is again an abelian monopole operator with two units of $\text{U}(1)_b$ charge but since $k = 2$, this requires turning on only one unit of flux for $\text{U}(1)_{\text{diag}}$ unlike in the $k = 1$ case above. Formally, we can assign such an operator the flux $\text{diag}(\frac{1}{2}, \frac{1}{2})$. This satisfies the fractional quantization condition $e^{2\pi i M} = -1 \in Z(\text{SU}(2))$.[8]

[8] An important modification of the quantization condition occurs when the gauge group has a non-trivial center under which all the matter transform trivially. The gauge group is then effectively $G/Z(G)$ where $Z(G)$ is the center and the quantization condition is then $e^{2\pi i M} \in Z(G)$. For example, in a $\text{SU}(N)$ gauge theory with adjoint matter, the $\mathbb{Z}_N$ subgroup decouples. When $N = 2$, this allows for example $M = \text{diag}(\frac{1}{2}, \frac{1}{2})$ when the gauge group is taken to be $\text{SU}(2)/\mathbb{Z}_2$ since $e^{2\pi i M} = -1$. This is in addition those $M$ satisfying $e^{2\pi i M} = 1$ allowed when the gauge group is $\text{SU}(2)$.

# Chapter 5

# Non-SUSY duality

We study perturbative and non-perturbative instabilities in a non-supersymmetric family of $AdS_4$ vacua of $M$-theory of the form $AdS_4 \times Y_7$. The spaces we study include orbifolds and orientation-reversed (or 'skew-whiffed') versions of SUSY spaces and the non-SUSY $Y_7 = M^{pqr}$ coset space family, part of which is known to satisfy the Breitenlohner-Friedman(BF) classical stability conditions. Global singlet marginal operators (GSMOs) in non-SUSY theories can destroy conformal fixed points due to $1/N$ corrections to their $\beta$ functions that cause them to run. We identify a small number of GSMOs that could potentially destabilize a non-SUSY background due to $1/N$ effects and show that such GSMOs are generically absent for non-SUSY spaces that are not orbifolds or skew-whiffed versions of SUSY theories. We then study possible tunneling decay of these spaces into a bubble of nothing due to the presence of a shrinking circle, studied earlier in arxiv:0709.4262. We find that the generic non-SUSY space is unstable towards such a non-perturbative decay channel. For completeness, we also demonstrate the tunneling decay of non-SUSY $S^7$ orbifolds which unlike the $M^{pqr}$ spaces also contain perturbative instabilities from marginal multi-trace operators.

## 5.1 Introduction

Extending the $AdS/CFT$ correspondence [9, 11, 12, 6, 7] to spacetimes with no supersymmetry is of obvious interest. From the gravity point of view, these would be examples of string theory backgrounds that break all SUSY and are yet stable in several senses to be discussed later. The corresponding gauge theory would be a non-supersymmetric theory with a non-trivial strongly coupled fixed point. The extension from $\mathcal{N} = 4$ $AdS/CFT$ to less symmetric $\mathcal{N} = 1$ theories has been a very fruitful venture [18,36,25] but the $\mathcal{N} = 0$ case has been harder due to various instabilities that arise in the absence of supersymmetry. The most basic stability check would be the Breitenlohner-Friedman (BF) stability conditions which are just positive mass criteria adapted to $AdS$ space. While SUSY often assures that these bounds are satisfied, one must check the BF stability of non-SUSY spaces on a case-by-case basis. In this paper, we work with spaces verified to be BF stable in works such as [141].

Beyond BF stability, the absence of SUSY allows the possibility of other instabilities such as tadpoles for SUGRA fields or equivalently $1/N$ corrections to the $\beta$ functions of the dual gauge theory that destroy conformality [142, 16, 143, 17, 144, 145]. This can be viewed as an example of the Dine-Seiberg phenomenon detailed in [146]. As pointed out in [142], for relevant and irrelevant operators, $1/N$ effects can only shift the zeroes of $\beta$ functions and thus there is a new conformal fixed point nearby in the space of coupling constants. For marginal operators, the instability is more dire as $1/N$ effects could potentially destroy the conformal fixed point by completely destroying the zeroes of the $\beta$ function.

For $AdS_5 \times X^5$, it has long been known that all such non-SUSY backgrounds have at least the marginal direction corresponding to the gauge coupling i.e $\frac{1}{g^2}\,\mathrm{tr}\, F_{\mu\nu}^2$. Equivalently, the dilaton can always develop a tadpole. We instead investigate the presence of marginal operators in two kinds of non-SUSY backgrounds of the form $AdS_4 \times X^7$. In $AdS_4$ backgrounds, there are no single-trace marginal operators like $\frac{1}{g^2}\,\mathrm{tr}\, F_{\mu\nu}^2$ above.

The first set can be characterized as non-SUSY spacetimes with SUSY spectrums. These are obtained from a SUSY spacetime by orbifolding, "skew-whiffing"[1] or some other such procedure which results in the non-SUSY spacetime inheriting the KK spectrum from the SUSY spacetime.The second set of non-SUSY $AdS_4 \times X^7$ we consider are those $X^7$ which are non-SUSY by construction, such as $M^{pqr}, Q^{pqr}$. We show that the former theories always contain marginal operators[2] while they are generically absent in members of the latter families. Hence $1/N$ effects destabilize non-SUSY backgrounds which have a SUSY-like spectrum while others like $M^{pqr}, Q^{pqr}$, if BF stable, are also likely generically stable against $1/N$ effects. A potentially more dire instability can afflict non-SUSY $AdS/CFT$ pairs as shown in [23]. The non-perturbative tunneling decay of Kaluza-Klein vacua with some compact directions into a bubble of nothing, first studied by Witten [147], was found to be present in several non-SUSY freely acting $Z_k$ orbifolds of $S^5$. Further, as is to be expected in the absence of any scale in a CFT, it was pointed out in [23] that the rate of such a decay is infinite. In a sense, the dual gauge theories are not merely unstable but simply do not exist as conformal theories.

In this paper, we investigate such tunneling instability in M-theory vacua of the type $AdS_4 \times X_7$. We work out two examples in detail – the non-SUSY orbifolds of $S^7$ and the family $M^{pqr}$. We study this instability by numerically investigating the existence of a tunneling ansatz symmetric on the internal space. We find that almost all non-SUSY backgrounds are unstable towards such a tunneling. (It is possible that the seeming exception also turns out to be unstable if a more general ansatz is considered.) In Section 5.3.1, we discuss orbifolds of $S^7$ while in Section 5.3.2, we consider the $M^{pqr}$ spaces (though we switch notation and use the equivalent $M^{mn}$ notation of [148]. The exact relation between the two notations are presented in the sections.) Details such as the Einstein equations involved are presented in appendices.

[1]Skew-whiffing (cf. [142,126]) involves reversing the orientation of a SUGRA background which in terms of vielbeins $e^a_\mu$ is the operation $e^a_\mu \to -e^a_\mu$, as explained in later sections.

[2]In [142], only potential single-trace marginal operators were considered and found absent in certain orbifolded and skew-whifed compactifications. However, these spaces are found to be destabilized by multi-trace marginal operators here.

## 5.2 Perturbative Stability

Supergravity backgrounds of the form $AdS_4 \times X^7$ are dual to gauge theories, in the $N \to \infty$ limit. In this case, the $AdS$ factor ensures that the dual gauge theory is conformal. For large but finite $N$, there could be $1/N$ corrections to the beta functions of the gauge theory and correspondingly, a tadpole for SUGRA fields that shifts the true vacuum [142, 16, 143, 17, 144]. While such corrections are known to be absent in supersymmetric cases, for non-supersymmetric compactifications, such effects could in principle destroy the conformal fixed point by lifting the zeroes of the $\beta$ functions for certain operators and causing the theory to flow to a far away point in coupling-constant space. In such cases, the duality would not exist even for large but finite $N$ but only at the formal point $N = \infty$.

In [142], it was pointed out that only operators marginal in the limit $N \to \infty$ are vulnerable to $1/N$ effects. It was argued that $1/N$ corrections to a relevant or irrelevant operator would only have the effect of shifting the zero of the $\beta$ function. For example, the $1/N$ perturbed $\beta$ function for the coupling constant $g$ of an operator $\mathcal{O}$ is of the form,

$$\beta(g) = \kappa(g - g_*) + \frac{a}{N} + \dots . \tag{5.2.1}$$

Here $g = g_*$ is the value of the coupling at the putative fixed point at $N = \infty$ and $\kappa = \Delta - d$ where $d$ is the dimensionality of spacetime and $\Delta$ is the dimension of the operator $\mathcal{O}$.

This perturbed $\beta$ function has a zero near $g_*$ at $g - g_* = -\frac{a}{N\kappa}$ as long as $\kappa \neq 0$ which is exactly the case for non-marginal operators. This is also demonstrated in Figure 5.1. Hence, for relevant and irrelevant operators, $1/N$ effects can only cause a small shift of the fixed point in coupling constant space. The corresponding effect in SUGRA is the generation of tadpoles, which can occur only for massless fields (which are related to marginal operators by duality). From the point of view of (gauged) SUGRA, one also realizes that such tadpoles can only afflict fields that are singlets under the gauge symmetry [142]. In the gauge theory, this translates to the realization that only global singlet marginal operators (GSMOs) are vulnerable to $1/N$ corrections.

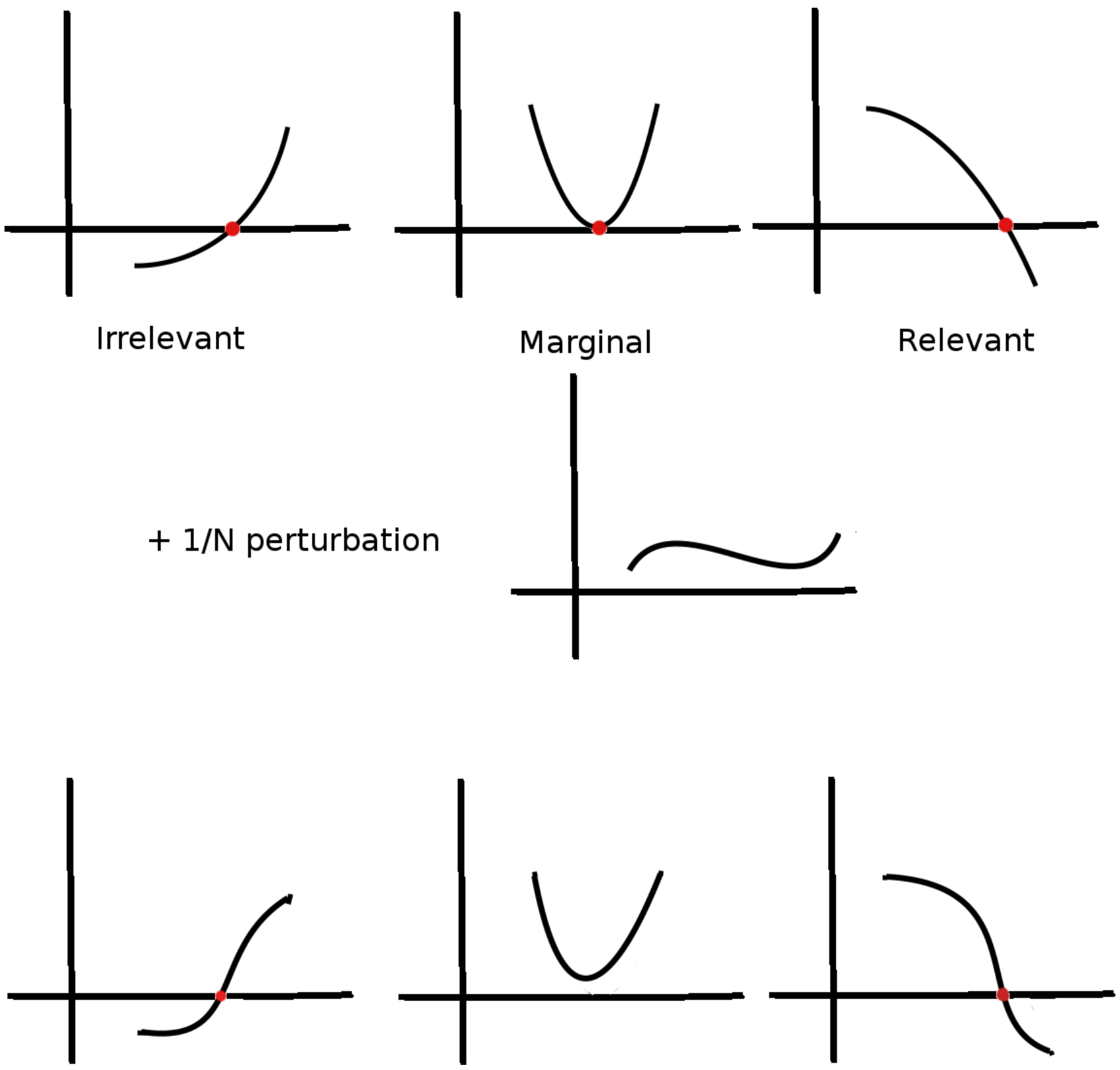

Figure 5.1: Effect of a $1/N$ contribution to the beta functions of generic irrelevant, marginal and relevant operators near a fixed point.

Hence if one can find a non-supersymmetric compactification $AdS_4 \times X^7$ with no marginal (possibly multi-trace) global-singlet operators, one can conclude that $1/N$ effects would not destabilize the conformal fixed point. It is important to note that the presence of such operators does not necessarily mean an instability but one cannot make any certain statements without detailed calculations in such a non-supersymmetric theory.

Dangerous GSMOs have always generically been present in various examples investigated to date. In compactifications of the form $AdS_5 \times X_5$, the dilaton is always unprotected against such a tadpole in the absence of SUSY and correspondingly, the $\mathrm{tr}\, F_{\mu\nu}^2$ operator related to the gauge coupling is always marginal in $3+1$ dimensions [16, 17, 149]. Further, in orbifolds of SUSY theories studied extensively at weak and strong coupling ( [21, 22]), marginal multi-trace are generically inherited from the parent SUSY theory which always contain such operators. Instabilities induced by such operators were studied in detail at weak coupling in [21, 22] and found to generically destroy non-SUSY orbifolds, even at leading order in $1/N$.

We avoid such instabilities by compactifying $M$-theory down to $AdS_4 \times X_7$. There is no generically-present massless field such as the dilaton dual to a marginal operator. We consider two types of non-supersymmetric $AdS_4$ vacua. First we consider breaking SUSY by 'skew-whiffing' or orbifolding SUSY spacetimes. We find that since the spectrum of such spaces is inherited from SUSY, GSMOs are always present and hence such spacetimes cannot be considered stable without non-trivial detailed calculations.

We then consider a family of spacetimes that are non-SUSY to begin with. We take $X_7 = M^{pqr}$ (sometimes known as $M^{mn}$), a family of non-SUSY coset spaces which are not orbifolds of any SUSY theory and hence do not inherit any marginal multi-trace operators. The spectrum needs to be investigated for marginal multi-trace operators on a case by case basis. We identify the small subset of operators that need to be studied carefully in any such non-SUSY backgrounds by studying the family of spaces $M^{pqr}$. We show that these dangerous operators are generically absent. We proceed identifying the marginal multi-trace operators in the unique SUSY space in the family, $M^{111}$, and show that the dimensions of these operators always change away from the marginal dimension 3 as we begin to change $p, q$ away from the SUSY point. Thus we argue for the $1/N$ perturbative stability of the generic non-SUSY member of the $M^{111}$ family. Our results are only partial since our analysis involves full explicit calculations of operator dimensions for the massless gauge multiplets but not for the special Betti multiplets which exist for backgrounds with non-trivial 2 cycles.

We expect similar results to hold for other non-SUSY families such as $Q^{pqr}$.

### 5.2.1 SUSY orbifolds and skew-whiffing

Orbifolding and skew-whiffing are two procedures that can produce non-SUSY spaces from SUSY spacetimes. One appeal of such constructions is due to the ease of construction since one does not need to write a whole new lagrangian or solve for a new SUGRA solution but can instead make global modifications alone. Another appeal is that it is easy to argue for the classical stability of such theories since the spectrum is inherited from the SUSY theory – either entirely in the case of skew-whiffing or a projection in the case of orbifolding. As the SUSY theory satisfies BF stability conditions, the non-SUSY theory is automatically BF stable. However, we show that precisely this fact leads to an $1/N$ instability due to GSMOs.

Orbifolding involves projecting to the singlet sector under the action of some discrete group $\Gamma$ on the fields of the theory. Generic orbifolds of SUSY spacetimes, even those of maximally SUSY spacetimes such as $AdS_4 \times S^7/\Gamma$, typically break all SUSY. The resulting spectrum is simply the portion of the SUSY spectrum that is invariant under the discrete group $\Gamma$.

| Spin | Name | Dim | $Y^W$ | KK expansion | Relevant spectrum |
|---|---|---|---|---|---|
| 1 | $A_\mu$ | 2 | 0 | $h_{\mu a} = A_\mu(x)Y_a(y) + \dots$ | $m_A^2 = M_{(1)(0)^2} + 16$ $-12\sqrt{M_{(1)(0)^2} + 16}$ |
| $\frac{1}{2}$ | $\lambda_L$ | 3/2 | $-1$ | $\psi_\mu = \gamma_\mu \lambda_L(x)\Xi(y) + \dots$ | $m_{\lambda_L} = -(M_{(1/2)^3} + 16)$ |
| $\frac{1}{2}$ | $\lambda_L$ | 3/2 | $+1$ | $\psi_\mu = \gamma_\mu \lambda_L(x)\Xi(y) + \dots$ | $m_{\lambda_L} = -(M_{(1/2)^3} + 16)$ |
| 0 | $\pi$ | 2 | 0 | $a_{abc} = \pi(x)Y_{[abc]} + \dots$ | $m_\pi^2 = 16M_{(1)^3}(M_{(1)^3} - 3)$ |
| 0 | $S$ | 1 | 0 | $h_{ab} = \delta_{ab}S(x)Y(y) + \dots$ | $m_S^2 = M_{(0)^3} + 144$ $-24\sqrt{M_{(0)^3} + 36}$ |

Table 5.1: Structure of the massless $AdS_4$ gauge multiplet. This multiplet always transforms in the adjoint representation of the global symmetry group $G$ of the dual field theory.

Skew-whiffing (cf. [142, 126]) involves reversing the orientation of a SUGRA background which in terms of vielbeins $e_\mu^a$ is the operation $e_\mu^a \to -e_\mu^a$. Such a background is still a solution of the equations of motion. The spectrum changes in that the positive and negative parts of the spectrum of fermionic laplacians are interchanged while the spectrum of bosonic operators does not change as they are quadratic in the vielbeins. This operation typically breaks all SUSY. Commonly studied examples include the squashed $S^7$ and orbifolds of the round $S^7$ [126, 142].

We will argue that any procedure such as orbifolding and skew-whiffing which inherit their spectrum from SUSY theories always contain at least one global singlet marginal operator (GSMO). This follows from the fact that any SUGRA theory based on a spacetime of the form $AdS_4 \times X^7$ contains an $AdS_4$ massless gauge multiplet, transforming in the adjoint representation of the symmetry group $G$ of $X^7$. Equivalently, the dual SUSY gauge theory must contain such a massless multiplet corresponding to the SUSY completion of the conserved global vector currents of the symmetry group $G$. For $\mathcal{N} > 2$, the multiplet occurs as part of a larger SUSY multiplet. The structure of this short multiplet is shown in Table 5.1 and is determined entirely by the properties of SUGRA on $AdS_4$ and is independent of $X^7$. This multiplet always transforms in the adjoint of the global symmetry group $G$ of $X^7$ and hence of the dual field theory. The highest component of this multiplet $A_\mu$ is a massless vector and is dual to the global conserved currents of $G$ in the field theory.

Such a multiplet survives intact through procedures like orbifolding or skew-whiffing since the top component $A_\mu$ owes its existence merely to the existence of a global symmetry group $G$ of the field theory. The superpartners of such an $A_\mu$ then retain their dimensions under orbifolding or skew-whiffing and the relations in Table 5.1 hold in the non-SUSY case as well. We focus on the two scalars $\pi, S$ found in the multiplet with dimensions 2 and 1 respectively. (Note that they have dimensions in a range where a choice of dressing allows one to pick $\Delta_+$ or $\Delta_- = 3 - \Delta_+$ for their dimension. See [13] for more details.) While each of $\pi$ and $S$ transform under $G$ (in the adjoint) and are single trace operators themselves, we can form the product double trace operator as a singlet under $G$,

$$\mathcal{O} \sim \mathrm{tr}\,(\pi S) \tag{5.2.2}$$

where tr is taken over the adjoint indices of $G$ carried by each of $\pi$ and $S$ (and not the gauge group). Hence $\mathcal{O}$ is a double trace operator which has dimension 3 and is hence marginal. It is also a singlet under $G$ by construction.

$\mathcal{O}$ is the GSMO we set out to demonstrate. It is present in any non-SUSY theory obtained from a SUSY theory by orbifolding, skew-whiffing or such procedure which does not change the dimensions of operators that survive the procedure. Hence the stability of such constructions is suspect until further explicit calculations of the $1/N$ corrections are done and found to not destroy the infinite $N$ fixed point.

### 5.2.2 Non-SUSY family $M^{pqr}$

We consider the tadpole stability of a family of non-supersymmetric $AdS_4$ compactifications and argue that generic members of such families do not contain GSMOs. Some commonly known examples are the $M^{pqr}$ and $Q^{pqr}$ families [126]. Such families are typically indexed by a set of integers and one member of the family is usually supersymmetric.

The strategy we adopt is to consider the dimensions of operators as functions on such families of non-supersymmetric spacetimes. The dimensions (or masses of the Kaluza-Klein states) often turn out to depend only on certain ratios of integers which index such families, such as for example $x = \frac{q}{3p}$ in the case of the $M^{pqr}$ manifolds. Further, for coset spaces such as $M^{pqr}$, the dimensions are algebraic functions (i.e only polynomials and fractional roots of various powers) of $x$. This is clear from the group-theoretic procedure used to find the Kaluza-Klein masses as a function of various Casimir-like quantities of the global symmetry group and using algebraic formulae typically of the form, $\Delta = \frac{d}{2} + \sqrt{\frac{d^2}{4} + m^2 L^2}$

We wish to investigate such non-supersymmetric families for GSMOs. We begin by focussing on the dimensions of operators $\Delta(x_0)$ at the point $x = x_0$ in the family which is supersymmetric. The claim is that operators which are *not* marginal at $x = x_0$ cannot be marginal in at least a small open neighborhood around $x_0$. Hence operators that are not GSMOs at the SUSY point $x_0$ cannot become GSMOs in at least a neighborhood of $x_0$.

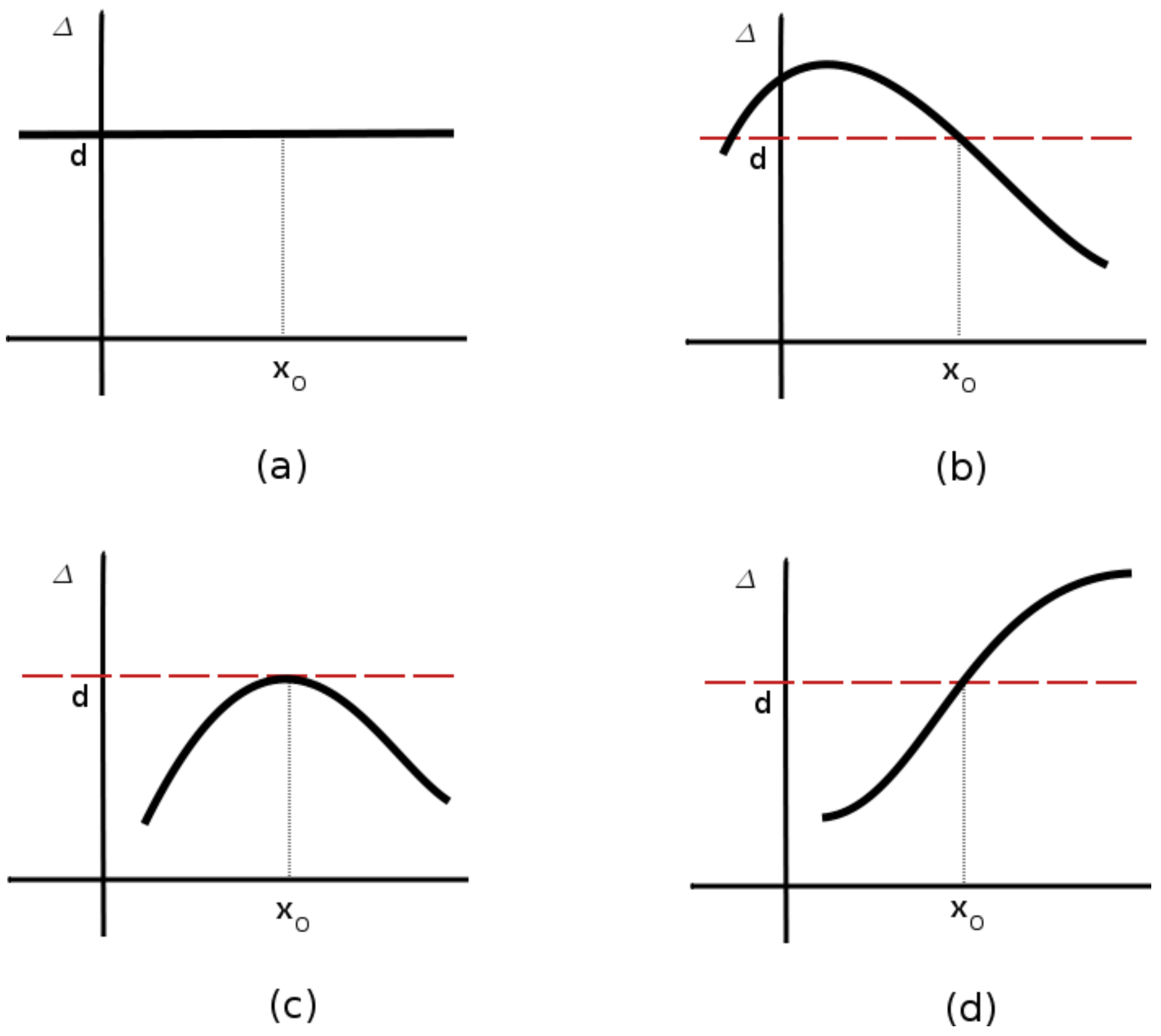

Figure 5.2: Possible behaviors of the dimension $\Delta(x)$ of an operator as a function of some $x$ parametrizing a non-SUSY family. $x = x_0$ is the SUSY member at which point the operator is marginal, $\Delta(x_0) = d$. Except in case (a), there is always a neighborhood of $x_0$ in which $\Delta(x_0) \neq d$. $\Delta(x)$ is a continuous analytic function of $x$ which might often be restricted to rational numbers.

On the other hand, there usually exist several operators that are GSMOs at $x_0$.

Since the dimension $\Delta(x)$ is an algebraic function of $x$, if such an operator stays marginal in a neighborhood around the SUSY point $x_0$, $\Delta(x)$ must in fact be a constant. This behavior is shown in Figure 5.2(a) and such special operators that stay marginal across the entire family would spell the end of perturbative tadpole stability for all of the members. It is precisely this possibility that must be investigated and ruled out. Any other behavior of $\Delta(x)$, as shown in Figures 5.2(b),(c),(d), correspond to operators that are GSMOs at the SUSY point but move away from marginality in the neighborhood of $x_0$.

In practice, the index $x$ of a family cannot be any real number. This does not pose a problem as long as the allowed set of $x$ is sufficiently dense, like for example the set of rational numbers as is the case with $M^{pqr}$.

We implement the above program with family $M^{pqr}$, first introduced by [148]. $M^{111}$ is the only supersymmetric member of this family and hence we first identify GSMOs of $M^{111}$ and consider their behavior as we move along the family. $M^{pqr}$ was constructed by Witten in the 1980s with an eye on Standard Model phenomenology. These manifolds were labeled as $M^{mn}$ by [148] and as $M^{pqr}$ in works such as [150], [125] among others. In either notation, the integers characterize the coset space,

$$\frac{SU(3)\times SU(2)\times U(1)}{SU(2)\times U(1)\times U(1)} \tag{5.2.3}$$

The exact definition of the coset in terms of the integers $p, q, r$ is given in Appendix 5.B. The integer $r$ is an orbifold[3] parameter – the space $M^{pqr}$ is the orbifold $M^{pq1}/Z_r$. As found in the appendices, the spectrum depends only on the ratio

$$x = \frac{q}{3p}. \tag{5.2.4}$$

[3]In the notation $M^{mn}$, we have $r = gcd(m, n)$ and $\frac{m}{n} = \frac{3p}{2q}$. We find it convenient to use the notation $M^{pqr}$ in this section of perturbative analysis and the notation $M^{mn}$ in the non-perturbative analysis of the next section. The relation between them is simple and one-to-one as indicated here.

The SUSY space is recovered at $x_0 = \frac{1}{3}$. The global symmetry group for all $x$ is,

$$G = SU(3) \times SU(2) \times U(1) \tag{5.2.5}$$

and hence operators transform in representations labeled by $(M_1, M_2)$,$J$ and $Y$ giving the representation under $SU(3), SU(2)$ and $U(1)$ of $G$ respectively.

For non-SUSY $M^{pqr}$, one needs to verify their BF stability before considering tadpole stability. This was done in [141] where it was found that classical stability exists in the interval,

$$\frac{29}{17\sqrt{66}} < x < \frac{9}{7\sqrt{6}} \tag{5.2.6}$$

**Marginal SUSY operators**

We will first identify all singlet marginal multi-trace operators of the SUSY case $M^{111}$. These are the only candidates for truly dangerous operators that are marginal for all $x$ (i.e for all $M^{pqr}$) if we assume that the dimensions are continuous functions of $x$.

To this end, we collect all the relevant operators for $M^{111}$ from the SUSY analysis of [125] in Table 5.2. For each SUSY multipllet, we work out the minimum energy of the Clifford vacuum and the resulting operators of least dimension. From this, we note the relevant operators and the multiplets in which they occur.

Using the table, we can write down the possible singlet marginal operators. We find that there are no single or triple trace or higher trace GSMOs. The only GSMOs that can be formed from the table above are infact double trace operators formed from the two scalars ($\mathrm{tr}\, S\ \pi$ and $\pi\ \phi$ ) or the two fermions ($\mathrm{tr}\, \lambda_L\ \lambda_L$ and $\mathrm{tr}\, \lambda_T\ \lambda_T$ ) of the massless gauge and Betti multiplets respectively. We investigate each of these in turn to check if these GSMOs remain marginal away from the SUSY point $M^{111}$ represented by $x = \frac{1}{3}$.

| Multiplet | $E_{0min}$ | Relevant fields | $(M_1, M_2), J$ | $E$ |
|---|---|---|---|---|
| Long graviton | 3 | $A_\mu$ | (1,1),0 | 3 |
| | | | (0,0),1 | 3 |
| Long gravitino | 2.5 | $\lambda_L, \lambda_T$ | (1,1),1 | 2.5 |
| | | $A_\mu, \pi, \phi$ | (1,1),1 | 3 |
| Short graviton | 4 | None | | |
| Short gravitino | 5/2 | $\lambda_{Ly=1}$ | (1,1),1 | 5/2 |
| | | $A_{\mu_{y=1,0}}$ | (1,1),1 | 3 |
| | 5/2 | $\lambda_{Ly=1}$ | (3,0),0 | 5/2 |
| | | $A_{\mu_{y=1,0}}$ | (3,0),0 | 3 |
| Short vector | 3 | S | (4,1),1 and (3,0),2 | 3 |
| Hypermultiplets | 2 | S | (3,0),1 | 2 |
| | | $\lambda_L$ | | 5/2 |
| | | $\pi$ | | 3 |
| Massless graviton | 2 | $A_\mu$ | (0,0),0 | 2 |
| | | $\chi^+_{y=-1}, \chi^+_{y=+1}$ | | 5/2 |
| | | $h_{\mu\nu}$ | | 3 |
| Massless vector | 1 | S | (1,1),0 and (0,0),1 | 1 |
| | | $\pi$ | | 2 |
| | | $\lambda_{Ly=\pm1}$ | | 3/2 |
| | | A | | 2 |
| Betti | 1 | $\pi$ | (0,0),0 | 1 |
| | | $\phi$ | | 2 |
| | | $\lambda_{Ty=\pm1}$ | | 3/2 |
| | | Z | | 2 |

Table 5.2: Relevant Operators for $M^{1,1,1}$. All complex representations above must be doubled to include the conjugate $(M_1, M_2) \to (M_2, M_1)$ and $y \to -y$

### GSMOs from the $SU(3) \times SU(2)$ Gauge multiplets

The gauge multiplets correspond to the global symmetry $G$ of the field theory and each component such as $S, \pi$ in Table 5.1 transforms in the adjoint of $G$ while the product is a singlet under $G$. We study the dimension of the two singlet marginal (at $x = \frac{1}{3}$) combinations $\mathrm{tr}\, S\pi$ and $\mathrm{tr}\, \lambda_L \lambda_L$ away from the SUSY point.

- $\mathrm{tr}\, \mathbf{S}\, \pi$

  In the AdS dual, this type of pair corresponds to operator pairs formed from an operator [151] of the type $Tr\phi\phi\phi\ldots$ by replacing an even number of the $\phi$s by fermions. [4] The energy of the scalar field $S$ is determined by the spectrum of the scalar laplacian $M_{(0)^3}$ worked out in Appendix 5.C while the pseudoscalar $\pi$s energy is determined by the 3-form laplacian $M_{(1)^3}$ found in Appendix 5.D. Some of these calculations were carried out earlier in the supersymmetric case, and partly in the non-supersymmetric case in [125, 150, 152]. We present a complete calculation in the non-SUSY case in the appendices.

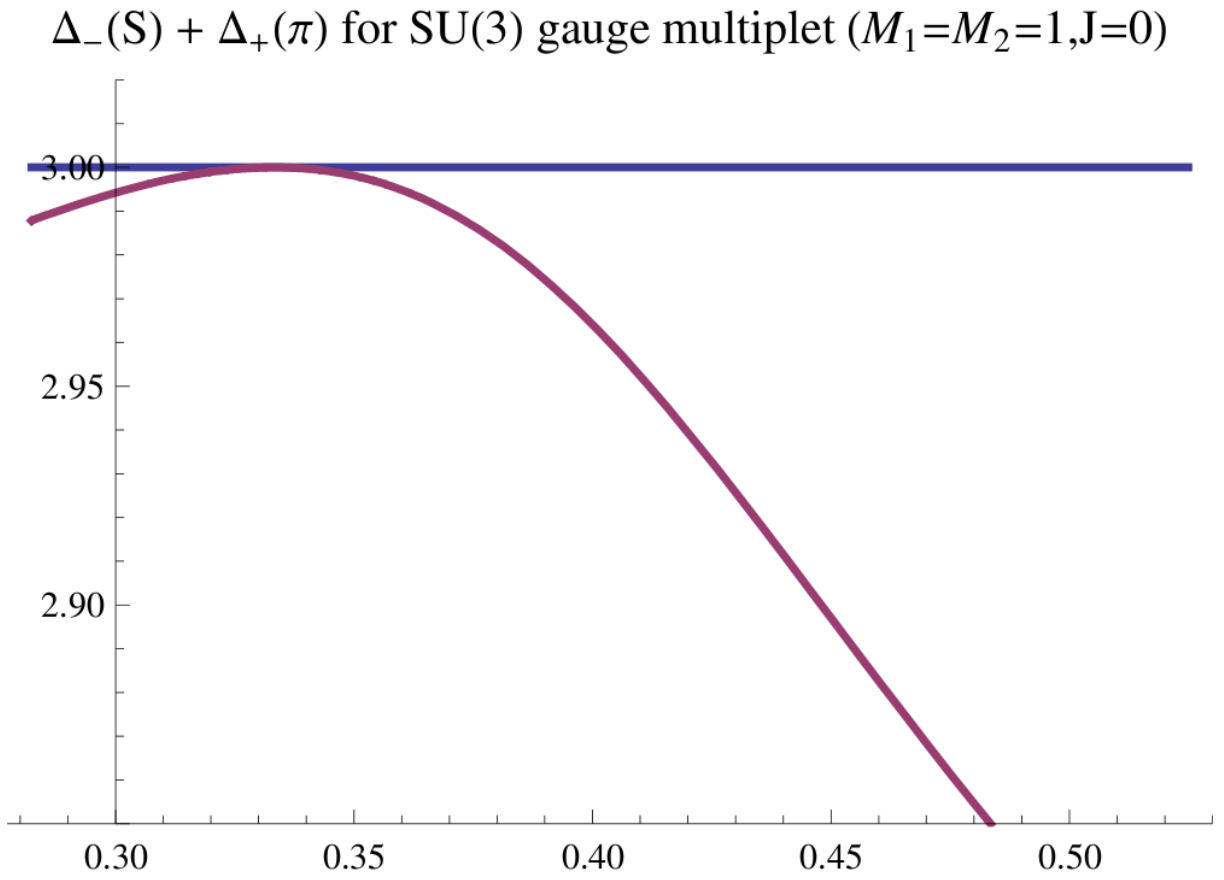

Figure 5.3: Plot of $\Delta_{-}(S) + \Delta_{+}(\pi)$ against $x$ for $SU(3)$ (i.e $M_1 = M_2 = 1, J = 0$) gauge multiplet. The operator is marginal only for $x = \frac{1}{3}$, the SUSY point.

The gauge groups in the theory are $SU(3) \times SU(2)$ and hence we could consider gauge multiplets of each of the groups. They have symmetry charges $(M_1, M_2), J = (1,1), 0$ and $(0,0), 1$ respectively. In figure 5.3, we plot the sum of the dimensions of $S$ and $\pi$ for the $SU(3)$ gauge multiplet while in figure 5.4, we plot the sum for the $SU(2)$ gauge multiplet. We have chosen to plot the combination $\Delta_{-}(S) + \Delta_{+}(\pi)$ in both cases – the combination $\Delta_{+}(S) + \Delta_{-}(\pi)$ is obtained by subtracting the shown graphs from the horizontal line at 6.

Thus we see that the dangerous marginal pair is marginal only at $x = 1/3$, the SUSY case. Further, the dimension remains strictly less (or greater) than 3 for all the non-SUSY compactifications. This non-trivial feature emerges when plotting the sum of the dimensions $S+\pi$ while each field in itself shows no special feature at $x = 1/3$. This provides a non-trivial check on our calculations.

[4] Such marginal operators are well-known in 3+1 D gauge theories such as $\mathcal{N} = 4$ and the $\mathcal{N} = \infty SU(N) \times SU(N)$ conifold theory [18]

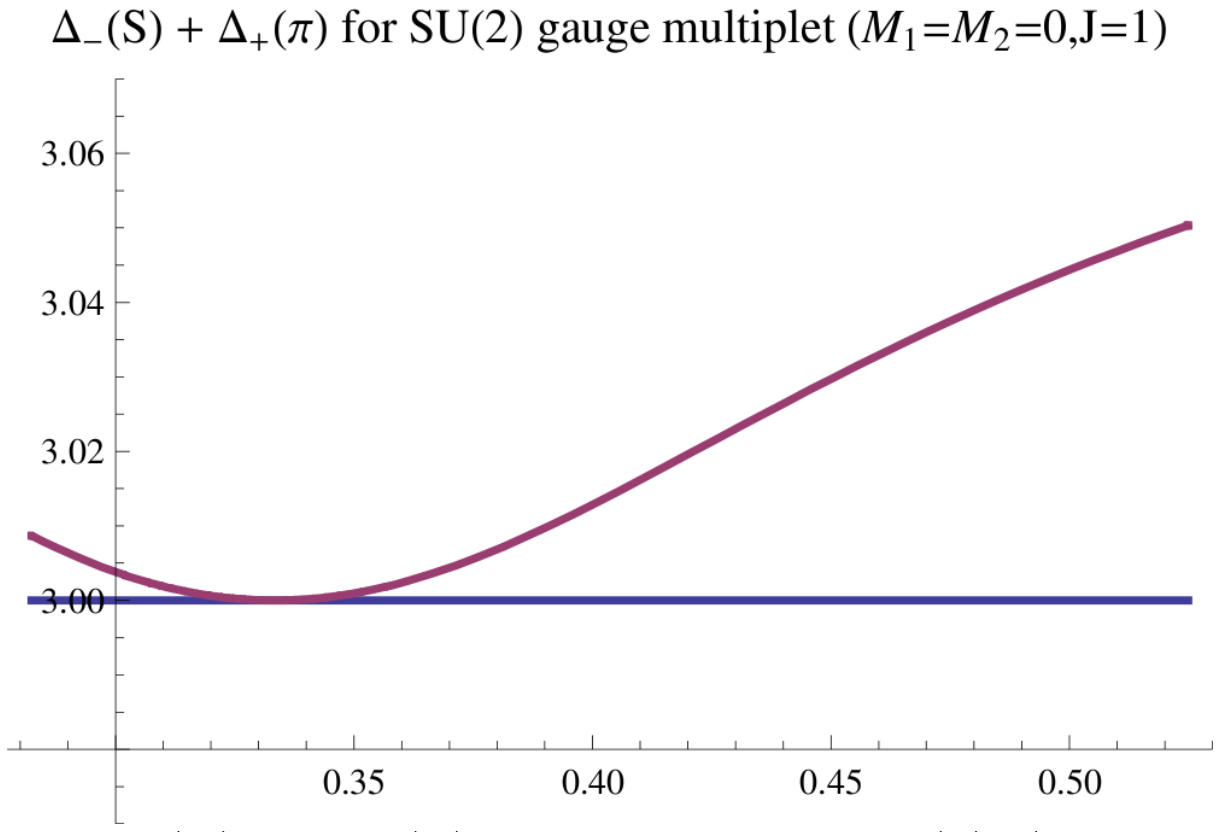

Figure 5.4: Plot of $\Delta_-(S) + \Delta_+(\pi)$ against $x$ for $SU(2)$ $(M_1 = M_2 = 0, J = 1)$ gauge multiplet. The operator is marginal only for $x = \frac{1}{3}$, the SUSY point.

While there is a unique mass given the symmetry charges for $S$, we have several eigenvalues $\lambda_1, \ldots, \lambda_8$ for $\pi$ corresponding (at the SUSY point) to the masses of $\pi$ in different SUSY multiplets. In the SUSY case, the mass of $\pi$ in the gauge multiplet is given by $\lambda_5$. The figures 5.3 and 5.4 were made using the mass $\lambda_5$. However it is possible that for some isolated non-SUSY $x \neq 1/3$, there is a marginal combination between $S$ of the gauge multiplet and $\pi$ in a different multiplet with the same symmetry charges with mass given by one of the other $\lambda_i$ of Appendix 5.D. We have carried out such checks and find no marginal combinations for any $x$.

- $\mathrm{tr}\, \lambda_{\mathbf{L}}\, \lambda_{\mathbf{L}}$

Another singlet pair can be formed by taking the two fermions $\lambda_L$ with opposite hypercharge. Since they only differ through $Y^W \to -Y^W$, they will have the same dimension which in the SUSY case is $3/2$, the minimal allowed dimension for spinors. Hence they form a marginal pair in the SUSY case.

To find their dimension for general $x$, we need the spectrum of the Dirac operator acting on 8 component Majorana spinors on $M^{pqr}$. This spectrum $M_{(1/2)^3}$ has been worked out in [153] and the cases of interest are reproduced in Appendix 5.E. From this, we compute and plot their dimensions for the $SU(2)$ (see figure 5.5) and $SU(3)$ (see figure 5.6). We find that the dimension always rises above 3/2 when $x \neq 1/3$ and hence the pair becomes irrelevant away from the SUSY point.

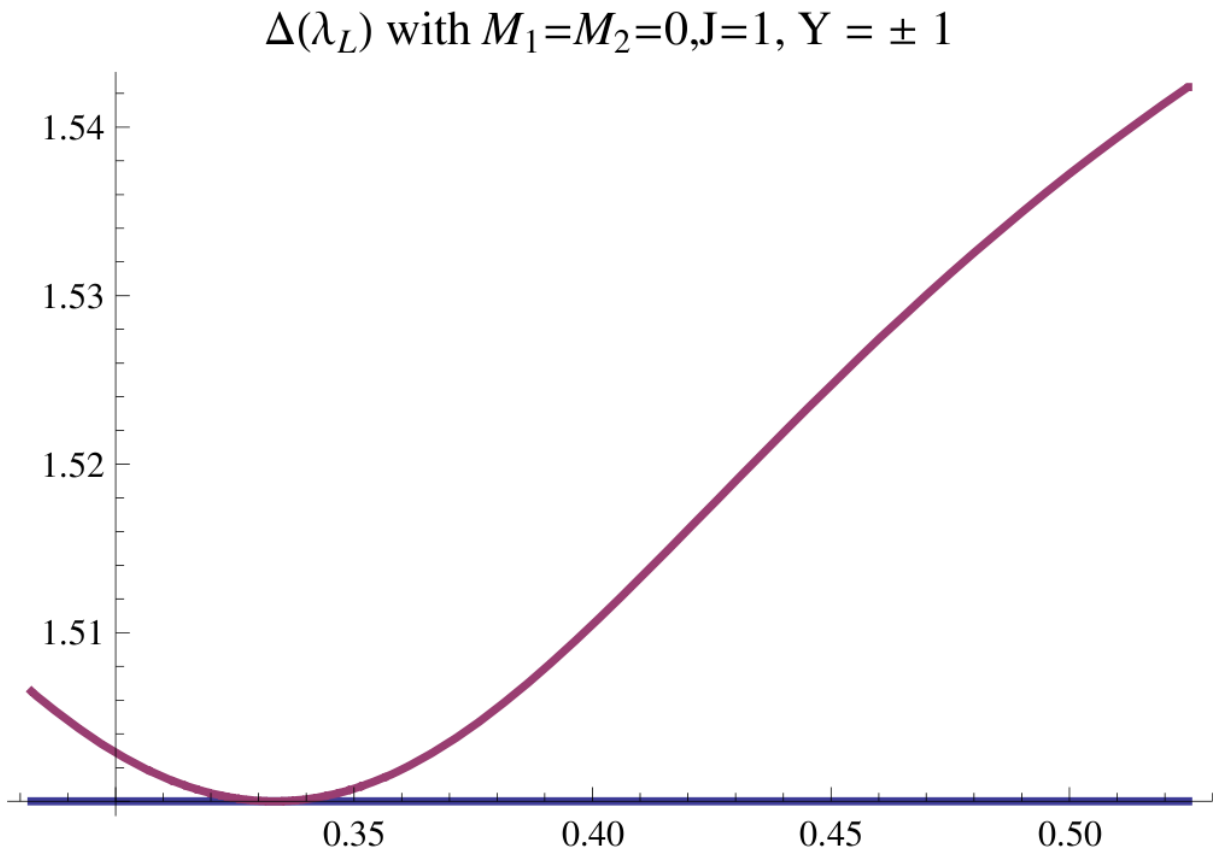

Figure 5.5: Plot of $\Delta(\lambda_L)$ against $x$ for $M_1 = M_2 = 0, J = 1$, the $SU(2)$ gauge multiplet. There are two such fermions with opposite $Y^w$ but same dimension. The product operator is marginal only for $x = \frac{1}{3}$, the SUSY point.

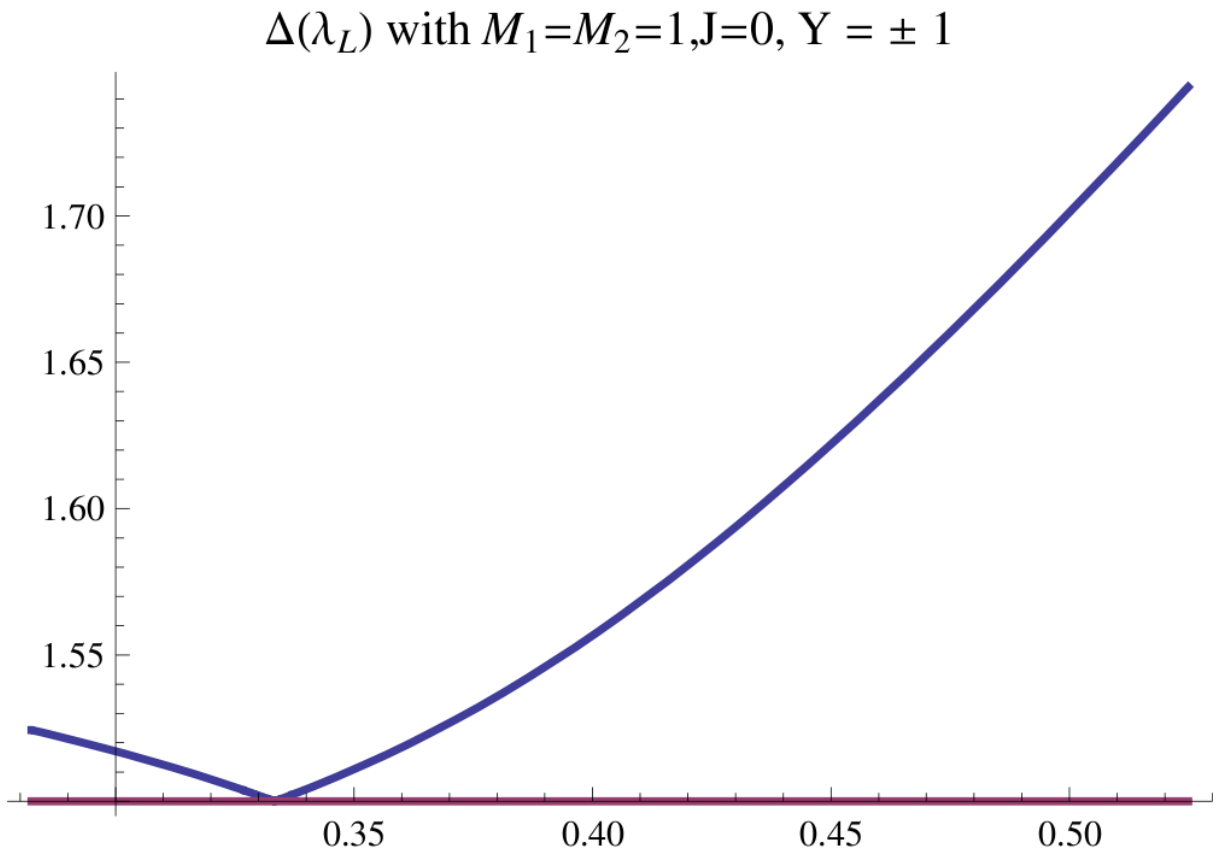

Figure 5.6: Plot of $\Delta(\lambda_L)$ against $x$ for $M_1 = M_2 = 1, J = 0$, the $SU(3)$ gauge multiplet. There are two such fermions with opposite $Y^w$ but same dimension. The product operator is marginal only for $x = \frac{1}{3}$, the SUSY point.

Hence we find that the GSMOs always found in the gauge multiplet for SUSY backgrounds, is generically not marginal for other members of the $M^{pqr}$ family. We expect similar results to hold for other non-SUSY families like $Q^{pqr}$ since our results here indicate that there is no fundamental reason that holds such operators fixed at marginality in the absence of SUSY.

### GSMOs from the Betti multiplet

The Betti multiplets, Table 5.3 are similar in structure to the gauge multiplets (shown in Table 5.1 earlier). They correspond to a 'hidden' global $U(1)$ symmetry – often called the baryonic symmetry [18] – that arises from the presence of non-trivial two cycles in the $M^{pqr}$ geometry. The Betti multiplet is naturally a singlet under $G$ and hence the double trace operators formed from components is also a singlet.

The pairs that can be formed here are similar as for the gauge multiplets i.e $\phi\pi$ and $\lambda_T\lambda_T$. But here the dimension of $\phi$ is given by Lichnerowicz operator and that of $\lambda_T$ by the Rarita-Schwinger operator. The spectrum of these two operators is much harder to calculate than those computed in the appendix and one usually avoids calculating them in SUSY cases by using SUSY mass relations instead. While we are able to make sufficient computations to show that $\phi\pi$ is not marginal away from the SUSY point $x = \frac{1}{3}$, our results are incomplete for the final operator, $\lambda_T\lambda_T$.

| Spin | Name | $\Delta$ | $Y^W$ | KK expansion | Relevant spectrum |
|---|---|---|---|---|---|
| 1 | $Z_\mu$ | 2 | 0 | $a_{\mu bc} = Z_\mu(x)Y_{bc}(y) + \dots$ | $m_Z^2 = M_{(1)^2(0)} - 32$ |
| $\frac{1}{2}$ | $\lambda_T$ | 3/2 | $-1$ | $\psi_a = \lambda_T(x)\Xi_a(y) + \dots$ | $m_{\lambda_T} = (M_{(\frac{3}{2})(\frac{1}{2})^2} + 8)$ |
| $\frac{1}{2}$ | $\lambda_T$ | 3/2 | $+1$ | $\psi_a = \lambda_T(x)\Xi_a(y) + \dots$ | $m_{\lambda_T} = (M_{(\frac{3}{2})(\frac{1}{2})^2} + 8)$ |
| 0 | $\phi$ | 2 | 0 | $h_{ab} = \phi(x)Y_{(ab)} + \dots$ | $m_\phi^2 = M_{(2)(0)^2} - 32$ |
| 0 | $\pi$ | 1 | 0 | $a_{abc} = \pi(x)Y_{[abc]} + \dots$ | $m_\pi^2 = 16M_{(1)^3}(M_{(1)^3} - 3)$ |

Table 5.3: Structure of the Betti multiplet. $\Delta$ is the SUSY dimension. $\mu$ are AdS indices, $a$ are $X_7$ indices

- $\phi\pi$

The dimension of $\pi$ has already been calculated in Appendix 5.D. There are several masses obtained for the symmetry charges $M_1 = M_2 = J = Y = 0$ corresponding to the Betti multiplet and only two of them $\lambda_8, \lambda_5$ are relevant and hence interesting. The resulting dimensions are plotted in Figure 5.7. We find that $\Delta_-(\pi) = 1$ only at the SUSY point.

The dimension of $\phi$ is given by the Lichnerowicz operator whose spectrum is difficult to calculate in general. Without any calculations, one can argue that since a VEV for $\phi$ should correspond to a blow up of the cone over $M^{pqr}$, the corresponding operator must have dimension of exactly 2. This should hold for the non-SUSY cases as well due to the non-trivial 2-cycle found in all $M^{pqr}$ as found in [152].[5] While only a direct computation of the singlet eigenvalue of the Lichnerowicz operator can confirm this argument, if we do assume that $\phi$ has fixed dimension 2, it does not form a marginal pair with $\phi$ above away from $x = 1/3$ as can be seen from Figure 5.7.

- $\boldsymbol{\lambda_T \lambda_T}$

  The dimension of $\lambda_T$ is determined by diagonalizing the Rarita-Schwinger operator, again a calculation always avoided in SUSY cases by using SUSY mass relations. However, since the smallest dimension a fermionic operator corresponding to $\lambda_T$ can have is $3/2$, the only way this pair could stay marginal for $x \neq 1/3$ is for $\Delta(\lambda_T)$ for each $\lambda_T$ to stay fixed at $3/2$. We are not able to discount this possibility without further computation.

[5] On the other hand, it is the Lichnerowicz operator that enters the BF bound criteria and violates it for some sufficiently large or small $x$ as found by [141]. At the threshold of violating the BF bound, a scalar has dimension $3/2$ in $2+1$ dimensions. Most likely the BF bound is violated not by the Betti $M_1 = M_2 = J = 0$ mode but a different mode of $\phi$.

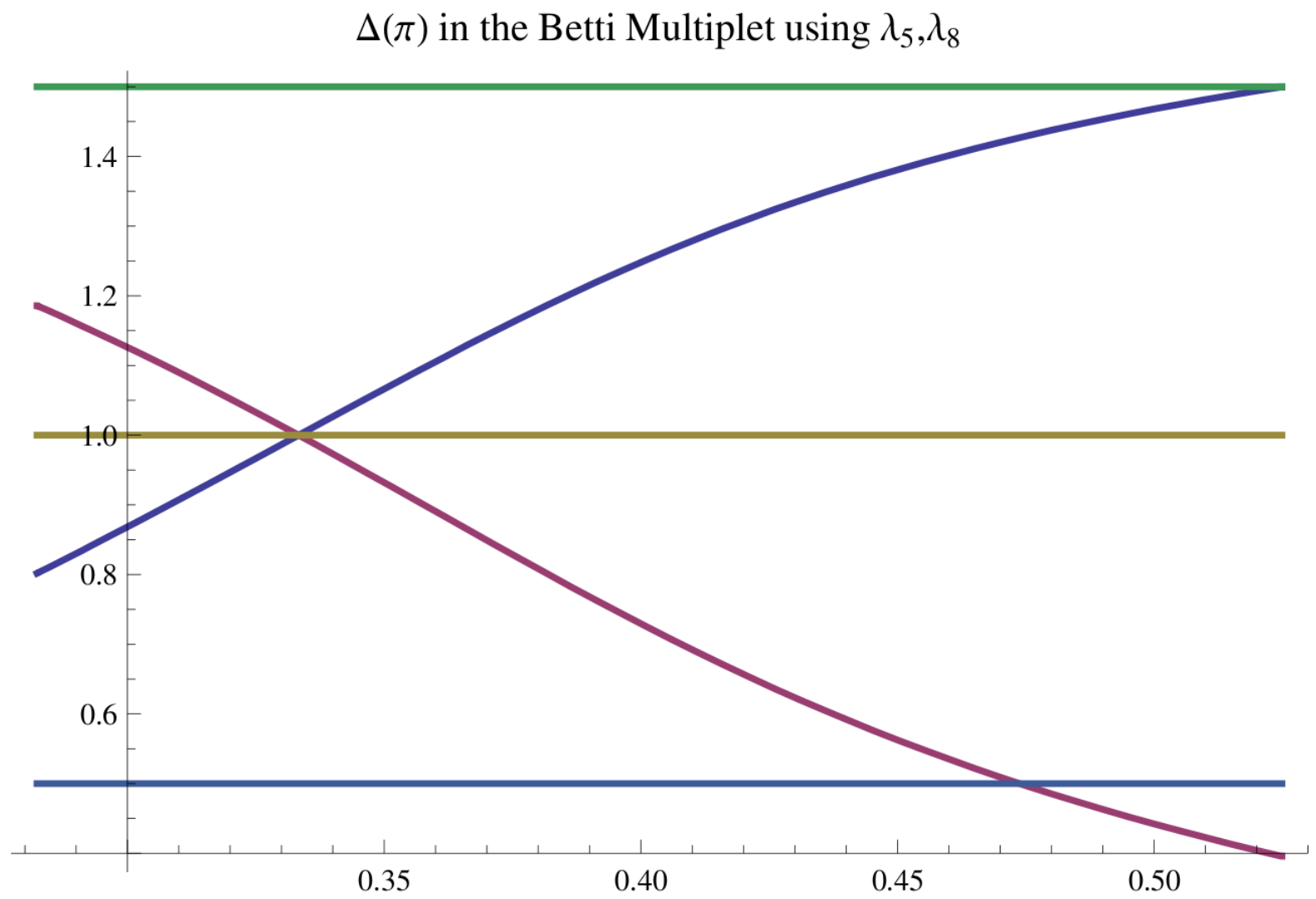

Figure 5.7: Plot of $\Delta(\pi)$ against $x$ for $M_1 = M_2 = J = 0$ i.e the Betti multiplet. The blue curve is $\lambda_8$ while the purple is $\lambda_5$. The other $\lambda$ are irrelevant.

### Conclusions on Perturbative stability

We have found that non-SUSY spaces obtained from SUSY spaces by procedures like orbifolding or skew-whiffing that do not change the spectrum inherit GSMOs from the SUSY space. In particular, there is always such an inherited GSMO formed from the two scalars in the massless gauge multiplet. Hence such constructions are suspect until computations of $1/N$ effects in such theories can establish otherwise.

For non-SUSY families such as $M^{pqr}$, we identified a small set of operators that one needs to study to rule out GSMOs. These double-trace operators are again formed from within the massless vector multiplets as above, by either combining the two scalars or the two fermions. Assuming an algebraic dependence of the dimensions $\Delta(x)$ on a parameter $x$ that parameterizes the non-SUSY family, we argued that we can restrict our study to the above operators since only they are marginal for the SUSY member of the family. By explicit computation, we were able to show that the double trace operators formed from the gauge multiplet, $\mathrm{tr}\, S\pi$ and $\mathrm{tr}\, \lambda_L \lambda_L$, are not marginal away from the SUSY point and hence not destabilizing. Our results remain incomplete since we have not performed a similar explicit (but more involved) computation for the Betti multiplet, which is a special multiplet with the structure of a gauge multiplet arising due to non-trivial 2 cycles in the topology of the manifold.

## 5.3 Non-pertubative Instability

Witten demonstrated [147] that a spacetime of the form $R^9 \times S^1$ without supersymmetric boundary conditions for fermions around the compact direction is non-perturbatively unstable. He exhibited a tunneling solution in which a 'bubble of nothing' – a sphere on which the $S^1$ shrinks to zero size forming a cigar shaped geometry and spacetime ends – nucleates at a point in spacetime. This solution is very closely related to the Euclidean continuation of the Schwarzschild black-hole geometry, where the size of the periodic time direction shrinks to zero size at the even horizon. Witten showed that such a solution can be interpreted as a bubble of nothing that nucleates expands out rapidly and wipes out all of spacetime.

It was realized in [23] that such an instability can also afflict $AdS$ compactifications, using the example of $AdS_5 \times S^5/Z_k$. The internal space, $S^5/Z_k$ in this case, can often be written as the fibration of a circle $S^1$ over a 4-dimensional base. Connecting such a structure with Witten's example, a tunneling instability was found to afflict non-SUSY orbifolds of the form $AdS_5 \times S^5/Z_k$ in [23].

We investigate if the non-SUSY $AdS_4$ compactifications studied in the previous sections, $AdS_4 \times M^{pqr}$ also exhibit such a tunneling instability. In addition, we work out the case of non-SUSY orbifolds $AdS_4 \times S^7/Z_k$ as a warm-up exercise. Our method for studying the tunneling process follows that in [23]. In what follows, we will switch to the notation $M^{mn}$ from $M^{pqr}$ of the last section. [6] Both $S^7$ and $M^{mn}$ can be viewed as $U(1)$ fibrations over a base manifold and this plays the role of the KK circle in

[6] The relation between these notations was explained earlier. We have $r = gcd(m,n)$ and $\frac{m}{n} = \frac{3p}{2q}$ and the relation is one-to-one.

Witten's [147] original set-up. The tunneling solution is obtained by ansatze similar to those in [23], differing from there primarily due to the different dimensionalities involved.

As in [23], we can show analyically that a tunneling instability exists for $Z_k$ orbifolds of $S^7$ and $M^{mn}$ when $k$ is sufficiently large (or consequently, the size of the $U(1)$ is small). Hence such spaces are unstable through such a tunneling decay. As we reduce $k$, a numerical solution of Einstein's equations is required and we find that typically the tunneling solution ceases to exist at some $k = k_{crit}$. Thus spaces for which a tunneling solution exists (i.e when $k > k_{crit}$) are unstable. On the other hand, spaces for which no solution exists ($k < k_{crit}$) can only said to be stable within the context of the ansatze we have used. Our ansatze are symmetric or smeared along the 6 dimensional base spaces over which the $S^1$ is fibered. It is possible that assuming a more general ansatz would reveal instabilities of these spaces as well.

Undertaking the numerical program outlined, we find that for the $M^{mn}$ family, all the non-SUSY members are unstable with the smeared ansatz but for $M^{11}$ and its $Z_2$ orbifold, $M^{22}$. For $S^7$, we find that $k_{crit} = 4$ and hence the only stable orbifolds are the SUSY ones, $S^7/Z_2$ and $S^7/Z_4$. One might expect similar results to hold for the $Q^{pqr}$ family which is similar in flavor to the $M^{mn}$ family.

### 5.3.1 $AdS_4 \times S^7/Z_k$

We consider freely acting $Z_k$ orbifolds of $S^7$ generated by the action,

$$z_i \to e^{2\pi i/k} z_i, \tag{5.3.1}$$

where $z_i, i = 1, \ldots, 4$ are complexified coordinates on $R^8$. The discussion here largely parallels that in [23] for $AdS_5 \times S^5/Z_k$ with some important differences in how the orbifold is defined. To find the action on the fermions, we need to consider the $\Gamma$ matrix, $\Gamma = \Gamma_{12} + \Gamma_{34} + \Gamma_{56} + \Gamma_{78}$, representing equal rotations in each of the 4 planes. The orbifold action on fermions is then generated by $g = e^{\frac{2\pi i}{k}\Gamma}$ and we look for spinors invariant under this action. It is possible to choose a basis in the $2^4 = 16$ dimensional space of spinors such that under the action of $\Gamma$, they have the 16 eigenvalues $s_1 + s_2 + s_3 + s_4$ , where $s_i = \pm\frac{1}{2}$ (c.f [154]). Under the action of $g$, we find that the 16 spinors transform with the eigenvalues $1, e^{\pm\frac{2\pi i}{k}}, e^{\pm\frac{4\pi i}{k}}$ with the following multiplicities,

$$\begin{aligned} 1 &\longrightarrow \text{6 of these} \\ e^{\pm\frac{2\pi i}{k}} &\longrightarrow \text{8 of these} \\ e^{\pm\frac{4\pi i}{k}} &\longrightarrow \text{2 of these} \end{aligned}$$

We see that for $SO(8)$, 6 spinors are always preserved by this kind of freely acting orbifold for any $k, \theta$. We do not want any SUSY to be preserved[7] for general $k$ and for this purpose, we include a factor $(-1)^F$ in the generator of the $Z_k$ orbifold,

$$g = (-1)^F e^{\frac{2\pi i}{k}\Gamma}. \tag{5.3.2}$$

Then we need one of the eigenvalues of $e^{\frac{2\pi i}{k}\Gamma}$ above to be $-1$ to preserve a fermionic spinor. This happens only for $k = 2$ (8 spinors) and $k = 4$ (2 spinors). For all other $k$, no SUSY is preserved. However, we restrict ourselves to even $k$ since when $k$ is odd, the $k$-th power of the generator $g^k = (-1)^{kF} = -1$ on fermions. This would project out all the fermions in M-theory and such a theory is beyond the scope of this publication. Hence we restrict to even $k$ and note that $k = 2, 4$ alone preserve some SUSY.

$S^7$ can be thought of as the Hopf $U(1)$ bundle over $CP^3$. Consider the complex

[7]When some SUSY is preserved, the space is not expected to be unstable even if a tunneling solution exists. This is because of the incompatibility of the unique spin structure on the tunneling solution and the SUSY spin structure [147].

variables $z_1, \ldots, z_4$ with $z_i\bar{z}_i = 1$. Defining the angular variables

$$z_1 = e^{i\chi}e^{i\phi_1}\cos\theta, \qquad z_2 = e^{i\chi}e^{i\phi_2}\sin\theta\cos\psi_1 \tag{5.3.3}$$

$$z_3 = e^{i\chi}e^{i\phi_3}\sin\theta\sin\psi_1\cos\psi_2, \qquad z_4 = e^{i\chi}\sin\theta\sin\psi_1\sin\psi_2, \tag{5.3.4}$$

$z_i$ provides coordinates on $S^7$ with $ds^2 = dz_i d\bar{z}_i$. Mod-ing out the $\chi$ rotations, we are left with $CP^3$. To obtain a metric on $CP^3$, we define the vielbeins,

$$e_1 = d\theta \tag{5.3.5}$$

$$e_2 = \sin\theta d\psi_1 \tag{5.3.6}$$

$$e_3 = \sin\theta\sin\psi_1 d\psi_2 \tag{5.3.7}$$

$$e_4 = \sin\theta\cos\theta(d\phi_1 - \cos^2\psi_1 d\phi_2 - \sin^2\psi_1\cos^2\psi_2 d\phi_3) \tag{5.3.8}$$

$$e_5 = \sin\theta\cos\psi_1\sin\psi_1(d\phi_2 - \cos^2\psi_2 d\phi_3) \tag{5.3.9}$$

$$e_6 = \sin\theta\sin\psi_1\cos\psi_2\sin\psi_2 d\phi_3 \tag{5.3.10}$$

and finally the $S^1$ fiber has the line element

$$e_7 = d\chi + \cos^2\theta d\phi_1 + \sin^2\theta\cos^2\psi_1 d\phi_2 + \sin^2\theta\sin^2\psi_1\cos^2\psi_2 d\phi_3 \equiv d\chi + A \tag{5.3.11}$$

In terms of the vielbeins, $CP^3$ has the metric $ds^2 = \sum_1^6 e_i e_i$ while on $S^7$, we have the metric $ds^2 = \sum_1^7 e_i e_i$. Here $\chi$ has period $2\pi/k$ for the $Z_k$ orbifold of $S^7$.

We make the following ansatz for the smeared bounce solution, similar to the one made in [23],

$$ds^2 = \rho(r)dr^2 + f(r)d\Omega_3^2 + g(r)ds^2_{CP^3} + h(r)(d\chi + A)^2 \tag{5.3.12}$$

where the first two terms represent a deformed $AdS_4$ space and the last two, a deformed $S^7$. As $h(r) \to 0$ at some $r_0$, the $S^1$ shrinks to zero size, giving the bubble of nothing.

We also have the $G_4$ flux of M-theory. In the appendix, it is shown that the equations of motion and the Bianchi identities give,

$$G_4 \sim \frac{R^6}{g^3 h^{1/2}} vol_{AdS_4} \tag{5.3.13}$$

Using the above form for the metric and the flux, we can write down the Einstein equations for this background. These equations are also presented in the appendix.

As in [23], we can study this ansatz in three regions – in the far UV (large $r$) where we recover $AdS_4 \times S^7$, in the tunneling region where the metric resembles the Euclidean continuation of a black hole solution and finally, very near the bubble of nothing where we have a singularity due to the smeared symmetric bubble solution (with a smeared source for the symmetric $G_4$ flux) we are considering. As explained in [23], these regions are cleanly separated only in the large $k$ limit.

In the first of the regions, the far UV ($r \sim R$), we expect the ansatz to reduce to,

$$\rho_I = 1, \qquad f_I = \frac{R^2}{4}\sinh^2(2r/R), \qquad g_I = h_I = R^2 \tag{5.3.14}$$

Using this in (5.3.12) gives a pure $AdS_4 \times S^7$ solution. Note that $R_{AdS} = R_{S^7}/2 = R/2$.

For $r \sim R/k$, we can ignore the curvature of $AdS_4 \times CP^3$ as explained in [23] and treat it as flat with a highly curved $S^1$ fibration of size $R/k$. Then we can use the known bubble solution on flat spacetime with 4 large dimensions of $AdS_4$ – the five-dimensional Euclidean black hole in the $AdS_4 + \chi$ directions.

$$\rho_{II} = \frac{1}{H(r)}, \qquad f_{II} = r^2 \qquad g_{II} = R^2 \qquad h_{II} = R^2 H(r), \tag{5.3.15}$$

$$H(r) = 1 - \frac{r_0^2}{r^2} \tag{5.3.16}$$

Defining $\tilde{r}$ as $r - r_0 = \frac{\tilde{r}^2}{2r_0}$ for $r \sim r_0$ i.e very close to the horizon at $r_0$, we find a metric of the form (ignoring the $CP^3$)

$$d\tilde{r}^2 + r_0^2 d\Omega_3^2 + \tilde{r}^2 \frac{R^2}{r_0^2}(d\chi + A)^2 \tag{5.3.17}$$

If $\chi$ is periodic with period $2\pi/k$, we need to have $r_0 = R/k$ to avoid a singularity.

Finally we have the smeared source of the $G_4$ flux which causes a singularity. Near this singularity, we can use the solution corresponding to several $M2$ branes wrapping the $S^3$ of $AdS_4$ (whose warping is $f$), smeared along the $CP^3$ directions (warped by $g$) and localized in the $r, \chi$ plane (warped by $\rho, h$). We expect log functions due to such a smearing. We find,

$$\rho_{III} = a(\log(r_*/r))^{1/3}, \qquad h_{III} = ak^2\tilde{r}^2(\log(r_*/r))^{1/3} \tag{5.3.18}$$

$$f_{III} = br_0^2 \frac{1}{(\log(r_*/r))^{2/3}} \tag{5.3.19}$$

$$g_{III} = cR^2(\log(r_*/r))^{1/3} \tag{5.3.20}$$

where $a, b, c, r_*$ are unknown constants.

As in [23], we can set $a = c$ using gauge freedom. We can show that Einstein's equations (Appendix 5.F) require,

$$a^3 = c^3 = \frac{6}{k}. \tag{5.3.21}$$

For large $k$, we can estimate the value of $b, r_*$ by matching the ansatz for region III above with region II for $r \sim r_0$. Using the values of $a, c$ obtained above, we find,

$$r_* \sim r_0 e^{\frac{k}{6}+\delta}, \qquad b \sim \frac{k^{2/3}}{36^{1/3}} + \beta \tag{5.3.22}$$

where $\beta, \delta$ are expected to be small when $k$ is large.

The above analysis is meaningful only for large $k$. For general $k$, we must resort to a numeric integration of Einstein's equations starting from the bubble region and integrating outwards to large $r$ towards the expected $AdS_4 \times S^7$ UV asymptotics.

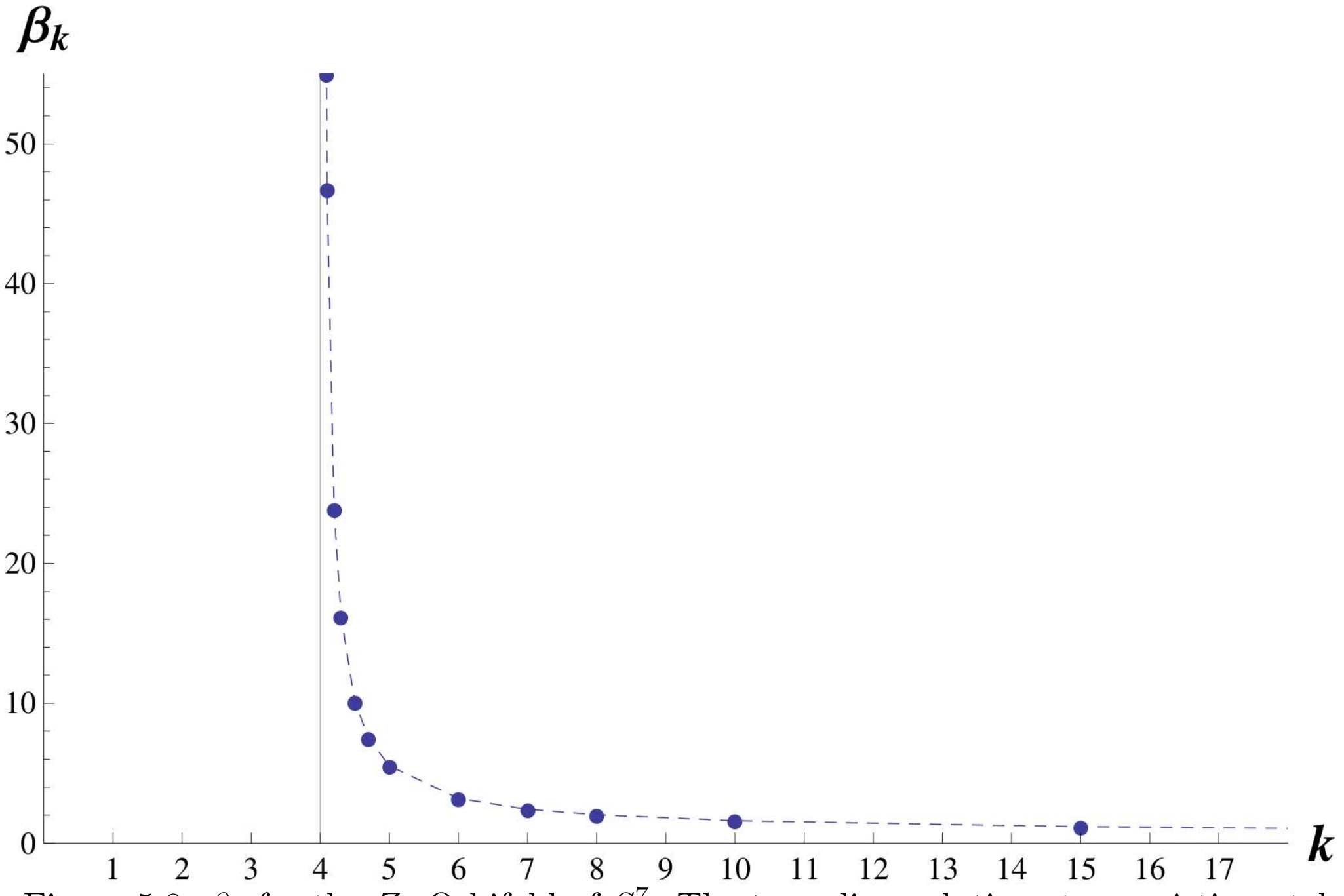

Figure 5.8: $\beta_k$ for the $Z_k$ Orbifold of $S^7$. The tunneling solution stops existing at $k = 4$.

We start with the ansatz of region III providing the initial conditions as in [23].[8] This ansatz has two unknown parameters $b, r_*$. We choose $b, r_*$ such that when we integrate outwards to large $r$, the solution asymptotes to $AdS_4 \times S^7$ of region I above. (For example, we need $h, g, \rho$ to approach 1 at large $r$. ) For general values of $b, r_*$, one can analyze the large $r$ asymptotics and find two blow-up modes $e^{6r/R}, e^{2r/R}$ in $h, g$. By tuning $b, r_*$, we can numerically obtain solutions with these modes absent up to some reasonable value of $r/R \sim 3$.

We determine $\beta, \delta$ (and hence $b, r_*$) for each $k$, using (5.3.22) as a guide for large $k$. The plot of $\beta_k$ against $k$ is shown in Fig.5.8.[9] Note that while we treat $k$ as a continuous parameter here, we use the orbifold interpretation only for even integer $k$. We find that as $k \to 4$, $\beta_k$ rapidly diverges as $\beta_k \sim \frac{1}{k-4}$. Hence we find a tunneling solution for all $k > 4$ and we conclude that all orbifolds $S^7/Z_k$ are unstable for $k > 4$.

[8] As was the case in [23], Einstein's equations presented in the Appendix 5.F are not all independent because of the Bianchi identity. Also, we can use gauge freedom to set for example $g(r) = R^2\rho(r)$, leaving us with three independent Einstein equations in the three functions $\rho, h, f$.

[9] The solid dots are the results of computation while the dashed line is a smooth interpolation.

Hence, only the $k = 2, 4$ orbifolds are seen to be stable. We have already seen that these are the only SUSY orbifolds (when the generator is defined with a factor of $(-1)^F$). Hence, in analogy with [23], we find that the SUSY solutions are dynamically stable, independent of any spin-structure argument. All the other orbifolds $k = 6, 8, \ldots$ are unstable.

### 5.3.2 $AdS_4 \times M^{pqr}$

In this section, we consider tunneling solutions for the $M^{pqr}$ manifolds, first introduced by [148]. These manifolds are labeled as $M^{pqr}$ in works such as [150], [125] among others. Here we use the notation $M^{mn}$ introduced by Witten where $m, n$ are two integers characterizing the coset space,

$$\frac{SU(3) \times SU(2) \times U(1)}{SU(2) \times U(1) \times U(1)} \tag{5.3.23}$$

As a point of connection with the $M^{pqr}$ notation used for the perturbative analysis, we note that $\frac{m}{n} = \frac{3p}{2q}$ and the SUSY manifold is $M^{111}$ or $M^{32}$ i.e with $\frac{m}{n} = \frac{3p}{2q} = \frac{3}{2}$.

This space can be viewed as a $U(1)$ bundle over the base space $CP^2 \times S^2$. Writing the metric of $CP^2$ in a way analogous to $CP^3$ of Section 5.3.1, the metric of the total space is given by,

$$\begin{aligned} ds^2 &= \frac{1}{\lambda_3^2}(d\chi + 2mA + nB)^2 \\ &+ \frac{1}{\lambda_1^2}(d\theta^2 + \sin^2\theta d\psi^2 + \cos^2\theta \sin^2\theta (d\phi_1 - \cos^2\psi d\phi_2)^2 + \sin^2\theta\cos^2\psi\sin^2\psi d\phi_2^2) \\ &+ \frac{1}{\lambda_2^2}(d\vartheta^2 + \sin^2\vartheta d\varphi^2) \\ A &= \cos^2\theta d\phi_1 + \sin^2\theta\cos^2\psi d\phi_2, \qquad B = \cos\vartheta d\varphi \end{aligned} \tag{5.3.24}$$

Here $\chi$ has period $4\pi$. $\lambda_1, \lambda_2, \lambda_3$ are the rescalings needed to make the above metric Einstein ($R_{\mu\nu} = 6g_{\mu\nu}$) for given $m, n$. $\lambda_1, \lambda_2$ depend only on $m/n$ while $\lambda_3$ transforms as $\lambda_3 \to \lambda_3 g$ when we pass from $M^{m,n} \to M^{m/g,n/g}$. For clarity, in the following, we will allow $m, n$ to be any two integers[10], and consider $Z_k$ orbifolds of such a space $M^{m,n}$ that reduce the period of $\chi$ to $4\pi/k$. The primary focus will be to investigate whether the tunneling instability which is always present for large $k$ continues to persist down to $k = 1$. If so, $M^{mn}$ is by itself unstable.

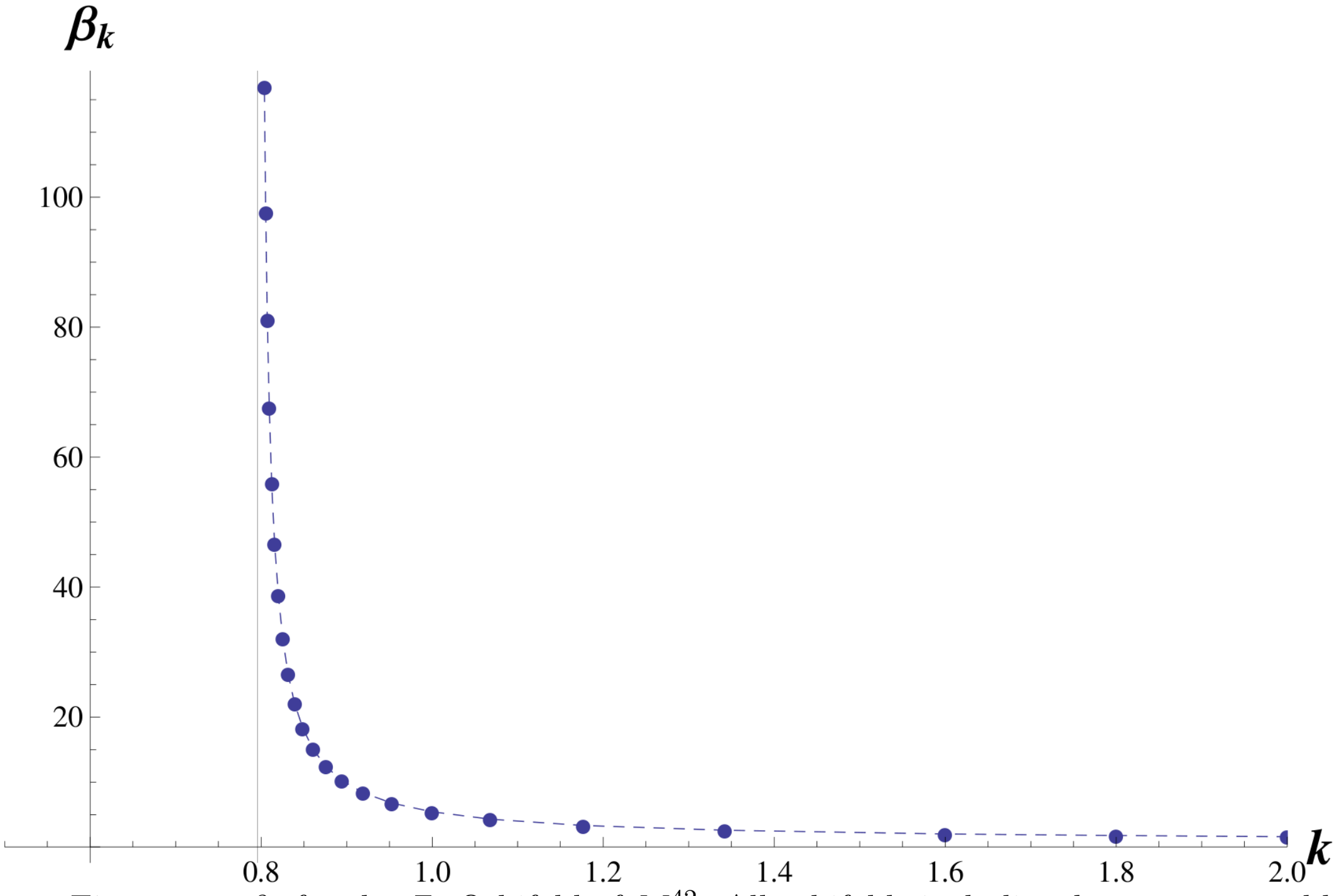

Figure 5.9: $\beta_k$ for the $Z_k$ Orbifold of $M^{42}$. All orbifolds including $k = 1$ are unstable.

The discussion of the bubble solution now parallels that in Section 5.3.1 (and in [23]) with some changes and we will be brief. We use a general ansatz for the bubble solution

[10]When $m, n$ are not co-prime and have a common factor $g$, we can pull out this factor of $g$ in the metric and redefine $\chi$ with period $4\pi/g$. This is just the $Z_g$ orbifold of the space $M^{m/g,n/g}$ and is identical to $M^{m,n}$

with individual squash factors $g_1, g_2$ for $CP^2$ and $S^2$.

$$ds^2 = \rho(r)dr^2 + f(r)d\Omega_3^2 + \frac{g_1(r)}{\lambda_1^2}ds^2_{CP^2} + \frac{g_2(r)}{\lambda_2^2}ds^2_{S^2} + \frac{h(r)}{\lambda_3^2}(d\chi + 2mA + nB)^2$$

As earlier, we have three separate regions where analytic expressions are possible when one considers the large $k$ limit. Matching up the expressions between the regions as in Section 5.3.1, we again look for numerical solutions at finite $k$ without blow-up modes up to some large $r/R \sim 3$ as a function of two parameters $\beta, \delta$. If such $\beta, \delta$ are found, a tunneling solution has been found and the space is unstable for such $k$.

For each space $M^{mn}$, we find that the tunneling solution always exists for large $k$ and hence these orbifolds are unstable. We also find that the tunneling solution stops existing for some value of $k = k_{crit}$ as in Section 5.3.1[11] and hence $Z_k$ orbifolds with $k < k_{crit}$ are stable.

Plotting $\beta_k$ against $k$ for the SUSY space $M^{32}$, we find that $\beta_k$ blows up at $k_{crit} = 1$ and hence there is no tunneling solution for the SUSY space $M^{32}$ while all its orbifolds are unstable. On the other hand, in Figure 5.9, the plot of $\beta_k$ for the non-SUSY space $M^{42}$ blows up at $k_{crit} \approx 0.795 < 1$. Hence, the tunneling solution exists at $k = 1$, showing that $M^{42}$ is unstable towards tunneling into a bubble of nothing.

In Figure 5.10, we plot[12] the value of $k_{crit}$ against $m$ for the spaces $M^{m2}$. We find that $k_{crit} < 1$ for $m > 3$, indicating unstable spaces $M^{m2}$ with behavior similar to Figure 5.9.

In fact, we find that $k_{crit} < 1$ for almost every non-SUSY member of the $M^{mn}$ family that is BF stable. The Breitenlohner-Friedman bound was worked out in [141] for these spaces. Within this BF-stable range, only three spaces are found to have $k_{crit} \geq 1$ and hence have no tunneling instability consistent with the ansatz we have used. These two spaces are the SUSY space $M^{32}$ and the non-SUSY spaces $M^{11}$ and $M^{22} = M^{11}/Z_2$, the latter of which is represented by the $m = 2$ point in Figure 5.10

[11] We treat $k$ as a continuous real number here at the level of the differential equations, though only integer $k$ have meaning as orbifold spaces.

[12] The exact numerical value of $k_{crit}$ was obtained by numerical solution through plots such as 5.9. This introduces some error in $k_{crit}$ that progressively grew as $m$ was reduced from 3. The origin of such systematic error is unclear to us but the error is small (say at most a few percent) and does not affect the important $k_{crit} > 1$ conclusion.

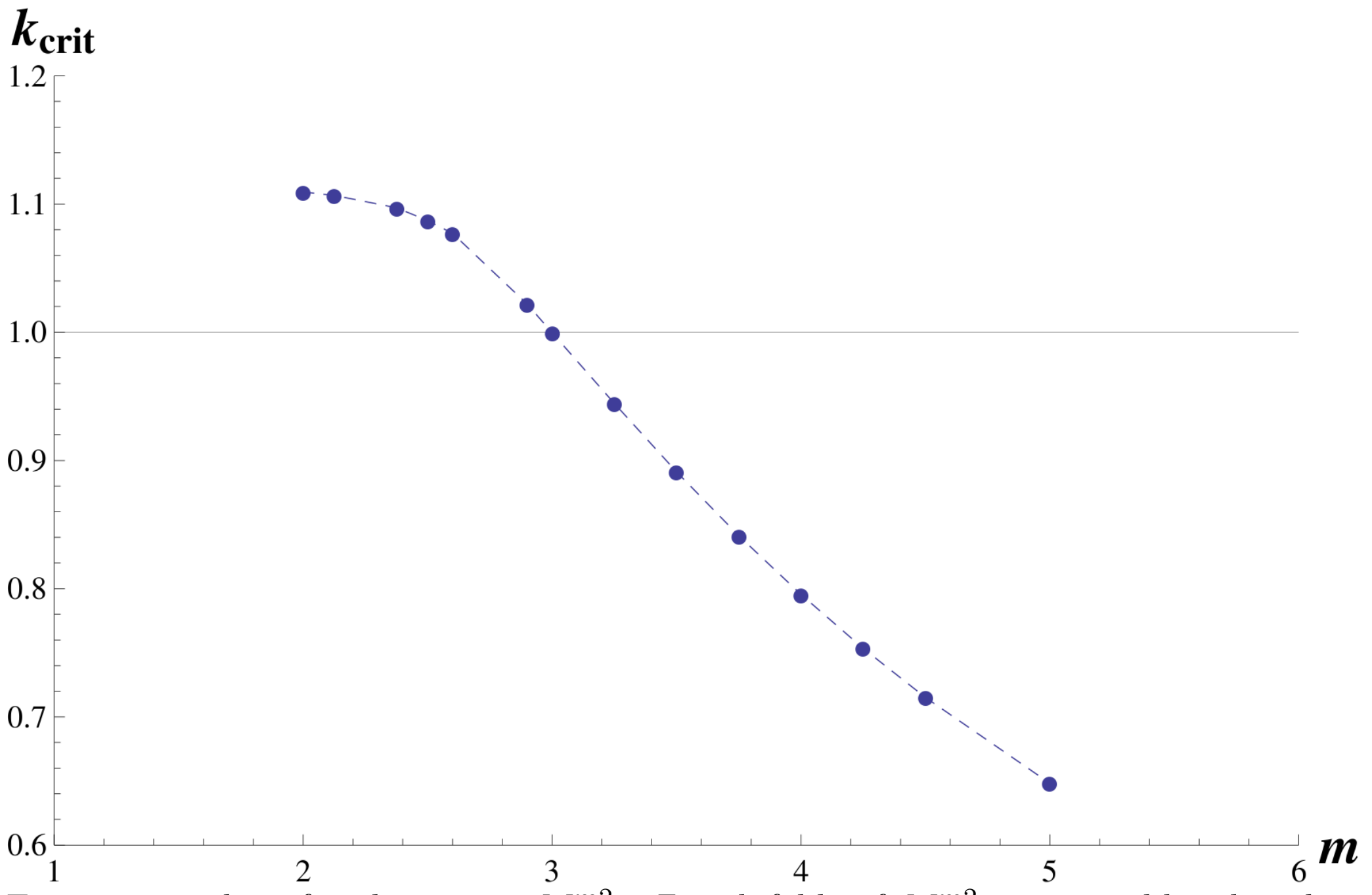

Figure 5.10: $k_{crit}$ for the spaces $M^{m2}$. $Z_k$ orbifolds of $M^{m2}$ are unstable when $k > k_{crit}$. Hence when $k_{crit} \geq 1$, the space $M^{m2}$ is stable, in itself.

## Conclusions on Non-perturbative Instability

We have investigated the non-perturbative stability of a series of M-theory backgrounds of the form $AdS_4 \times X_7$ and have found most non-supersymmetric spaces to be unstable. The instability under study involves the tunneling into a bubble of nothing due to a $S^1$ (called the KK $S^1$) in the compact Einstein manifold $X_7$, a process first studied by Witten [147]. The existence of the instability can be phrased in terms of the size of the KK $S^1$ being larger than a certain critical size.

When $X_7 = S^7/Z_k$, we find that orbifolds with $k > 4$ are unstable while $k = 2, 4$ are stable. These orbifolds preserve some SUSY and hence are expected to be stable.

However, as found in [23], it appears that the reason for stability is dynamical and the SUSY spin-structure does not play a crucial role in warding off the tunneling decay.

When $X_7 = M^{mn}$, a well-known family of non-SUSY compactifications (a part of which has long been known to be Breitenlohner-Friedman (BF) stable), the results seem intriguing. Only one member of this family, $M^{32}$, is supersymmetric and it is found to be stable against the tunneling decay. Most of the remaining non-SUSY spaces are destabilized by the tunneling since the size of the KK $S^1$ shrinks for large $m, n$. But there appear to be non-SUSY spaces - $M^{11}$ and its $Z_2$ orbifold - which do not decay by tunneling and also satisfy the BF stability bounds. We have assumed a symmetric ansatz in which the bubble solution is smeared on the base $CP^2 \times S^2$. It is possible that if we localize this bubble solution on $CP^2 \times S^2$ by using a less symmetric (and more complicated) ansatz, the spaces $M^{11}, M^{11}/Z_2$ will turn out to be unstable as well.

In conclusion, the tunneling instability is a rather generic problem in non-SUSY backgrounds of the form we have studied. Further, the instability appears related to the size of the fibered $S^1$ which becomes just large enough to ensure stability for the SUSY member of the families we studied. Hence the geometry of non-SUSY spacetimes such as $AdS_4 \times M^{mn}$ appears to generically invite such a non-perturbative instability. On the other hand, procedures (such as orbifolding or 'skew-whiffing') that generate non-SUSY spacetimes with the same (non-perturbatively stable) geometry as SUSY spacetimes were argued to have perturbative instabilities due to GSMOs. Hence requiring perturbative and non-perturbative stability appears to impose duelling demands on the geometry with little room for error.

# 5.A Spectrum, masses and dimensions, normalizations

## 5.A.1 Metric and curvature

We take $AdS_4$ to have radius $R$, i.e $ds^2 = \frac{R^2}{z^2} ds^2_{mink}$. We find that $\mathrm{Ricci}^{AdS}_{\mu\nu} = \frac{-3}{R^2} g^{AdS}_{\mu\nu}$. We set $R = 1/4$. Hence we have,

$$\mathrm{Ricci}^{AdS}_{\mu\nu} = -48 g^{AdS}_{\mu\nu}, \quad \mathrm{Ricci}^{X_7}_{mn} = 24 g^{X_7}_{mn} \tag{5.A.1}$$

Here *Ricci* is computed for example by GRTensor. It uses the GR convention which is twice the tensor used for group theory.

In [126], they set $R^{AdS}_{\mu\nu} = -12\lambda^2 g^{AdS}_{\mu\nu}$ and $R^{X_7}_{mn} = 6\lambda^2 g^{X_7}_{mn}$. This gives the identification $R^2 = \frac{1}{4\lambda^2}$ and we are led to set $\lambda = 2$. In [125, 124], they set $R = \frac{1}{4e}$ and hence we set $e = 1$. In Chapter 3 and 4, [124] uses the non-GR convention for the curvature tensor.

## 5.A.2 Definition of *AdS* masses

We define the scalar and vector masses through,

$$(\mathit{\Delta}_{AdS} + m^2_{scalar})\phi = 0 \tag{5.A.2}$$

$$\mathit{\Delta} B_\mu + \nabla_\mu \nabla^\rho B_\rho + (m^2_{vector} + 32) B_\mu = 0 \tag{5.A.3}$$

These definitions differ by 32 from that in [126] and implicitly used in [124].

From [124], (2.2.14), we find the mass-energy relations for particles of different spin,

$$m^2_{scalar} = 16E(E-3), \qquad E = \frac{3}{2} \pm \sqrt{\frac{9}{4} + \frac{m_s^2}{16}} \tag{5.A.4}$$

$$|m_{1/2}| = 4E - 6, \qquad E = \frac{1}{4}|m_{1/2} + 2| + 1 \tag{5.A.5}$$

$$m^2_{vector} = 16E(E-3), \qquad E = \frac{3}{2} \pm \sqrt{\frac{9}{4} + \frac{m_v^2}{16}} \tag{5.A.6}$$

$$|m_{3/2} + 4| = 4E - 6, \qquad E = \frac{1}{4}|m_{3/2} + 6| + 1 \tag{5.A.7}$$

Relating $AdS$ masses to the $X_7$ spectrum,

$$m^2_{\Sigma} = M_{(0)^3} + 144 + 24\sqrt{M_{(0)^3} + 36} \tag{5.A.8}$$

$$m^2_{S} = M_{(0)^3} + 144 - 24\sqrt{M_{(0)^3} + 36} \tag{5.A.9}$$

$$m^2_{\pi} = 16M_{(1)^3}(M_{(1)^3} - 3) \tag{5.A.10}$$

$$m^2_{\phi} = M_{(2)(0)^2} - 32 \tag{5.A.11}$$

$$m^2_{A} = M_{(1)(0)^2} + 16 - 12\sqrt{M_{(1)(0)^2} + 16} \tag{5.A.12}$$

$$m^2_{Z} = M_{(1)^2(0)} - 32 \tag{5.A.13}$$

$$m_{\lambda_L} = -(M_{(\frac{1}{2})^3} + 16) \tag{5.A.14}$$

$$m_{\lambda_T} = M_{(\frac{3}{2})(\frac{1}{2})^2} + 8 \tag{5.A.15}$$

where $M_{(i)(j)(k)}$ are as defined in [124] with $e = 1$.

## 5.B $M^{p,q,r}$ geometry

$$M^{pqr} = \frac{G}{H} = \frac{SU(3)^c \times SU(2)^w \times U(1)}{SU(2)^c \times U(1)' \times U(1)''} \tag{5.B.1}$$

We set,

$$\boxed{x = \frac{q}{3p}, \qquad y = \frac{r}{3p}} \tag{5.B.2}$$

Writing $g = h + k$, we pick the generators of $k$ as

$$SU(3)^c \quad : \quad \lambda_4, \ldots \lambda_7 \quad \ldots (4) \tag{5.B.3}$$

$$SU(2)^w \quad : \quad \sigma_1, \sigma_2 \quad \ldots (2) \tag{5.B.4}$$

$$U(1) \quad : \quad Z = p\frac{\sqrt{3}}{2}i\lambda_8 + \frac{q}{2}i\sigma_3 + riY \quad \ldots (1) \tag{5.B.5}$$

This lets us take for $Z', Z''$ the orthogonal complement of $Z$,

$$Z' \quad = \quad 2pr\frac{i\sqrt{3}}{2}\lambda_8 + 2rq\frac{i}{2}\sigma_3 - (3p^2 + q^2)iY \tag{5.B.6}$$

$$Z'' \quad = \quad -q\frac{i\sqrt{3}}{2}\lambda_8 + 3p\frac{i}{2}\sigma_3 \tag{5.B.7}$$

We also label $J_3 = \sigma_3/2$.

Let $\Omega(x) = L^{-1}(y)dL(y) = \Omega^h T_h + \Omega^k T_k$ be the Maurer-Cartan form which can be expanded in a basis of generators $h + k$. We can rescale the $K$ vielbeins with one scale factor per irrep of $H$ acting on $K$. In our case, this allows three scalings,

$$B^A \quad = \quad \frac{1}{a}\Omega^A, \quad A = 1, \ldots, 4 \tag{5.B.8}$$

$$B^m \quad = \quad \frac{1}{b}\Omega^m, \quad m = 1, 2 \tag{5.B.9}$$

$$B^3 \quad = \quad \frac{1}{c}(\sqrt{3}p\Omega^8 + q\Omega^{\bar{3}} + 2r\Omega^Y) = \frac{1}{c}(3p^2 + q^2 + 2r^2)\Omega^Z \tag{5.B.10}$$

The notation above is consistent[13] with that in [150] and [124]. We use the same generators as in [124] and set ,

$$T_i = \frac{1}{2}i\lambda_i, \frac{1}{2}i\sigma_m, iY \tag{5.B.11}$$

[13]We have $\Omega = \Omega^3 T_3 + \Omega^8 T_8 + \Omega^Y T_y + \ldots = \Omega^Z Z + \ldots$. Now $Z = \sqrt{3}pT_8 + qT_3 + rT_Y$ and hence $TrZZ = \frac{3p^2}{2} + \frac{q^2}{2} + r^2$. Hence $TrZ\Omega = \Omega^Z TrZZ = -\Omega^Z(\frac{3p^2}{2} + \frac{q^2}{2} + r^2)$. On the other hand, $TrZ\Omega = TrZ(\Omega^3 T_3 + \Omega^8 T_8 + \Omega^Y T_y) = -\frac{1}{2}\left(\sqrt{3}p\Omega^8 + q\Omega^3 + 2r\Omega^y\right)$.

From the rescaling of the vielbeins (5.B.8), we read off the components of $D^H$,

$$D^H_A = -\frac{a}{2} i\lambda_A \tag{5.B.12}$$

$$D^H_m = -\frac{b}{2} i\sigma_m \tag{5.B.13}$$

$$D^H_A = -\frac{c}{3p^2+q^2+2r^2} Z \tag{5.B.14}$$

We can choose $a, b, c$ such that $g$, the metric on $G/H$ is Einstein with $R^{ac}_{bc} = 12e^2\delta^a_b$.Castellani et al [150] do this to find,

$$a = 3x\gamma\sqrt{\frac{2}{3}\alpha}, \quad b = \gamma\sqrt{2\beta}, \quad c = q\gamma \tag{5.B.15}$$

$$4\beta^3 - 6\beta^2 + 9(x^2+\frac{1}{4})\beta - \frac{9}{2}x^2 = 0 \tag{5.B.16}$$

$$\alpha = \frac{1}{9x^2}(3\beta - 4\beta^2) \tag{5.B.17}$$

$$\gamma = \pm\sqrt{\frac{12e^2}{\beta(1-\beta)}} \tag{5.B.18}$$

We find (as they find in [124,125]) for $M^{1,1,1}$ ($x = 1/3$) that $a = 8/\sqrt{3}, b = 4\sqrt{2}, c = 8$. Note that $a, b, c, \alpha, \beta, \gamma$ depend only on the ratio $x = \frac{q}{3p}$. As a consequence, the mass spectrum depends only on this ratio as well and hence one can view the dimensions of operators as functions of $x$ alone and not $p, q$ individually.

## 5.C Scalar laplacian spectrum $M^3_{(0)}$

The action of the scalar laplacian can be reduced to the action of the generators $K$. Using (5.B.12),

$$\varDelta = D^H_a D^{Ha} = \frac{a^2}{4}\lambda_A\lambda_A + \frac{b^2}{4}\sigma_m\sigma_m - \frac{c^2}{(3p^2+q^2+2r^2)^2} Z^2 \tag{5.C.1}$$

For the scalar spectrum, we expand in harmonics of $G$ which contain the trivial representation of $H$ i.e $J_c = Z' = Z'' = 0$. By evaluating on the Young diagrams, this gives,

$$
\begin{aligned}
J_3 &= \frac{q}{3p}(M_2 - M_1), \qquad Y = \frac{2r}{3p}(M_2 - M_1) && \text{(5.C.2)}\\
Z &= i\frac{3p^2 + q^2 + 2r^2}{3p}(M_2 - M_1) && \text{(5.C.3)}\\
\lambda_A \lambda_A &= 4(M_1 + M_2 + M_1 M_2) && \text{(5.C.4)}\\
\sigma_m \sigma_m &= 4(J(J+1) - J_3^2) && \text{(5.C.5)}
\end{aligned}
$$

Substituting the above into the laplacian, we find,

$$
\begin{aligned}
H_0 &= a^2(x)(M_1 + M_2 + M_1 M_2) + b^2(x)J(J+1) + x^2(\gamma^2(x) - b^2(x))(M_1 - M_2)^2 \\
&= \gamma^2 \left(6x^2\alpha(M_1 + M_2 + M_1 M_2) + 2\beta J(J+1) + x^2(1 - 2\beta)(M_1 - M_2)^2\right)
\end{aligned}
$$

where $\alpha, \beta, \gamma$ are as defined in (5.B.15).[14]

We find that $\gamma^2 - b^2 \geq 0$ for all $x$ in the BF range.

Since $J_3 \in \mathbf{Z}/2$, we find the following constraints on the representations of $G$ that contribute to the scalar spectrum,

- $M_2 - M_1 \in \frac{3p}{2q} Z$ (and also $M_2, M_1 \in Z$)
- $J \geq |J_3| = |\frac{q}{3p}(M_2 - M_1)|$

# 5.D 3-form laplacian spectrum $M^3_{(1)}$

We use the results of [152] who worked out the $15 \times 15$ matrix that needs to be diagonalized. Correcting what appear to be a few mistakes in the entries, we obtain the correct matrix presented below.

[14] The last term of this result disagrees with the spectrum derived in [152] when $r \neq 1$. It appears that $r$ has been set to 1 implicitly in [152]

The first 5 columns are,

$$\left(\begin{matrix}
0 & CiY & -\frac{1}{2}BqY & 0 & 0 \\
-CiY & 0 & BiJ & 0 & 0 \\
2BqY & -4Bi(1+J) & \Omega & 0 & 0 \\
0 & 0 & 0 & 0 & 0 \\
0 & 0 & 0 & 0 & 0 \\
0 & 0 & -\frac{1}{4}Ai(4+\mathrm{M1}+\mathrm{M2}) & CiY & 0 \\
0 & 0 & -\frac{1}{4}A(\mathrm{M1}-\mathrm{M2}) & 0 & CiY \\
\frac{1}{4}A(4+\mathrm{M1}+\mathrm{M2}) & \frac{1}{4}Ai(\mathrm{M1}-\mathrm{M2}) & 0 & -BJ & -\frac{1}{2}BiqY \\
-\frac{1}{4}Ai(\mathrm{M1}-\mathrm{M2}) & \frac{1}{4}A(4+\mathrm{M1}+\mathrm{M2}) & 0 & \frac{1}{2}BiqY & -BJ \\
-\frac{1}{4}A(4+\mathrm{M1}+\mathrm{M2}) & \frac{1}{4}Ai(\mathrm{M1}-\mathrm{M2}) & 0 & BJ & -\frac{1}{2}BiqY \\
\frac{1}{4}Ai(\mathrm{M1}-\mathrm{M2}) & \frac{1}{4}A(4+\mathrm{M1}+\mathrm{M2}) & 0 & \frac{1}{2}BiqY & BJ \\
0 & 0 & 0 & 0 & 0 \\
0 & 0 & 0 & 0 & 0 \\
0 & 0 & 0 & 0 & 0 \\
0 & 0 & 0 & 2Ai(\mathrm{M1}-\mathrm{M2}) & 2A(4+\mathrm{M1}+\mathrm{M2})
\end{matrix}\right.$$

and the next 5 are,

$$\begin{matrix}
0 & 0 & A(\mathrm{M1}+\mathrm{M2}) & -Ai(\mathrm{M1}-\mathrm{M2}) & -A(\mathrm{M1}+\mathrm{M2}) \\
0 & 0 & Ai(\mathrm{M1}-\mathrm{M2}) & A(\mathrm{M1}+\mathrm{M2}) & Ai(\mathrm{M1}-\mathrm{M2}) \\
2Ai(\mathrm{M1}+\mathrm{M2}) & 2A(\mathrm{M1}-\mathrm{M2}) & 0 & 0 & 0 \\
-CiY & 0 & -2B(1+J) & BiqY & 2B(1+J) \\
0 & -CiY & -BiqY & -2B(1+J) & -BiqY \\
0 & 0 & 0 & 0 & 0 \\
0 & 0 & 0 & 0 & 0 \\
0 & 0 & \varDelta & 0 & 0 \\
0 & 0 & 0 & \varDelta & 0 \\
0 & 0 & 0 & 0 & -\varDelta \\
0 & 0 & 0 & 0 & 0 \\
0 & 0 & \frac{1}{2}Ai(\mathrm{M1}-\mathrm{M2}) & \frac{1}{2}A(4+\mathrm{M1}+\mathrm{M2}) & \frac{1}{2}Ai(\mathrm{M1}-\mathrm{M2}) \\
0 & 0 & -\frac{1}{2}A(4+\mathrm{M1}+\mathrm{M2}) & \frac{1}{2}Ai(\mathrm{M1}-\mathrm{M2}) & \frac{1}{2}A(4+\mathrm{M1}+\mathrm{M2}) \\
A(4+\mathrm{M1}+\mathrm{M2}) & -Ai(\mathrm{M1}-\mathrm{M2}) & 0 & 0 & 0 \\
0 & 0 & 0 & 0 & 0
\end{matrix}$$

and the last 5 columns are,

$$
\left.\begin{array}{ccccc}
Ai(\mathrm{M1}-\mathrm{M2}) & 0 & 0 & 0 & 0 \\
A(\mathrm{M1}+\mathrm{M2}) & 0 & 0 & 0 & 0 \\
0 & 0 & 0 & 0 & 0 \\
BiqY & 0 & 0 & 0 & \frac{1}{2}Ai(\mathrm{M1}-\mathrm{M2}) \\
2B(1+J) & 0 & 0 & 0 & \frac{1}{2}A(\mathrm{M1}+\mathrm{M2}) \\
0 & 0 & 0 & \frac{1}{2}A(\mathrm{M1}+\mathrm{M2}) & 0 \\
0 & 0 & 0 & -\frac{1}{2}Ai(\mathrm{M1}-\mathrm{M2}) & 0 \\
0 & \frac{1}{2}Ai(\mathrm{M1}-\mathrm{M2}) & -\frac{1}{2}A(\mathrm{M1}+\mathrm{M2}) & 0 & 0 \\
0 & \frac{1}{2}A(\mathrm{M1}+\mathrm{M2}) & \frac{1}{2}Ai(\mathrm{M1}-\mathrm{M2}) & 0 & 0 \\
0 & \frac{1}{2}Ai(\mathrm{M1}-\mathrm{M2}) & \frac{1}{2}A(\mathrm{M1}+\mathrm{M2}) & 0 & 0 \\
-\varDelta & \frac{1}{2}A(\mathrm{M1}+\mathrm{M2}) & -\frac{1}{2}Ai(\mathrm{M1}-\mathrm{M2}) & 0 & 0 \\
\frac{1}{2}A(4+\mathrm{M1}+\mathrm{M2}) & 0 & -CiY & BJ & 0 \\
-\frac{1}{2}Ai(\mathrm{M1}-\mathrm{M2}) & CiY & 0 & -\frac{1}{2}BiqY & 0 \\
0 & 4B(1+J) & -2BiqY & -\Omega & \varDelta \\
0 & 0 & 0 & 2\varDelta & 0
\end{array}\right)
$$

In the above matrix, we have used the notation of [152] which is related to our notation by,

$$A = \frac{a}{8}, \quad B = \frac{b}{8}, \quad C = \frac{c}{8}, \quad \varDelta = \frac{1}{4}\gamma\alpha, \quad \Omega = \frac{\gamma\beta}{2} \tag{5.D.1}$$

where $a, b, c, \alpha, \beta$ are defined in Appendix 5.B. One can show that $2\varDelta^2 + \Omega^2 = 3$.

The matrix can be diagonalized to find the eigenvalues. For simplicity, we perform this task assuming $M_1 = M_2$ since this case will be sufficient for all purposes of this paper.

We find the eigenvalues,

$$\lambda_1, \lambda_2 = \frac{1}{4}\sqrt{H + 16\varDelta^2} \tag{5.D.2}$$

$$\lambda_3, \lambda_4 = -\frac{1}{4}\sqrt{H + 16\varDelta^2} \tag{5.D.3}$$

$$\lambda_9 = \lambda_{10} = \ldots = \lambda_{15} = 0 \tag{5.D.4}$$

and $\lambda_5, \lambda_6, \lambda_7, \lambda_8$ are the roots of the fourth degree equation,

$$\begin{aligned} x^4 + \left(-3 - \frac{H}{8}\right) x^2 &+ \left(3\Omega - \Omega^3\right) x \\ &+ \left(\frac{H^2}{256} - \frac{1}{16} H \left(-3 + \Omega^2\right) + 6A^2(-1 + \Omega^2) m(2+m)\right) = 0 \end{aligned}$$

The equation has a general analytic solution which we do not reproduce because of the complexity. When the last term proportional to $(-1+\Omega^2)m(2+m)$ is zero as when $m = 0$, this equation factorizes simply and the roots are,

$$\lambda_5 = -\frac{\Omega}{2} + \frac{1}{4}\sqrt{H + 48 - 12\Omega^2} \quad (5.D.5)$$

$$\lambda_6 = -\frac{\Omega}{2} - \frac{1}{4}\sqrt{H + 48 - 12\Omega^2} \quad (5.D.6)$$

$$\lambda_7 = \frac{\Omega}{2} - \frac{1}{4}\sqrt{H + 4\Omega^2} \quad (5.D.7)$$

$$\lambda_8 = \frac{\Omega}{2} + \frac{1}{4}\sqrt{H + 4\Omega^2} \quad (5.D.8)$$

When $x = 1/3$, we have $\Delta = 1, \Omega = 1$ and the above result agrees with the SUSY spectrum found in [124].

## 5.E Dirac operator spectrum $M_{(1/2)^3}$

This was completely worked out for general $M^{pqr}$ manifolds in [153]. We will primarily be interested in the spinors occuring in the gauge multiplets. These are determined by the spinor spectrum with the symmetry charges $M_1 = M_2 = 1, J = 0$ and $M_1 = M_2 = 0, J = 1$, i.e in the adjoints of the $SU(3)$ and $SU(2)$ gauge groups. Looking this up in [153], we find that these occur in the exceptional representation "2" and the regular representation "3 + 4" in the notation of Sec 10 there.

### 5.E.1 Exceptional representation "2"

The relevant matrix to diagonlize is[15],

$$\gamma \begin{pmatrix} \frac{-q}{2}Y + \mathit{\Delta}_{1/2} & \frac{q}{p}\sqrt{\frac{2}{3}\alpha} M_1 \\ \frac{q}{p}\sqrt{\frac{2}{3}\alpha}\,(2+M_2) & \frac{q}{2}Y + \mathit{\Delta}_- \end{pmatrix} - \begin{pmatrix} 7 & 0 \\ 0 & 7 \end{pmatrix} \tag{5.E.1}$$

where

$$\mathit{\Delta}_2 = -\frac{1}{3p^2+q^2}\left(\frac{\alpha+\beta}{2}q^2 + 2p^2\beta^2\right) = -\frac{\beta}{2} \tag{5.E.2}$$

$$\mathit{\Delta}_+ = -\frac{1}{3p^2+q^2}\left(\frac{\alpha+\beta}{2}q(3p-q) - 2p(p+q)\beta^2\right) = \frac{\beta}{2} - \frac{q}{2p}\alpha \tag{5.E.3}$$

$$\mathit{\Delta}_- = \frac{1}{3p^2+q^2}\left(\frac{\alpha+\beta}{2}q(3p+q) + 2p(p-q)\beta^2\right) = \frac{\beta}{2} + \frac{q}{2p}\alpha \tag{5.E.4}$$

$$\lambda_{1,2} = -7 + \frac{\gamma(x)}{2}\left(\mathit{\Delta}_{1/2} + \mathit{\Delta}_-\right) \pm \sqrt{\frac{\gamma(x)^2}{4}(\mathit{\Delta}_- - \mathit{\Delta}_{1/2} + 3x)^2 + H} \tag{5.E.5}$$

$$H = 6x^2\gamma(x)^2\alpha(x)m(m+2) \tag{5.E.6}$$

### 5.E.2 Regular representation "3 + 4"

The matrix is,

$$\gamma \begin{pmatrix} \frac{q}{2r}Y + \mathit{\Delta}_+ & \sqrt{2\beta}(J + \frac{1}{2} + \frac{q}{2r}Y) \\ \sqrt{2\beta}(J + \frac{1}{2} - \frac{q}{2r}Y) & -\frac{q}{2r}Y + \mathit{\Delta}_- \end{pmatrix} - \begin{pmatrix} 7 & 0 \\ 0 & 7 \end{pmatrix}$$

[15]The operator diagonlized in [153] is $D$ while we are interested in $D-7$. Hence we have modified the matrix presented here appropriately.

giving the eigenvalues,

$$\begin{aligned}
\lambda_{1,2} &= -7 + \frac{\gamma(x)}{2}\left(\Delta_+ + \Delta_-\right) \\
&\pm\sqrt{\frac{\gamma(x)^2}{4}(\Delta_- - \Delta_+ + 3x)^2 + \frac{\gamma(x)^2}{2}\left(1-9x^2\right)\beta(x) + H} \\
H &= 2\gamma(x)^2\beta(x)J(J+1) \qquad (5.E.7)
\end{aligned}$$

## 5.F Einstein's equations

Einstein's equations for 11 D SUGRA are,

$$R_{MN} = \frac{1}{12}\left(F_M^{PQR}F_{NPQR} - \frac{1}{12}G_{MN}F^2\right) \qquad (5.F.1)$$

Here $F_4$ is the 4-form M-theory flux. Using its equations of motion and the Bianchi identity, we find that it must scale like

$$F_4 \sim \frac{R^6}{g_1^2 g_2 h^{1/2}} vol_{AdS_4} \qquad (5.F.2)$$

for an ansatz of the type (5.3.25) of Section 5.3.2 where $R$ is the radius of $X_7$ and twice the radius of $R_{AdS}$.

We can work out the normalization and write out Einstein's equations for each part $AdS_4$ and $X_7$ of the geometry.

$$\begin{aligned}
R_{MN} &= -12\frac{R^{12}}{g_1^4 g_2^2 h}G_{MN}, \qquad (M,N \in AdS_4) \qquad (5.F.3) \\
R_{MN} &= 6\frac{R^{12}}{g_1^4 g_2^2 h}G_{MN}, \qquad (M,N \in X_7) \qquad (5.F.4) \\
& \qquad (5.F.5)
\end{aligned}$$

For $AdS_4 \times S^7$ with the ansatz (5.3.12) of Section 5.3.1, we get for the Einstein equations with flux,

$$\begin{aligned}
\frac{12R^{12}p}{g^6h} + \frac{3f'^2}{4f^2} + \frac{3g'^2}{2g^2} + \frac{h'^2}{4h^2} + \frac{3f'p'}{4fp} + \frac{3g'p'}{2gp} + \frac{h'p'}{4hp} - \frac{3f''}{2f} - \frac{3g''}{g} - \frac{h''}{2h} &= 0 \\
2 + \frac{12R^{12}f}{g^6h} - \frac{f'^2}{4fp} - \frac{3f'g'}{2gp} - \frac{f'h'}{4hp} + \frac{f'p'}{4p^2} - \frac{f''}{2p} &= 0 \\
8 - \frac{6R^{12}}{g^5h} - \frac{2h}{g} - \frac{3f'g'}{4fp} - \frac{g'^2}{gp} - \frac{g'h'}{4hp} + \frac{g'p'}{4p^2} - \frac{g''}{2p} &= 0 \\
-\frac{6R^{12}}{g^6} + \frac{6h^2}{g^2} - \frac{3f'h'}{4fp} - \frac{3g'h'}{2gp} + \frac{h'^2}{4hp} + \frac{h'p'}{4p^2} - \frac{h''}{2p} &= 0
\end{aligned}$$

For $AdS_4 \times M^{mn}$ with the ansatz (5.3.25) of Section 5.3.2, we get for the Einstein equations with flux,

$$\begin{aligned}
\frac{3f'^2}{4f^2} + \frac{g_1'^2}{g_1^2} + &\quad \ldots \qquad (5.F.6) \\
+\frac{g_2'^2}{2g_2^2} + \frac{h'^2}{4h^2} + \frac{3f'p'}{4fp} + \frac{g_1'p'}{g_1p} + \frac{g_2'p'}{2g_2p} + \frac{h'p'}{4hp} - \frac{3f''}{2f} - \frac{2g_1''}{g_1} - \frac{g_2''}{g_2} - \frac{h''}{2h} &= -12\frac{R^{12}}{g_1^4g_2^2h}\rho \\
2 - \frac{f'^2}{4fp} - \frac{f'g_1'}{g_1p} - \frac{f'g_2'}{2g_2p} - \frac{f'h'}{4hp} + \frac{f'p'}{4p^2} - \frac{f''}{2p} &= -12\frac{R^{12}}{g_1^4g_2^2h}f \\
\frac{4a^4m^2h^2}{c^2g_1^2} + \frac{b^4n^2h^2}{2c^2g_2^2} - \frac{3f'h'}{4fp} - \frac{g_1'h'}{g_1p} - \frac{g_2'h'}{2g_2p} + \frac{h'^2}{4hp} + \frac{h'p'}{4p^2} - \frac{h''}{2p} &= 6\frac{R^{12}}{g_1^4g_2^2h}h \\
6a^2 - \frac{2a^4m^2h}{c^2g_1} - \frac{3f'g_1'}{4fp} - \frac{g_1'^2}{2g_1p} - \frac{g_1'g_2'}{2g_2p} - \frac{g_1'h'}{4hp} + \frac{g_1'p'}{4p^2} - \frac{g_1''}{2p} &= 6\frac{R^{12}}{g_1^4g_2^2h}g_1 \\
b^2 - \frac{b^4n^2h}{2c^2g_2} - \frac{3f'g_2'}{4fp} - \frac{g_1'g_2'}{g_1p} - \frac{g_2'h'}{4hp} + \frac{g_2'p'}{4p^2} - \frac{g_2''}{2p} &= 6\frac{R^{12}}{g_1^4g_2^2h}g_2 \\
&\ldots
\end{aligned}$$

In the above equations, $a = 1/\lambda_1, b = 1/\lambda_2, c = 1/\lambda_3$ of (5.3.25). These scale factors must be chosen so that $R_{ab}^{X_7} = 6g_{ab}^{X_7}$.